\documentclass{aa}  

\usepackage{graphicx}
\usepackage{txfonts}
\usepackage{hyperref}
\usepackage{adjustbox}
\usepackage{url}
\begin{document} 

\def\teff{${T}_{\rm eff}$}
\def\kms{{km\,s}$^{-1}$}
\def\logg{$\log g$}
\def\micro{$\xi_{\rm t}$}
\def\macro{$\zeta_{\rm RT}$}
\def\rad{$v_{\rm r}$}
\def\vsini{$v\sin i$}
\def\ebv{$E(B-V)$}
\def\kepler{\textit{Kepler}}

   \title{A plethora of new, magnetic chemically peculiar stars from LAMOST DR4}

\author{S.~H{\"u}mmerich\inst{1,2}
        \and E.~Paunzen\inst{3}
        \and K.~Bernhard\inst{1,2}
        }
\institute{Bundesdeutsche Arbeitsgemeinschaft f{\"u}r Ver{\"a}nderliche Sterne e.V. (BAV), D-12169 Berlin, Germany, \email{ernham@rz-online.de}
\and American Association of Variable Star Observers (AAVSO), 49 Bay State Rd, Cambridge, MA 02138, USA
\and Department of Theoretical Physics and Astrophysics, Masaryk University, Kotl\'a\v{r}sk\'a 2, 611\,37 Brno, Czech Republic}

\date{}

 
  \abstract
   {Magnetic chemically peculiar (mCP) stars are important to astrophysics because their complex atmospheres lend themselves perfectly to the investigation of the interplay between such diverse phenomena as atomic diffusion, magnetic fields, and stellar rotation. The most up-to-date catalogue of these objects was published a decade ago. Since then, no large scale spectroscopic surveys targeting this group of objects have been carried out. An increased sample size of mCP stars, however, is important for statistical studies.}
   {The present work is aimed at identifying new mCP stars using spectra collected by the Large Sky Area Multi-Object Fiber Spectroscopic Telescope (LAMOST).}
   {Suitable candidates were selected by searching LAMOST DR4 spectra for the presence of the characteristic 5200\,\AA\ flux depression. Spectral classification was carried out with a modified version of the MKCLASS code and the accuracy of the classifications was estimated by comparison with results from manual classification and the literature. Using parallax data and photometry from Gaia DR2, we investigated the space distribution of our sample stars and their properties in the colour-magnitude diagram.}
   {Our final sample consists of 1002 mCP stars, most of which are new discoveries (only 59 common entries with the Catalogue of Ap, HgMn and Am stars). Traditional mCP star peculiarities have been identified in all but 36 stars, highlighting the efficiency of the code's peculiarity identification capabilities. The derived temperature and peculiarity types are in agreement with manually derived classifications and the literature. Our sample stars are between 100\,Myr and 1\,Gyr old, with the majority having masses between 2\,$M_\odot$ and 3\,$M_\odot$. Our results could be considered as strong evidence for an inhomogeneous age distribution among low-mass ($M$\,$<$\,3\,M$_\odot$) mCP stars; however, we caution that our sample has not been selected on the basis of an unbiased, direct detection of a magnetic field. We identified several astrophysically interesting objects: the mCP stars LAMOST J122746.05+113635.3 and LAMOST J150331.87+093125.4 have distances and kinematical properties in agreement with halo stars; LAMOST J034306.74+495240.7 is an eclipsing binary system ($P$\textsubscript{orb}\,=\,5.1435$\pm$0.0012\,d) hosting an mCP star component; and LAMOST J050146.85+383500.8 was found to be an SB2 system likely comprising of an mCP star and a supergiant component.}
	 {With our work, we significantly increase the sample size of known Galactic mCP stars, paving the way for future in-depth statistical studies.}


   \maketitle

%
\section{Introduction} \label{introduction}

The chemically peculiar (CP) stars of the upper main sequence (spectral types early B to early F) are traditionally characterised by the presence of certain absorption lines of abnormal strength or weakness that indicate peculiar surface abundances \citep{preston74}. For most groups of CP stars, current theories ascribe the observed chemical peculiarities to the interplay between radiative levitation and gravitational settling (atomic diffusion) \citep{michaud70,richer00}: whereas most elements sink under the force of gravity, those with numerous absorption lines near the local flux maximum are radiatively accelerated towards the surface. Because CP stars are generally slow rotators and boast calm radiative atmospheres, atomic diffusion processes are able to significantly influence the chemical composition of the outer stellar layers.

Following \citet{preston74}, CP stars are traditionally divided into the following four main groups: CP1 stars (the metallic-line or Am/Fm stars), CP2 stars (the magnetic Bp/Ap stars), CP3 stars (the Mercury-Manganese or HgMn stars), and CP4 stars (the He-weak stars). Although the chemical composition within a group may vary considerably, each group is characterised by a distinct set of peculiarities. The CP1 stars show underabundances of Ca and Sc and overabundances of the iron-peak and heavier elements. CP2 stars exhibit excesses of elements such as Si, Sr, Eu, or the rare-earth elements. The CP3 stars are characterised by enhanced lines of Hg and Mn and other heavy elements, whereas the main characteristic of the CP4 stars is anomalously weak He lines. Further classes of CP stars have been described, such as the He strong stars -- early B stars with anomalously strong He lines --, the $\lambda$ Bootis stars \citep{murphy17}, which boast unusually low surface abundances of iron-peak elements, or the barium stars, which are characterised by enhancements of the \textit{s}-process elements Ba, Sr, Y, and C \citep{bidelman51}. Generally, with regard to the strength of chemical peculiarities, a continuous transition from normal to peculiar stars is observed \citep{loden87}.

Most of the CP2 and He-peculiar stars possess stable and globally organised magnetic fields with strengths of up to several tens of kG \citep{babcock47,auriere07}, the origin of which is still a matter of some controversy \citep{moss04}. However, a body of evidence has been built up that strongly favours the fossil field theory, which states that the magnetic field is a relic of the 'frozen-in' interstellar magnetic field (e.g. \citealt{braithwaite04}). These stars are often referred to as magnetic chemically peculiar (mCP) stars in the literature -- a convention which we will adhere to throughout this paper. The magnetic field affects the diffusion processes in such a way that mCP stars show a non-uniform distribution of chemical elements (chemical spots or belts) on their surfaces, which can be studied in detail via the technique of Doppler imaging \citep{kochukhov17}. As the magnetic axis is oblique to the rotation axis (oblique rotator model; \citealt{stibbs50}), mCP stars show strictly periodic light, spectral, and magnetic variations with the rotation period. The photometric variability arises because flux is redistributed in the abundance patches (e.g. \citealt{wolff71,molnar73,krticka13}).

The mCP stars show vastly differing abundance patterns. Some of the most peculiar objects belong to this group, such as the extreme lanthanide star HD\,51418 \citep{jones74} or Przybylski's star HD\,101065 (\citealt{przybylski66}), which is widely regarded as the most peculiar star known. Excesses of several orders of magnitude are commonly observed in these objects. \citet{morgan33} already recognised a relationship between a CP2 star's temperature and the predominant spectral peculiarities and showed that the CP2 stars can thus be sorted into subgroups. Since then, many authors have proposed corresponding classification schemes with varying levels of detail (cf. the discussions in \citealt{wolff83} and \citealt{gray09}). It is generally useful to at least differentiate between the 'cool' CP2 stars mostly characterised by Sr, Cr and Eu peculiarities and the 'hot' CP2 stars that generally show Si overabundances, although considerable overlap exists.

The mCP stars are important to astrophysics in several respects. Their complex atmospheres lend themselves perfectly to the investigation of such diverse phenomena as atomic diffusion, magnetic fields, stellar rotation and their interplay. They furthermore provide important testing grounds for the evaluation of model atmospheres \citep{krticka09} and, through their characteristic light variability, allow the derivation of rotational periods with great accuracy and comparatively little effort.

\begin{figure}
        \includegraphics[width=\columnwidth]{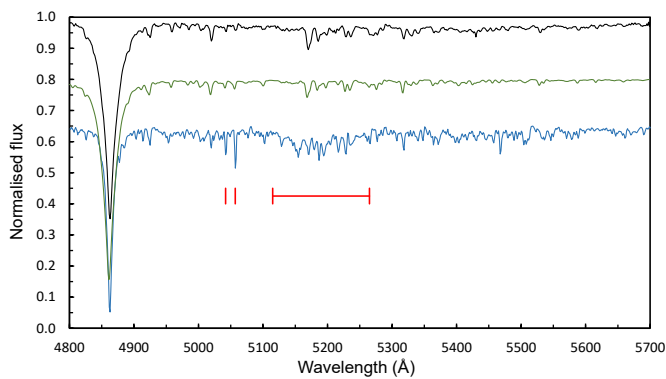}
    \caption{4800\,\AA\ to 5700\,\AA\ region of (from top to bottom) the non-CP A0 V star LAMOST\,J194655.00+402559.5 (HD\,225785), a synthetic spectrum with ${T}_{\rm eff}$\,=\,9750\,K, $\log g$\,=\,4.0, [M/H]\,=\,0.0 and a microturbulent velocity of 2\,km\,s$^{-1}$, and the newly-identified Si-strong mCP star LAMOST\,J025951.09+540337.5 (\#78; TYC\,3701-157-1). The position of the characteristic 5200\,\AA\ depression and the Si II lines at 5041\,\AA\ and 5055/56\,\AA\ are indicated. LAMOST spectra have been taken from DR4.}
    \label{fig_5200A_depression}
\end{figure}

The most up-to-date collection of CP stars -- the most recent version of the General Catalogue of CP Stars -- was published a decade ago \citep{RM09}. It contains about 3500 mCP stars or candidates ($\sim$2000 confirmed mCP stars and $\sim$1500 candidate mCP stars). Since then, several studies have identified new mCP stars on a minor scale (e.g. \citealt{huemmerich18,Scholz2019,sikora19}) but no large scale spectroscopic surveys have been conducted during the past recent decades that aim specifically at the identification of new mCP stars 'en masse'.

The works of \citet{hou15} and \citet{qin19} warrant special mention as they exploited spectra collected by the Large Sky Area Multi-Object Fiber Spectroscopic Telescope (LAMOST) of the Chinese Academy of Science. \citet{hou15} presented a list of 3537 candidate CP1 stars from LAMOST Data Release (DR) 1. Building on this work, \citet{smalley17} investigated pulsational properties versus metallicism in this subgroup of CP stars. \citet{qin19} searched for CP1 stars in the low-resolution spectra of LAMOST DR5 and compiled a catalogue of 9372 CP1 stars. They identified CP2 stars as a contaminant and searched for corresponding candidates in their sample of CP1 candidates, identifying 1131 candidate CP2 stars in this process.

Here we present our efforts aimed at identifying new mCP stars using spectra from the publicly available LAMOST DR4, which have led to the discovery of 1002 mCP stars. With this work, we significantly increase the sample size of known Galactic mCP stars, paving the way for future in-depth statistical studies. Spectroscopic data and target selection process are discussed in Section \ref{dataanalysis}. Spectral classification workflow and results are detailed in Section \ref{spectral_classification} and discussed, together with other interesting information on our sample of stars, in Section \ref{discussion}. We conclude our findings in Section \ref{conclusion}.

\section{Spectroscopic data and target selection} \label{dataanalysis}

This section contains a description of the employed spectral archive and the MKCLASS code and details the process of target selection.

\subsection{The Large Sky Area Multi-Object Fiber Spectroscopic Telescope (LAMOST)} \label{LAMOST}

The LAMOST telescope \citep{lamost1,lamost2}, also called the Guo Shou Jing\footnote{Guo Shou Jing (1231–1316) was a Chinese astronomer, hydraulic engineer, mathematician, and politician of the Yuan Dynasty.} Telescope, is a reflecting Schmidt telescope located at the Xinglong Observatory in Beijing, China. It boasts an effective aperture of 3.6$-$4.9\,m and a field of view of 5$\degr$. Thanks to its unique design, LAMOST is able to take 4000 spectra in a single exposure with spectral resolution R\,$\sim$\,1800, limiting magnitude r\,$\sim$\,19\,mag and wavelength coverage from 3700 to 9000\,\AA. LAMOST is therefore particularly suited to survey large portions of the sky and is dedicated to a spectral survey of the entire available northern sky. LAMOST data products are released to the public in consecutive data releases and can be accessed via the LAMOST spectral archive.\footnote{\url{http://www.lamost.org}} With about 10 million stellar spectra contained in DR6, the LAMOST archive constitutes a real treasure trove for researchers.

\begin{figure}
        \includegraphics[width=\columnwidth]{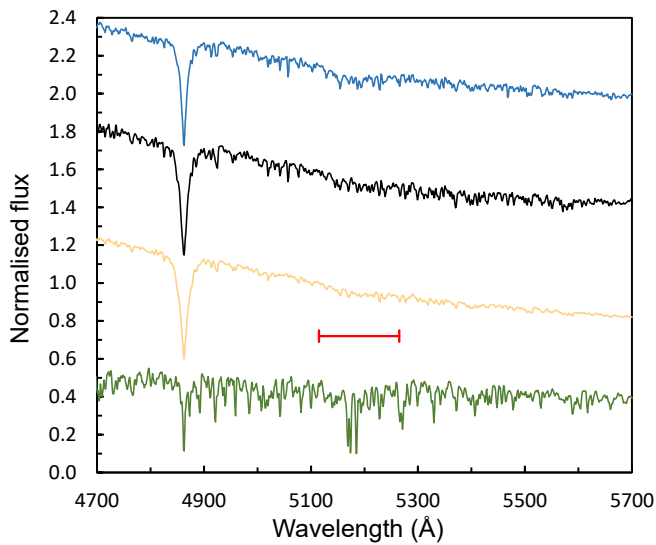}
    \caption{4700\,\AA\ to 5700\,\AA\ region of (from top to bottom) the LAMOST DR4 spectra of the mCP stars LAMOST J034458.31+464848.7 (\#139; TYC\,3313-1279-1), LAMOST J040642.34+454640.8 (\#180; HD\,25706), LAMOST J072118.92+223422.7 (\#792; TYC\,1909-1687-1), and the late-type 'impostor' LAMOST J001159.88+435908.5 (GSC\,02794-00977). The position of the characteristic 5200\,\AA\ depression is indicated.}
    \label{fig_5200A_depression2}
\end{figure}

\subsection{The MKCLASS code} \label{MKCLASS}

MKCLASS is a computer program written to classify stellar spectra on the Morgan-Keenan-Kellman (MKK) system \citep{gray14}. It has been designed to emulate the process of classifying by a human classifier. First, a rough spectral type is assigned, which is then refined by direct comparison with spectra from standard star libraries.

Currently, MKCLASS is able to classify spectra in the violet-green region (3800-5600\,\AA) in either rectified or flux-calibrated format. Several studies (e.g. \citealt{gray14,gray16,huemmerich18}) have shown that, providing input spectra of sufficient signal-to-noise (S/N), the results of MKCLASS compare well with the results of manual classification (precision of 0.6 spectral subclass and half a luminosity class according to \citealt{gray14}).

MKCLASS comes with two libraries of MKK standard spectra, which have been acquired with the Gray/Miller (GM) spectrograph on the 0.8\,m reflector of the Dark Sky Observatory in North Carolina, USA. \textit{libr18} contains rectified spectra in the spectral range from 3800–4600\,\AA\ and a resolution of 1.8\,\AA\ that were obtained with a 1200\,g\,mm$^{-1}$ grating. \textit{libnor36} boasts flux-calibrated and normalised spectra in the spectral range from 3800–5600\,\AA\ and a resolution of 3.6\,\AA\ obtained with a 600\,g\,mm$^{-1}$ grating. MKCLASS allows for the use of additional spectral libraries tailored to the specific needs of the researcher.

An interesting feature of the MKCLASS code is its ability to identify a set of spectral peculiarities, such as found in CP1 and CP2 stars, barium stars, carbon-rich giants etc. For more information on the MKCLASS code, we refer the reader to \citet{gray14} and the corresponding website.\footnote{\url{http://www.appstate.edu/~grayro/mkclass/}}

\subsection{Target selection criteria} \label{target}

To select suitable mCP star candidates, we specifically searched for the presence of the tell-tale 5200\,\AA\ depression in the LAMOST DR4 spectra of early-type stars. In the following, we provide background information and detail our selection criteria and the methods employed in the construction of the present sample of stars.

\subsubsection{The flux depressions in mCP stars} \label{5200A_depression}

The first to notice significant flux depressions at 4100\,\AA, 5200\,\AA, and 6300\,\AA\ in the spectrum of the mCP star HD\,221568 was \citet{1969ApJ...157L..59K}. Similar features in the ultraviolet region at 1400\,\AA, 1750\,\AA, and 2750\,\AA\ were later identified and investigated \citep{1977A&A....56..413J,1978A&A....70..379J}. It was found that these spectral features solely occur in mCP stars. \citet{khan07} showed that Fe is the principal contributor to the 5200\,\AA\ depression for the whole range of $T_{\rm eff}$ of mCP stars, while Cr and Si play a role primarily in the low $T_{\rm eff}$ region. Figure \ref{fig_5200A_depression} shows the 4800\,\AA\ to 5700\,\AA\ region of the spectra of a non-CP star, a corresponding synthetic spectrum and the newly-identified mCP star LAMOST J025951.09+540337.5 (\#78\footnote{The numbers given behind the identifiers refer to the internal identification number and facilitate easy identification in the tables.}; TYC\,3701-157-1), illustrating the 5200\,\AA\ depression in the latter object.

\begin{figure*}
        \includegraphics[width=\textwidth]{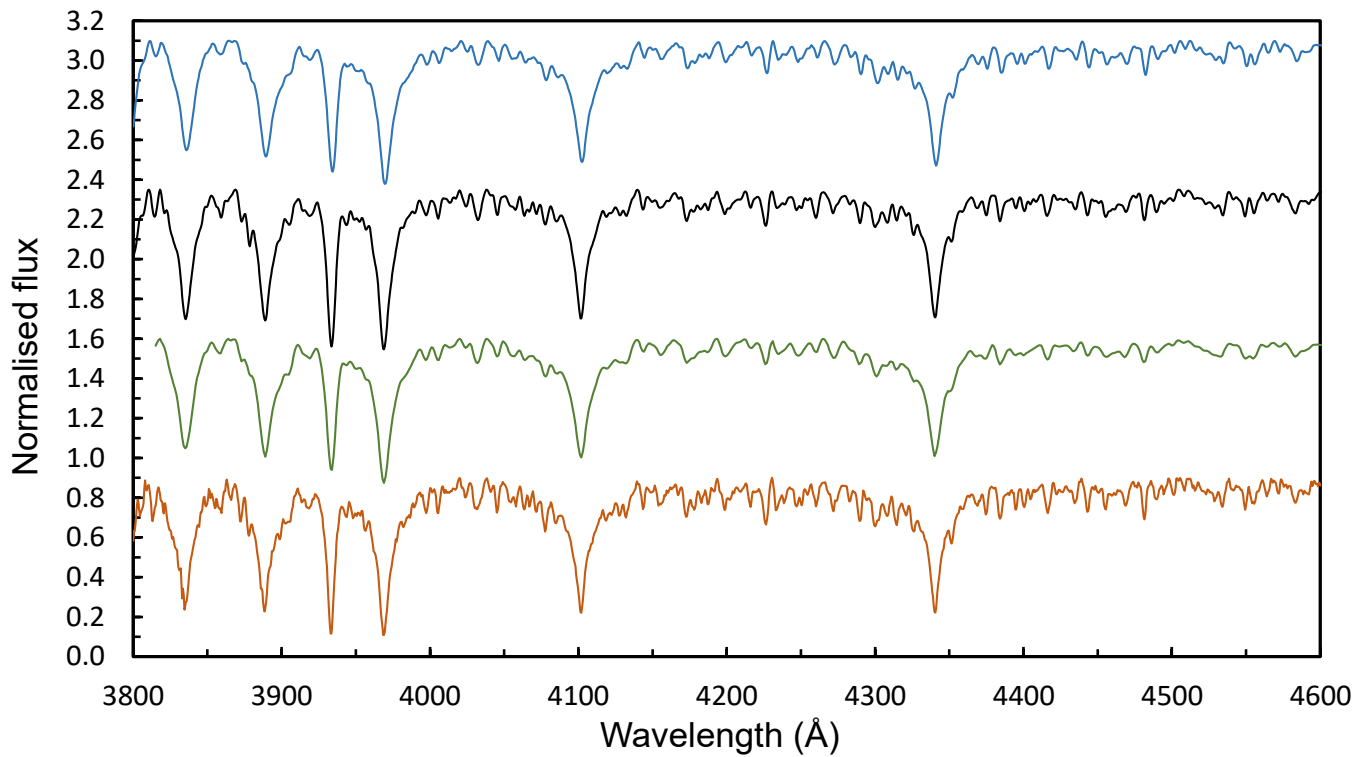}
    \caption{Blue-violet region of (from top to bottom) the F0 V standard spectra from the \textit{liblamost}, \textit{libsynth}, \textit{libnor36}, and \textit{libr18} standard libraries.}
    \label{fig_F0V_standards}
\end{figure*}

To investigate the flux depression at 5200\,\AA, \citet{1976A&A....51..223M} introduced the narrow-band three-filter $\Delta a$ system, which samples the depth of this depression by comparing the flux at the center (5220~{\AA}, $g_{\rm 2}$) with the adjacent regions (5000~{\AA}, $g_{\rm 1}$ and 5500~{\AA}, $y$) using a band-width of 130~{\AA} for $g_{\rm 1}$ and $g_{\rm 2}$ and 230~{\AA} for the Str{\"o}mgren $y$ filter. The index was introduced as: $$ a = g_{\rm 2} - (g_{\rm 1} + y)/2. $$ This quantity is slightly dependent on temperature in the sense that it increases towards lower temperatures. Therefore, the intrinsic peculiarity index had to be defined as: $$ \Delta a = a - a_{\rm 0}(g_{\rm 1} - y), $$ that is, the difference between the individual $a$ value and the $a$ value of non-peculiar stars of the same colour. The locus of the $a_{\rm 0}$ values for non-peculiar objects was termed the normality line. Virtually all mCP stars have positive $\Delta a$ values up to $+75$\,mmag \citep{2005A&A...441..631P}. Only extreme cases of CP1 and CP3 stars can exhibit marginally positive $\Delta a$ values. Be/Ae, B[e] and $\lambda$ Bootis stars exhibit significant negative values. In summary, it has been shown that the $\Delta a$ system is an efficient and reliable means of identifying mCP stars.

\subsubsection{Sample selection} \label{sample_selection}

In the present study, we concentrated on the publicly-available spectra from LAMOST DR4 \citep{lamost1,DR4}. As first step, the complete catalogue was cross-matched with the $Gaia$ DR2 catalogue \citep{2018A&A...616A...1G}. To identify suitable targets, we exploited the $G$ versus $(BP-RP)$ diagram to set a corresponding limit on the investigated spectral type range (hotter than mid F, i.e. $(BP-RP)$\,$<$\,0.45\,mag). From the remaining objects, apparent supergiants were excluded. As this approach is bound to miss highly-reddened hot objects, we searched the spectral types listed in the DR4 VizieR online catalogue\footnote{\url{http://cdsarc.u-strasbg.fr/viz-bin/cat/V/153}} \citep{DR4} for additional early-type (B-, A-, and F-type) targets, which were also included into the analysis.

Only spectra with a S/N of more than 50 in the Sloan $g$ band were considered for further analysis. This cut was deemed necessary because a lower S/N renders the detection of mCP star features difficult \citep{2011AN....332...77P}. If more than one spectrum was available for a single object, only the spectrum with the highest Sloan $g$ band S/N was included into the analysis.

From the remaining spectra, suitable candidates were selected by the presence of the tell-tale 5200\,\AA\ depression. To calculate synthetic $\Delta a$ values, all spectra were normalised to the flux at 4030\,\AA. This guarantees that the large absolute flux differences introduced by the apparent visual magnitude do not cause any numerical biases in the final magnitudes. The filter curves of $g_1$, $g_2$, and $y$ as defined in \citet{2003MNRAS.341..849K} were then folded with the spectra and the corresponding magnitudes calculated. All objects with a significant positive $\Delta a$ index were visually inspected for the presence of a 5200\,\AA\ depression in order to sort out glitches in the spectra or contamination by other objects such as cool stars with strong features in the 5200\,\AA\ range.

Figure \ref{fig_5200A_depression2} illustrates this process by providing sample LAMOST DR4 spectra of mCP stars showing 5200\,\AA\ flux depressions of various strengths. Also shown is the 'impostor' LAMOST J001159.88+435908.5 (GSC\,02794-00977). This object is actually a mid to late K star whose 5200\,\AA\ region is dominated by absorption lines of the \ion{Mg}{i} triplet at 5167\,\AA, 5173\,\AA, and 5184\,\AA, which leads to a significantly positive $\Delta a$ value and highlights the need for setting a limit on the investigated spectral type range via the above described colour-colour cut.

In this way, a list of 1002 mCP star candidates was compiled. This collection of stars is referred to in the following as the 'final sample'. We here emphasise that our sample is obviously biased towards mCP stars with conspicuous flux depressions at 5200\,\AA. However, not all mCP stars show significant 5200\,\AA\ depressions, in particular in low-resolution spectra, and such objects will have been missed by the imposed selection criteria. On the other hand, early-type stars with significant 5200\,\AA\ depressions are nearly always mCP stars; therefore, the chosen approach should be well suited to collecting a pure sample of mCP stars.

\begin{figure*}
        \includegraphics[width=\textwidth]{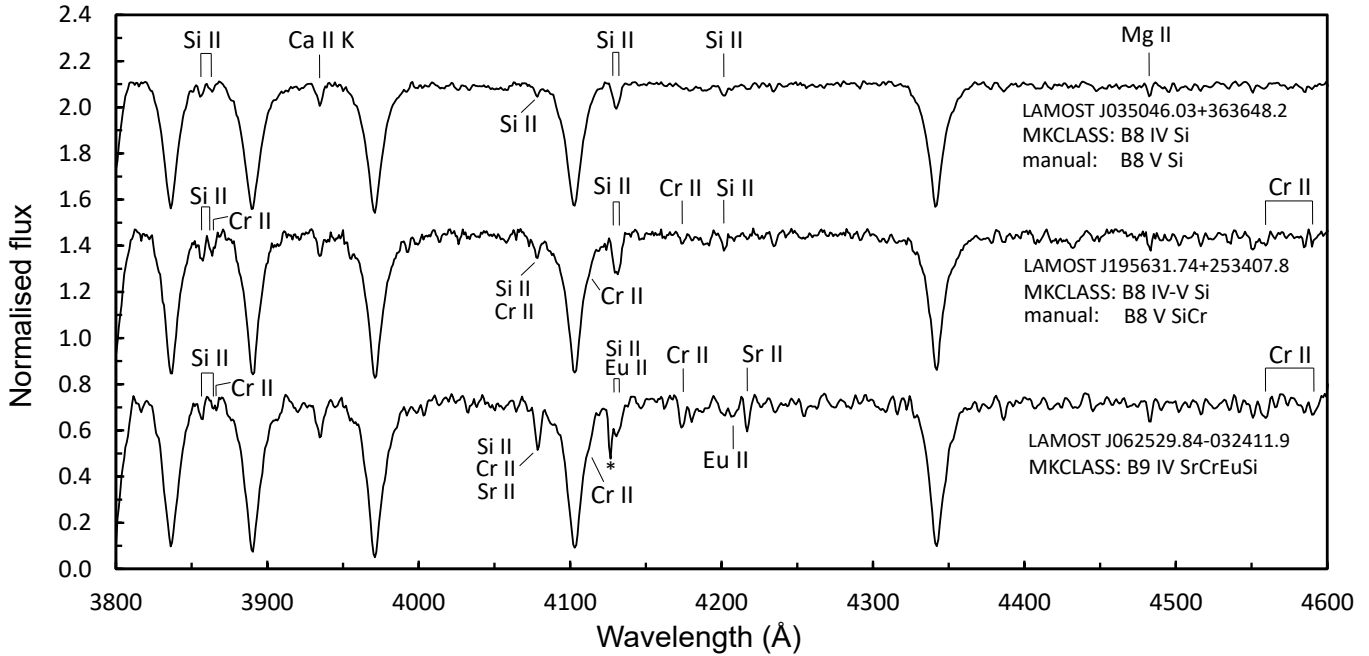}
    \caption{Showcase of three newly identified 'hot' mCP stars, illustrating the blue-violet region of the LAMOST DR4 spectra of (from top to bottom) LAMOST J035046.03+363648.2 (\#151; Gaia DR2 220081859486642816), LAMOST J195631.74+253407.8 (\#929; Gaia DR2 2026771741029840128), and LAMOST J062529.84-032411.9 (\#576; TYC 4789-2924-1). MKCLASS final types and, where available, manual types derived in the present study are indicated. Some prominent lines of interest are identified. The asterisk marks the position of a 'glitch' in the spectrum of LAMOST J062529.84-032411.9.}
    \label{spectra_early_types}
\end{figure*}

\section{Spectral classification} \label{spectral_classification}

Spectral classification is an important tool in astrophysics, which allows for the easy identification of astrophysically interesting objects. Furthermore, it enables to place stars in the Hertzsprung-Russell diagram, thus enabling the derivation of physical parameters. However, in the era of large survey projects such as LAMOST, RAVE, or SDSS/SEGUE, which produce a multitude of stellar spectra, human manual classification is no longer able to cope with the amount of data and the need for automatic classification has arisen. For the present study, we chose to employ a modified version of the MKCLASS code (cf. Section \ref{MKCLASS}) that has been specifically tailored to the needs of our project. More details are provided in this section, which details the spectral classification workflow.

\subsection{Spectral classification with the MKCLASS code} \label{spectral_classification_MKCLASS}

As deduced from an investigation of the program code, the current version of the MKCLASS code (v1.07) is able to identify the following spectral features, which are important in the detection of CP2 stars: the blend at 4077\,\AA\ (which may contain contributions from \ion{Si}{ii} 4076\,\AA, \ion{Sr}{ii} 4077\,\AA, and \ion{Cr}{ii} 4077\,\AA), the blend at 4130\,\AA\ (due to enhanced \ion{Si}{ii} 4128/30\,\AA\ and/or \ion{Eu}{ii} 4130\,\AA), and the \ion{Eu}{ii} 4205\,\AA\ line. On a significant detection of these features, the following output is created: 'Sr' (4077\,\AA), 'Si' (4130\,\AA), 'Eu' (4205\,\AA). As other elements besides Sr and Si contribute to the blends at 4077\,\AA\ and 4130\,\AA, the output may be misleading in some cases. Nevertheless, this allows a robust detection of mCP stars (e.g. \citealt{huemmerich18}) and is a good starting point for further investigations. 

To suit the special needs of our project, which is solely concerned with the identification and classification of mCP stars among a sample of early-type candidate stars, we opted to refine the MKCLASS peculiarity identification routine. The code was therefore altered to probe several additional lines, with the advantage that the new version is now able to more robustly identify traditional mCP star peculiarities. In addition, we enabled the identification of Cr peculiarities and, to some extent, He peculiarities, which are relevant to the classification of mCP stars. The choice of lines was dictated by the resolution and quality of the input material (i.e. LAMOST DR4 spectra), in particular concerning the availability of neighbouring continuum flux to probe a certain line and the numerous line blends due to the low resolution. We therefore stress that the resulting version of the MKCLASS code, which is referred to hereafter as MKCLASS\_mCP, has been created specifically for identifying and classifying mCP stars in LAMOST low-resolution spectra. Applying the code to spectra of other resolutions will require a corresponding update of the peculiarity classification routine and, perhaps, an update and enlargement of the employed standard star libraries (see below). Table \ref{table_lines} lists the lines and blends identified by MKCLASS\_mCP, as well as the spectral range in which the corresponding features are searched for. We note that at the resolution of the employed LAMOST spectra, all these lines are, to some extent, blended with other absorption lines. Nevertheless, the listed ions generally constitute the main contributors to these blends in mCP stars.

A sample output of MKCLASS\_mCP is provided in column five of Table \ref{table_MKCLASS_manual}. Further information on the interpretation of this output is provided below; a discussion of the accuracy of the derived classifications is provided in Section \ref{evaluation}.

\begin{table}
\caption{Absorption lines and blends identified by the modified version of the MKCLASS code (MKCLASS\_mCP) and used in the identification and classification of mCP stars in the present study. The columns denote: (1) Blend/line. (2) Wavelength (\AA). (3) Spectral range in which the corresponding feature was probed.}
\label{table_lines}
\begin{center}
\begin{tabular}{cll}
\hline
\hline
(1) & (2) & (3) \\
\textbf{Blend/line} & \textbf{Wavelength (\AA)} & \textbf{SpT\_range} \\
\hline
(\ion{Si}{ii}/\ion{Cr}{ii}/\ion{Sr}{ii}) & 4076/77 & B7$-$F5 \\
\hline
(\ion{Si}{ii}/\ion{Eu}{ii}) & 4128/30 & B3$-$F2 \\
\hline
\ion{Si}{ii} & 3856 & B3$-$F2 \\
\ion{Si}{ii} & 4200 & B3$-$A2 \\
\ion{Si}{ii} & 5041 & B3$-$F2 \\
\ion{Si}{ii} & 5056 & B3$-$F2 \\
\ion{Si}{ii} & 6347 & B3$-$F2 \\
\ion{Si}{ii} & 6371 & B3$-$F2 \\
\hline
\ion{Cr}{ii} & 3866 & B7$-$F2 \\
\ion{Cr}{ii} & 4172 & B7$-$F2 \\
\hline
\ion{Sr}{ii} & 4216 & B7$-$F2 \\
\hline
\ion{Eu}{ii} & 4205 & B7$-$F2 \\
\hline
\ion{He}{i} & 4009 & B0$-$A0 \\
\ion{He}{i} & 4026 & B0$-$A0 \\
\ion{He}{i} & 4144 & B0$-$A0 \\
\ion{He}{i} & 4387 & B0$-$A0 \\
\hline
\end{tabular}
\end{center}
\end{table}

We note that the \ion{Si}{ii} line at 5041\,\AA\ increases significantly with temperature type; therefore, different detection limits were applied depending on the investigated temperature range. Furthermore, the \ion{Si}{ii} 6347/71\,\AA\ lines were found to show a significant scatter in MKK standard stars. However, at the resolution of the LAMOST spectra, the red \ion{Si}{ii} lines are a readily detectable and outstanding feature of strong Si stars and contribute significantly to an unambiguous detection of Si peculiarity, in particular as the corresponding lines in the blue-violet region are difficult to detect because of continuum flux issues (3856/62\,\AA) and blending issues (4076\,\AA\ and 4128/30\,\AA).

\begin{figure*}
        \includegraphics[width=\textwidth]{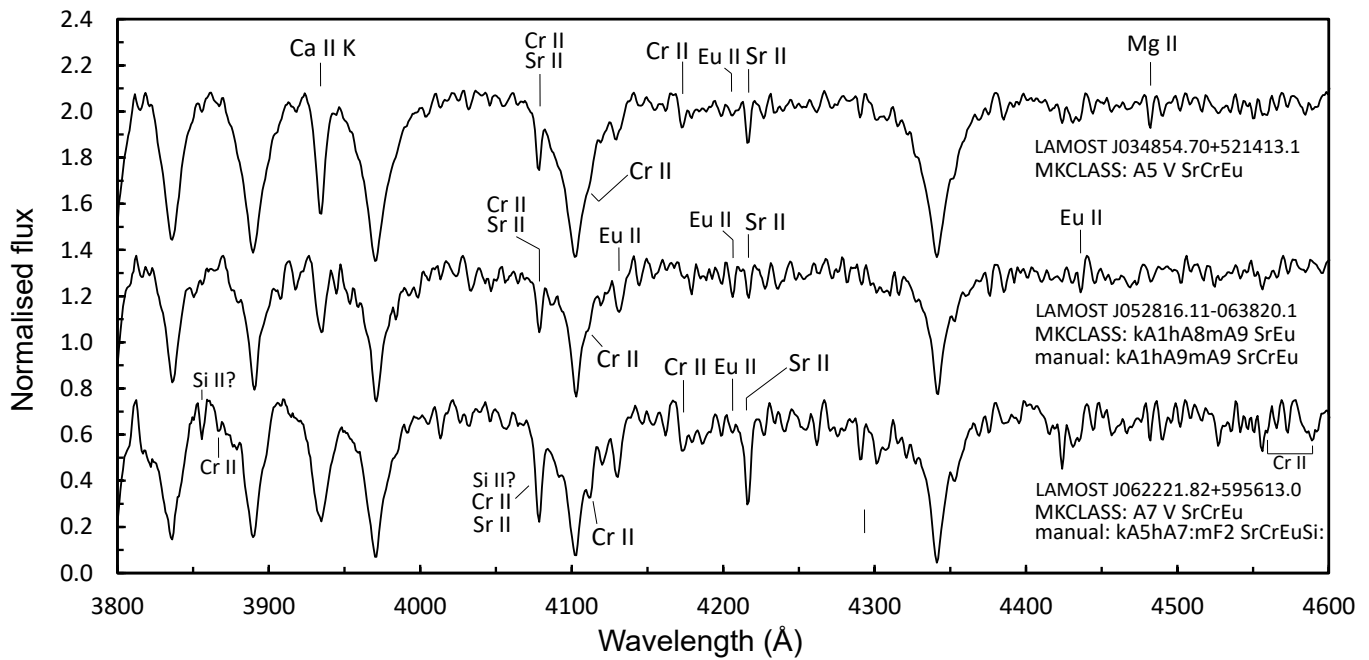}
    \caption{Showcase of three newly identified 'cool' mCP stars, illustrating the blue-violet region of the LAMOST DR4 spectra of (from top to bottom) LAMOST J034854.70+521413.1 (\#150; Gaia DR2 251609324623302400), LAMOST J052816.11-063820.1 (\#344; TYC 4765-708-1), and LAMOST J062221.82+595613.0 (\#561; TYC 3776-269-1). MKCLASS final types and, where available, manual types derived in the present study are indicated. Some prominent lines of interest are identified.}
    \label{spectra_late_types}
\end{figure*}

As has been desribed in Section \ref{MKCLASS}, MKCLASS is a computer program that emulates the workflow of a human classifier in the traditional MKK spectral classification process, which involves comparing the input spectrum to a set of standard star spectra. It is therefore imperative to carefully select standard star libraries that match the input spectra in resolution and calibrationwise. MKCLASS comes with the two standard libraries \textit{libr18} and \textit{libnor36} (cf. Section \ref{MKCLASS}), which, unfortunately, do not match the spectral resolution of the LAMOST low-resolution spectra. Furthermore, as far as we are aware of, a standard library based on LAMOST spectra does not exist.

\begin{table*}
\caption{Spectral classifications derived by manual classification and the MKCLASS\_mCP code. The columns denote: (1) Internal identification number. (2) LAMOST identifier. (3) Spectral type derived by manual classification. (4) MKCLASS\_mCP final type. (5) MKCLASS\_mCP output using the standard star libraries \textit{libr18}, \textit{libnor36}, \textit{libsynth}, and \textit{liblamost}.}
\label{table_MKCLASS_manual}
\scriptsize{
\begin{center}
\begin{tabular}{|l|l|l|l|l|}
\hline
(1) & (2) & (3) & (4) & (5) \\
\hline
\textbf{No.}	&	\textbf{LAMOST ID}	&	\textbf{SpT\_manual}	&	\textbf{SpT\_MKCLASS\_mCP}	&	\textbf{Output using \textit{libr18}/\textit{libnor36}/\textit{libsynth}/\textit{liblamost}} \\
\hline
142	&	J034541.53+275631.8	&	kA3hA6mA6 SrCrEu	&	A6 V SrCrEu 	&	kA4hA4mA8  bl4077 bl4130 Sr4216 Cr4172 Eu4205	\\
	&		&		&		&	A6 V  bl4077 bl4130 Sr4216 Eu4205	\\
	&		&		&		&	A6 V  bl4077 bl4130 Sr4216 Eu4205	\\
	&		&		&		&	A5 IV$-$V  bl4077 bl4130 Sr4216	\\
	\hline
151	&	J035046.03+363648.2	&	B8 V Si	&	B8 IV Si 	&	B6 IV$-$V  Si5041 Si5056 Si6347 bl4130	\\
	&		&		&		&	B8 IV  Si6347 bl4130	\\
	&		&		&		&	B8 IV  Si5041 Si5056 Si6347 bl4130	\\
	&		&		&		&	B7 IV  Si5041 Si5056 Si6347 bl4077 bl4130	\\
	\hline
232	&	J043201.64+471447.8	&	A2 V SrCrEu(Si)	&	A1 V SrCrEu 	&	kA2hA6mA8  bl4077 Sr4216 Cr4172 Eu4205	\\
	&		&		&		&	A1 IV$-$V  bl4077 bl4130 Sr4216	\\
	&		&		&		&	kA2hA3mA6  bl4077 bl4130 Sr4216 Cr4172	\\
	&		&		&		&	A1 V  bl4077 bl4130 Sr4216	\\
	\hline
306	&	J051844.95+380605.3	&	B9 V SrCrEuSi	&	B9 IV SrCr 	&	kB9hA0mA2  bl4077 bl4130 Sr4216 Cr4172	\\
	&		&		&		&	B9.5 III$-$IV  bl4077 bl4130 Sr4216 Cr4172	\\
	&		&		&		&	kB9.5hA1mA3  bl4077 bl4130 Sr4216 Cr3866 Cr4172	\\
	&		&		&		&	B9 IV  bl4077 bl4130 Sr4216 Cr4172	\\
	\hline
344	&	J052816.11-063820.1	&	kA1hA9mA9 SrCrEu	&	kA1hA8mA9 SrEu 	&	A1 II$-$III  bl4077 bl4130 Sr4216 Eu4205	\\
	&		&		&		&	kA2hA5mA9  bl4077 bl4130 Sr4216 Eu4205	\\
	&		&		&		&	kA1hA9mA8  bl4077 bl4130 Sr4216 Eu4205	\\
	&		&		&		&	kA1hA8mA9  Si3856 bl4077 bl4130 Sr4216 Eu4205	\\
	\hline
561	&	J062221.82+595613.0	&	kA5hA7:mF2 SrCrEuSi:	&	A7 V SrCrEu 	&	A8 mA4 metal weak	\\
	&		&		&		&	kA6hA8mF3  bl4077 bl4130 Sr4216 Cr3866	\\
	&		&		&		&	kA5hA7mF1  bl4077 bl4130 Sr4216 Cr3866 Cr4172 Eu4205	\\
	&		&		&		&	A7 V  bl4077 bl4130 Sr4216 Cr4172	\\
	\hline
596	&	J062909.51+023823.8	&	B8 V Si	&	B8 IV$-$V Si 	&	B8 IV$-$V  Si5041 Si5056 Si6347 Si6371 bl4130	\\
	&		&		&		&	B8 IV$-$V  Si5041 Si5056 Si6347 Si6371 bl4130	\\
	&		&		&		&	B8 IV  Si5041 Si5056 Si6347 Si6371 bl4130	\\
	&		&		&		&	B8 IV$-$V  Si5041 Si5056 Si6347 Si6371 bl4077 bl4130	\\
	\hline
724	&	J065511.76+115158.3	&	A7 V SrCrEu	&	A7 V SrEu 	&	A8 V  bl4077 bl4130 Sr4216 Eu4205	\\
	&		&		&		&	A9 V  bl4077 bl4130 Sr4216 Eu4205	\\
	&		&		&		&	A7 V  Si6347 bl4077 bl4130 Sr4216 Eu4205	\\
	&		&		&		&	A7 V  bl4077 bl4130 Sr4216	\\
	\hline
732	&	J065647.94+242958.8	&	A0 V SiSrCr	&	B9.5 V Sr	&	kB9hA7mA5  Sr4216 Cr3866 Eu4205	\\
	&		&		&		&	B9.5 IV$-$V bl4077 bl4130	\\
	&		&		&		&	kB9.5hA3mA3  bl4077 Sr4216	\\
	&		&		&		&	B9.5 V bl4077 bl4130	\\
	\hline
929	&	J195631.74+253407.8	&	B8 V SiCr	&	B8 IV$-$V Si 	&	B8 V  Si5041 Si5056 Si6347 Si6371 bl4077 bl4130	\\
	&		&		&		&	B8 IV$-$V  Si5056 Si6347 Si6371 bl4077 bl4130	\\
	&		&		&		&	B8 IV$-$V  Si5041 Si5056 Si6347 Si6371 bl4077 bl4130	\\
	&		&		&		&	A0 II$-$III  Si5056 Si6347 Si6371 bl4077 bl4130	\\
	\hline
\end{tabular}                                                                                                                                                                
\end{center}                                                                                                                                                                    
}                                                                                                                                                                       
\end{table*}

In the framework of the LAMOST-$Kepler$ project, \citet{gray16} presented MKK spectral classifications of more than 100\,000 LAMOST spectra of about 80\,000 objects situated in the $Kepler$ field. The authors solved the above-mentioned issue by degrading the flux-calibrated LAMOST spectra to a resolution of R\,$\sim$\,1100 and truncating them to the 3800$-$5600\,\AA\ region in order to enable the use of MKCLASS with the flux-calibrated \textit{libnor36} library (cf. also the MKCLASS documentation). Here we follow their approach, but we also searched for alternative methods, as degrading spectra obviously results in loss of information. This, however, is detrimental to the identification of the often subtle chemical peculiarities of our sample stars.

We synthesised a library of spectra using the program SPECTRUM\footnote{\url{http://www.appstate.edu/~grayro/spectrum/spectrum.html}} \citep{SPECTRUM} and ATLAS9 model atmospheres \citep{ATLAS9}, which is referred to hereafter as the \textit{libsynth} library. Only dwarf spectra (luminosity class V) were synthesised because no models were available to reproduce the subtle differences in surface gravity among early-type giant stars. Furthermore, we collected a set of LAMOST standard star spectra (the \textit{liblamost} library) by carefully choosing a grid of suitable high S/N spectra from the list presented by \citet{gray16}. Only dwarf and giant spectra were chosen, as not enough suitable spectra of higher luminosity class objects were available to build up a corresponding grid. We note, however, that it has been well confirmed that mCP stars are generally main-sequence objects (cf. \citealt{netopil17}, and references therein, and Sections \ref{introduction}, \ref{evaluation}, and \ref{evolutionary_status}); therefore, we do not expect that the absence of spectra of higher luminosity class objects in these two libraries significantly affects our results -- in particular as the \textit{libr18} and \textit{libnor36} libraries boast corresponding spectra of all luminosity classes.

The stars and corresponding LAMOST spectrum identifiers of the \textit{liblamost} library are given in the Appendix in Table \ref{table_liblamost}. Although it contains a very suitable grid of dwarf spectra, we note that the \textit{liblamost} library is far from being a perfect set of standard star spectra (a corresponding quality flag that estimates the suitability of a spectrum as a standard is also provided in Table \ref{table_liblamost}). It contains a fast rotator and some spectra show 'impurities' not expected in MKK standards. These shortcomings will lead to increased uncertainties in the derivation of the temperature and luminosity classes. However, the library consists of spectra obtained with the same instrument -- and hence, importantly, of the same resolution as our input spectra -- and was extremely valuable in the identification of chemical peculiarities. We nevertheless explicitly caution against using the \textit{liblamost} library as a standard star library out of the context of the present investigation. We also emphasise the need for a standard star library based on LAMOST low-resolution spectra, which will greatly facilitate further research based on this excellent data source. The \textit{liblamost} library may very well serve as a starting point; this, however, is beyond the scope of the present investigation.

In the following, an overview over the employed spectral libraries is presented. As an example, Figure \ref{fig_F0V_standards} illustrates the F0V standard spectra from the corresponding libraries. We note that the \textit{libsynth} and \textit{liblamost} libraries only contain spectra in the approximate spectral type range of our sample stars.

\begin{itemize}
	\item \textbf{\textit{libr18}}: spectral range from 3800–4600\,\AA, resolution of 1.8\,\AA\ (R\,$\sim$\,2200), normalised and rectified spectra; all luminosity classes (Ia-V)
	\item \textbf{\textit{libnor36}}: spectral range from 3800–5600\,\AA, resolution of 3.6\,\AA\ (R\,$\sim$\,1100), flux-calibrated and normalised spectra; all luminosity classes (Ia-V)
	\item \textbf{\textit{libsynth}}: spectral range from 3800–4600\,\AA, smoothed to a resolution of 3.0\,\AA\ and an output spacing of 0.5\,\AA, flux-calibrated and normalised synthetic spectra; only dwarf spectra (luminosity class V); spectral types B5 to F5
	\item \textbf{\textit{liblamost}}: spectral range from 3800–5600\,\AA, resolution R\,$\sim$\,1800, flux-calibrated and normalised spectra; only dwarf and giant spectra (luminosity classes V and III); spectral types B3 to G0
\end{itemize}

mCP stars may exhibit peculiar \ion{Ca}{ii} K profiles and line strengths \citep{faraggiana87,gray09,ghazaryan18} as well as generally enhanced metal-lines. Several authors have therefore adopted a notation that indicates separate spectral types as derived from the \ion{Ca}{ii} K line (the k-type), the hydrogen lines (the h-type), and the general strength of the metal-lines (the m-type), in the same way as is usually done for CP1 stars. The MKCLASS code also assigns k/h/m-types in cases where discrepant spectral types are derived from the corresponding features. As mCP stars are prone to exhibiting marked Ca and He deficiencies (e.g. \citealt{gray09,ghazaryan18}) and often enhanced metal-lines, the hydrogen-line profile is a better indicator of the actual effective temperature \citep{gray09}. Where they have been derived by the code (or by manual classification), k/h/m types are listed in the present study.

For most stars, only minor differences in temperature and luminosity types were found between the results from the different spectral libraries. In cases where the same spectral type was derived more than once, the most common spectral type was adopted (cf. Table \ref{table_MKCLASS_manual}). If no common classifications existed, spectral types were favoured in the order \textit{liblamost} > \textit{libsynth} > \textit{libnor36} > \textit{libr18}. In the case of strong differences between the derived classifications, the corresponding spectra were visually inspected and the best fitting type was chosen. 

To determine significant chemical peculiarities from the 'raw' MKCLASS\_mCP output, the number of detections $N$\textsubscript{det} of the peculiar strength of a given line with the different standard star libraries (0\,$\le$\,$N$\textsubscript{det}\,$\le$\,4) was counted, which provides an estimation of significance. Obviously, $N$\textsubscript{det}\,=\,4 is a very robust detection; $N$\textsubscript{det}\,<\,2 detections, on the other hand, have to be viewed with caution. Furthermore, we required that the identification of overabundances cannot be based on a single strong line (with the exception of the \ion{Cr}{ii} 4172\,\AA\ line in the identification of a Cr peculiarity; see below).

To come up with an approach that forms a compromise between spurious detections and overly high thresholds required a good amount of experimentation and experience in comparing the results of manual and automatic classification. Table \ref{table_peculiarities} lists the conditions found to work best with the input material and our methodological approach. $N$\textsubscript{det}($\lambda$) is the number of detections of a peculiarly strong line at the specified wavelength (\AA). For instance, a Cr peculiarity was flagged when (a) a strong \ion{Cr}{ii} 4172\,\AA\ line was detected with a least two different standard star libraries \textbf{or} (b) a strong \ion{Cr}{ii} 3866\,\AA\ line \textbf{and} a strong \ion{Cr}{ii} 4172\,\AA\ line were detected at least once \textbf{or} (c) a strong blend at 4077\,\AA\ was detected at least twice \textbf{and} a strong \ion{Cr}{ii} 4172\,\AA\ line was detected at least once.

Following the conventions of the MKK system, the peculiarity types 'Si', 'Cr', 'Sr', and 'Eu' were flagged according to the conditions given in Table \ref{table_peculiarities} and attached to the temperature and luminosity types in the final spectral classification. Several stars in our sample that were not assigned any of the above mentioned peculiarity types nevertheless show strong blends at 4077\,\AA\ and/or 4130\,\AA. In these cases, we decided to add the non-standard suffixes 'bl4077' and 'bl4130' to the derived spectral types (e.g. 'B8 IV bl4130') if the corresponding blends had been detected at least twice. In these objects, apart from the strong blends, the peculiarities are either too subtle to have passed our significance criteria, no other significant features are present, or the code failed to identify them for some reason. Manual classification is necessary to throw more light on this matter (cf. Section \ref{peculiarity_types}).

In addition, we opted to probe the \ion{He}{i} lines at 4009\,\AA, 4026\,\AA, 4144\,\AA, and 4387\,\AA\ to identify CP2 stars with weak \ion{He}{i} lines and He-peculiar objects. The corresponding detection thresholds for all four standard star libraries were determined using the He lines of 626 apparently chemically-normal B stars with spectra boasting S/N\,$>$\,100. The number of detections as 'weak' or 'strong' of the He lines with the different standard star libraries was counted and the results across all lines and libraries were added up to yield $N$\textsubscript{det}(He-wk) and $N$\textsubscript{det}(He-st). He-weakness and He-overabundance were assumed when $N$\textsubscript{det}(He-wk)\,$>$\,2 and $N$\textsubscript{det}(He-st)\,$>$\,2, respectively. In this way, we identified 55 mCP stars with weak \ion{He}{i} lines and three mCP stars with apparently strong \ion{He}{i} lines. As expected, these are mostly B7$-$B9 Si CP2 stars, which are notorious for their weak He lines, and mid-B type stars (likely He-peculiar objects). Interestingly, in three mid-B type stars, both weak and strong \ion{He}{i} lines were identified, which strongly suggests He peculiarity. The corresponding suffixes 'He-wk' and 'He-st' were added to the derived spectral types. Several He-peculiar objects are discussed in Section \ref{Hepec}.

In this way, peculiarities were identified in all but 36 stars from our sample, which highlights the efficiency of the chosen approach. The (mostly low S/N) spectra of the remaining objects were investigated manually and searched for the presence of chemical peculiarities. Most of these objects show subtle or complicated peculiarities that failed to meet the imposed significance criteria. Corresponding peculiarity types were manually added to the final spectral types. The remaining objects are a 'mixed bag' containing stars with enhanced metal-lines and strong flux depressions that nevertheless lack the traditional Si, Cr, Sr, Eu peculiarities and several He-peculiar objects. Appropriate comments were added to the tables, in which manually-derived peculiarity types and all further additions that are not directly based on the MKCLASS\_mCP output are highlighted.

\begin{table}
\caption{Conditions employed to flag the presence of an overabundance from the 'raw' MKCLASS\_mCP output. The columns denote: (1) Overabundant ion. (2) Condition(s) required to be met for a detection to be deemed significant.}
\label{table_peculiarities}
\begin{tabular}{cll}
\hline
\hline
(1) & (2) & \\
\textbf{Ion} & \textbf{condition(s)} & \\
\hline
\textbf{\ion{Si}{ii}} & ($N$\textsubscript{det}(3856)\,$+$\,$N$\textsubscript{det}(4200)\,$+$\,$N$\textsubscript{det}(5041)\,$+$ & \\
   & $N$\textsubscript{det}(5056)\,$+$\,$N$\textsubscript{det}(6347)\,$+$\,$N$\textsubscript{det}(6371))\,$>$\,1 & \\
\hline
\textbf{\ion{Cr}{ii}} & $N$\textsubscript{det}(4172)\,$>$\,1 & OR \\
   & ($N$\textsubscript{det}(3866)\,$>$\,0\,AND\,$N$\textsubscript{det}(4172)\,$>$\,0) & OR \\
	 & ($N$\textsubscript{det}(4077)\,$>$\,1 AND $N$\textsubscript{det}(4172)\,$>$\,0) & \\
\hline
\textbf{\ion{Sr}{ii}} & ($N$\textsubscript{det}(4077)\,$>$\,0 AND $N$\textsubscript{det}(4216)\,$>$\,0) & \\
\hline
\textbf{\ion{Eu}{ii}} & ($N$\textsubscript{det}(4130)\,$>$\,2 AND $N$\textsubscript{det}(4205)\,$>$\,0) & \\
\hline
\end{tabular}
\end{table}

\subsection{Evaluation} \label{evaluation}

As a test of the validity of our approach, we manually classified a sample of ten randomly chosen stars and compared our results with the final spectral types derived in the described manner from the MKCLASS\_mCP output, which, for convenience, are termed hereafter the 'MKCLASS final types'. In addition, we visually inspected the spectra of about 100 further stars to check for the presence of peculiarly strong lines and evaluate the reliability of the classifications. The results from the manual classification are shown in Table \ref{table_MKCLASS_manual} and highlight the good agreement between the manually- and automatically-derived (hydrogen-line) temperature types. In general, we estimate the uncertainty of the derived temperature types to be $\pm$1\,subclass. This, however, increases to about $\pm$2\,subclasses towards later and more peculiar mCP stars for which the classification is notoriously difficult and, for the most extreme objects, unreliable. Figures \ref{spectra_early_types} and \ref{spectra_late_types} showcase, respectively, three 'hot' mCP stars and three 'cool' mCP stars, which have been newly identified as such in the present study. MKCLASS final types and, where available, manual types from Table \ref{table_MKCLASS_manual} are indicated.

\begin{figure}
				\includegraphics[width=\columnwidth]{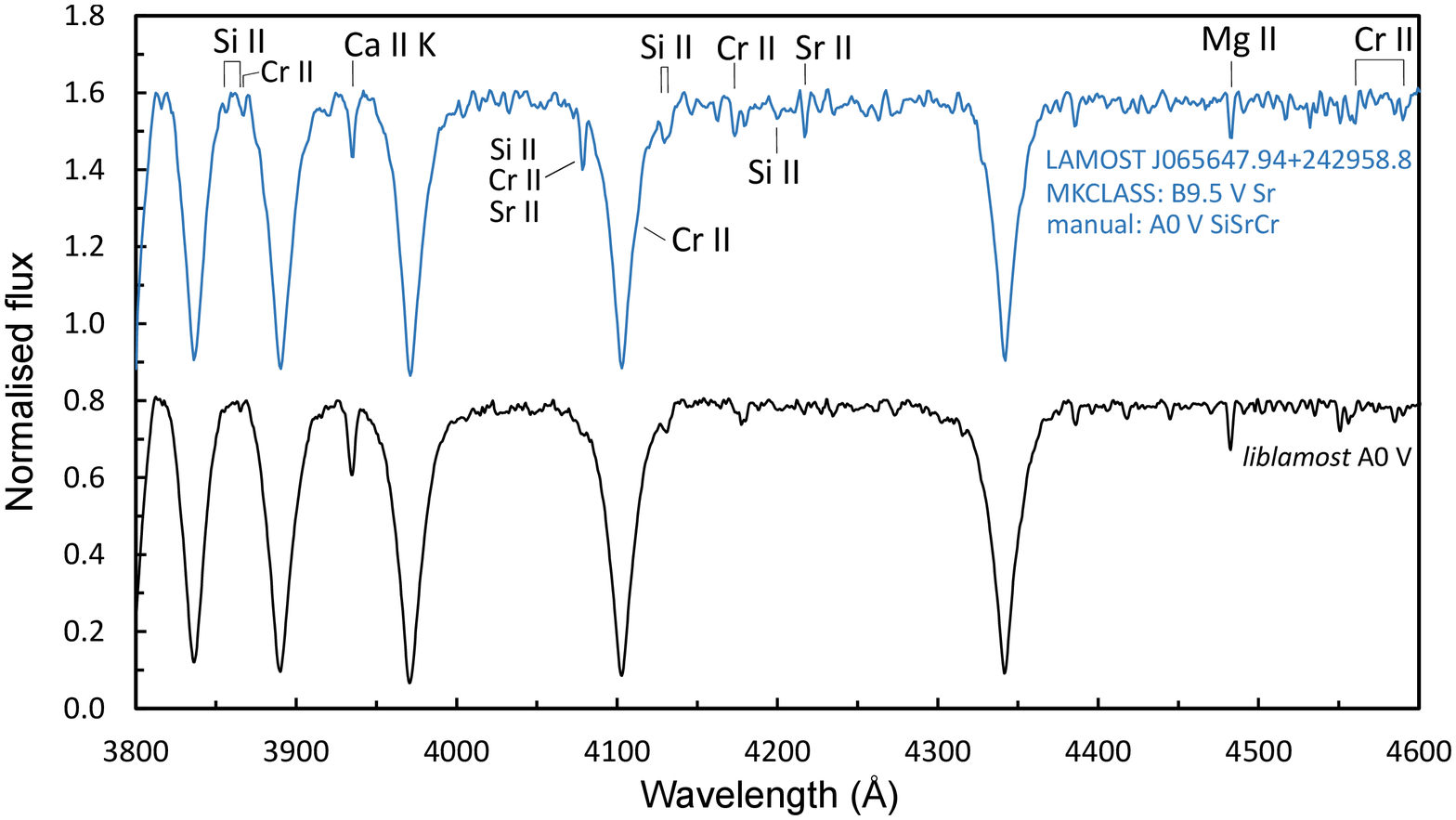} \\
				\includegraphics[width=\columnwidth]{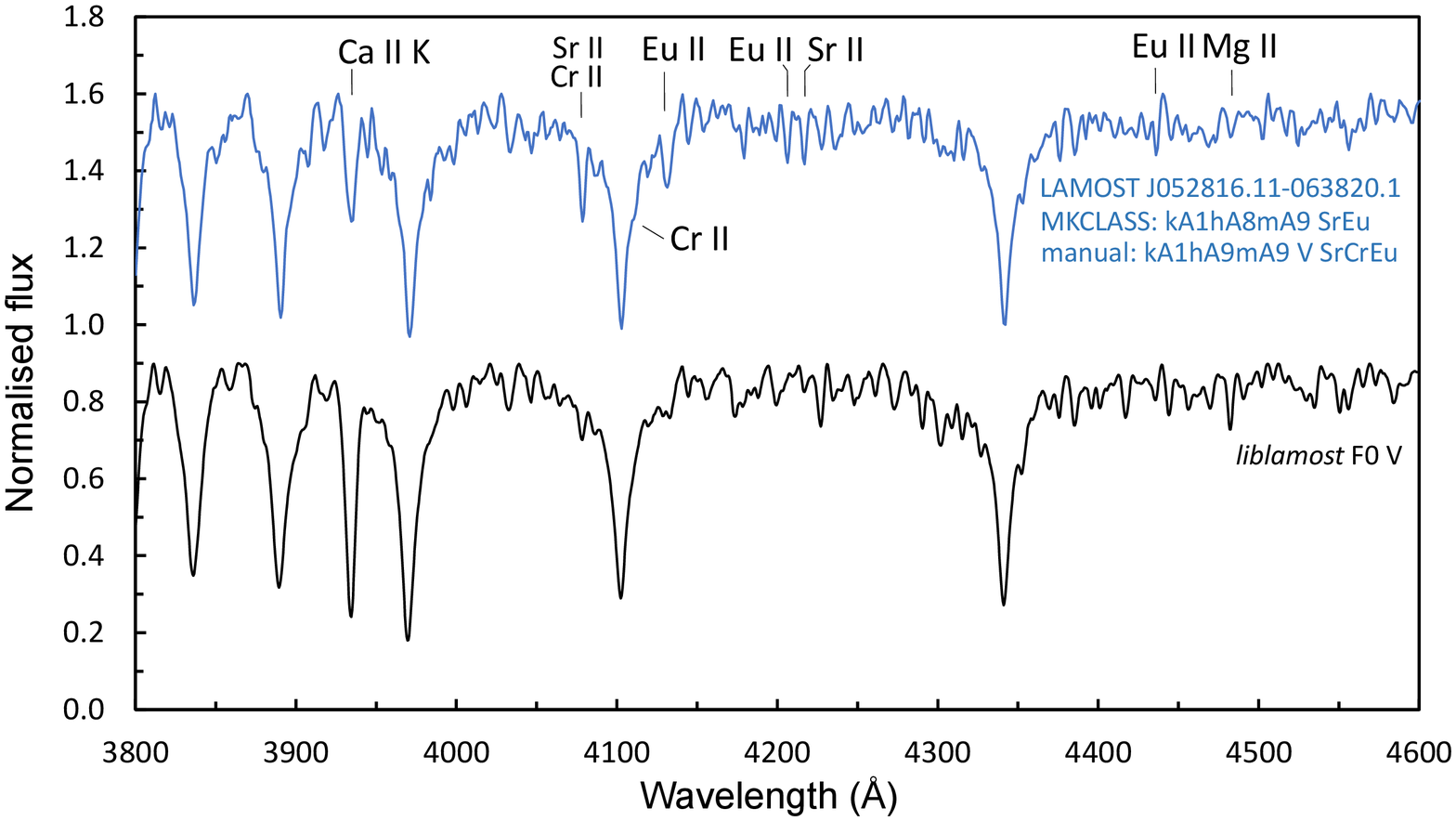}
    \caption{Upper panel shows a comparison of the blue-violet spectra of the mCP star LAMOST J065647.94+242958.8 (\#732; TYC\,1898-1408-1; manual type: A0 V SiSrCr; MKCLASS final type: B9.5 V Sr) to the \textit{liblamost} A0 V standard spectrum. Lower panel illustrates a comparison of the blue-violet spectra of the mCP star LAMOST\,J052816.11-063820.1 (\#344; TYC\,4765-708-1; manual type: kA1hA9mA9 SrCrEu; MKCLASS final type: kA1hA8mA9 SrEu) to the \textit{liblamost} F0 V standard spectrum. Some prominent lines of interest are indicated. We note the weak \ion{Ca}{ii} K line with a peculiar profile and the unusual profile of the H$\epsilon$ line in LAMOST\,J052816.11-063820.1.}
    \label{spectra_comparison}
\end{figure}

There is also a generally good agreement in regard to the derived peculiarity types. However, with our approach, we obviously missed the presence of peculiarities in several objects (Table \ref{table_MKCLASS_manual}). A further investigation of these stars shows that this has been mostly due to either the imposed significance criteria, weak or complicated peculiarities, or the absence of continuum flux to probe certain lines (or a combination thereof). A good case in point is LAMOST J065647.94+242958.8 (\#732; TYC\,1898-1408-1), whose blue-violet spectrum is shown in the upper panel of Figure \ref{spectra_comparison}. It shows enhanced \ion{Si}{II} lines at 3856/62\,\AA, 4128/31\,\AA, and 4200\,\AA. Furthermore, strong \ion{Cr}{II} lines are present at 3866\,\AA, 4111\,\AA, 4172\,\AA, 4559\,\AA, and 4588\,\AA. The blend at around 4077\,\AA\ is notoriously difficult to interpret and can contain contributions from \ion{Si}{II}, \ion{Cr}{II}, and \ion{Sr}{II}. The strong \ion{Sr}{II} line at 4216\,\AA, however, indicates the presence of a Sr peculiarity. Consequently, we have classified this star as A0 V SiSrCr. The MKCLASS\_mCP code missed the rather subtle Si and Cr peculiarities (MKCLASS final type: B9.5 V Sr).

The bottom panel of Figure \ref{spectra_comparison} illustrates the case of LAMOST J052816.11-063820.1 (\#344; TYC\,4765-708-1; also shown in Figure \ref{spectra_late_types}), a cool CP2 star that may serve as a warning and an example illustrating the difficulty of classifying the more extreme mCP stars, which may possess peculiarities that render their spectra difficult to match to any standard star \citep{gray09}. The star exhibits a weak \ion{Ca}{ii} K line with a peculiar profile. While the hydrogen-line profile points to a late A-type star, the broad K line is that of an early A-type star and reasonably matched by that of an A1 V standard. As CP2 stars are prone to exhibiting marked Ca deficiencies (e.g. \citealt{gray09,ghazaryan18}), the hydrogen-line profile is a better indicator of the actual effective temperature than the K line strength \citep{gray09}, hence LAMOST J052816.11-063820.1 is obviously a late-type mCP star. However, while the H$\gamma$, H$\delta$, H8, and H9 lines are reasonably well matched by that of an A9 V star, the H$\epsilon$ line exhibits an unusual profile, which makes it difficult to match the hydrogen-line profile to that of any standard star. We also note the unusually weak Mg II line at 4481\,\AA. The MKCLASS\_mCP code duly assigned final k/h/m-types of A1, A8, and A9; we prefer k/h/m-types of A1, A9, and A9.

The spectrum of LAMOST J052816.11-063820.1 shows strong blends at 4077\,\AA\ and 4130\,\AA, which -- judging from the strong lines at 4216\,\AA\ (\ion{Sr}{II}), 4205\,\AA\ (\ion{Eu}{II}), and 4435\,\AA\ (\ion{Eu}{II}) -- are mainly caused by overabundances of Sr and Eu. The strong \ion{Cr}{II} lines at 3866\,\AA\ and 4111\,\AA\ (the bump in the red wing of H$\delta$) indicate a Cr overabundance; a corresponding enhanced line at 4172\,\AA, however, is notably absent. We have classified this star as kA1hA9mA9 V SrCrEu. The MKCLASS\_mCP code duly identified the main peculiarities (MKCLASS final type: kA1hA8mA9 SrEu).

The examples show that the peculiarity types derived in the present investigation are in many cases not exhaustive but rather denote the main peculiarities present. They are still very useful for first orientation and an excellent starting point for more detailed investigations; they are furthermore suited to statistical studies (cf. Section \ref{peculiarity_types}).

Some cautionary words are necessary in regard to luminosity classification. It is well known that problems with the luminosity classification may arise by the confusion of luminosity criteria and mCP star characteristica or peculiarities. In the early A-type stars, luminosity classification is primarily based on hydrogen-line profiles. The mCP stars, in general, are slow rotators that display narrow lines and hydrogen-line profiles that are easily misinterpreted as belonging to stars of higher luminosity. Indeed, the hydrogen-line profiles of many late B- and early A-type Si stars of our sample are best matched by standards of luminosity class III although there is no further indication that these star are in fact giant stars. Additional confusion may arise due to peculiarly strong lines that are also used in luminosity classification. \ion{Si}{ii} lines, for example, are enhanced in giants and supergiants as well as in several types of mCP stars, which might lead to corresponding misclassifications \citep{loden89}. This holds especially true for classifications based on photographic plates or low S/N spectra. In regard to CP1 stars, the term 'anomalous luminosity effect' has been coined, which describes the perplexing situation that luminosity criteria from different regions of the spectrum indicate different luminosities. This also applies at least partly to mCP stars, which may show strong general enhancements of metal-lines in their spectra that are reminiscent of much cooler stars. Furthermore, while obtaining the hydrogen-line type is fairly straightforward for most mCP stars, the more peculiar objects show distorted atmospheres and unusual and peculiar hydrogen-line profiles that may not match any standard star \citep{gray09}, which may result in classification errors. Abnormal hydrogen-line profiles are especially observed in cool mCP stars \citep{kochukhov02}.

\begin{figure}
        \includegraphics[width=\columnwidth]{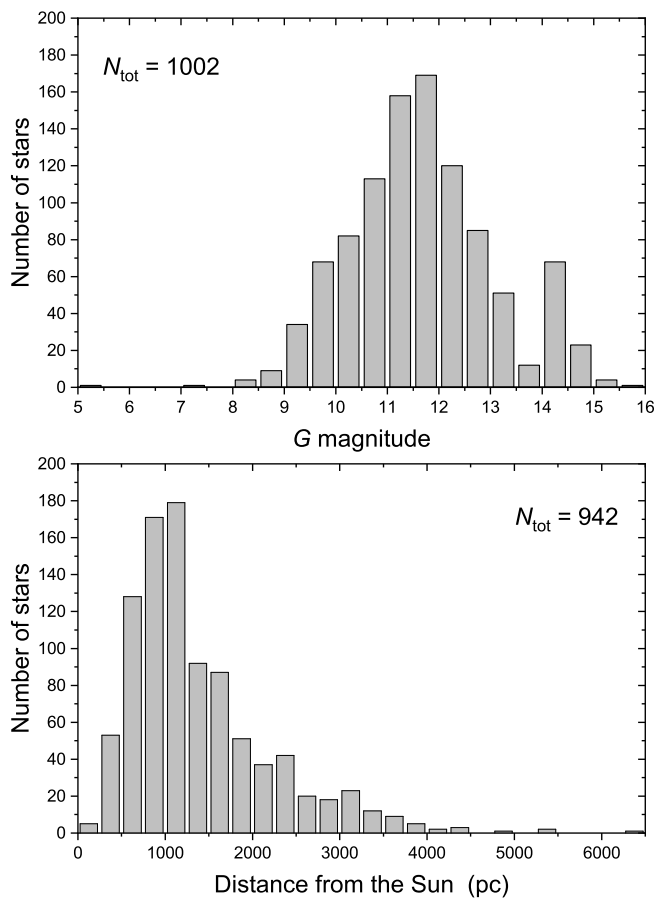}
    \caption{Histograms of the $G$ magnitudes (upper panel) and distances from the Sun (lower panel). For the construction of the latter, only stars with absolute parallax errors less than 25\,\% were employed.}
		\label{histograms}
\end{figure}

These issues also impact automatic classification routines such as the MKCLASS code and will surely be at the root of the high luminosity classification of many of our sample stars, which should be regarded with caution. It has been well confirmed that mCP stars are generally main-sequence objects (e.g. \citealt{netopil17} and references therein) and the results from the colour-magnitude diagram (CMD) (Section \ref{evolutionary_status}) fully support this finding. A detailed analysis of this issue is necessary but beyond the scope of the present investigation.

At this point, we would like to also recall that at least half of the mCP stars are spectroscopic variables, that is, the observed line strengths may vary considerably over the rotation period \citep[e.g.][]{gray09}, which should always be kept in mind when working with mCP star spectra.

Table \ref{table_master1} in the Appendix presents the MKCLASS final types along with essential data for our sample stars.

\section{Discussion} \label{discussion}

This section discusses properties of our final sample such as magnitude distribution, distances from the Sun, evolutionary status, and their distribution in space, and discusses interesting objects such as the eclipsing binary system LAMOST J034306.74+495240.7.

\subsection{Magnitudes and distances from the Sun} \label{magnitudes_and_distances}

In Figure \ref{histograms}, we present the histograms of the $G$ magnitudes and the distances from the Sun of our sample stars. The magnitude distribution peaks between 11th and 12th magnitude, which corresponds to a distance of about 1\,kpc for the investigated range of spectral types. While our sample contains only a few new mCP stars within 500\,pc around the Sun, there is a significant number of objects beyond 2\,kpc, which might help to shed more light on the Galactic radial metallicity gradient \citep{2016A&A...585A.150N} and its influence on the formation and evolution of CP stars.

In summary, our sample is a perfect extension to the mCP stars listed in the catalogue of \citet{RM09}, which are on the average closer and brighter, peaking at 9th magnitude.

\subsection{Evolutionary status} \label{evolutionary_status}

In the following sections, we investigate the evolutionary status of our sample stars in the $(BP - RP)_0$ versus $M_{\mathrm{G,0}}$ and mass versus fractional age on the main sequence spaces. We caution that, due to the imposed selection criteria (cf. Section \ref{sample_selection}), our sample is biased towards stars showing a conspicuous 5200\,\AA\ flux depression in the low-resolution LAMOST spectra. Therefore, our results, while being based on a statistically significant sample size, have to be viewed with caution and their general validity needs to be tested by a more extended sample selected via different methodological approaches.

\subsubsection{Colour-magnitude diagram} \label{evolutionary_status_CMD}

To investigate the astrophysical properties of our sample stars in a CMD, we employed the homogeneous Gaia DR2 photometry from \citet{gaia3}. Most of our sample stars are situated within the Galactic disk farther than 500\,pc from the Sun; therefore, interstellar reddening (absorption) cannot be neglected. Because hardly any objects have Str{\"o}mgren-Crawford indices available \citep{paunzen15b}, we relied on the published reddening map by \citet{Green2018}. To interpolate within this map, parallaxes were directly converted to distances. To limit the error of the absorption values to 0.1\,mag, only objects with relative parallax errors of at most 25\,\% (942 in total) were used. The transformation of the reddening values was performed using the relations:
\begin{equation}
E(B - V) = 0.76E(BP - RP) = 0.40A_G.
\end{equation}
These relations already take into account the conversion to extinction in different bands using the coefficients as listed in \citet{Green2018}.

\begin{figure}
\begin{center}
\includegraphics[width=80mm, clip]{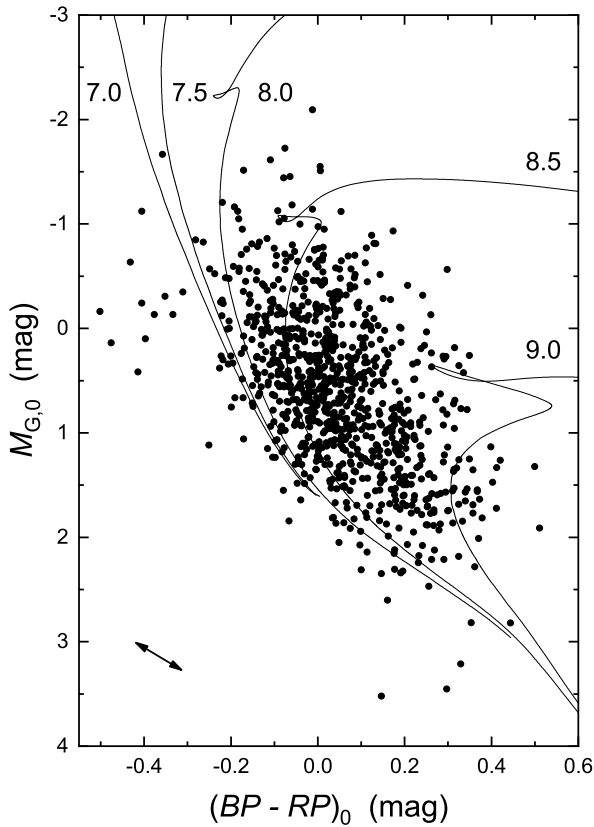}
\caption{$(BP - RP)_0$ versus $M_{\mathrm{G,0}}$ diagram of our sample stars, together with PARSEC isochrones for solar metallicity [Z]\,=\,0.020 (listed are the logarithmic ages). The arrow indicates the reddening vector for the maximum expected error due to the employed reddening map and the parallax error.}
\label{CMD} 
\end{center} 
\end{figure}

In Figure \ref{CMD}, we present the CMD of our sample stars together with PARSEC isochrones \citep{bressan12} for solar metallicity [Z]\,=\,0.020. We favour this value because it has been shown to be consistent with recent results of Helioseismology \citep{Vagnozzi2019}. Also included is the reddening vector according to an uncertainty of 0.1\,mag for $E(B-V)$. About 20 stars are situated below the zero-age main sequence (ZAMS) to such an extent that it cannot be explained by errors in the reddening estimation. Inconsistent photometry or binarity might possibly have led to the observed positions but, with the available data, we are unable to shed more light on this matter.

The stars LAMOST J061609.42+265703.2 (\#537; $M_{\mathrm{G,0}}$\,=\,+7.78\,mag, $(BP-RP)_0$\,=\,+0.840\,mag), LAMOST J064757.48+105648.2 (\#678; $M_{\mathrm{G,0}}$\,=\,+7.75\,mag, $(BP-RP)_0$\,=\,+1.721\,mag), and LAMOST J202943.73+384756.6 (\#943; $M_{\mathrm{G,0}}$\,=\,+3.61\,mag, $(BP-RP)_0$\,=\,+1.180\,mag) are not plotted in the CMD because they lie outside the chosen boundaries. The available spectra clearly confirm that they are mCP stars. We double-checked the identifications in the Gaia DR2 and LAMOST catalogues and searched for nearby objects on the sky that might have influenced the photometry, albeit with negative results. In the case of LAMOST J202943.73+384756.6, we strongly suspect that binarity might be at the root of the observed outlying position. Its spectrum has an unusual black-body curve, and its spectral energy distribution (SED) looks like the superposition of two objects, with a clearly visible infrared excess. However, we are unable to explain the reasons behind the observed inconsistent photometry for the other two stars.

\begin{figure}
\begin{center}
\includegraphics[width=\columnwidth]{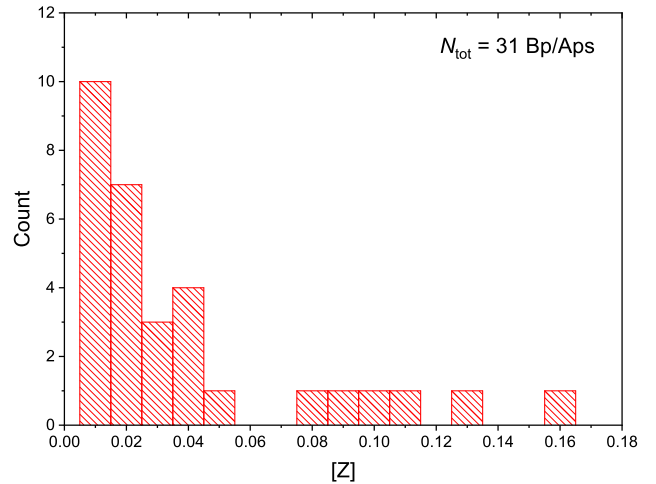}
\caption{Distribution of [Z] values for the CP2 stars from the \citet{ghazaryan18} catalogue with a least three measurements of C, N, O and S.}
\label{plot_Z}
\end{center} 
\end{figure}

From the distribution of stars in Figure \ref{CMD}, we conclude that the majority of our sample stars is between 100\,Myr and 1\,Gyr old. There are only a few very young stars and an accumulation of objects older than 300\,Myr.

The here employed isochrones have been calculated for stars of standard composition, that is, chemically normal stars, assuming solar metallicity [Z]\,=\,0.020. As the chemical bulk composition of mCP stars is unknown, however, the choice of the right chemical composition for the theoretical tracks remains an open question (cf. e.g. the discussion in \citealt{bagnulo06}). The main question is whether the apparent overabundances encountered at the surface are representative for the whole stellar interior. If diffusion is assumed as the main mechanism (in line with most theoretical studies), the overall abundance will be close to solar because corresponding underabundances are expected in the stellar interior. All current isochrone calculations are based on assuming a [Z] value for the whole star; it is currently not possible to consider different [Z] values for different layers of the stellar atmosphere. A detailed discussion of the influence of [Z] on the determination of age on the main sequence is provided in the following section.

\begin{figure}
\includegraphics[width=\columnwidth]{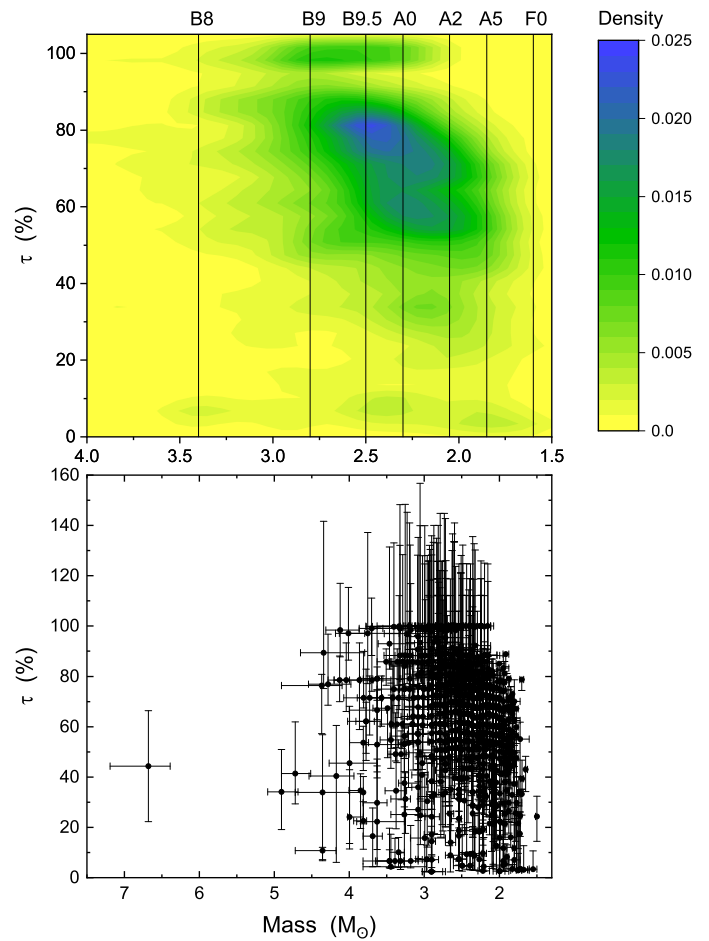}
\caption{Mass versus fractional age on the main sequence ($\tau$) distribution assuming solar metallicity [Z]\,=\,0.020 for the 903 sample stars fulfilling the imposed accuracy criteria. Upper panel shows a density plot for masses up to 4\,M$_\odot$. The position of the spectral types has been based on the information given in \citet{pecaut13}.}
\label{plot_age_mass}
\end{figure}

\begin{figure}
\includegraphics[width=\columnwidth]{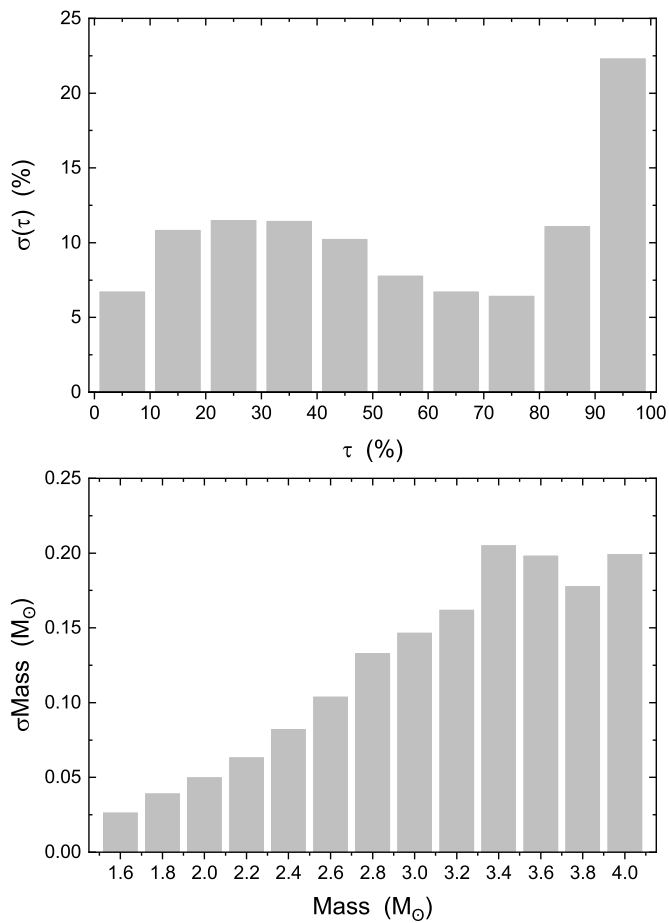}
\caption{Distributions of errors for the derived fractional ages on the main sequence ($\tau$; upper panel) and masses (lower panel) assuming solar metallicity [Z]\,=\,0.020.}
\label{plot_age_mass_errors}
\end{figure}

\subsubsection{Mass versus age on the main sequence} \label{evolutionary_status_mass_age}

To examine the evolutionary status of our sample stars in more detail, we investigated the distribution of mass ($M$) versus age on the main sequence. Age on the main sequence ($\tau$) is here defined as the fraction of life spent on the main sequence, with the ZAMS corresponding to $\tau$\,=\,0\,\% and the terminal-age main sequence (TAMS) to $\tau$\,=\,100\,\%. Only objects with absolute parallax errors less than 25\,\% were considered in this process. We again used PARSEC isochrones \citep{bressan12} for solar metallicity [Z]\,=\,0.020 and 7.0\,$<$\,log\,$t$\,$<$\,10.0 (step size: 0.025). Bearing in mind the observational errors, the chosen step size is sufficient not to run into numerical inaccuracies due to the grid being too coarse. We did not interpolate within the grid but always selected the point with minimal Euclidean distance to the observed value in the $(BP-RP){_0}$ versus $M_{\mathrm{G,0}}$ space. Only grid points representing the main sequence (flagged ``1'' in the corresponding isochrones) were used. As next step, we discarded all data points with a distance of more than 0.05\,mag between observed value and theoretical grid point, which guarantees the exclusion of points below the ZAMS and above the TAMS. In this way, masses and ages were derived for 903 sample stars fulfilling the imposed accuracy criteria.

As final step, the lifetime on the main sequence was calculated for a given mass using the upper envelope of the isochrone grid. With this parameter, the fractional lifetime of a star on the main sequence can be easily calculated. To compute upper and lower limits for $\tau$ and $M$, the full error ellipse was taken into account. This procedure is described in more detail in \citet{kochukhov06}.

Table \ref{table_age_mass} lists the derived masses and fractional ages on the main sequence for solar metallicity [Z]\,=\,0.020, which are graphically represented in Figure \ref{plot_age_mass}. The density plot in the upper panel clearly shows that there are only very few stars in our sample younger than $\tau$\,<\,20\,\%. Most stars have a relative age of $\tau$\,$\ge$\,60\,\%.

To check the reliability of our results, we have investigated the error distribution of the age and mass estimates in detail. This analysis was done for solar metallicity. We have investigated the influence of the assumed metallicity in a second step. For this, masses were binned in 0.2\,M$_\odot$ and ages in 10\,\% intervals. Sizes have been chosen to guarantee the availability of a significant number of data points in each bin. Figure \ref{plot_age_mass_errors} illustrates the corresponding histograms. The absolute errors of the masses increase linearly until $M$\,=\,3.4\,M$_\odot$ and then flatten out, which means that the relative error stays constant over the whole investigated mass range. The situation is different for the derived ages; up to $\tau$\,$\le$\,90\,\%, absolute errors remain almost constant. Relative errors obtained for younger stars, therefore, are significantly larger than for old ones. This, however, does not impact our conclusions (see below). The significant increase of the errors for the last age bin is due to the higher density of isochrones in this region; furthermore, taking the error ellipse into account, some stars may be located above the TAMS.

In order to evaluate the effect of the chosen metallicity on our results, we have investigated the [Z] distribution for CP2 stars from the \citet{ghazaryan18} catalogue with a least three measurements of C, N, O, and S. These light elements were chosen because they contribute the most to the derived [Z] values. They appear significantly underabundant in most CP2 stars, which therefore show lower [Z] values than chemically normal stars (Figure \ref{plot_Z}). For most objects, we find [Z] values in the range from about 0.008 to 0.060. From the reference source, we were not able to estimate the errors of the derived [Z] values, which mainly depend on the errors of the individual abundance determinations; these, however, are mostly not available.

We emphasise that all available isochrones use scaled abundances according to the abundance pattern of the Sun. Whether this approximation can also be applied to CP stars remains at the present time unknown (cf. Figure 1 of \citealt{ghazaryan18}). On the basis of open cluster members, \citet{bagnulo06} investigated the influence of the overall metallicity on the error of the age determination for CP2 stars, using metallicities of [Z]\,=\,0.020 (solar) and [Z]\,=\,0.008, as derived from the corresponding host clusters. As clusters with [Z]\,>\,0.020 are very rare in the Milky Way \citep{2016A&A...585A.150N}, no isochrones for metallicities exceeding solar metallicity were considered in their study.

\begin{figure}
\includegraphics[width=\columnwidth]{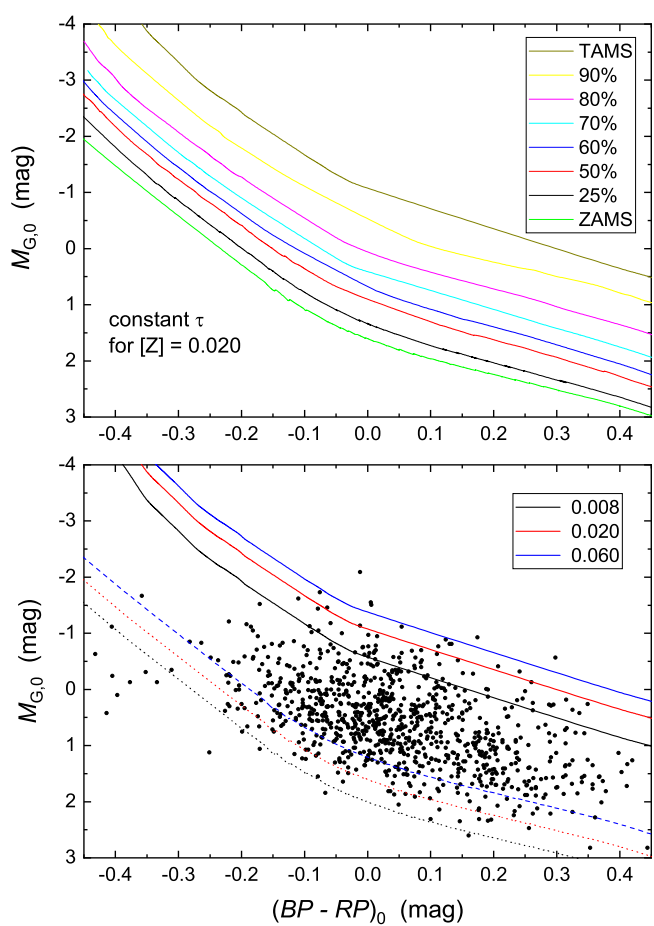}
\caption{Lines of constant fractional ages on the main sequence ($\tau$) for solar metallicity [Z]\,=\,0.020 (upper panel). Lower panel shows the positions of the ZAMS and TAMS for isochrones with [Z]\,=\,0.008, 0.020, and 0.060. Values have been chosen to cover the main range of [Z] values found for CP2 stars (Figure \ref{plot_Z}).}
\label{plots_sequences_Zs}
\end{figure}

When estimating $\tau$, we have to consider two effects, which are illustrated in Figure \ref{plots_sequences_Zs}. As is well known, lines of constant $\tau$ are not distributed equally across the main sequence (Figure \ref{plots_sequences_Zs}, upper panel). For a constant value of $(BP-RP){_0}$, they are much tighter in terms of $M_{\mathrm{G,0}}$ for the first $\sim$70\,\% of the main-sequence lifetime. The total interval of $M_{\mathrm{G,0}}$ from ZAMS to TAMS spans about 2.6\,mag, whereas the intervals covered to $\tau$\,=\,25\,\% and $\tau$\,=\,50\,\% amount to only 0.3\,mag and 0.7\,mag, respectively.

The lower panel of Figure \ref{plots_sequences_Zs} explores the impact of isochrones of different metallicity on the positions of the ZAMS and TAMS. The magnitude differences between the positions of the ZAMS are nearly constant and amount to 0.4\,mag, which corresponds to an age difference of about 30\,\% at the ZAMS. This means
that a ZAMS star of solar metallicity ([Z]\,=\,0.020) would have already spent 30\,\% of its main-sequence lifetime for [Z]\,=\,0.008 but would be situated 0.4\,mag below the ZAMS for [Z]\,=\,0.060. This illustrates the dilemma ellicited by the lack of knowledge of the overall metallicity of mCP stars and the resulting loss of accuracy, as clearly demonstrated by \citet{bagnulo06}. The uncertainty is largest for stars near the ZAMS and has to be considered together with the distribution of errors for a given distinct metallicity (Figure \ref{plot_age_mass_errors}). However, because individual [Z] values and their errors are unavailable for our sample stars, we are not able to provide reliable estimations of the contribution to the computed errors.

\begin{figure}
\includegraphics[width=\columnwidth]{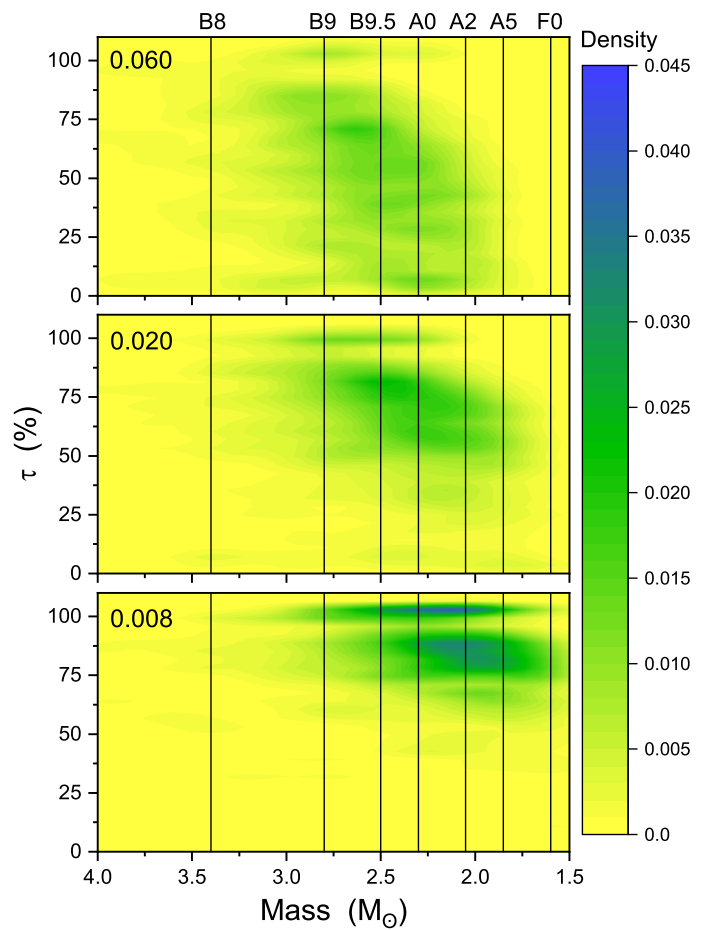}
\caption{Mass versus fractional age on the main sequence ($\tau$) distributions for isochrones with [Z]\,=\,0.008, 0.020, and 0.060, illustrating the differences in the derived mass and age distributions.}
\label{plot_different_metallicities}
\end{figure}

Figure \ref{plot_different_metallicities} illustrates the mass versus $\tau$ distributions for all three investigated isochrones. The majority of our sample stars is situated in the rather narrow spectral range from B8 to A0 (cf. Section \ref{peculiarity_types}). However, there is also a lesser but significant amount of stars with spectral types between A5 and F0. Any calibrated mass distribution should represent these results to some extent. The lower-metallicity isochrone ([Z]\,=\,0.008) yields mainly old ($\tau$\,$>$75\%) and cooler (later than spectral type A0) stars; no young stars are present. On the other hand, adopting the isochrone for [Z]\,=\,0.060 results in a quite homogeneous age and mass distribution. However, there is a lack of stars cooler than A5, in conflict with the observations. Overall, as expected, none of the employed isochrones is suitable to reproduce the observed distribution of spectral types. Nevertheless, assuming solar metallicity offers the best compromise, with most stars situated in the late B-type realm and a tail of objects extending down to spectral type F0.

To further tackle this important problem, a modern and detailed abundance analysis of the light elements is needed. The current available data are rare and mainly based
on the assumption of local thermodynamical equilibrium \citep{1990ApJS...73...67R}. For the relevant spectral type domain, almost all suitable spectral lines (i.e. lines of sufficient strength) are situated in the spectral region redwards of 6000\,\AA. Unfortunately, the medium-resolution spectra of the LAMOST survey, which should be sufficient in terms of resolution, do not cover a significant amount of the specified spectral region \citep{zhang20}.

Finally, diffusion calculations for light elements are needed to estimate the influence of the magnetic field and to what extent the observed surface abundances are representative of the stellar composition. Until now, however, because of the lack of corresponding observations, such calculations are not available \citep{2012MNRAS.425.2715S}. Therefore, in the following, we have adopted the results for solar metallicity ([Z]\,=\,0.020) as best approximation. Assuming isochrones for lower [Z] values would lead to the derivation of older fractional ages (cf. Figure \ref{plot_different_metallicities}).

\citet{hubrig00} put forth the hypotheses that mCP stars with masses $M$\,$<$\,3\,M$_\odot$ are concentrated towards the centre of the main-sequence band and that magnetic fields only appear in stars that have completed about 30\,\% of their lifetime on the main sequence. In their investigation of the evolutionary status of mCP stars, \citet{kochukhov06} demonstrated that mCP stars with $M$\,$>$\,3\,$M_\odot$ are distributed homogeneously among the main sequence. They further identified 22 young ($\tau$\,$<$\,30\,\%) mCP stars among their sample stars with $M$\,$\le$\,3\,$M_\odot$, thereby rejecting the proposal of \citet{hubrig00} that all observably magnetic low-mass CP stars have completed a significant fraction of their main-sequence lifetime. That very young (ZAMS to 25\% on the main sequence) mCP stars do exist has been unambiguously demonstrated by several studies on the basis of members of open clusters (cf. e.g. \citealt{bagnulo03}, \citealt{poehnl03}, \citealt{landstreet07}, and \citealt{landstreet08}).

Nevertheless, \citet{kochukhov06} also find an uneven distribution for mCP stars with masses of $M$\,$<$\,3\,M$_\odot$, in particular for stars with $M$\,$\le$\,2\,$M_\odot$, which tend to cluster in the centre of the main-sequence band. Confirming their previous results, \citet{Hubrig2007} again found that mCP stars with $M$\,$<$\,3\,M$_\odot$ are concentrated towards the centre of the main-sequence band.

\begin{figure}
\includegraphics[width=\columnwidth]{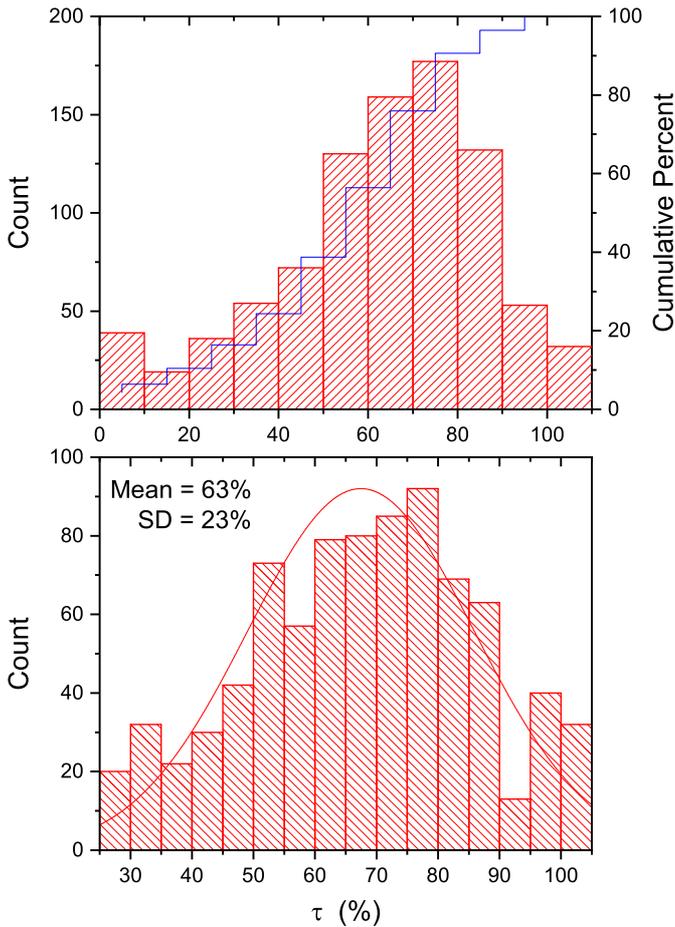}
\caption{Distribution of fractional ages on the main sequence ($\tau$) among the 903 sample stars fulfilling our accuracy criteria.}
\label{plots_age}
\end{figure}

As is obvious from Figure \ref{plot_age_mass}, most of our sample stars are situated in the 2\,$\le$\,$M_\odot$\,$\le$3 bin. In agreement with the results of the aformentioned studies, we also find an uneven distribution of the fractional lifetime; however, our sample stars boast a mean fractional age of $\tau$\,=\,63\,\% (standard deviation of 23\,\%; cf. Figure \ref{plots_age}). Young mCP stars, while undoubtedly present, are conspicuously underrepresented in our sample.

In summary, our results strongly suggest an inhomogeneous age distribution among low-mass ($M$\,$<$\,3\,M$_\odot$) mCP stars as hinted at by previous studies. However, we stress that our sample is biased towards mCP stars showing a conspicuous 5200\,\AA\ flux depression in the low-resolution LAMOST spectra and has not been selected on the basis of an unbiased, direct detection of a magnetic field. Therefore, our results have to be viewed with caution and their general validity needs to be tested by a more extended sample selected via different methodological approaches. It remains to be sorted out in what way the occurrence of the 5200\,\AA\ depression is connected with this result, in particular why this feature is apparently much more prominent in older stars. Several studies have shown that the 5200\,\AA\ depression increases with magnetic field strength and the atmospheric metal content (e.g. \citealt{2005A&A...433..671K}; \citealt{2006A&A...448.1153K}).

Our analysis has been based on a statistically significant sample of mCP stars. Furthermore, due to the applied methods, it is not impacted by potential error sources that have been proposed to influence the results of former studies, such as a displacement of stars from the ZAMS by the application of negative Lutz-Kelker corrections or incorrect ${T}_{\rm eff}$ values caused by the anomalous flux distributions of mCP stars (cf. e.g. the discussions in \citealt{kochukhov06} and \citealt{netopil08}). Even if the here derived error margins had been significantly underestimated, which we see no reason to believe, the general conclusion would hold. However, we caution that individual [Z] values and their errors are not available for our sample stars and that the influence of [Z] on the derived fractional ages is large.

\subsection{Space distribution} \label{space_distribution}

\begin{figure}
        \includegraphics[width=\columnwidth]{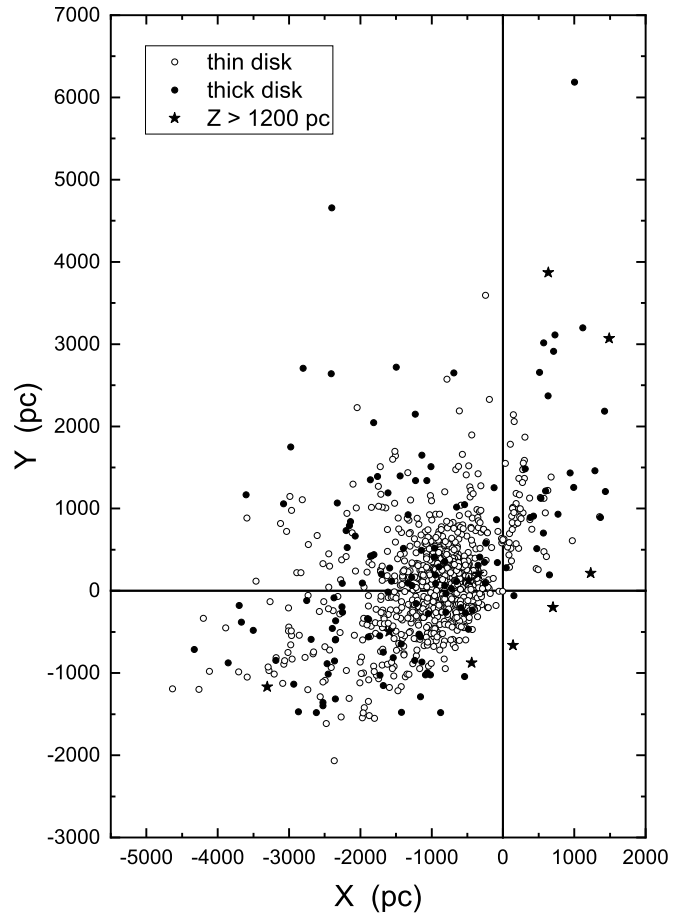}
    \caption{Distribution of the 942 stars with absolute parallax errors less than 25\%	in the [XY] plane. Stars were divided in probable members of the thin and thick disk according to the scale heights given in \citet{2001MNRAS.322..426O,2017MNRAS.470.2113A}. Ten stars have Z values larger than 1200\,pc and might be Halo objects.}
		\label{map_3D}
\end{figure}

\begin{table*}
\caption{Kinematic and astrometric data for the ten stars of our sample with a height larger than 1200\,pc above the Galactic plane. The columns denote: (1) Internal identification number. (2) LAMOST ID. (3) MKCLASS final type. (4) X-coordinate towards the Galactic centre. (5) Y-coordinate in direction of Galactic rotation. (6) Z-coordinate towards the north Galactic pole. (7) Radial velocity ($R_{\mathrm{V}}$) taken from LAMOST DR4. (8) Standard deviation of $R_{\mathrm{V}}$. (9) Total spatial velocity ($v_{\mathrm{tot}}$). (10) Standard deviation of $v_{\mathrm{tot}}$.}
\label{space_3D}
\begin{center}
\begin{adjustbox}{max width=\textwidth}
\begin{tabular}{llllllllll}
\hline
\hline
(1) & (2) & (3) & (4) & (5) & (6) & (7) & (8) & (9) & (10) \\
No.	&	LAMOST\_ID & SpT\_final	&	X (pc) &	Y (pc)	&	Z (pc)	&	$R_{\mathrm{V}}$ (km\,s$^{-1}$)	&	$\sigma R_{\mathrm{V}}$ (km\,s$^{-1}$) & 
$v_{\mathrm{tot}}$ (km\,s$^{-1}$) & $\sigma v_{\mathrm{tot}}$ (km\,s$^{-1}$) \\
\hline
816	&	J073950.01+201812.7 & B9 IV$-$V SrCrEu &	$-$3305	&	$-$1170	&	+1239	&	+71	&	4	&	74	&	7	\\
859	&	J091053.70+285032.7 & A9 V SrCrEu  &	$-$1592	&	$-$496	&	+1487	&	$-$24	&	9	&	44	&	11	\\
870	&	J104323.95+045214.3 & A1 IV$-$V CrEu &	$-$440	&	$-$875	&	+1265	&	+23	&	3	&	30	&	4	\\
873	&	J114130.23+403822.7 & B9 IV$-$V CrEuSi &	$-$480	&	+125	&	+1380	&	$-$43	&	5	&	98	&	16	\\
875	&	J122139.23+383309.5 & kA2hA4mA7 bl4077 bl4130 &	$-$346	&	+202	&	+1732	&	$-$42	&	1	&	133	&	10	\\
876	&	J122746.05+113635.3 & B8 IV Si (He-wk) &	+142	&	$-$665	&	+2298	&	+182	&	5	&	292	&	39	\\
879	&	J140422.54+044357.9 & kA4hA7mF0 SrCrEu & +703	&	$-$202	&	+1355	&	$-$55	&	15	&	200	&	23	\\
880	&	J150331.87+093125.4 & A8 V SrCrEu &	+1233	&	+213	&	+1736	&	$-$38	&	5	&	312	&	33	\\
881	&	J155549.85+401144.4 & B8 IV Si &	+1491	&	+3070	&	+4063	&	+90	&	5	&	96	&	10	\\
895	&	J181156.38+523411.4 & B7 IV$-$V Si &	+635	&	+3869	&	+2002	&	$-$6	&	5	&	147	&	31	\\
\hline
\end{tabular}
\end{adjustbox}
\end{center}
\end{table*}

To investigate the location of our sample stars in the Galactic [XYZ] plane, the corresponding coordinates were obtained from a conversion of spherical Galactic coordinates (latitude and longitude) to Cartesian coordinates using the distance $d$ from \citet{2018AJ....156...58B}. In this work, the positive X-axis points towards the Galactic centre, the Y-axis is positive in the direction of Galactic rotation and the positive Z-axis points towards the north Galactic pole. Only objects with absolute parallax errors less than 25\,\% were considered in this process. 942 stars satisfied this criterion.

We divided our sample in candidate members of the thin disk (scale height of 350\,pc) and the thick disk (1200\,pc). The scale heights were taken from \citet{2001MNRAS.322..426O} and \citet{2017MNRAS.470.2113A}. Our results are shown in Figure \ref{map_3D}. From the sample of 942 stars, 797 objects likely belong to the thin disk and 135 objects to the thick disk. The remaining ten stars qualify for being members of the halo population and are thus worth a closer look.

As a first step, we checked the spectra of the halo star candidates and confirmed that all objects exhibit the typical spectral features of mCP stars and a clearly visible flux depression. For the calculation of the total spatial velocity $v_{\mathrm{tot}}$, the radial velocity (RV) is needed. Data from the LAMOST survey include automatically measured RV information of the spectra \citep{anguiano18}. We checked the reliability of these values for mCP stars, that is, the spectral range from late B- to early F-type objects, by searching for common entries with the RV catalogue of \citet{2007AN....328..889K}. In total, 11 stars were found that are common to both our sample and this catalogue. Some of these objects boast more than one spectrum in LAMOST DR4; in these cases, mean RV values were calculated. Comparing the RV values from both sources, we find a mean difference of +2.4\,km\,s$^{-1}$, which lends confidence that the LAMOST RVs are useful in a statistical sense. However, we caution that an external uncertainty we cannot account for is introduced by the spot-induced RV variations of mCP stars that can reach up to $\pm$50\,km\,s$^{-1}$ \citep{1999A&A...351..283P}.

The space velocities were calculated following the formulae of \citet{1987AJ.....93..864J}. The final values are listed in Table \ref{space_3D}. Stars of the halo population show $v_{\mathrm{tot}}$\,>\,180\,km\,s$^{-1}$ as compared to the local standard of rest \citep{2004AJ....128.1177V}. We therefore conclude that the stars LAMOST J122746.05+113635.3 (\#876; Gaia DR2 3907547639444408064) and LAMOST J150331.87+093125.4 (\#880; Gaia DR2 1167894108493926016) are kinematically true halo objects, which is of considerable interest as no halo CP2 stars have been discovered so far. Considering the error, the star LAMOST J140422.54+044357.9 (\#879) does not satisfy this criterion. 

\begin{figure}
        \includegraphics[width=\columnwidth]{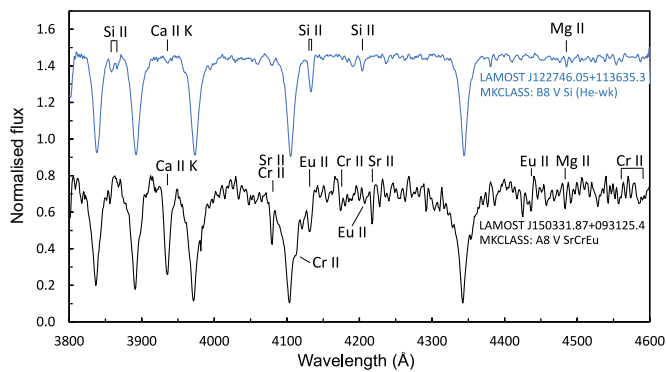}
    \caption{Blue-violet spectra of the proposed halo stars LAMOST\,J122746.05+113635.3 (\#876; MKCLASS final type B8 IV Si (He-wk); upper spectrum) and LAMOST\,J150331.87+093125.4 (\#880; MKCLASS final type A8 V SrCrEu; lower spectrum). Some prominent lines of interest are indicated.}
    \label{comparison_halo}
\end{figure}

\citet{1996ApJS..103..433B} identified LAMOST J122746.05+113635.3 as candidate field horizontal-branch star. Its spectrum, however, is that of a Si CP2 star (Figure \ref{comparison_halo}). There are very strong \ion{Si}{ii} lines at 3856/62\,\AA, 4128/31\,\AA, 4200\,\AA, 5041/56\,\AA, and 6347/71\,\AA. In addition, the \ion{He}{i} lines are weak, which is commonly observed in CP2 stars. It has consequently been classified as B8 IV Si (He-wk) by the MKCLASS\_mCP code. Almost all blue horizontal-branch stars, on the other hand, are metal-weak and their spectra rather resemble that of $\lambda$ Bootis stars \citep{gray09}. We feel therefore safe in rejecting the proposed horizontal-branch classification.

In summary, according to the available evidence, LAMOST J122746.05+113635.3 and LAMOST J150331.87+093125.4 are bona-fide CP2 stars whose distances and kinematical properties are in agreement with halo stars. If confirmed, they would be the first CP2 halo objects known and therefore of great interest.

LAMOST J155549.85+401144.4 (\#881; Gaia DR2 1382933122321062912) is another interesting object because it is listed in the catalogue of hot subdwarfs by \citet{2017A&A...600A..50G}. The location in the $(BP - RP)_0$ versus $M_{\mathrm{G,0}}$ diagram (Figure \ref{CMD}), however, does not support this classification. The same is true for the spectrum, which is that of a classical Si CP2 star (MKCLASS final type: B8 IV Si). We note that the occurrence of abundance anomalies in hot subdwarfs has been well established; for example, \citet{2018MNRAS.473.4021W} identified two hot subdwarfs with effective temperatures of about 37\,000\,K and enrichments of 1.5 to 3\,dex in heavy metals. This, however, is very different from what we see in LAMOST J155549.85+401144.4, which is significantly cooler than that ($\sim$14\,000\,K) and shows the abundance pattern of a CP2 star. The current evidence, therefore, points to it being no subdwarf but a Si CP2 star.

Although the LAMOST survey is avoiding dense regions such as star clusters, we searched for possible cluster members among our sample stars. To this end, the positions, diameter, proper motions, distances and their errors of star clusters from \citet{2013A&A...558A..53K} and \citet{2018A&A...618A..93C} were employed and we searched for matches within 3\,$\sigma$ of these parameters. In total, seven matches in six open clusters were found, which are listed in Table \ref{OCs}. Judging from a comparison of the ages derived from Figure \ref{CMD} and the cluster ages, all objects seem to be true cluster members.

\begin{table}
\caption{Open cluster members among our sample stars. The columns denote: (1) Internal identification number. (2) LAMOST ID. (3) Open cluster.}
\label{OCs}
\begin{center}
\begin{tabular}{lll}
\hline
\hline
(1) & (2) & (3) \\
No.	&	LAMOST\_ID & Open cluster	\\
\hline
203 & J041641.15+511253.2 & NGC 1528 \\
269 & J045933.63+395715.8 & Alessi 2 \\
317 & J052059.29+351123.5 & Gulliver 8 \\
497 & J060815.12+045107.6 & NGC 2168 \\
502 & J060827.84+204832.0 & NGC 2168 \\
675 & J064741.02+072458.8 & Collinder 115 \\
868 & J095855.77-044413.8 & Collinder 359 \\
\hline
\end{tabular}
\end{center}
\end{table}

\subsection{Peculiarity type distribution} \label{peculiarity_types}

\begin{figure*}
        \includegraphics[width=\textwidth]{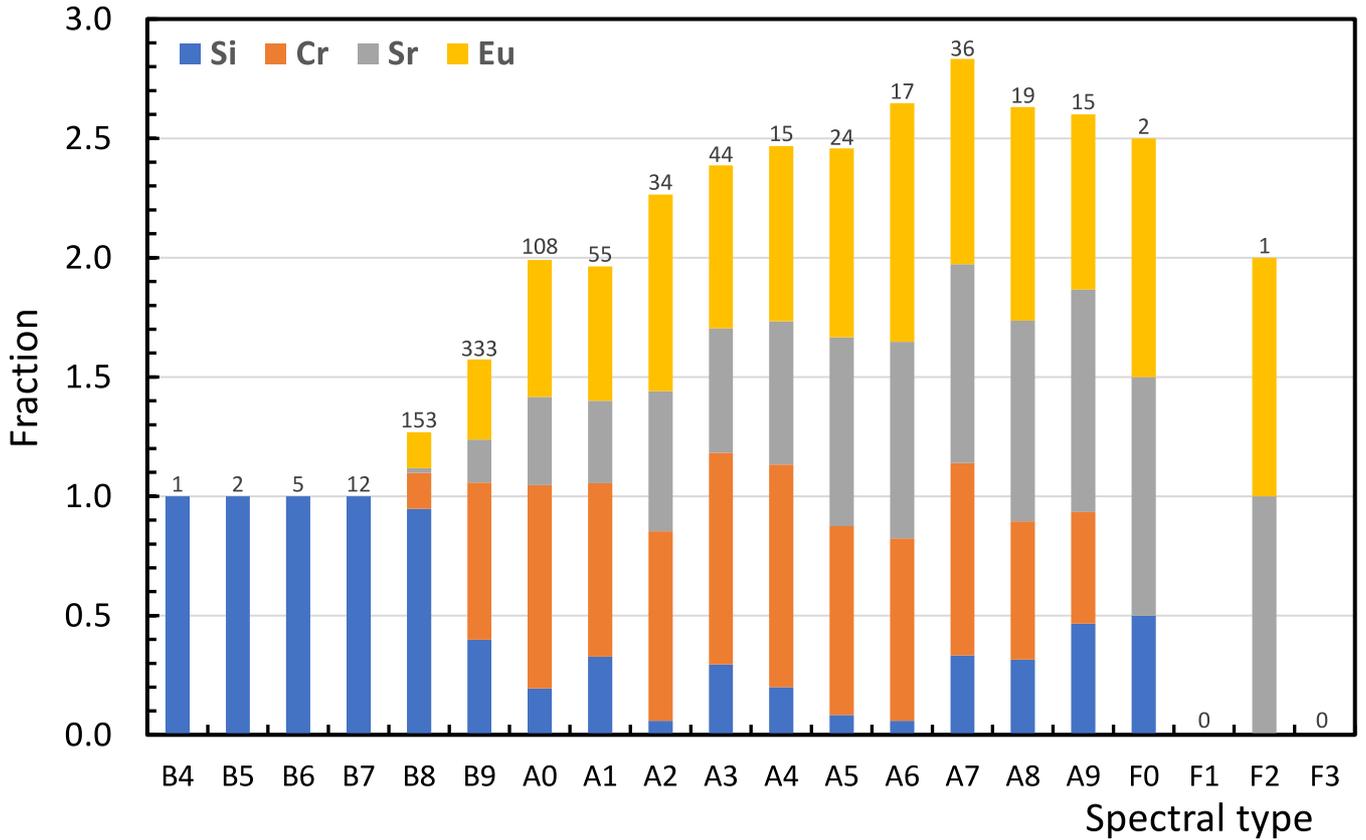}
    \caption{Fractional distribution of chemical peculiarities versus hydrogen-line spectral type for the 876 stars with unambiguous peculiarity type identifications. The numbers above the bars indicate the number of objects in the corresponding spectral type bin. Because a single object may have multiple peculiarities, fractions may exceed 1.}
    \label{peculiarity_type_distribution}
\end{figure*}

Figure \ref{peculiarity_type_distribution} explores the distribution of Si, Cr, Sr, and Eu peculiarities versus hydrogen-line spectral type for the 876 stars of our sample with unambiguous identifications (i.e. without the stars in which only strong 4077\,\AA\ and/or 4130\,\AA\ blends or no traditional peculiarities were identified). Stars with hydrogen-line types of B9.5 are not listed separately but included under spectral type B9. The number of stars per spectral type bin varies considerably, with the B8$-$A0 stars forming the vast majority of our sample. Nevertheless, some tentative trends can be identified, although the interpretation towards the low temperature end is severely hampered by the small number statistics for objects later than A9:

\begin{itemize}
	\item Si peculiarities are present between spectral types B4 and F0. They play a dominant role in stars with spectral types earlier than B9, strongly decreasing in importance in later-type objects. Except for He peculiarites (which are not shown in the plot), Si peculiarities are the only chemical peculiarities identified in objects earlier than B8.
	\item Cr peculiarities set in at spectral type B8 and form an important part of the peculiarity mix between spectral types B9 and A9.
	\item Sr and Eu peculiarities set in at spectral type B8 and increase in strength towards later types.
\end{itemize}

This is in good agreement with the expectations and the literature. It is well known that Si peculiarities are present throughout a wide range of effective temperatures in mCP stars \citep[e.g.][]{RM09}. The hottest stars with Eu peculiarities from the \citet{RM09} catalogue are of spectral type B8, which holds true also for the vast majority of stars with Cr and Sr peculiarities, with the exception of only three objects (HD 35502, spectral type B6 Sr Cr Si; HD 167288, spectral type B7 Si Cr; HD 213918, spectral type B7 Si Sr). Likewise, the work of \citet{ghazaryan18} contains atomic data for Eu, Cr, and Sr from effective temperatures of, respectively, 12\,900\,K ($\sim$B8), 14\,700\,K ($\sim$B6), and 13\,300\,K ($\sim$B8) downwards. The good agreement of the peculiarity type 'blue borders' between the present work and the literature provides independent proof of the reliability of the here derived spectral types.

\begin{figure}
        \includegraphics[width=0.5\textwidth]{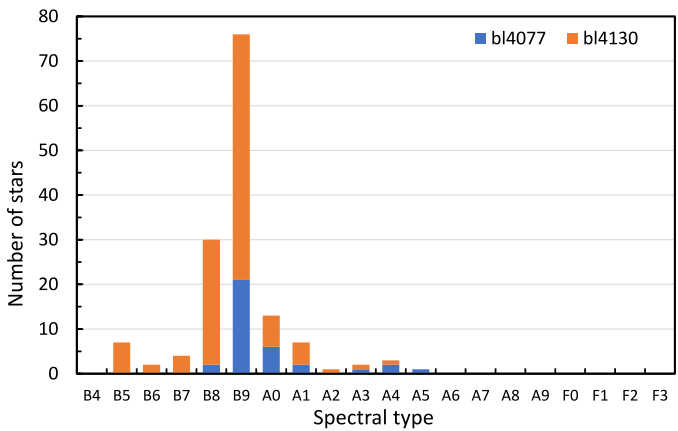}
    \caption{Distribution of stars in which only strong blends at 4077\,\AA\ and/or 4130\,\AA\ were identified.}
    \label{bl4077bl4130}
\end{figure}

Figure \ref{bl4077bl4130} illustrates the distribution of stars in which only strong blends at 4077\,\AA\ and/or 4130\,\AA\ were identified. Again, stars with hydrogen-line types of B9.5 are included under spectral type B9. These stars were not assigned Si, Cr, Sr, and Eu types with the here employed workflow because, apart from the strong blends, the peculiarities are either too subtle to have passed our significance criteria, no other significant features are present or the code failed to identify them for some reason (cf. Section \ref{spectral_classification_MKCLASS}).

Manual classification is necessary to throw more light on what elements contribute to the observed blends. Nevertheless, the distribution of the 4130\,\AA\ blend identifications, in particular for objects earlier than B9, is in general agreement with the distribution of Si peculiarities. We therefore expect that most of the 'bl4130' stars in our sample will turn out to be Si stars. No similar predictions can be made for the 'bl4077' stars from the available data.

\subsection{Comparison with samples from the literature} \label{discussion_literature}

The following sections compare our results with the works of \citet{RM09}, \citet{Skiff}, and \citet{qin19}. We further note that 22 of our sample stars are contained in the sample of strongly magnetic Ap stars of \citet{Scholz2019}. As the authors do not list spectral types, a direct comparison of results was not possible. The stars common to both samples are identified in Table \ref{table_master1}.

\subsubsection{Comparison with the compilations of \citet{RM09} and \citet{Skiff}} \label{discussion_RM09}

Our final sample contains 59 mCP stars or candidates that are also included in the catalogue of \citet{RM09}. This low level of coincidence (6.65\,\%) is expected because the \citet{RM09} sample mostly consists of bright stars (peaking at around 9th magnitude) for which there are no LAMOST spectra available.

Table \ref{table_comparison_RM09} gives a comparison of the spectral types from the present study, the RM09 catalogue, and the compilation of \citet{Skiff}. For most stars, there is a good general agreement between the different sources. For example, for the 46 stars that have at least one other detailed literature classification listing the temperature subtype, the determined hydrogen-line types agree within $\pm$2\,subclasses, which seems reasonable considering the inhomogeneous source material behind the literature classifications and the inherent difficulties in classifying mCP stars. For several stars, our results provide a first detailed classification; furthermore, we confirm several doubtful objects as mCP stars and show that some suspected CP1 stars are in fact mCP stars. A more detailed investigation into this matter will be the topic of an upcoming study that will be concerned with a new classification of stars in the RM09 catalogue based on homogeneous spectroscopic material.

\subsubsection{Comparison with the sample of \citet{qin19}} \label{discussion_Qin}

\citet{qin19} searched for CP1 stars in low resolution spectra of early-type stars from LAMOST DR5 and compiled a catalogue of 9372 CP1 stars. Because cooler CP2 stars may exhibit similar spectral features (Ca deficiency, overabundance of Fe-group elements), the authors expect a contamination of their sample by these objects. To identify potential CP2 stars among the CP1 star candidates, they used the 4077\,\AA\ blend as reference line, which may contain contributions from \ion{Si}{ii}, \ion{Cr}{ii}, and \ion{Sr}{ii}. Synthetic spectra with overabundances of Sr, Cr, Eu, and Si of 2.0 dex were computed for different effective temperatures and the equivalent widths of the 4077\,\AA\ blend were calculated and compared for both the templates and the observed spectra. If the equivalent width of the observed 4077\,\AA\ feature ($EW$\textsubscript{4077\_obs}) exceeded that of the corresponding templates ($EW$\textsubscript{4077\_temp}), a star was flagged as a CP2 star candidate. In this way, \citet{qin19} flagged 1131 stars within their sample of CP1 star candidates as CP2 star candidates. From a cursory investigation of about 20 randomly chosen objects, several bona-fide CP2 stars have indeed been found among the objects with high values of ($EW$\textsubscript{4077\_obs}\,$-$\,$EW$\textsubscript{4077\_temp}), in line with the expectations of \citet{qin19}. However, the incidence of CP2 stars seems to drop rather sharply towards lower values of ($EW$\textsubscript{4077\_obs}\,$-$\,$EW$\textsubscript{4077\_temp}). We assume that this is because a strong 4077\,\AA\ feature alone, while often helpful, is an insufficient criterion for identifying CP2 stars.

Only 45 objects are common to both our sample and the \citet{qin19} catalogue. Of these stars, about 70\,\% (31 objects) were flagged as CP2 star candidates. Table \ref{table_comparison_Qin} compares the \citet{qin19} k/h/m spectral types to the final types derived in the present study for all objects common to both lists. In general, the agreement between the derived spectral types is poor. With the here employed methodology, different k/h/m-types were assigned to only 21 of these stars. A detailed investigation of the stars common to both lists, in particular a comparison of the automatically derived classifications to manually derived spectral types and an investigation of the source(s) of the observed discrepancies, is beyond the scope of the present paper but might be beneficial to both studies and help with the refinement of the employed algorithms.

On first glance, the low level of coincidence between the \citet{qin19} catalogue and the here presented sample of mCP seems surprising. We have identified several reasons for this, which are related to the different approaches and goals of both studies.

\citet{qin19} explicitly searched for CP1 stars. CP2 stars were identified as a contaminant and corresponding candidates were subsequently identified. Because they searched for CP2 star candidates within their sample of CP1 star candidates, that is, among objects with pronounced differences between k and h spectral types, their sample will not contain any CP2 stars that do not share these characteristics. However, early-type CP2 stars generally do not show significant (if any) differences between k and h types; this phenomenon is mostly restricted to late-type CP2 stars. Thus, we assume that the \citet{qin19} subsample of CP2 star candidates is biased towards late-type CP2 stars.\footnote{Although not statistically significant, it is interesting to note that all stars from the \citet{qin19} candidate sample we were able to confirm as CP2 stars were indeed late-type CP2 stars.} This is further corroborated by the fact that, in agreement with the expectations for identifying CP1 stars, \citet{qin19} constrained their search to objects with 6500\,K\,<\,${T}_{\rm eff}$\,<\,11\,000\,K. In fact, their final catalogue contains only 11 objects with ${T}_{\rm eff}$\,>\,10\,000\,K. Thus, only very few stars hotter than spectral type A0 are present in their sample.

Our study follows a very different approach and is concerned with the identification of mCP stars that were selected among early-type targets by the presence of a significant 5200\,\AA\ flux depression. The spectra of CP1 stars, on the other hand, generally do not show this feature \citep{2005A&A...441..631P}. However, not all mCP stars distinctly show a 5200\,\AA\ depression either, in particular in low-resolution spectra. Therefore, we will have missed any such mCP stars, which might have found their way into the \citet{qin19} candidate sample. Most important, however, is that the majority of our sample stars is situated in the spectral range of B8 to A0 (9700\,K\,<\,${T}_{\rm eff}$\,<\,12\,500\,K), which renders them mostly incompatible with the candidate sample of \citet{qin19}.

In summary, the above mentioned issues, combined with the fact that the incidence of CP2 stars seems to drop rather sharply towards lower values of ($EW$\textsubscript{4077\_obs}\,$-$\,$EW$\textsubscript{4077\_temp}) in the CP2 star candidate subsample of \citet{qin19} and the different source material ($\sim$9.0 million spectra in DR5 vs. $\sim$7.6 million spectra in DR4), illustrate that no significant overlap between both samples is to be expected. We would like to again stress that it never was the intention of \citet{qin19} to collect a pure sample of CP2 stars. Their subsample of candidates, however, is a valuable starting point for further investigations. As a spin-off of the present study, we intend to investigate the \citet{qin19} CP2 star candidates to confirm or reject their status as mCP stars. It is clear that investigations based on a broader analysis of LAMOST early-type spectra (i.e. also including stars without conspicuous 5200\,\AA\ flux depressions) will lead to the discovery of many more mCP stars in the future.

\begin{table*}
\caption{Comparison of the spectral types derived in this study, the RM09 catalogue, and the compilation of \citet{Skiff}. The columns denote: (1) Internal identification number. (2) Identification number from the RM09 catalogue. (3) LAMOST identifier. (4) Spectral type, as derived in this study. (5) Spectral type from the RM09 catalogue. An asterisk (*) or a question mark (?) in parantheses behind the spectral type denote, respectively, well-known CP2 stars and CP2 stars of doubtful nature. (6) Spectral type from the compilation of \citet{Skiff}. (7) Corresponding reference from the compilation by \citet{Skiff}.}
\label{table_comparison_RM09}
\begin{center}
\begin{adjustbox}{max width=0.9\textwidth}
\begin{tabular}{cclllll}
\hline
\hline
(1) & (2) & (3) & (4) & (5) & (6) & (7) \\
\textbf{No.}	&	\textbf{ID\_RM09}	&	\textbf{ID\_LAMOST}	&	\textbf{SpT\_final}	&	\textbf{SpT\_RM09}	&	SpT\_Skiff	&	Ref\_Skiff	\\
\hline
14	&	1236	&	J004947.16+525208.2	&	B9 V CrEuSi  (He-wk) 	&	B9 Si	&	ApSi	&	\citet{1998PASP..110..270B}	\\
19	&	1543	&	J010007.84+045339.3	&	kB8hA3mA6 Cr   	&	A0 Si Cr (?)	&	A0/2	&	\citet{hou15}	\\
22	&	1680	&	J010651.35+154426.9	&	kA1hA7mA7 SrCrEu   	&	A3	&	A3p	&	\citet{1955ApJ...121...32N}	\\
33	&	2210	&	J013055.87+452155.7	&	kA1hA3mA4 SrCr   	&	A0 Sr	&	A0pSrCr?	&	\citet{1975AAS...21...25F}	\\
38	&	2750	&	J015059.58+540259.1	&	B8 IV  bl4130  	&	B8 Si	&	B8IIIpSi	&	\citet{1999AAS..137..451G}	\\
61	&	3660	&	J022120.02+280415.6	&	A8 V SrEuSi   	&	A2 Sr Eu	&	ApSrEu	&	\citet{1983AJ.....88.1182B}	\\
77	&	4580	&	J025945.31+541941.5	&	kB7hA7mA6 CrEuSi   	&	A2 Si Cr Eu	&	B9pSi	&	\citet{1999AAS..137..451G}	\\
80	&	4740	&	J030350.21+463718.3	&	A8 V SrCrEuSi   	&	Sr Eu	&	ApSrEu Sr vstrong	&	\citet{1985AJ.....90..341B}	\\
83	&	4820	&	J030633.66+025615.7	&	B9.5 IV$-$V CrEu   	&	A2 Cr Eu (?)	&	A2pCrEu Sr 4077A weak	&	\citet{1973AJ.....78...47D}	\\
130	&	5750	&	J034000.57+444858.4	&	kA0hA1mA3 (Si)    	&	A0	&	n/a	&	n/a	\\
132	&	5800	&	J034112.38+453031.7	&	A1 V SrCrEu   	&	Sr Eu	&	ApSrEu	&	\citet{1985AJ.....90..341B}	\\
134	&	5850	&	J034229.41+353820.9	&	A0 IV$-$V bl4077 bl4130  	&	A1 Sr	&	A1pSr	&	\citet{1977AJ.....82..598G}	\\
138	&	5860	&	J034417.13+494336.6	&	B8 IV$-$V CrSi   	&	A0 Cr Eu (?)	&	ApCrEu:	&	\citet{1985AJ.....90..341B}	\\
180	&	6530	&	J040642.34+454640.8	&	kB9hA2mA6 CrSi   	&	A0 Si	&	ApSi	&	\citet{1985AJ.....90..341B}	\\
228	&	7210	&	J042736.18+063643.1	&	A2 IV$-$V SrCrEu   	&	A0 Sr Cr Eu	&	ApSrCrEu	&	\citet{1973AJ.....78..687B}	\\
241	&	7740	&	J044407.32-005639.0	&	F0 V SrEuSi   	&	F0 Sr Eu	&	FpSrEu	&	\citet{1973AJ.....78..687B}	\\
252	&	7960	&	J045121.11+093555.8	&	A0 V SrCrEuSi   	&	A0 Sr Cr Eu	&	ApSrCrEu	&	\citet{1973AJ.....78..687B}	\\
275	&	8140	&	J050210.72+464600.0	&	B8 III$-$IV Si   	&	Si	&	ApSi	&	\citet{1985AJ.....90..341B}	\\
318	&	8872	&	J052118.97+320805.7	&	B4 Vpn   	&	B8 Si Sr	&	B9	&	\citet{1931AnHar.100...61C}	\\
325	&	8938	&	J052259.54+343944.9	&	B9.5 V Cr   	&	B9 Si Cr Sr	&	A0/2	&	\citet{hou15}	\\
334	&	9082	&	J052616.48+331544.2	&	B8 IV$-$V  bl4130  	&	A0 Si Sr	&	A2	&	\citet{1931AnHar.100...61C}	\\
343	&	9200	&	J052812.16+415006.4	&	kA0hA3mA7 Si   	&	B5 Si	&	ApSi	&	\citet{1983AJ.....88.1182B}	\\
351	&	9295	&	J053025.32+332639.6	&	A6 V SrEu   	&	A1-	&	A7	&	\citet{1931AnHar.100...61C}	\\
360	&	9350	&	J053239.91+434307.5	&	B8 IV Si   	&	B9 Si	&	B8pSi	&	\citet{1975AAS...21...25F}	\\
371	&	9740	&	J053504.75-012406.5	&	kA0hA2mA4 CrEu   	&	A2 Cr Eu	&	A0pSi	&	\citet{1983AA...118..102G}	\\
393	&	10107	&	J054102.12+332331.1	&	B7 IV Si  (He-wk) 	&	B9 Si (?)	&	n/a	&	n/a	\\
403	&	10323	&	J054630.44+273518.1	&	B9 V CrEu   	&	B8 Si (?)	&	n/a	&	n/a	\\
411	&	10375	&	J054819.92+333516.9	&	kA4hA9mF1 SrCrEuSi   	&	A5 Si Sr	&	A7	&	\citet{1931AnHar.100...61C}	\\
409	&	10385	&	J054757.12+235011.8	&	B9.5 II$-$III EuSi   	&	B9 Si Sr	&	B8	&	\citet{1931AnHar.100...61C}	\\
418	&	10447	&	J055002.80+234023.9	&	B9 IV  bl4130  	&	B9 Si Cr	&	B9	&	\citet{1931AnHar.100...61C}	\\
428	&	10460	&	J055121.05+420610.5	&	B8 IV Si   	&	A Si	&	ApSi	&	\citet{1983AJ.....88.1182B}	\\
436	&	10536	&	J055237.95+274922.8	&	A6 V SrCrEu   	&	A2-	&	n/a	&	n/a	\\
445	&	10602	&	J055422.76+305401.8	&	B8 III  bl4130  	&	A0-	&	A2	&	\citet{1931AnHar.100...61C}	\\
470	&	10900	&	J060045.94-035344.3	&	A0 IV$-$V Si   	&	A0 Cr Eu	&	A0VpSi	&	\citet{1999AAS..137..451G}	\\
476	&	10915	&	J060227.33+282943.9	&	B8 III$-$IV EuSi  (He-wk) 	&	A0 Si	&	A3	&	\citet{1931AnHar.100...61C}	\\
475	&	10917	&	J060225.92+244628.5	&	B8 IV Si   	&	B9 Si	&	B8/9.5IV/VSi:	&	\citet{1979RA......9..479C}	\\
560	&	11800	&	J062155.55+001812.2	&	B8 III Si   	&	A0 Si Sr	&	A0pSi	&	\citet{1999AAS..137..451G}	\\
564	&	11810	&	J062257.61+231625.8	&	A1 IV$-$V SrCrEu   	&	A2 Sr	&	ApSr	&	\citet{1983AJ.....88.1182B}	\\
597	&	12200	&	J062914.34+004257.0	&	B7 III$-$IV Si   	&	B9 Si	&	ApSi	&	\citet{1973AJ.....78..687B}	\\
604	&	12300	&	J063035.50+035245.3	&	A1 IV$-$V SrCr   	&	A2 Sr Eu	&	ApSrEu	&	\citet{1973AJ.....78..687B}	\\
611	&	12360	&	J063218.45+032146.3	&	B9 III$-$IV Si  (He-wk) 	&	B9 Si	&	ApSi	&	\citet{1973AJ.....78..687B}	\\
627	&	12630	&	J063744.29+195655.1	&	kA1hA7mF4 SrCrSi   	&	A Sr Eu	&	ApSrEu	&	\citet{1983AJ.....88.1182B}	\\
630	&	12690	&	J063752.90+091516.7	&	B8 IV Si   	&	B9 Si	&	ApSi	&	\citet{1973AJ.....78..687B}	\\
715	&	13790	&	J065403.63+221545.2	&	A6 IV SrCrEu   	&	A2 Sr Eu	&	ApSrEu	&	\citet{1972PASP...84..446B}	\\
721	&	13980	&	J065458.31+040826.9	&	kA1hA3mA6 SrCrEu   	&	A2 Sr Cr Eu	&	FpSrCrEu	&	\citet{1973AJ.....78..687B}	\\
753	&	14520	&	J070252.77+023700.0	&	kB9hA9mA7 SrSi   	&	A2 Si	&	ApSi	&	\citet{1973AJ.....78..687B}	\\
763	&	14740	&	J070617.23+101601.6	&	B9 IV EuSi   	&	A0 Si	&	ApSi	&	\citet{1973AJ.....78..687B}	\\
774	&	15123	&	J071337.30+040720.7	&	B8 IV CrEuSi   	&	B9 Si	&	n/a	&	n/a	\\
792	&	15650	&	J072118.92+223422.7	&	B9 V bl4077   	&	Sr (?)	&	ApSr	&	\citet{1983AJ.....88.1182B}	\\
827	&	17490	&	J074851.40+001619.1	&	kB8hA3mA3 CrEu   	&	Cr Eu	&	ApCrEu	&	\citet{1973AJ.....78..687B}	\\
829	&	17540	&	J074959.61+013517.8	&	kA1hA3mA7 SrCrEuSi   	&	Sr Cr Eu	&	ApSrCrEu	&	\citet{1973AJ.....78..687B}	\\
830	&	17630	&	J075041.80-060338.3	&	A2 IV SrCrEu   	&	A0 Cr Eu	&	FpCrEu	&	\citet{1973AJ.....78..687B}	\\
838	&	18380	&	J080339.87-082141.0	&	kB9hA3mA8 SrCrEu   	&	A0 Cr Eu	&	ApCrEu	&	\citet{1973AJ.....78..687B}	\\
867	&	24620	&	J095644.95-021719.5	&	kA2hA3mA6 SrCrEu   	&	A2 Sr Eu Cr	&	ApSrCrEu	&	\citet{1981AJ.....86..553B}	\\
873	&	29280	&	J114130.23+403822.7	&	B9 IV$-$V CrEuSi   	&	A0- (?)	&	A0m:	&	\citet{1959AAHam...5..105S}	\\
877	&	31550	&	J122855.36+255446.3	&	B9 V CrEu   	&	A0 Sr Cr Eu (*)	&	B8pSiCrSr	&	\citet{1990AAS...85.1069S}	\\
967	&	59010	&	J222549.96+343851.0	&	B8 IV EuSi   	&	A0 Si (?)	&	ApSi:	&	\citet{1985AJ.....90..341B}	\\
980	&	60185	&	J230905.79+523711.2	&	B8 IV EuSi   	&	B8 Si	&	ApSi	&	\citet{1998PASP..110..270B}	\\
1001	&	61520	&	J235740.51+470001.7	&	B9 IV CrSi   	&	B8 (?)	&	n/a	&	n/a	\\
\hline
\end{tabular}
\end{adjustbox}
\end{center}
\end{table*}

\begin{table}
\caption{Comparison of the spectral types derived in this study to the catalogue of \citet{qin19}. The columns denote: (1) Internal identification number. (2) LAMOST identifier. (3) Spectral type, as derived in this study. (4) Spectral type from \citet{qin19}. (5) Ap\_flag from \citet{qin19}. A value of 1 indicates that the star is a CP2 star candidate.}
\label{table_comparison_Qin}
\begin{center}
\begin{adjustbox}{max width=0.5\textwidth}
\begin{tabular}{clllc}
\hline
\hline
(1) & (2) & (3) & (4) & (5) \\
\hline
\textbf{No.} &	\textbf{ID\_LAMOST}	&	\textbf{SpT\_final}	&	\textbf{SpT\_Qin}	&	\textbf{Ap\_flag}	\\
2	&	J000834.32+321205.4	&	A2 IV SrCrEu   	&	kA2hA7mA7	&	1	\\
12	&	J003425.61+452108.7	&	kA1hA7mA9 SrEuSi   	&	kA2hA3mA5	&	1	\\
25	&	J011435.00+534757.2	&	A1 IV$-$V SrCrEuSi   	&	kA1hA5mA5	&	1	\\
29	&	J012028.54+480545.6	&	kA4hA7mA8 SrCrEu   	&	kA5hA6mA7	&	1	\\
30	&	J012122.31+442826.2	&	kA3hA7mA8 CrEu   	&	kA5hA5mA7	&	1	\\
35	&	J014508.87+322430.3	&	kA1hA6mA7 SrCrEu   	&	kA1hA5mA5	&	1	\\
68	&	J023936.01+540059.5	&	kA2hA5mA7 CrEu   	&	kA6hA5mA5	&	0	\\
70	&	J024243.57+453720.4	&	A1 V SrCrEuSi   	&	kA6hA6mA7	&	1	\\
106	&	J032333.76+510336.6	&	A2 V CrEuSi   	&	kA3hA6mA7	&	0	\\
218	&	J042235.95+411448.9	&	A1 IV SrCrEuSi   	&	kA6hA5mA7	&	1	\\
235	&	J043752.46+533259.6	&	B9.5 V CrSi   	&	kA0hA1mA5	&	1	\\
244	&	J044446.24+513129.2	&	A9 V SrEu   	&	kA5hA5mF0	&	0	\\
246	&	J044713.47+540515.5	&	A2 IV CrEu   	&	kA6hA6mF0	&	0	\\
252	&	J045121.11+093555.8	&	A0 V SrCrEuSi   	&	kA2hA1mA6	&	1	\\
259	&	J045508.24+204943.7	&	kA2hA4mA7 SrCrEu   	&	kA6hA5mA7	&	1	\\
286	&	J050925.66+512444.9	&	A3 III$-$IV CrEu   	&	kA5hA6mA7	&	0	\\
330	&	J052459.98-064651.8	&	A0 IV$-$V Si   	&	kA3hA5mA7	&	1	\\
332	&	J052552.24+344817.3	&	A0 V CrEu   	&	kA1hA1mA5	&	1	\\
344	&	J052816.11-063820.1	&	kA1hA9mA9 SrCrEu   	&	kA2hA7mA6	&	1	\\
387	&	J053840.17+413754.0	&	kA1hA7mA9 SrCrEuSi   	&	kA3hA6mA7	&	1	\\
423	&	J055045.30+372809.0	&	kA2hA4mA7 SrCrEu   	&	kA6hA5mA7	&	1	\\
427	&	J055110.67+360517.0	&	B9.5 V SrCr   	&	kA1hA1mA5	&	1	\\
441	&	J055333.92+180202.9	&	F2 V SrEu   	&	kA7hA7mF5	&	0	\\
483	&	J060356.06+213033.2	&	A8 V SrEu   	&	kA5hA6mF0	&	0	\\
542	&	J061743.05+595315.3	&	A1 IV Eu   	&	kA3hA6mA7	&	1	\\
570	&	J062449.08+190854.0	&	kA2hA3mA7 SrCrEu   	&	kA3hA5mF0	&	1	\\
571	&	J062449.21+161325.7	&	A0 V Cr   	&	kA1hA1mA5	&	0	\\
627	&	J063744.29+195655.1	&	kA1hA7mF4 SrCrSi   	&	kA2hA6mA7	&	1	\\
643	&	J064158.41+223927.2	&	A5 IV SrCrEu   	&	kA5hA6mA7	&	0	\\
689	&	J064907.51+114600.1	&	kA1hA7mA8 SrCr   	&	kA2hA6mA5	&	0	\\
690	&	J064947.96+202510.8	&	kA3hA5mA7 bl4077 bl4130  	&	kA7hA7mA7	&	0	\\
770	&	J070907.08+441114.7	&	B9.5 V CrEu   	&	kA1hA6mA5	&	1	\\
773	&	J071258.59+065952.3	&	A2 IV SrCrEu   	&	kA3hA6mA7	&	1	\\
775	&	J071413.88+142449.5	&	A3 III$-$IV SrCrEu   	&	kA6hA5mA7	&	1	\\
779	&	J071550.38+063655.9	&	A1 IV SrCrEuSi   	&	kA2hA5mA7	&	1	\\
814	&	J073548.83+123225.1	&	kA1hA9mA8 SrEu   	&	kA6hA7mA7	&	1	\\
828	&	J074919.49+051551.8	&	kA1hA9mA9 SrCrEu   	&	kA2hA7mA5	&	1	\\
831	&	J075220.93+113710.6	&	kA5hA6mA9 SrCrEu   	&	kA5hA6mA7	&	0	\\
848	&	J082137.47+064401.0	&	A1 IV SrCrEu   	&	kA2hA5mA5	&	1	\\
850	&	J082326.74+072116.4	&	kA3hA7mF1 SrCrEu   	&	kA2hA6mA7	&	0	\\
853	&	J083539.12+002150.4	&	kA2hA3mA6 SrCrEu   	&	kA3hA1mA7	&	0	\\
861	&	J092233.23+072519.4	&	kA3hA6mA9 SrCrEu   	&	kA6hA5mA7	&	1	\\
879	&	J140422.54+044357.9	&	kA4hA7mF0 SrCrEu   	&	kA6hA6mA7	&	1	\\
899	&	J185127.99+262012.1	&	A0 IV$-$V Cr   	&	kA0hA1mA7	&	1	\\
909	&	J192524.14+431911.5	&	kA3hA7mA9 SrCrEu   	&	kA6hA5mA5	&	1	\\
\hline
\end{tabular}
\end{adjustbox}
\end{center}
\end{table}

\begin{figure*}
        \includegraphics[width=\textwidth]{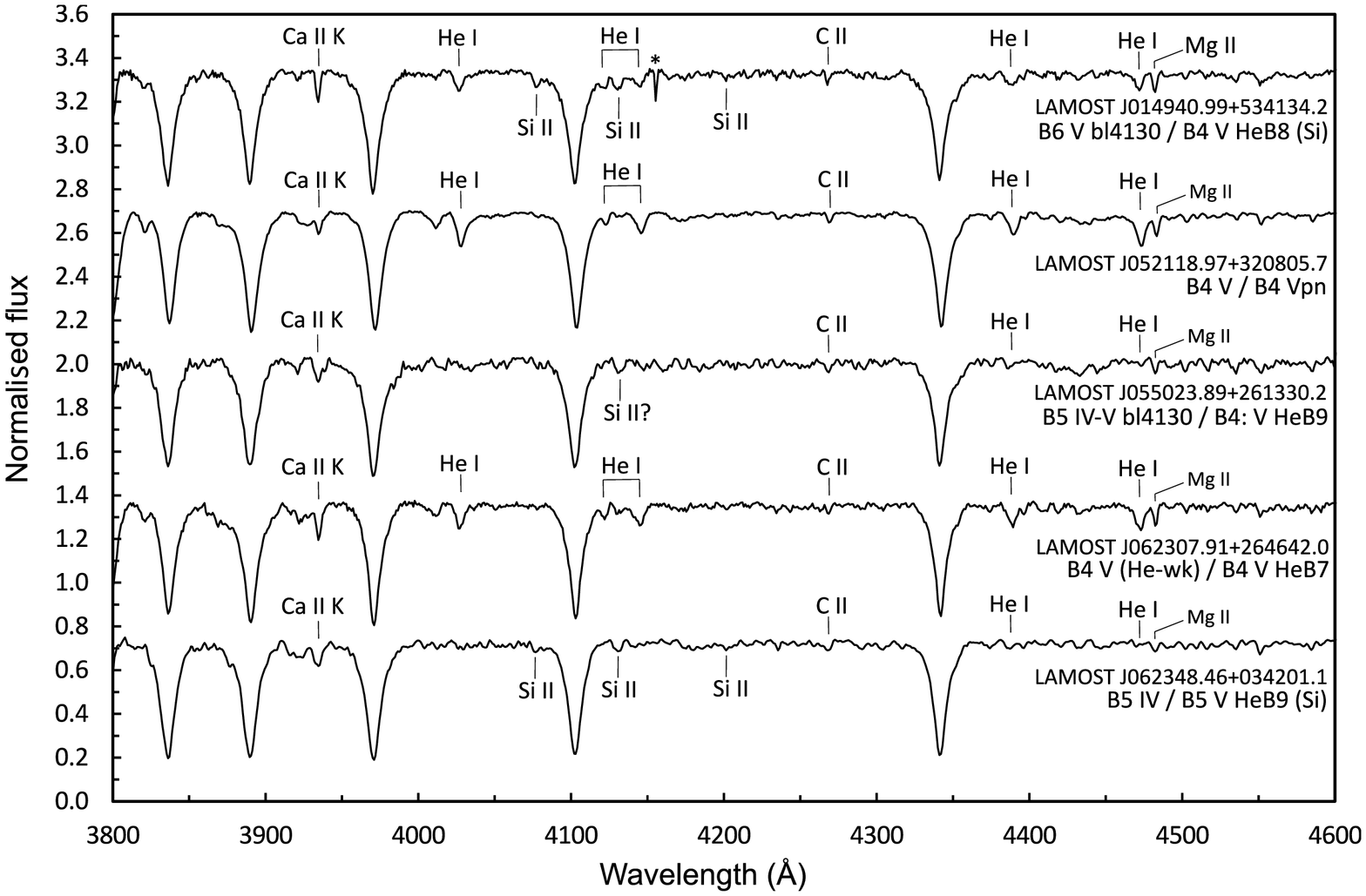}
    \caption{Showcase of five newly identified peculiar mid-B type stars, illustrating the blue-violet region of the LAMOST DR4 spectra of (from top to bottom) LAMOST J014940.99+534134.2 (\#37; TYC 3684-1139-1), LAMOST J052118.97+320805.7 (\#318; HD 242764), LAMOST J055023.89+261330.2 (\#421; TYC 1866-861-1), LAMOST J062307.91+264642.0
 (\#565; Gaia DR2 3432273606513132544), and LAMOST J062348.46+034201.1 (\#567; HD 256582). MKCLASS final types and manual types derived in the present study are indicated. Some prominent lines of interest are identified. The asterisk marks the position of a 'glitch' in the spectrum of LAMOST J014940.99+534134.2.}
    \label{spectra_hepec}
\end{figure*}

\subsection{The mid-B type mCP stars - He-peculiar objects?} \label{Hepec}

23 stars of our sample have MKCLASS final types earlier than B7. More than half of these objects were classified as showing peculiarly weak \ion{He}{i} lines ('He-wk'). Interestingly, three stars were identified as showing both weak and strong ('He-st') \ion{He}{i} lines, which strongly suggests He peculiarity. Besides that, only Si overabundances and strong 4130\,\AA\ blends were identified in several of these objects. As He-rich stars are generally hotter than spectral type B4 \citep{gray09} and therefore not expected to contribute to our sample, we consider these objects good candidates for He-weak (CP4) stars.

Figure \ref{spectra_hepec} showcases the spectra of five mid-B type stars. The hydrogen-line profiles and the prominent \ion{C}{ii} 4267\,\AA\ lines corroborate the classifications, although we manually derived slightly different temperature types in two cases. LAMOST J014940.99+534134.2 (\#37; TYC 3684-1139-1) and LAMOST J062348.46+034201.1 (\#567; HD 256582) are He-weak stars with Si overabundances. LAMOST J062307.91+264642.0 (\#565; Gaia DR2 3432273606513132544) is also certainly He-weak but does not fit any of the standard subclasses of the He-weak stars (the 'hot' Si stars, i.e. Si stars with hotter temperatures than the classical Si CP2 stars; the P Ga stars; the Sr Ti stars; cf. \citealt{gray09}). It is here classified as B4 V HeB7 (R. O. Gray, personal communication).

LAMOST J055023.89+261330.2 (\#421; TYC 1866-861-1) boasts a rather noisy spectrum ($g$ band S/N of 79) that is, apart from the hydrogen lines, basically a 'smattering' of metal-lines, without any particularly outstanding features -- except for the strong C II 4267\,\AA\ and the weak Mg II 4481\,\AA\ lines that support its classification as a mid B-type object. This is also supported by the colour index $(BP-RP){_0}$\,=\,$-0.182$\,mag. The He I lines, then, seem to be curiously absent from its spectrum, so the star may be related to the CP4 stars, although the metal-lines seem way too strong to support this interpretation. We here tentatively classify it as B4: V HeB9. The star shows a strong flux depression at 5200\,\AA, and, according to data from the SuperWASP archive \citep{butters10}, is a periodic photometric variable with a period of about 11.5\,d. It certainly merits a closer look -- this, however, is beyond the scope of the present investigation.

The He lines of LAMOST J052118.97+320805.7 (\#318; HD 242764) do not look weak for its temperature type but have broad profiles suggesting rapid rotation. This, however, is not supported by the hydrogen-line profile, which almost exactly matches that of the B4 V standard. The presence of a number of unidentified metal-lines (which do show evidence for rotational broadening, but not to the extent the He I lines imply) and the conspicuous 5200\,\AA\ depression suggest that this star is indeed chemically peculiar, although it also does not fit into any of the standard mid-B peculiarity subtypes. It has been classified as B4 Vpn in the present study (R. O. Gray, personal communication).

Additional proof that this star is indeed chemically peculiar comes from its periodic photometric variability with a period of about 5.1\,d in SuperWASP data and the conspicuous 5200\,\AA\ flux depression in its spectrum. Incidentally, the star is listed with a spectral type of B8 Si Sr in the \citet{RM09} catalogue, which is not supported by the available LAMOST spectrum. We were unable to get at the root of this classification; however, the star is listed as B9 in the Henry Draper extension \citep{1931AnHar.100...61C} and was classified as B5p by \citet{chargeishvili88}. Further He-peculiar objects and candidates can be gleaned from Table \ref{table_master1}.

\subsection{The eclipsing binary system LAMOST J034306.74+495240.7} \label{eclipser}

\begin{figure}
        \includegraphics[width=\columnwidth]{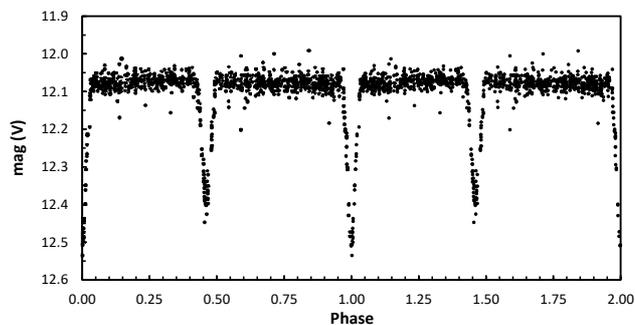}
    \caption{ASAS-SN light curve of the eclipsing binary system LAMOST J034306.74+495240.7 (\#135; TYC\,3321-881-1). The data have been folded with the orbital period of $P$\textsubscript{orb}\,=\,5.1435$\pm$0.0012\,d.}
    \label{eclipser_lightcurve}
\end{figure}

The star LAMOST J034306.74+495240.7 (\#135; TYC\,3321-881-1) was identified as an eclipsing binary system in ASAS-SN data \citep{jayasinghe19}. It is listed in the International Variable Star Index of the AAVSO (VSX; \citealt{VSX}) under the designation ASASSN-V J034306.74+495240.8 and with a period of 5.1431\,d. We have analyzed the available ASAS-SN data for this star and derive a period of 5.1435$\pm$0.0012\,d and an epoch of primary minimum at HJD 2457715.846$\pm$0.002. The light curve is shown in Figure \ref{eclipser_lightcurve} and illustrates that the orbit is slightly eccentric, the secondary minimum occurs at an orbital phase of $\varphi$\,=\,0.46. In addition, there is evidence for out-of-eclipse variability in agreement with rotational modulation on one component of the system.

\begin{figure}
        \includegraphics[width=\columnwidth]{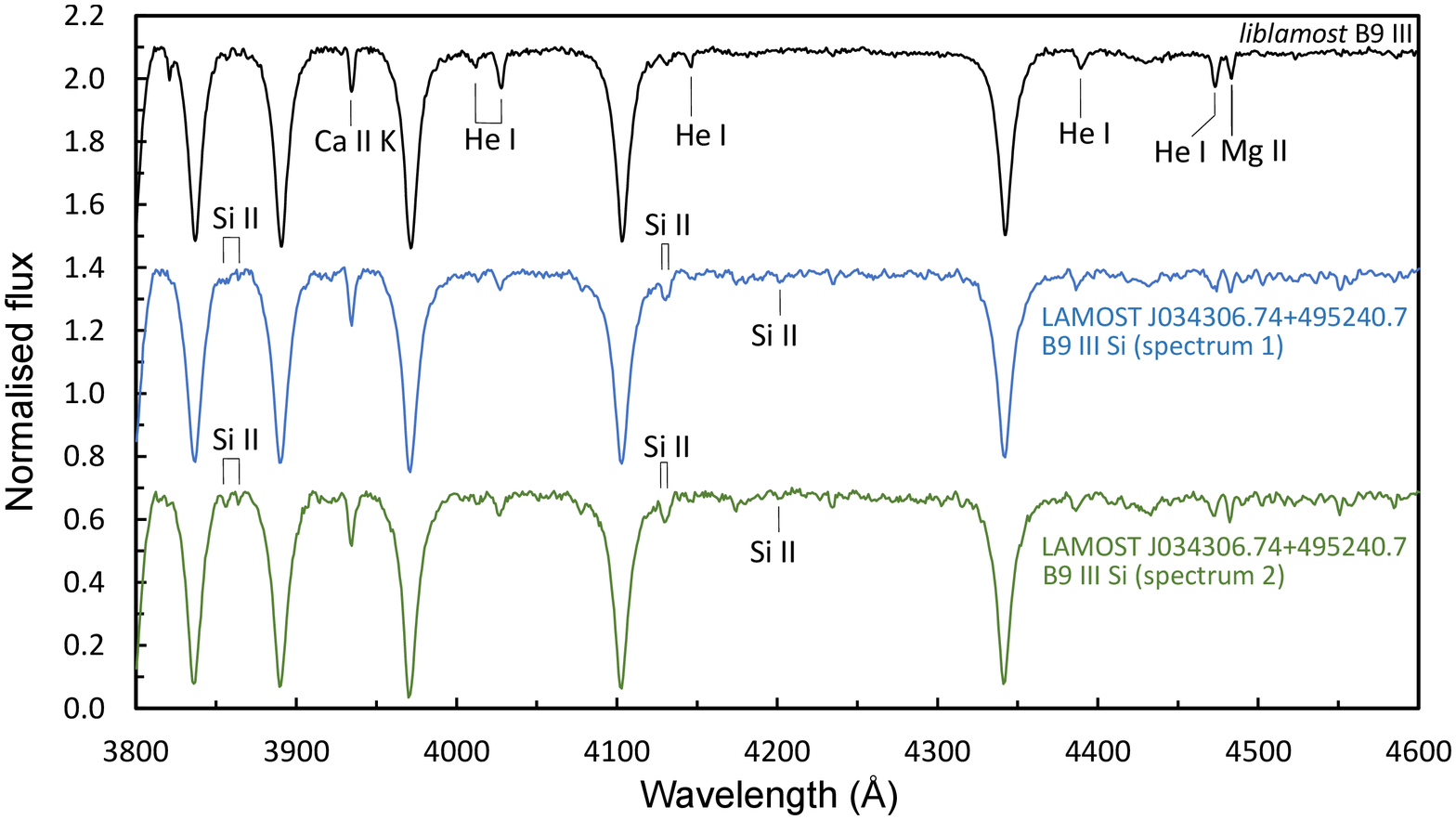}
    \caption{Comparison of the blue-violet spectra of the eclipsing binary system LAMOST J034306.74+495240.7 (\#135; TYC\,3321-881-1) to the \textit{liblamost} B9 III standard spectrum (upper spectrum). Some prominent lines of interest are indicated.}
    \label{eclipser_spectra}
\end{figure}

Two spectra are available for this star in LAMOST DR4. The first spectrum ('spectrum1') was obtained on 23 October 2015 (MJD 57318; observation median UTC 17:33:00; $g$ band S/N: 258), which corresponds to an orbital phase of $\varphi$\,=\,0.890. The second spectrum ('spectrum2') was taken on 19 January 2016 (MJD 57406; observation median UTC 11:46:00; $g$ band S/N: 207), which corresponds to an orbital phase of $\varphi$\,=\,0.952. Therefore, both spectra were taking during maximum light, and we find no significant difference between them. Both show a strong flux depression at 5200\,\AA\ and enhanced \ion{Si}{ii} lines at 3856/62\,\AA, 4128/31\,\AA, 4200\,\AA, 5041/56\,\AA, and 6347/71\,\AA. We have analyzed the spectrum with the highest S/N (spectrum1) and derive a spectral type of B9 III Si. Figure \ref{eclipser_spectra} compares the blue-violet part of both spectra to the liblamost B9 III standard, whose hydrogen-line profile provides a good fit to the observed ones.

In summary, we conclude that at least one component of the LAMOST J034306.74+495240.7 system is a Si CP2 star. It is, therefore, of great interest because mCP stars in eclipsing binaries are exceedingly rare \citep{RM09,niemczura17,kochukhov18,skarka19} and accurate parameters for the components can be derived via an orbital solution of the system. We strongly encourage further studies of this interesting object.

\subsection{The SB2 system LAMOST J050146.85+383500.8} \label{SB2_system}

\begin{figure}
        \includegraphics[width=\columnwidth]{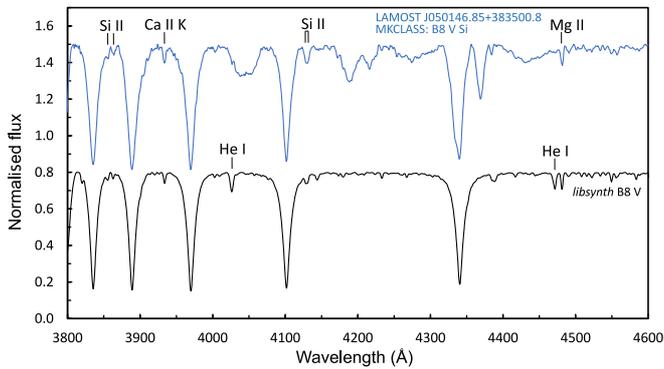}
    \caption{Comparison of the blue-violet spectra of the propsed SB2 system LAMOST J050146.85+383500.8 (\#272; HD 280281; MKCLASS final type B8 V Si) to the \textit{libsynth} B8 V standard spectrum (upper spectrum). Some prominent lines of interest are indicated.}
    \label{SB2}
\end{figure}

\begin{figure}
        \includegraphics[width=0.5\textwidth]{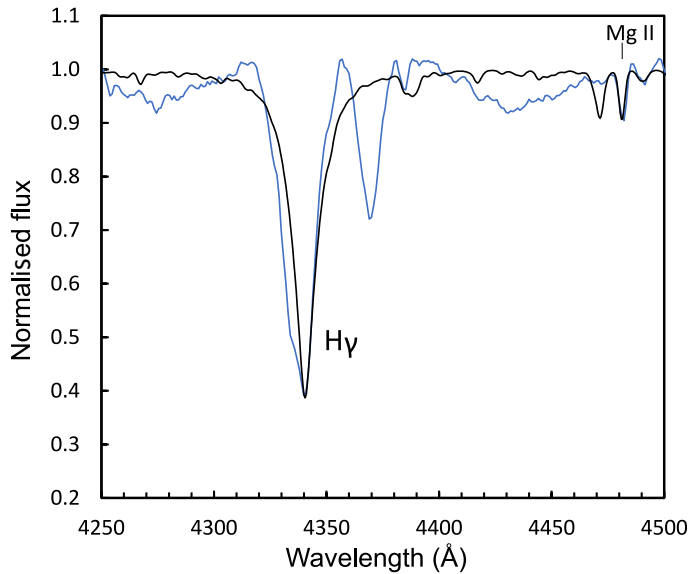}
    \caption{Close-up view of the H$\gamma$ region of the proposed SB2 system LAMOST J050146.85+383500.8 (blue spectrum) and the \textit{libsynth} B8 V standard (black spectrum), illustrating the peculiar profile of the H$\gamma$ line indicative of binarity.}
    \label{SB2_Hgam}
\end{figure}

Figure \ref{SB2} illustrates the peculiar spectrum of LAMOST J050146.85+383500.8 (\#272; HD 280281), which we suspect to be a blend of two different stars. This becomes especially obvious in the profile of the H$\gamma$ line (Figure \ref{SB2_Hgam}).

\begin{figure}
\begin{center}
\includegraphics[width=85mm]{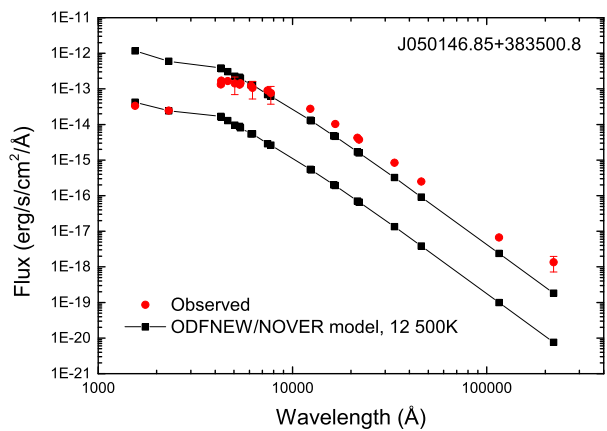}
\caption[]{Comparison of the SED of LAMOST J050146.85+383500.8 (red dots) to a Kurucz ODFNEW/NOVER model with $T_\mathrm{eff}$\,=\,12\,500\,K (black squares). The model was forced to either fit the ultraviolet (lower model) or optical flux (upper model). The discrepancies are clearly visible, the star's SED cannot be fitted with a single star flux model.}
\label{fig_SED}
\end{center}
\end{figure}

To further investigate this matter, we employed the VO Sed Analyzer tool VOSA\footnote{\url{http://svo2.cab.inta-csic.es/theory/vosa/}} v6.0 \citep{2008A&A...492..277B} to fit the SED to the available photometry. For comparison, we used a Kurucz ODFNEW/NOVER model \citep{1997A&A...318..841C} with $T_\mathrm{eff}$\,=\,12\,500\,K, which corresponds to a spectral type of B8. We emphasise that a change of $T_\mathrm{eff}$ of about 2000\,K in either direction will not impact our conclusions. Figure \ref{fig_SED} illustrates the results of the fitting process. The flux model was fitted to either match the ultraviolet or the optical wavelength region. In any case, the discrepancies are readily visible and it is obvious that the observed flux distribution of LAMOST J050146.85+383500.8 cannot be fitted with a single star flux model. We note that it is well known that CP2 stars show a 'blueing' effect \citep{1980A&A....89..230M}, which leads to observed flux discrepancies due to stronger absorption in the ultraviolet than in chemically normal stars. However, a slight shift in the ultraviolet region will not alter our conclusions.

Because the features of the companion star are readily visible in the available LAMOST spectrum, we conclude that its absolute magnitude must be similar to the B-type main-sequence component. We therefore assume that it is a supergiant star with a progenitor of higher mass. Such a combination of components is quite unusual among mCP stars; in order to put further constraints on this spectroscopic binary system, orbital elements or the analysis of light-travel time effects are needed. LAMOST J050146.85+383500.8, therefore, is an interesting target for follow-up studies.

\section{Conclusions} \label{conclusion}

We carried out a search for mCP stars in the publicly available spectra of LAMOST DR4. Suitable candidates were selected by searching for the presence of the characteristic 5200\,\AA\ flux depression. In consequence, our sample is biased towards mCP stars with conspicuous flux depressions at 5200\,\AA. Spectral classification was carried out with a modified version of the MKCLASS code (MKCLASS\_mCP) and, for a subsample of stars, by manual classification. We evaluated our results by spot-checking with manually derived spectral types and comparison to samples from the literature.

The main findings of the present investigation are summarised in the following:

\begin{itemize}
	\item We identified 1002 mCP stars, most of which are new discoveries. There are only 59 common entries with the catalogue of \citealt{RM09}. With our work, we significantly increase the sample size of known Galactic mCP stars, paving the way for future in-depth statistical studies.
	\item To suit the special needs of our project, we updated the current version (v1.07) of the MKCLASS code to probe several additional lines, with the advantage that the new version (here termed MKCLASS\_mCP) is now able to more robustly identify traditional mCP star peculiarities, including Cr peculiarities and, to some extent, He peculiarities. 
	\item mCP star peculiarities (Si, Cr, Sr, Eu, strong blends at 4077\,\AA\ and/or 4130\,\AA) were identified in all but 36 stars of our sample, highlighting the efficiency of the chosen approach and the peculiarity identification routine. The remaining objects (mostly mCP stars with weak or complicated peculiarities and He-peculiar objects) were manually searched to locate the presence of peculiarities.
	\item Comparisons between manually derived spectral types and the MKCLASS\_mCP final types indicate a good agreement between the derived temperature and peculiarity types. This is further corroborated by a comparison with spectral types from the \citet{RM09} and \citet{Skiff} catalogues and the good agreement of the peculiarity type versus spectral type distribution between this study and the literature. However, with our approach, we missed the presence of certain peculiarities in several objects. The peculiarity types presented here are therefore not exhaustive. They nevertheless form a sound basis for statistical and further studies.
	\item Our sample stars are between 100\,Myr and 1\,Gyr old, with the majority having masses between 2\,$M_\odot$ and 3\,$M_\odot$. We investigated the evolutionary status of 903 mCP stars, deriving a mean fractional age on the main sequence of $\tau$\,=\,63\,\% (standard deviation of 23\,\%). Young mCP stars, while undoubtedly present, are conspicuously underrepresented in our sample. Our results could be considered as strong evidence for an inhomogeneous age distribution among low-mass ($M$\,$<$\,3\,M$_\odot$) mCP stars, as hinted at by previous studies. However, we caution that our sample has not been selected on the basis of an unbiased, direct detection of a magnetic field. Therefore, our results have to be viewed with caution and their general validity needs to be tested by a more extended sample selected via different methodological approaches.
	\item The mCP stars LAMOST J122746.05+113635.3 (\#876) and LAMOST J150331.87+093125.4 (\#880) boast distances and kinematical properties in agreement with halo stars. If confirmed, they would be the first CP2 halo objects known and therefore of special interest.
	\item We identified LAMOST J034306.74+495240.7 (\#135; TYC\,3321-881-1) as an eclipsing binary system ($P$\textsubscript{orb}\,=\,5.1435$\pm$0.0012\,d) hosting a Si CP2 star component (spectral type B9 III Si). This is of great interest because mCP stars in eclipsing binaries are exceedingly rare.
	\item The star LAMOST J050146.85+383500.8 was identified as an SB2 system likely comprising of a Si CP2 star and a supergiant.
\end{itemize}
	
Future investigations will be concerned with an in-depth study of the new mCP stars identified in this work, particularly with regard to their photometric variability, along with further development and refinement of the approach for identifying and classifying mCP stars in large spectroscopic databases using the MKCLASS code.

\begin{acknowledgements}
We thank the referee for his thoughtful report that helped to significantly improve the paper. This work has been supported by the DAAD (project No. 57442043). The Guo Shou Jing Telescope (the Large Sky Area Multi-Object Fiber Spectroscopic Telescope, LAMOST) is a National Major Scientific Project built by the Chinese Academy of Sciences. Funding for the project has been provided by the National Development and Reform Commission. LAMOST is operated and managed by National Astronomical Observatories, Chinese Academy of Sciences. This work presents results from the European Space Agency (ESA) space mission Gaia. Gaia data are being processed by the Gaia Data Processing and Analysis Consortium (DPAC). Funding for the DPAC is provided by national institutions, in particular the institutions participating in the Gaia MultiLateral Agreement (MLA). The Gaia mission website is https://www.cosmos.esa.int/gaia. The Gaia archive website is https://archives.esac.esa.int/gaia.
\end{acknowledgements}

\clearpage

%
%

\bibliographystyle{aa}
\bibliography{lamost_apstars}

\appendix

\section{Essential data for our sample stars}

Table \ref{table_master1} lists essential data for our sample stars. It is organised as follows:

\begin{itemize}
\item Column 1: Internal identification number.
\item Column 2: LAMOST identifier.
\item Column 3: Alternativ identifier (HD number, TYC identifier, or GAIA DR2 number).
\item Column 4: Right ascension (J2000). Positional information was taken from GAIA DR2 \citep{2018A&A...616A...1G,gaia3}.
\item Column 5: Declination (J2000).
\item Column 6: MKCLASS final type, as derived in this study.\footnote{We note that, as in the \citet{RM09} catalogue, the ’p’ denoting peculiarity was omitted from the spectral classifications.} All further additions to the spectral type that are not directly based on the MKCLASS\_mCP output are highlighted using italics. For an easy identification, manually altered spectral types are indicated by asterisks.
\item Column 7: Sloan $g$ band S/N of the analysed spectrum.
\item Column 8: $G$\,mag (GAIA DR2).
\item Column 9: $G$\,mag error.
\item Column 10: Parallax (GAIA DR2).
\item Column 11: Parallax error.
\item Column 12: Dereddened colour index $(BP-RP){_0}$ (GAIA DR2).
\item Column 13: Colour index error.
\item Column 14: Absorption in the $G$ band, $A_G$.
\item Column 15: Intrinsic absolute magnitude in the $G$ band, $M_{\mathrm{G,0}}$.
\item Column 16: Absolute magnitude error.
\end{itemize} 

Tables \ref{table_master1}, \ref{table_age_mass}, and \ref{table_liblamost} are available at the CDS.\footnote{\url{http://cdsarc.u-strasbg.fr/viz-bin/cat/J/A+A/640/A40}}

\setcounter{table}{0}  
\begin{sidewaystable*}
\caption{Essential data for our sample stars, sorted by increasing right ascension. The columns denote: (1) Internal identification number. (2) LAMOST identifier. (3) Alternativ identifier (HD number, TYC identifier or GAIA DR2 number). (4) Right ascension (J2000; GAIA DR2). (5) Declination (J2000; GAIA DR2). (6) Spectral type, as derived in this study. (7) Sloan $g$ band S/N ratio of the analysed spectrum. (8) $G$\,mag (GAIA DR2). (9) $G$\,mag error. (10) Parallax (GAIA DR2). (11) Parallax error. (12) Dereddened colour index $(BP-RP){_0}$ (GAIA DR2). (13) Colour index error. (14) Absorption in the $G$ band, $A_G$. (15) Intrinsic absolute magnitude in the $G$ band, $M_{\mathrm{G,0}}$. (16) Absolute magnitude error.}
\label{table_master1}
\begin{center}
\begin{adjustbox}{max width=\textwidth}
\begin{tabular}{lllcclcccccccccc}
\hline
\hline
(1) & (2) & (3) & (4) & (5) & (6) & (7) & (8) & (9) & (10) & (11) & (12) & (13) & (14) & (15) & (16) \\
No.	&	ID\_LAMOST	&	ID\_alt	&	RA(J2000) 	&	 Dec(J2000)    	&	SpT\_final	&	S/N\,$g$	&	$G$\,mag	&	e\_$G$\,mag	&	pi (mas)	&	e\_pi	&	$(BP-RP){_0}$	&	e\_$(BP-RP){_0}$	&	$A_G$	&	$M_{\mathrm{G,0}}$	&	e\_$M_{\mathrm{G,0}}$	\\
\hline
1	&	J000700.42+053333.6	&	TYC 7-1183-1       	&	00 07 00.30 	&	 +05 33 33.64	&	A0 IV$-$V SrCrEuSi   	&	369	&	10.5972	&	0.0005	&	+1.579	&	0.056	&	+0.238	&	0.002	&	0.15	&	+1.44	&	0.09	\\
2	&	J000834.32+321205.4	&	TYC 2263-195-1     	&	00 08 34.32 	&	 +32 12 05.48	&	A2 IV SrCrEu   	&	166	&	11.6306	&	0.0005	&	+1.355	&	0.046	&	+0.325	&	0.003	&	0.11	&	+2.18	&	0.09	\\
3	&	J000840.99+580213.2	&	Gaia DR2 422492913555750912	&	00 08 41.00 	&	 +58 02 13.21	&	A0 Ib$-$II bl4077   	&	105	&	14.1363	&	0.0025	&	+0.152	&	0.023	&	$-$0.012	&	0.012	&	1.18	&	$-$1.14	&	0.33	\\
4	&	J001137.65+473203.2	&	TYC 3251-1665-1    	&	00 11 37.73 	&	 +47 32 02.75	&	B9 V Cr   	&	173	&	10.7570	&	0.0008	&	+0.955	&	0.058	&	+0.054	&	0.003	&	0.17	&	+0.49	&	0.14	\\
5	&	J001345.60+562953.5	&	TYC 3661-1153-1    	&	00 13 45.84 	&	 +56 29 53.60	&	B9 IV$-$V CrSi  (He-wk) 	&	241	&	10.3974	&	0.0006	&	+1.176	&	0.031	&	$-$0.030	&	0.002	&	0.62	&	+0.13	&	0.08	\\
6	&	J001407.53+553052.8	&	TYC 3657-780-1     	&	00 14 07.53 	&	 +55 30 52.89	&	B9 IV$-$V Eu   	&	264	&	11.0914	&	0.0011	&	+0.912	&	0.046	&	$-$0.016	&	0.003	&	0.62	&	+0.27	&	0.12	\\
7	&	J002343.42+535755.9	&	Gaia DR2 419523991644430208	&	00 23 43.43 	&	 +53 57 55.92	&	B9 IV$-$V bl4077 bl4130  	&	111	&	12.8617	&	0.0004	&	+0.283	&	0.033	&	+0.010	&	0.002	&	0.77	&	$-$0.65	&	0.26	\\
8	&	J002454.57+385413.2	&	Gaia DR2 379258467075146112	&	00 24 54.58 	&	 +38 54 13.24	&	A8 V SrCrEu   	&	122	&	12.5838	&	0.0003	&	+0.745	&	0.064	&	+0.239	&	0.002	&	0.16	&	+1.78	&	0.19	\\
9	&	J002616.49+562735.1	&	TYC 3661-142-1     	&	00 26 16.50 	&	 +56 27 35.14	&	kA3hA7mF2 Si   	&	59	&	11.7673	&	0.0004	&	+0.803	&	0.040	&	$-$0.048	&	0.002	&	0.48	&	+0.81	&	0.12	\\
10	&	J003312.88+543141.3	&	TYC 3658-79-1      	&	00 33 12.64 	&	 +54 31 41.33	&	B8 V Si   	&	560	&	10.2213	&	0.0006	&	+1.081	&	0.051	&	$-$0.085	&	0.002	&	0.37	&	+0.02	&	0.11	\\
11	&	J003318.47+571616.8 $^{a}$	&	TYC 3662-378-1     	&	00 33 18.47 	&	 +57 16 16.79	&	B9 IV$-$V Si   	&	338	&	10.6583	&	0.0009	&	+1.105	&	0.053	&	$-$0.082	&	0.003	&	0.61	&	+0.26	&	0.12	\\
12	&	J003425.61+452108.7	&	Gaia DR2 389018591278583552	&	00 34 25.58 	&	 +45 21 09.21	&	kA1hA7mA9 SrEuSi   	&	64	&	14.1081	&	0.0005	&	+0.347	&	0.039	&	+0.222	&	0.015	&	0.17	&	+1.64	&	0.25	\\
13	&	J004555.55+553553.8	&	TYC 3659-1537-1    	&	00 45 55.53 	&	 +55 35 53.73	&	kA0hA3mA6 SrEu   	&	332	&	10.5123	&	0.0004	&	+1.602	&	0.055	&	+0.058	&	0.002	&	0.24	&	+1.30	&	0.09	\\
14	&	J004947.16+525208.2	&	HD 232285  	&	00 49 46.83 	&	 +52 52 08.24	&	B9 V CrEuSi  (He-wk) 	&	285	&	9.3967	&	0.0004	&	+2.214	&	0.064	&	+0.049	&	0.001	&	0.45	&	+0.67	&	0.08	\\
15	&	J005002.58+434300.5	&	TYC 2810-568-1     	&	00 50 02.59 	&	 +43 43 00.55	&	B9 V SrEu   	&	232	&	11.8448	&	0.0005	&	+0.752	&	0.048	&	$-$0.029	&	0.003	&	0.13	&	+1.10	&	0.15	\\
16	&	J005535.39+433430.0	&	TYC 2810-1634-1    	&	00 55 35.40 	&	 +43 34 30.04	&	kB9.5hA1mA3 Eu   	&	296	&	11.6205	&	0.0013	&	+0.475	&	0.073	&	$-$0.048	&	0.005	&	0.17	&	$-$0.17	&	0.34	\\
17	&	J005604.18+453914.9	&	TYC 3262-1253-1    	&	00 56 04.19 	&	 +45 39 14.95	&	A5 V SrEu   	&	164	&	11.6601	&	0.0004	&	+0.837	&	0.051	&	+0.222	&	0.002	&	0.23	&	+1.05	&	0.14	\\
18	&	J005937.83+532354.0	&	TYC 3668-724-1     	&	00 59 37.83 	&	 +53 23 53.93	&	B9 IV$-$V Si   	&	199	&	11.7537	&	0.0007	&	+0.649	&	0.066	&	+0.052	&	0.002	&	0.43	&	+0.39	&	0.22	\\
19	&	J010007.84+045339.3	&	HD 5844    	&	01 00 07.85 	&	 +04 53 39.46	&	kB8hA3mA6 Cr   	&	293	&	9.7105	&	0.0009	&	+1.189	&	0.118	&	+0.106	&	0.003	&	0.11	&	$-$0.02	&	0.22	\\
20	&	J010159.03+505149.9	&	Gaia DR2 404326915576637184	&	01 01 59.02 	&	 +50 51 49.99	&	kA3hA7mA8 SrCrEu   	&	190	&	13.4949	&	0.0005	&	+0.457	&	0.032	&	+0.267	&	0.003	&	0.37	&	+1.43	&	0.16	\\
21	&	J010522.41+501031.3  $^{a}$	&	TYC 3271-1597-1    	&	01 05 22.42 	&	 +50 10 29.60	&	B9 V Cr   	&	367	&	10.4967	&	0.0006	&	+0.968	&	0.054	&	+0.038	&	0.002	&	0.31	&	+0.12	&	0.13	\\
22	&	J010651.35+154426.9	&	HD 6590    	&	01 06 51.17 	&	 +15 44 26.92	&	kA1hA7mA7 SrCrEu   	&	416	&	9.9956	&	0.0009	&	+1.277	&	0.067	&	+0.262	&	0.004	&	0.29	&	+0.24	&	0.12	\\
23	&	J011154.28+563254.7	&	TYC 3677-2219-1    	&	01 11 54.20 	&	 +56 32 54.70	&	B9.5 IV$-$V CrEu   	&	142	&	10.9232	&	0.0004	&	+1.285	&	0.035	&	$-$0.086	&	0.002	&	0.48	&	+0.99	&	0.08	\\
24	&	J011430.72+432537.2	&	TYC 2812-530-1     	&	01 14 30.73 	&	 +43 25 37.19	&	A9 V Sr   	&	187	&	11.0312	&	0.0007	&	+1.098	&	0.044	&	+0.398	&	0.003	&	0.10	&	+1.13	&	0.10	\\
25	&	J011435.00+534757.2	&	Gaia DR2 410353407527493504	&	01 14 35.00 	&	 +53 47 57.22	&	A1 IV$-$V SrCrEuSi   	&	115	&	12.1234	&	0.0003	&	+1.025	&	0.046	&	+0.205	&	0.002	&	0.68	&	+1.50	&	0.11	\\
26	&	J011524.23+515128.5	&	TYC 3276-520-1     	&	01 15 24.26 	&	 +51 51 28.85	&	A0 V SrCr   	&	452	&	11.2142	&	0.0012	&	+1.519	&	0.072	&	+0.068	&	0.003	&	0.40	&	+1.72	&	0.11	\\
27	&	J011627.87+250924.1	&	TYC 1750-2062-1    	&	01 16 27.88 	&	 +25 09 24.15	&	B8 III$-$IV Si   	&	140	&	12.5959	&	0.0007	&	+0.210	&	0.057	&	$-$	&	$-$	&	$-$	&	$-$	&	$-$	\\
28	&	J011816.48+514639.8	&	TYC 3276-500-1     	&	01 18 16.41 	&	 +51 46 40.53	&	B9 IV$-$V CrEuSi   	&	409	&	10.7842	&	0.0006	&	+1.368	&	0.063	&	$-$0.012	&	0.003	&	0.34	&	+1.13	&	0.11	\\
29	&	J012028.54+480545.6	&	TYC 3269-867-1     	&	01 20 28.55 	&	 +48 05 45.63	&	kA4hA7mA8 SrCrEu   	&	300	&	11.3762	&	0.0006	&	+0.935	&	0.055	&	+0.214	&	0.002	&	0.29	&	+0.94	&	0.14	\\
30	&	J012122.31+442826.2	&	Gaia DR2 397601443468210944	&	01 21 22.30 	&	 +44 28 26.12	&	kA3hA7mA8 CrEu   	&	152	&	12.2442	&	0.0014	&	+0.520	&	0.050	&	+0.316	&	0.007	&	0.14	&	+0.69	&	0.21	\\
31	&	J012220.76+532354.4	&	TYC 3670-871-1     	&	01 22 20.76 	&	 +53 23 54.43	&	B9 IV Si   	&	298	&	11.0429	&	0.0032	&	+0.985	&	0.054	&	+0.069	&	0.013	&	0.52	&	+0.49	&	0.13	\\
32	&	J013031.12+454021.1	&	TYC 3278-454-1     	&	01 30 31.13 	&	 +45 40 21.16	&	B9 IV CrEuSi   	&	233	&	11.1469	&	0.0025	&	+0.853	&	0.082	&	$-$0.088	&	0.006	&	0.10	&	+0.70	&	0.21	\\
33	&	J013055.87+452155.7	&	TYC 3278-614-1     	&	01 30 55.88 	&	 +45 21 55.77	&	kA1hA3mA4 SrCr   	&	734	&	9.2923	&	0.0006	&	+1.037	&	0.077	&	+0.112	&	0.003	&	0.14	&	$-$0.77	&	0.17	\\
34	&	J014016.31+422946.5	&	TYC 2823-1986-1    	&	01 40 16.31 	&	 +42 29 46.52	&	B9 V Cr   	&	113	&	12.5656	&	0.0007	&	+0.476	&	0.067	&	$-$0.081	&	0.005	&	0.13	&	+0.82	&	0.31	\\
35	&	J014508.87+322430.3	&	Gaia DR2 305411761460144896	&	01 45 08.88 	&	 +32 24 30.39	&	kA1hA6mA7 SrCrEu   	&	109	&	14.2946	&	0.0005	&	+0.382	&	0.054	&	+0.241	&	0.003	&	0.13	&	+2.08	&	0.31	\\
36	&	J014905.86+543820.9	&	Gaia DR2 408536189683896960	&	01 49 05.87 	&	 +54 38 21.00	&	kA0hA5mA6 Cr   	&	187	&	12.4605	&	0.0005	&	+0.615	&	0.045	&	+0.125	&	0.004	&	0.23	&	+1.17	&	0.17	\\
37	&	J014940.99+534134.2	&	TYC 3684-1139-1    	&	01 49 41.00 	&	 +53 41 34.26	&	\textit{B4} V \textit{HeB8 (Si)}* 	&	154	&	12.4184	&	0.0007	&	+0.327	&	0.038	&	$-$0.010	&	0.005	&	0.45	&	$-$0.46	&	0.26	\\
38	&	J015059.58+540259.1	&	HD 11140   	&	01 50 59.78 	&	 +54 03 02.16	&	B8 IV  bl4130  	&	691	&	8.5498	&	0.0006	&	+1.015	&	0.071	&	$-$0.110	&	0.004	&	0.20	&	$-$1.62	&	0.16	\\
39	&	J015458.39+541849.0	&	TYC 3684-2153-1    	&	01 54 58.39 	&	 +54 18 49.09	&	B9 V Cr   	&	367	&	11.3345	&	0.0014	&	+0.872	&	0.064	&	+0.116	&	0.005	&	0.20	&	+0.84	&	0.17	\\
40	&	J015545.33+543748.2	&	TYC 3688-1164-1    	&	01 55 45.33 	&	 +54 37 48.21	&	B9.5 III$-$IV Si   	&	152	&	11.8995	&	0.0005	&	+0.940	&	0.048	&	$-$0.004	&	0.003	&	0.41	&	+1.35	&	0.12	\\
41	&	J015559.12+563558.0	&	Gaia DR2 504860387608173184	&	01 55 59.13 	&	 +56 35 58.06	&	B9.5 II$-$III Si   	&	139	&	14.0054	&	0.0004	&	+0.408	&	0.027	&	+0.031	&	0.004	&	0.53	&	+1.53	&	0.15	\\
42	&	J015628.29+534409.9	&	TYC 3684-1671-1    	&	01 56 28.30 	&	 +53 44 09.94	&	B9.5 IV  bl4130  	&	336	&	11.8104	&	0.0005	&	+0.413	&	0.040	&	$-$0.052	&	0.003	&	0.26	&	$-$0.38	&	0.22	\\
43	&	J015654.63+543533.0	&	Gaia DR2 408394047746449280	&	01 56 54.64 	&	 +54 35 33.06	&	A1 II$-$III  bl4130  	&	177	&	12.9211	&	0.0003	&	+0.239	&	0.034	&	+0.001	&	0.003	&	0.40	&	$-$0.59	&	0.32	\\
44	&	J015729.58+542137.2	&	TYC 3684-165-1     	&	01 57 29.77 	&	 +54 21 36.51	&	B8 V Si   	&	496	&	10.5459	&	0.0012	&	+0.642	&	0.083	&	$-$0.054	&	0.007	&	0.32	&	$-$0.74	&	0.29	\\
45	&	J020004.86+560852.5	&	TYC 3689-200-1     	&	02 00 04.61 	&	 +56 08 51.37	&	B9 IV Si   	&	538	&	10.1176	&	0.0011	&	+4.935	&	0.314	&	+0.146	&	0.003	&	0.07	&	+3.52	&	0.15	\\
46	&	J020024.73+544319.0	&	Gaia DR2 504415600792977920	&	02 00 24.73 	&	 +54 43 19.06	&	A1 IV$-$V SrCr   	&	253	&	12.3203	&	0.0005	&	+0.477	&	0.037	&	+0.154	&	0.003	&	0.51	&	+0.21	&	0.17	\\
47	&	J020034.64+451914.2	&	TYC 3280-1768-1    	&	02 00 34.65 	&	 +45 19 14.24	&	kA1hA3mA6 CrEu   	&	215	&	11.4997	&	0.0008	&	+0.853	&	0.074	&	+0.071	&	0.004	&	0.13	&	+1.03	&	0.19	\\
48	&	J020228.03+565629.3	&	Gaia DR2 505090907095341824	&	02 02 28.03 	&	 +56 56 29.45	&	kB9.5hA5mA6 Cr   	&	103	&	14.0315	&	0.0005	&	+0.298	&	0.028	&	+0.143	&	0.003	&	0.58	&	+0.82	&	0.21	\\
49	&	J020338.02+533204.2	&	TYC 3685-1684-1    	&	02 03 38.02 	&	 +53 32 04.26	&	B8 V Si   	&	426	&	11.1758	&	0.0033	&	+0.742	&	0.058	&	$-$0.018	&	0.015	&	0.27	&	+0.26	&	0.18	\\
50	&	J020415.91+503019.4	&	HD 12532   	&	02 04 16.21 	&	 +50 30 19.66	&	B6 III Si   	&	687	&	9.5114	&	0.0011	&	+1.084	&	0.065	&	$-$0.097	&	0.006	&	0.29	&	$-$0.61	&	0.14	\\
\hline
\end{tabular}                                                                                                                                                                   
\end{adjustbox}
\end{center}                                                                                                                                             
\end{sidewaystable*}
\setcounter{table}{0}  
\begin{sidewaystable*}
\caption{Essential data for our sample stars, sorted by increasing right ascension. The columns denote: (1) Internal identification number. (2) LAMOST identifier. (3) Alternativ identifier (HD number, TYC identifier or GAIA DR2 number). (4) Right ascension (J2000; GAIA DR2). (5) Declination (J2000; GAIA DR2). (6) Spectral type, as derived in this study. (7) Sloan $g$ band S/N ratio of the analysed spectrum. (8) $G$\,mag (GAIA DR2). (9) $G$\,mag error. (10) Parallax (GAIA DR2). (11) Parallax error. (12) Dereddened colour index $(BP-RP){_0}$ (GAIA DR2). (13) Colour index error. (14) Absorption in the $G$ band, $A_G$. (15) Intrinsic absolute magnitude in the $G$ band, $M_{\mathrm{G,0}}$. (16) Absolute magnitude error.}
\label{table_master2}
\begin{center}
\begin{adjustbox}{max width=\textwidth}
\begin{tabular}{lllcclcccccccccc}
\hline
\hline
(1) & (2) & (3) & (4) & (5) & (6) & (7) & (8) & (9) & (10) & (11) & (12) & (13) & (14) & (15) & (16) \\
No.	&	ID\_LAMOST	&	ID\_alt	&	RA(J2000) 	&	 Dec(J2000)    	&	SpT\_final	&	S/N\,$g$	&	$G$\,mag	&	e\_$G$\,mag	&	pi (mas)	&	e\_pi	&	$(BP-RP){_0}$	&	e\_$(BP-RP){_0}$	&	$A_G$	&	$M_{\mathrm{G,0}}$	&	e\_$M_{\mathrm{G,0}}$	\\
\hline
51	&	J020417.01+553439.5	&	Gaia DR2 456542864521078144	&	02 04 17.04 	&	 +55 34 39.35	&	B9.5 IV Si   	&	169	&	12.4943	&	0.0072	&	$-$	&	$-$	&	$-$	&	$-$	&	$-$	&	$-$	&	$-$	\\
52	&	J020425.19+561142.4	&	TYC 3689-1567-1    	&	02 04 25.19 	&	 +56 11 42.54	&	B8 V Si   	&	305	&	11.4011	&	0.0026	&	+0.835	&	0.052	&	$-$0.092	&	0.011	&	0.48	&	+0.53	&	0.14	\\
53	&	J020842.44+544442.1	&	Gaia DR2 456423464428927360	&	02 08 42.45 	&	 +54 44 42.11	&	B9 III$-$IV Si   	&	122	&	13.3649	&	0.0008	&	+0.417	&	0.029	&	+0.074	&	0.004	&	0.54	&	+0.93	&	0.16	\\
54	&	J020921.65+471008.4	&	HD 13090   	&	02 09 21.45 	&	 +47 10 10.99	&	B9 III$-$IV Si   	&	298	&	9.0616	&	0.0006	&	+1.413	&	0.070	&	$-$0.167	&	0.003	&	0.26	&	$-$0.45	&	0.12	\\
55	&	J021106.70+492548.2	&	TYC 3289-2552-1    	&	02 11 06.70 	&	 +49 25 48.16	&	B9 III$-$IV Si   	&	181	&	12.3874	&	0.0005	&	+0.192	&	0.046	&	+0.005	&	0.004	&	0.36	&	$-$1.55	&	0.52	\\
56	&	J021147.71+515247.3	&	TYC 3293-1775-1    	&	02 11 47.77 	&	 +51 52 49.09	&	B9 III$-$IV bl4077 bl4130  	&	204	&	10.6664	&	0.0015	&	+0.717	&	0.037	&	+0.075	&	0.006	&	0.32	&	$-$0.38	&	0.12	\\
57	&	J021429.02+553451.3	&	HD 13592   	&	02 14 28.73 	&	 +55 34 53.10	&	B8 IV$-$V Si   	&	266	&	9.5172	&	0.0005	&	+1.757	&	0.051	&	$-$0.091	&	0.002	&	0.33	&	+0.41	&	0.08	\\
58	&	J021626.36+442611.8	&	TYC 2842-681-1     	&	02 16 26.36 	&	 +44 26 11.89	&	B9.5 III$-$IV EuSi   	&	882	&	9.3855	&	0.0008	&	+1.569	&	0.074	&	+0.035	&	0.005	&	0.17	&	+0.20	&	0.11	\\
59	&	J021927.75+420707.6	&	TYC 2838-1789-1    	&	02 19 27.75 	&	 +42 07 07.67	&	B8 IV$-$V $^{b}$	&	983	&	9.4303	&	0.0005	&	+1.109	&	0.083	&	$-$0.091	&	0.003	&	0.12	&	$-$0.47	&	0.17	\\
60	&	J022113.98+381906.4	&	Gaia DR2 331697064392371584	&	02 21 13.99 	&	 +38 19 06.36	&	A0 V Cr   	&	136	&	14.1572	&	0.0015	&	+0.369	&	0.050	&	+0.042	&	0.008	&	0.13	&	+1.87	&	0.30	\\
61	&	J022120.02+280415.6	&	HD 14522   	&	02 21 20.03 	&	 +28 04 15.86	&	A8 V SrEuSi   	&	198	&	8.7493	&	0.0005	&	+3.525	&	0.055	&	+0.206	&	0.003	&	0.23	&	+1.26	&	0.06	\\
62	&	J022252.38+485816.0	&	Gaia DR2 355271624484226944	&	02 22 52.39 	&	 +48 58 16.02	&	kA0hA3mA6 Cr   	&	134	&	13.3035	&	0.0004	&	+0.823	&	0.135	&	+0.160	&	0.003	&	0.28	&	+2.60	&	0.36	\\
63	&	J023335.11+530446.3	&	TYC 3687-654-1     	&	02 33 35.12 	&	 +53 04 46.41	&	B9.5 II$-$III Si   	&	120	&	11.6941	&	0.0007	&	+0.690	&	0.039	&	$-$0.065	&	0.003	&	0.58	&	+0.31	&	0.13	\\
64	&	J023534.25+520926.0	&	Gaia DR2 452192028288916096	&	02 35 34.25 	&	 +52 09 26.09	&	B9.5 II$-$III Si   	&	169	&	13.0338	&	0.0008	&	+0.402	&	0.037	&	+0.002	&	0.003	&	0.67	&	+0.38	&	0.21	\\
65	&	J023800.44+374400.0	&	Gaia DR2 333718035483431040	&	02 38 00.45 	&	 +37 44 00.04	&	A0 IV$-$V bl4077 bl4130  	&	114	&	12.5399	&	0.0006	&	+0.537	&	0.071	&	$-$0.020	&	0.004	&	0.15	&	+1.04	&	0.29	\\
66	&	J023805.20+513946.1	&	TYC 3308-1326-1    	&	02 38 05.20 	&	 +51 39 46.15	&	B8 IV Si   	&	147	&	11.8472	&	0.0013	&	+0.722	&	0.039	&	$-$0.011	&	0.006	&	0.67	&	+0.47	&	0.13	\\
67	&	J023844.12+510644.1	&	TYC 3308-1621-1    	&	02 38 44.13 	&	 +51 06 44.16	&	B9.5 IV bl4077 bl4130  	&	110	&	12.6169	&	0.0005	&	+0.551	&	0.049	&	+0.017	&	0.003	&	0.74	&	+0.58	&	0.20	\\
68	&	J023936.01+540059.5	&	Gaia DR2 453949185310404736	&	02 39 36.02 	&	 +54 00 59.65	&	kA2hA5mA7 CrEu   	&	163	&	12.3750	&	0.0009	&	+1.085	&	0.030	&	+0.511	&	0.007	&	0.64	&	+1.91	&	0.08	\\
69	&	J024028.73+473922.8	&	TYC 3300-2476-1    	&	02 40 28.74 	&	 +47 39 22.83	&	B9 V Cr   	&	116	&	11.7478	&	0.0006	&	+0.823	&	0.045	&	+0.089	&	0.004	&	0.27	&	+1.05	&	0.13	\\
70	&	J024243.57+453720.4	&	TYC 3296-281-1     	&	02 42 42.76 	&	 +45 37 20.73	&	A1 V SrCrEuSi   	&	142	&	11.9910	&	0.0006	&	+1.019	&	0.037	&	+0.311	&	0.003	&	0.17	&	+1.86	&	0.09	\\
71	&	J024542.78+555858.9	&	Gaia DR2 454542577989769472	&	02 45 42.78 	&	 +55 58 59.00	&	kA0hA4mA6 CrSi   	&	163	&	11.8698	&	0.0004	&	+1.010	&	0.033	&	+0.061	&	0.002	&	1.02	&	+0.87	&	0.09	\\
72	&	J025123.15+455720.9	&	TYC 3297-693-1     	&	02 51 23.16 	&	 +45 57 20.94	&	B9.5 IV$-$V \textit{Cr}*	&	318	&	11.1158	&	0.0006	&	+0.625	&	0.047	&	+0.090	&	0.002	&	0.33	&	$-$0.23	&	0.17	\\
73	&	J025317.44+465342.6	&	Gaia DR2 437431973738031744	&	02 53 17.45 	&	 +46 53 42.66	&	kA3hA3mA7 CrEu   	&	135	&	12.7318	&	0.0004	&	+0.458	&	0.039	&	+0.208	&	0.003	&	0.25	&	+0.78	&	0.19	\\
74	&	J025507.91+463730.1	&	Gaia DR2 434413749897657728	&	02 55 07.91 	&	 +46 37 30.19	&	A0 V SrCr   	&	96	&	14.5096	&	0.0004	&	+0.394	&	0.034	&	+0.060	&	0.002	&	0.63	&	+1.86	&	0.19	\\
75	&	J025708.41+241901.3	&	TYC 1782-825-1     	&	02 57 08.41 	&	 +24 19 01.31	&	A9 V SrEu   	&	170	&	11.5772	&	0.0007	&	+0.938	&	0.042	&	+0.237	&	0.003	&	0.22	&	+1.22	&	0.11	\\
76	&	J025716.49+572715.7	&	TYC 3709-175-1     	&	02 57 16.36 	&	 +57 27 15.73	&	B9 V Eu   	&	221	&	10.7272	&	0.0007	&	+0.870	&	0.155	&	$-$0.840	&	0.003	&	3.06	&	$-$2.64	&	0.39	\\
77	&	J025945.31+541941.5 $^{a}$	&	HD 18410   	&	02 59 45.28 	&	 +54 19 44.96	&	kB7hA7mA6 CrEuSi   	&	770	&	9.0794	&	0.0012	&	+2.858	&	0.035	&	+0.167	&	0.006	&	0.41	&	+0.95	&	0.06	\\
78	&	J025951.09+540337.5	&	TYC 3701-157-1     	&	02 59 51.20 	&	 +54 03 39.14	&	B8 III$-$IV Si   	&	409	&	10.5057	&	0.0011	&	+1.183	&	0.045	&	0.000	&	0.004	&	1.33	&	$-$0.46	&	0.10	\\
79	&	J030339.08+472125.5	&	Gaia DR2 435816451857186048	&	03 03 39.08 	&	 +47 21 25.54	&	kA0hA3mA5 CrSi   	&	160	&	12.1342	&	0.0006	&	+0.501	&	0.032	&	+0.123	&	0.005	&	0.48	&	+0.15	&	0.15	\\
80	&	J030350.21+463718.3	&	TYC 3310-1808-1    	&	03 03 50.21 	&	 +46 37 18.37	&	A8 V SrCrEuSi   	&	653	&	9.8183	&	0.0003	&	+1.688	&	0.040	&	+0.282	&	0.002	&	0.33	&	+0.63	&	0.07	\\
81	&	J030459.32+351024.5	&	Gaia DR2 138608975579548544	&	03 04 59.33 	&	 +35 10 24.58	&	A7 V SrCrEu   	&	130	&	12.3721	&	0.0003	&	+0.831	&	0.052	&	+0.340	&	0.002	&	0.40	&	+1.57	&	0.14	\\
82	&	J030614.98+485615.0	&	TYC 3318-18-1      	&	03 06 14.91 	&	 +48 56 17.16	&	kB9hA5mA3 SrCrEu   	&	431	&	10.1998	&	0.0007	&	+1.257	&	0.044	&	+0.039	&	0.002	&	0.81	&	$-$0.11	&	0.09	\\
83	&	J030633.66+025615.7	&	TYC 58-1131-1      	&	03 06 33.67 	&	 +02 56 15.77	&	B9.5 IV$-$V CrEu   	&	301	&	11.4566	&	0.0008	&	+0.808	&	0.089	&	+0.019	&	0.002	&	0.25	&	+0.75	&	0.24	\\
84	&	J030655.59+492941.0	&	Gaia DR2 436280858083105024	&	03 06 55.59 	&	 +49 29 41.04	&	A0 IV$-$V Cr   	&	111	&	12.2319	&	0.0003	&	+1.184	&	0.032	&	+0.252	&	0.002	&	0.89	&	+1.71	&	0.08	\\
85	&	J030708.46+460837.7	&	TYC 3310-1240-1    	&	03 07 08.47 	&	 +46 08 37.72	&	B9.5 IV bl4077 bl4130  	&	222	&	11.0603	&	0.0012	&	+1.249	&	0.048	&	+0.004	&	0.006	&	0.42	&	+1.12	&	0.10	\\
86	&	J030709.75+535142.4	&	TYC 3702-136-1     	&	03 07 09.72 	&	 +53 51 42.60	&	B8 III$-$IV  bl4130  	&	194	&	10.1437	&	0.0010	&	+1.434	&	0.125	&	$-$0.142	&	0.003	&	1.39	&	$-$0.47	&	0.20	\\
87	&	J030759.91+452730.7	&	TYC 3310-129-1     	&	03 07 59.79 	&	 +45 27 30.71	&	B9 IV  bl4130  	&	411	&	10.6727	&	0.0009	&	$-$	&	$-$	&	$-$	&	$-$	&	$-$	&	$-$	&	$-$	\\
88	&	J030837.10+364000.8	&	Gaia DR2 139337264594236288	&	03 08 37.11 	&	 +36 40 00.81	&	A0 V Cr   	&	104	&	14.3423	&	0.0008	&	+0.094	&	0.052	&	$-$	&	$-$	&	$-$	&	$-$	&	$-$	\\
89	&	J030915.26+262955.5	&	TYC 1791-480-1     	&	03 09 15.27 	&	 +26 29 55.58	&	A3 V Sr   	&	217	&	11.2401	&	0.0013	&	+1.735	&	0.054	&	+0.162	&	0.002	&	0.45	&	+1.99	&	0.08	\\
90	&	J030937.10+451010.4	&	TYC 3310-2246-1    	&	03 09 37.11 	&	 +45 10 10.51	&	\textit{A0 V Cr}*	&	136	&	11.8616	&	0.0008	&	+0.594	&	0.042	&	+0.074	&	0.005	&	0.45	&	+0.28	&	0.16	\\
91	&	J031043.71+480727.8	&	Gaia DR2 435920218270165504	&	03 10 43.72 	&	 +48 07 27.87	&	A0 IV$-$V SrCrEu   	&	160	&	12.7285	&	0.0002	&	+0.663	&	0.032	&	+0.181	&	0.001	&	0.66	&	+1.18	&	0.11	\\
92	&	J031054.55+462725.9	&	TYC 3310-1345-1    	&	03 10 54.42 	&	 +46 27 25.89	&	kA1hA4mA6 SrCrEu   	&	421	&	10.6114	&	0.0011	&	+1.043	&	0.067	&	+0.171	&	0.004	&	0.52	&	+0.19	&	0.15	\\
93	&	J031100.46+443818.2	&	TYC 2860-145-1     	&	03 11 00.46 	&	 +44 38 18.20	&	A1 II$-$III Si   	&	246	&	11.5547	&	0.0009	&	$-$0.513	&	0.191	&	$-$	&	$-$	&	$-$	&	$-$	&	$-$	\\
94	&	J031111.67+461637.6	&	TYC 3310-2115-1    	&	03 11 11.68 	&	 +46 16 37.67	&	B9.5 IV$-$V CrEu   	&	185	&	12.2008	&	0.0014	&	+0.725	&	0.038	&	+0.111	&	0.006	&	0.47	&	+1.03	&	0.12	\\
95	&	J031142.28+080707.6	&	HD 19846   	&	03 11 42.05 	&	 +08 07 07.60	&	B9 IV$-$V Eu   	&	695	&	8.5098	&	0.0007	&	+3.773	&	0.057	&	$-$0.414	&	0.004	&	0.98	&	+0.42	&	0.06	\\
96	&	J031244.57+113233.7	&	TYC 651-963-1      	&	03 12 44.56 	&	 +11 32 34.00	&	B9.5 IV Cr   	&	154	&	10.0352	&	0.0009	&	+1.951	&	0.047	&	+0.024	&	0.003	&	0.67	&	+0.81	&	0.07	\\
97	&	J031315.28+331930.4	&	Gaia DR2 137397859224918016	&	03 13 15.29 	&	 +33 19 30.48	&	A0 V CrEu   	&	190	&	12.0185	&	0.0005	&	+1.345	&	0.057	&	+0.232	&	0.003	&	0.42	&	+2.24	&	0.11	\\
98	&	J031536.60+430106.0	&	TYC 2856-451-1     	&	03 15 36.63 	&	 +43 01 03.49	&	B8 IV \textit{Si}*	&	297	&	10.0378	&	0.0007	&	+1.417	&	0.068	&	$-$0.080	&	0.003	&	0.35	&	+0.44	&	0.12	\\
99	&	J031714.31+565041.4	&	TYC 3710-812-1     	&	03 17 14.32 	&	 +56 50 41.40	&	B8 IV CrSi  (He-wk) 	&	247	&	11.0567	&	0.0005	&	+0.903	&	0.038	&	$-$0.185	&	0.002	&	1.96	&	$-$1.12	&	0.10	\\
100	&	J031722.65+490836.3	&	TYC 3319-464-1     	&	03 17 22.36 	&	 +49 08 36.59	&	B9.5 IV$-$V CrEu   	&	276	&	9.6825	&	0.0009	&	+2.436	&	0.041	&	+0.048	&	0.004	&	0.66	&	+0.96	&	0.06	\\
\hline
\end{tabular}                                                                                                                                                                   
\end{adjustbox}
\end{center}                                                                                                                                             
\end{sidewaystable*}
\setcounter{table}{0}  
\begin{sidewaystable*}
\caption{Essential data for our sample stars, sorted by increasing right ascension. The columns denote: (1) Internal identification number. (2) LAMOST identifier. (3) Alternativ identifier (HD number, TYC identifier or GAIA DR2 number). (4) Right ascension (J2000; GAIA DR2). (5) Declination (J2000; GAIA DR2). (6) Spectral type, as derived in this study. (7) Sloan $g$ band S/N ratio of the analysed spectrum. (8) $G$\,mag (GAIA DR2). (9) $G$\,mag error. (10) Parallax (GAIA DR2). (11) Parallax error. (12) Dereddened colour index $(BP-RP){_0}$ (GAIA DR2). (13) Colour index error. (14) Absorption in the $G$ band, $A_G$. (15) Intrinsic absolute magnitude in the $G$ band, $M_{\mathrm{G,0}}$. (16) Absolute magnitude error.}
\label{table_master3}
\begin{center}
\begin{adjustbox}{max width=\textwidth}
\begin{tabular}{lllcclcccccccccc}
\hline
\hline
(1) & (2) & (3) & (4) & (5) & (6) & (7) & (8) & (9) & (10) & (11) & (12) & (13) & (14) & (15) & (16) \\
No.	&	ID\_LAMOST	&	ID\_alt	&	RA(J2000) 	&	 Dec(J2000)    	&	SpT\_final	&	S/N\,$g$	&	$G$\,mag	&	e\_$G$\,mag	&	pi (mas)	&	e\_pi	&	$(BP-RP){_0}$	&	e\_$(BP-RP){_0}$	&	$A_G$	&	$M_{\mathrm{G,0}}$	&	e\_$M_{\mathrm{G,0}}$	\\
\hline 
101	&	J032020.62+485347.2	&	TYC 3319-1872-1    	&	03 20 20.63 	&	 +48 53 47.20	&	B5 III$-$IV  bl4130  	&	306	&	11.3562	&	0.0012	&	+0.956	&	0.041	&	+0.089	&	0.003	&	0.99	&	+0.27	&	0.11	\\
102	&	J032111.38+530600.5	&	TYC 3703-715-1     	&	03 21 11.38 	&	 +53 06 00.52	&	B8 IV$-$V EuSi   	&	154	&	11.3153	&	0.0015	&	+0.467	&	0.199	&	$-$	&	$-$	&	$-$	&	$-$	&	$-$	\\
103	&	J032122.27+445902.7	&	TYC 2873-3205-1    	&	03 21 22.28 	&	 +44 59 00.12	&	B9.5 V bl4077 bl4130  	&	209	&	9.9483	&	0.0006	&	+2.086	&	0.058	&	+0.001	&	0.002	&	0.32	&	+1.22	&	0.08	\\
104	&	J032224.63+441318.6	&	TYC 2873-1033-1    	&	03 22 24.64 	&	 +44 13 18.62	&	A8 V SrCr   	&	165	&	11.4434	&	0.0012	&	+1.370	&	0.077	&	+0.301	&	0.002	&	0.39	&	+1.74	&	0.13	\\
105	&	J032252.44+320514.7	&	Gaia DR2 124329274471999232	&	03 22 52.45 	&	 +32 05 14.71	&	A0 V SrCrEuSi   	&	220	&	11.8053	&	0.0003	&	+0.718	&	0.067	&	+0.221	&	0.002	&	0.60	&	+0.49	&	0.21	\\
106	&	J032333.76+510336.6	&	Gaia DR2 442651497173870080	&	03 23 33.76 	&	 +51 03 36.63	&	A2 V CrEuSi   	&	107	&	12.9872	&	0.0006	&	+0.880	&	0.031	&	+0.313	&	0.002	&	1.34	&	+1.37	&	0.09	\\
107	&	J032343.27+442245.4	&	Gaia DR2 241472445889902080	&	03 23 43.27 	&	 +44 22 45.50	&	A0 Ib$-$II EuSi   	&	108	&	14.6248	&	0.0041	&	+0.243	&	0.036	&	+0.010	&	0.02	&	0.41	&	+1.14	&	0.33	\\
108	&	J032413.45+525937.5	&	TYC 3703-669-1     	&	03 24 13.45 	&	 +52 59 37.59	&	B7 V  bl4130  	&	124	&	11.5487	&	0.0016	&	+0.747	&	0.034	&	$-$0.074	&	0.006	&	1.56	&	$-$0.64	&	0.11	\\
109	&	J032611.50+495758.8	&	TYC 3320-16-1      	&	03 26 11.50 	&	 +49 57 58.83	&	B8 V Si   	&	136	&	12.2755	&	0.0006	&	+0.591	&	0.040	&	$-$0.056	&	0.003	&	0.84	&	+0.30	&	0.16	\\
110	&	J032719.03+142144.4	&	Gaia DR2 41984099888203392	&	03 27 19.04 	&	 +14 21 44.25	&	B9 V CrEu   	&	186	&	12.7108	&	0.0004	&	+0.473	&	0.045	&	$-$0.018	&	0.004	&	0.61	&	+0.47	&	0.21	\\
111	&	J032747.71+453527.0	&	TYC 3312-1515-1    	&	03 27 47.72 	&	 +45 35 27.10	&	B9 III$-$IV Cr   	&	104	&	11.9996	&	0.0010	&	+0.712	&	0.039	&	+0.024	&	0.004	&	0.51	&	+0.75	&	0.13	\\
112	&	J032823.89+542139.1	&	TYC 3703-2-1       	&	03 28 23.90 	&	 +54 21 39.08	&	B9.5 V SrCr   	&	179	&	11.5554	&	0.0006	&	+1.048	&	0.059	&	+0.024	&	0.002	&	1.58	&	+0.08	&	0.13	\\
113	&	J032856.39+475618.4	&	TYC 3316-892-1     	&	03 28 56.69 	&	 +47 56 20.09	&	B9 IV$-$V  bl4130  	&	463	&	9.6703	&	0.0005	&	+1.448	&	0.048	&	$-$0.037	&	0.002	&	0.69	&	$-$0.22	&	0.09	\\
114	&	J032939.33+060540.8	&	TYC 63-463-1       	&	03 29 39.35 	&	 +06 05 40.82	&	B9 V CrEuSi   	&	303	&	11.5663	&	0.0007	&	+1.036	&	0.038	&	$-$0.038	&	0.003	&	0.71	&	+0.93	&	0.09	\\
115	&	J033003.45+563459.0	&	Gaia DR2 448609132207300096	&	03 30 03.45 	&	 +56 34 58.99	&	A1 IV$-$V CrEu   	&	148	&	11.9160	&	0.0004	&	+1.603	&	0.042	&	+0.074	&	0.002	&	1.39	&	+1.55	&	0.08	\\
116	&	J033053.47+550051.2	&	TYC 3707-124-1     	&	03 30 53.18 	&	 +55 00 51.27	&	B9 IV$-$V Si   	&	258	&	9.9624	&	0.0015	&	+2.276	&	0.055	&	$-$0.101	&	0.006	&	0.86	&	+0.89	&	0.07	\\
117	&	J033221.41+472318.3	&	TYC 3316-1385-1    	&	03 32 21.19 	&	 +47 23 17.08	&	B9 V  bl4130  	&	216	&	10.2552	&	0.0006	&	+2.000	&	0.043	&	$-$0.191	&	0.002	&	1.10	&	+0.66	&	0.07	\\
118	&	J033228.57+433038.0	&	TYC 2874-2420-1    	&	03 32 28.58 	&	 +43 30 38.26	&	B8 III$-$IV EuSi   	&	146	&	11.1063	&	0.0026	&	+0.635	&	0.066	&	+0.048	&	0.008	&	0.43	&	$-$0.31	&	0.23	\\
119	&	J033325.39+452124.0	&	TYC 3312-694-1     	&	03 33 25.32 	&	 +45 21 24.29	&	A0 V SrCr   	&	324	&	11.2030	&	0.0009	&	+1.128	&	0.051	&	+0.179	&	0.002	&	0.60	&	+0.87	&	0.11	\\
120	&	J033341.86+542708.3	&	TYC 3720-422-1     	&	03 33 41.87 	&	 +54 27 08.35	&	B8 III$-$IV EuSi   	&	180	&	11.4739	&	0.0006	&	+0.979	&	0.034	&	$-$0.006	&	0.003	&	1.50	&	$-$0.07	&	0.09	\\
121	&	J033348.98+334153.1	&	HD 278822  	&	03 33 48.99 	&	 +33 41 53.05	&	B6 III$-$IV Si   	&	256	&	9.9195	&	0.0005	&	+1.706	&	0.066	&	$-$0.004	&	0.002	&	1.21	&	$-$0.13	&	0.10	\\
122	&	J033359.91+564753.9	&	Gaia DR2 448596934500298880	&	03 33 59.92 	&	 +56 47 53.88	&	A0 II$-$III EuSi   	&	145	&	12.1883	&	0.0040	&	+0.515	&	0.204	&	$-$	&	$-$	&	$-$	&	$-$	&	$-$	\\
123	&	J033512.23+490936.7	&	Gaia DR2 249686244426935296	&	03 35 12.24 	&	 +49 09 36.76	&	B9.5 II$-$III CrSi   	&	120	&	12.7989	&	0.0007	&	+0.487	&	0.038	&	+0.106	&	0.003	&	0.76	&	+0.48	&	0.18	\\
124	&	J033546.83+512723.7	&	TYC 3325-152-1     	&	03 35 46.83 	&	 +51 27 23.73	&	B9 V Cr   	&	111	&	11.6677	&	0.0008	&	+1.249	&	0.037	&	$-$0.761	&	0.003	&	3.16	&	$-$1.01	&	0.08	\\
125	&	J033611.64+582252.1	&	TYC 3728-990-1     	&	03 36 11.65 	&	 +58 22 52.22	&	B8 III$-$IV Si   	&	109	&	12.0052	&	0.0006	&	+0.854	&	0.066	&	$-$0.352	&	0.003	&	1.97	&	$-$0.31	&	0.17	\\
126	&	J033620.54+575723.8	&	TYC 3724-886-1     	&	03 36 20.55 	&	 +57 57 23.90	&	B8 IV$-$V Si   	&	160	&	11.9796	&	0.0003	&	+1.021	&	0.036	&	+0.189	&	0.001	&	1.30	&	+0.72	&	0.09	\\
127	&	J033648.54+571248.6	&	Gaia DR2 448982588204437120	&	03 36 48.54 	&	 +57 12 48.62	&	B9.5 V Cr   	&	142	&	11.9229	&	0.0004	&	+1.106	&	0.044	&	+0.239	&	0.002	&	1.50	&	+0.64	&	0.10	\\
128	&	J033726.54+523401.6	&	TYC 3716-278-1     	&	03 37 26.55 	&	 +52 34 01.65	&	A6 V SrCrEu   	&	149	&	11.5672	&	0.0007	&	+1.715	&	0.033	&	+0.373	&	0.002	&	1.10	&	+1.63	&	0.07	\\
129	&	J033742.76+494930.0	&	Gaia DR2 249729468977400832	&	03 37 42.77 	&	 +49 49 30.06	&	B9 III$-$IV Si   	&	104	&	13.3761	&	0.0003	&	+0.373	&	0.025	&	$-$0.034	&	0.002	&	1.18	&	+0.05	&	0.15	\\
130	&	J034000.57+444858.4	&	TYC 2875-2489-1    	&	03 40 00.58 	&	 +44 48 58.26	&	kA0hA1mA3 \textit{(Si)}*	&	771	&	9.8025	&	0.0007	&	+1.479	&	0.769	&	$-$	&	$-$	&	$-$	&	$-$	&	$-$	\\
131	&	J034036.63+455233.0	&	TYC 3313-1894-1    	&	03 40 36.45 	&	 +45 52 32.95	&	kA0hA2mA5 Cr   	&	216	&	10.5868	&	0.0007	&	+0.318	&	0.458	&	$-$	&	$-$	&	$-$	&	$-$	&	$-$	\\
132	&	J034112.38+453031.7	&	TYC 3313-1037-1    	&	03 41 12.10 	&	 +45 30 31.79	&	A1 V SrCrEu   	&	336	&	9.7342	&	0.0004	&	+2.202	&	0.054	&	+0.036	&	0.001	&	0.56	&	+0.88	&	0.07	\\
133	&	J034114.89+483916.7	&	TYC 3317-108-1     	&	03 41 14.90 	&	 +48 39 16.70	&	A0 IV$-$V Sr   	&	118	&	11.9607	&	0.0008	&	+0.933	&	0.035	&	+0.155	&	0.003	&	0.81	&	+1.00	&	0.10	\\
134	&	J034229.41+353820.9	&	HD 22961   	&	03 42 29.42 	&	 +35 38 20.93	&	A0 IV$-$V bl4077 bl4130  	&	800	&	9.5145	&	0.0003	&	+3.312	&	0.039	&	$-$0.251	&	0.001	&	1.00	&	+1.12	&	0.06	\\
135	&	J034306.74+495240.7	&	TYC 3321-881-1     	&	03 43 06.74 	&	 +49 52 40.80	&	\textit{B9 III Si}* &	258	&	11.9711	&	0.0006	&	+0.475	&	0.038	&	+0.085	&	0.002	&	1.06	&	$-$0.70	&	0.18	\\
136	&	J034312.35+581724.9	&	Gaia DR2 449154489978207488	&	03 43 12.35 	&	 +58 17 25.02	&	A1 IV SrCr   	&	154	&	11.8213	&	0.0005	&	+1.151	&	0.029	&	$-$0.058	&	0.002	&	1.44	&	+0.68	&	0.07	\\
137	&	J034411.65+500117.8	&	TYC 3321-374-1     	&	03 44 11.34 	&	 +50 01 17.86	&	B7 IV$-$V Si   	&	481	&	9.6221	&	0.0005	&	+1.555	&	0.039	&	$-$0.218	&	0.002	&	0.82	&	$-$0.24	&	0.07	\\
138	&	J034417.13+494336.6	&	TYC 3321-1539-1    	&	03 44 17.38 	&	 +49 43 34.76	&	B8 IV$-$V CrSi   	&	457	&	9.6305	&	0.0006	&	+1.238	&	0.041	&	$-$0.101	&	0.002	&	0.86	&	$-$0.77	&	0.09	\\
139	&	J034458.31+464848.7	&	TYC 3313-1279-1    	&	03 44 58.31 	&	 +46 48 48.78	&	B9 III$-$IV Si   	&	389	&	11.3576	&	0.0011	&	+1.024	&	0.052	&	+0.116	&	0.004	&	0.65	&	+0.76	&	0.12	\\
140	&	J034514.32+294335.4	&	HD 281193  	&	03 45 14.32 	&	 +29 43 35.50	&	A4 IV CrEuSi   	&	136	&	9.9949	&	0.0005	&	+2.384	&	0.085	&	+0.115	&	0.002	&	0.45	&	+1.43	&	0.09	\\
141	&	J034525.09+523642.2	&	TYC 3716-571-1     	&	03 45 25.09 	&	 +52 36 42.26	&	A1 IV$-$V CrEu   	&	237	&	11.8307	&	0.0005	&	+1.304	&	0.040	&	+0.255	&	0.002	&	1.01	&	+1.40	&	0.08	\\
142	&	J034541.53+275631.8	&	TYC 1807-159-1     	&	03 45 41.41 	&	 +27 56 31.86	&	\textit{kA3hA6mA6} SrCrEu* $^{d}$  	&	259	&	10.5867	&	0.0008	&	+1.279	&	0.088	&	+0.195	&	0.003	&	0.22	&	+0.90	&	0.16	\\
143	&	J034543.16+583801.1	&	Gaia DR2 473184522761462400	&	03 45 43.17 	&	 +58 38 01.09	&	A0 IV  bl4130  	&	129	&	14.3359	&	0.0002	&	+0.434	&	0.025	&	+0.228	&	0.002	&	1.19	&	+1.33	&	0.13	\\
144	&	J034550.78+592754.7	&	Gaia DR2 474041145398913024	&	03 45 50.78 	&	 +59 27 54.74	&	B9 IV Eu   	&	129	&	12.2978	&	0.0017	&	+0.964	&	0.030	&	+0.299	&	0.008	&	1.08	&	+1.13	&	0.08	\\
145	&	J034614.32+514056.5	&	Gaia DR2 251478379660219904	&	03 46 14.33 	&	 +51 40 56.51	&	B8 V Si  (He-wk) 	&	152	&	12.3500	&	0.0009	&	+0.794	&	0.036	&	+0.097	&	0.005	&	1.15	&	+0.70	&	0.11	\\
146	&	J034704.76+305136.2	&	HD 281171  	&	03 47 04.77 	&	 +30 51 36.04	&	A4 IV$-$V Sr   	&	164	&	11.0769	&	0.0009	&	+1.616	&	0.099	&	+0.314	&	0.002	&	0.77	&	+1.35	&	0.14	\\
147	&	J034731.69+394858.2	&	HD 275801  	&	03 47 31.70 	&	 +39 48 58.31	&	B9 IV$-$V Cr   	&	238	&	11.4844	&	0.0007	&	+0.800	&	0.056	&	+0.073	&	0.003	&	0.57	&	+0.43	&	0.16	\\
148	&	J034828.34+460128.5	&	TYC 3326-2034-1    	&	03 48 28.35 	&	 +46 01 28.77	&	A0 II$-$III CrEu   	&	395	&	11.1605	&	0.0024	&	$-$0.534	&	0.518	&	$-$	&	$-$	&	$-$	&	$-$	&	$-$	\\
149	&	J034834.07+391150.8	&	TYC 2863-759-1     	&	03 48 34.07 	&	 +39 11 50.92	&	A0 III$-$IV Eu   	&	194	&	11.7411	&	0.0008	&	+1.146	&	0.043	&	+0.101	&	0.004	&	0.81	&	+1.23	&	0.10	\\
150	&	J034854.70+521413.1	&	Gaia DR2 251609324623302400	&	03 48 54.70 	&	 +52 14 13.13	&	A5 V SrCrEu   	&	147	&	12.8362	&	0.0003	&	+0.999	&	0.031	&	+0.267	&	0.002	&	1.07	&	+1.76	&	0.08	\\
\hline
\end{tabular}                                                                                                                                                                   
\end{adjustbox}
\end{center}                                                                                                                                             
\end{sidewaystable*}
\setcounter{table}{0}  
\begin{sidewaystable*}
\caption{Essential data for our sample stars, sorted by increasing right ascension. The columns denote: (1) Internal identification number. (2) LAMOST identifier. (3) Alternativ identifier (HD number, TYC identifier or GAIA DR2 number). (4) Right ascension (J2000; GAIA DR2). (5) Declination (J2000; GAIA DR2). (6) Spectral type, as derived in this study. (7) Sloan $g$ band S/N ratio of the analysed spectrum. (8) $G$\,mag (GAIA DR2). (9) $G$\,mag error. (10) Parallax (GAIA DR2). (11) Parallax error. (12) Dereddened colour index $(BP-RP){_0}$ (GAIA DR2). (13) Colour index error. (14) Absorption in the $G$ band, $A_G$. (15) Intrinsic absolute magnitude in the $G$ band, $M_{\mathrm{G,0}}$. (16) Absolute magnitude error.}
\label{table_master4}
\begin{center}
\begin{adjustbox}{max width=\textwidth}
\begin{tabular}{lllcclcccccccccc}
\hline
\hline
(1) & (2) & (3) & (4) & (5) & (6) & (7) & (8) & (9) & (10) & (11) & (12) & (13) & (14) & (15) & (16) \\
No.	&	ID\_LAMOST	&	ID\_alt	&	RA(J2000) 	&	 Dec(J2000)    	&	SpT\_final	&	S/N\,$g$	&	$G$\,mag	&	e\_$G$\,mag	&	pi (mas)	&	e\_pi	&	$(BP-RP){_0}$	&	e\_$(BP-RP){_0}$	&	$A_G$	&	$M_{\mathrm{G,0}}$	&	e\_$M_{\mathrm{G,0}}$	\\
\hline
151	&	J035046.03+363648.2	&	Gaia DR2 220081859486642816	&	03 50 46.04 	&	 +36 36 48.27	&	B8 \textit{V} Si* $^{d}$   	&	129	&	12.4742	&	0.0004	&	+0.384	&	0.048	&	$-$0.119	&	0.003	&	0.51	&	$-$0.11	&	0.27	\\
152	&	J035222.92+325344.1	&	HD 279137  	&	03 52 22.93 	&	 +32 53 44.13	&	\textit{B9 IV (Cr)}*	&	138	&	10.8883	&	0.0006	&	+1.439	&	0.055	&	$-$0.031	&	0.002	&	0.47	&	+1.21	&	0.10	\\
153	&	J035238.56+284808.0	&	Gaia DR2 167014411808069504	&	03 52 38.57 	&	 +28 48 08.01	&	B9.5 III CrEu   	&	84	&	14.3013	&	0.0008	&	+0.070	&	0.049	&	$-$	&	$-$	&	$-$	&	$-$	&	$-$	\\
154	&	J035323.41+511945.5	&	Gaia DR2 251273492539058560	&	03 53 23.42 	&	 +51 19 45.51	&	B9 III  bl4130  	&	136	&	12.4213	&	0.0007	&	+0.455	&	0.590	&	$-$	&	$-$	&	$-$	&	$-$	&	$-$	\\
155	&	J035328.65+401211.4	&	Gaia DR2 230000141564434432	&	03 53 28.65 	&	 +40 12 11.46	&	B9 IV$-$V CrSi   	&	248	&	11.7017	&	0.0008	&	+1.274	&	0.033	&	+0.058	&	0.003	&	1.08	&	+1.15	&	0.08	\\
156	&	J035402.92+483618.9	&	Gaia DR2 246983193871427072	&	03 54 02.92 	&	 +48 36 18.93	&	B8 IV$-$V EuSi   	&	124	&	12.0396	&	0.0007	&	+0.886	&	0.036	&	+0.070	&	0.005	&	1.19	&	+0.59	&	0.10	\\
157	&	J035458.95+524755.5	&	TYC 3717-914-1     	&	03 54 58.84 	&	 +52 47 54.37	&	B9 IV Si   	&	370	&	10.7480	&	0.0005	&	+1.066	&	0.056	&	$-$0.264	&	0.002	&	1.71	&	$-$0.83	&	0.13	\\
158	&	J035502.04+422421.6	&	Gaia DR2 231448438897036800	&	03 55 02.05 	&	 +42 24 21.67	&	A0 II$-$III bl4077 bl4130  	&	117	&	14.2345	&	0.0006	&	+0.355	&	0.029	&	+0.017	&	0.003	&	1.00	&	+0.98	&	0.18	\\
159	&	J035508.23+444208.1	&	TYC 2876-1121-1    	&	03 55 08.23 	&	 +44 42 08.23	&	B9.5 II$-$III  bl4130  	&	203	&	12.1215	&	0.0003	&	+0.401	&	0.042	&	+0.013	&	0.002	&	0.92	&	$-$0.78	&	0.23	\\
160	&	J035601.03+473843.6	&	TYC 3330-2807-1    	&	03 56 01.04 	&	 +47 38 43.68	&	B8 IV Si  (He-wk) 	&	457	&	10.1658	&	0.0019	&	+1.490	&	0.120	&	$-$0.065	&	0.005	&	0.93	&	+0.10	&	0.18	\\
161	&	J035642.77+513747.2	&	TYC 3339-770-1     	&	03 56 42.77 	&	 +51 37 47.19	&	A0 IV$-$V Cr   	&	343	&	11.2870	&	0.0009	&	+1.030	&	0.050	&	+0.150	&	0.002	&	0.94	&	+0.41	&	0.12	\\
162	&	J035700.35+240805.7	&	HD 283172  	&	03 57 00.36 	&	 +24 08 05.79	&	B9 V CrSi   	&	243	&	11.3804	&	0.0009	&	+0.928	&	0.079	&	$-$0.052	&	0.004	&	0.31	&	+0.91	&	0.19	\\
163	&	J035716.27+344839.6	&	HD 279178  	&	03 57 16.27 	&	 +34 48 39.65	&	A7 V SrCrEu   	&	250	&	11.4612	&	0.0006	&	+1.230	&	0.041	&	+0.198	&	0.002	&	0.48	&	+1.43	&	0.09	\\
164	&	J035738.10+522745.9	&	TYC 3339-300-1     	&	03 57 38.11 	&	 +52 27 45.89	&	B9.5 III$-$IV Si   	&	297	&	11.2969	&	0.0008	&	+0.825	&	0.048	&	$-$0.047	&	0.002	&	1.37	&	$-$0.49	&	0.14	\\
165	&	J035752.06+580323.1	&	TYC 3725-1418-1    	&	03 57 52.06 	&	 +58 03 23.11	&	B9.5 II$-$III Si   	&	136	&	12.4059	&	0.0004	&	+0.453	&	0.040	&	+0.041	&	0.003	&	1.27	&	$-$0.58	&	0.20	\\
166	&	J035828.16+555117.3	&	Gaia DR2 468737788503660288	&	03 58 28.16 	&	 +55 51 17.35	&	A0 III$-$IV CrEu   	&	176	&	12.2127	&	0.0005	&	+1.135	&	0.032	&	+0.148	&	0.002	&	1.09	&	+1.40	&	0.08	\\
167	&	J035945.29+331757.7	&	TYC 2361-392-1     	&	03 59 45.30 	&	 +33 17 57.72	&	B8 IV Si   	&	141	&	12.6797	&	0.0003	&	+0.409	&	0.039	&	+0.025	&	0.002	&	0.71	&	+0.03	&	0.21	\\
168	&	J040009.29+584106.1	&	Gaia DR2 470056824500164608	&	04 00 09.29 	&	 +58 41 06.16	&	B9 IV Cr   	&	117	&	12.4413	&	0.0002	&	+0.845	&	0.032	&	+0.082	&	0.001	&	1.59	&	+0.49	&	0.10	\\
169	&	J040023.62+462840.7	&	HD 25010   	&	04 00 23.62 	&	 +46 28 40.69	&	B8 IV$-$V Si   	&	358	&	10.3137	&	0.0006	&	+1.381	&	0.051	&	$-$0.144	&	0.002	&	1.08	&	$-$0.06	&	0.09	\\
170	&	J040207.61+511231.9	&	Gaia DR2 250653299261292800	&	04 02 07.62 	&	 +51 12 31.91	&	B9 IV EuSi   	&	112	&	12.5781	&	0.0004	&	+0.873	&	0.040	&	+0.124	&	0.002	&	1.24	&	+1.04	&	0.11	\\
171	&	J040213.87+370957.1 $^{a}$	&	HD 279277  	&	04 02 13.74 	&	 +37 09 57.15	&	B9 V CrEuSi   	&	311	&	10.6138	&	0.0004	&	+1.863	&	0.047	&	+0.003	&	0.001	&	0.88	&	+1.09	&	0.07	\\
172	&	J040217.66+340724.2	&	Gaia DR2 170944650483159040	&	04 02 17.66 	&	 +34 07 24.23	&	A2 V SrCrEu   	&	82	&	14.6761	&	0.0005	&	+0.207	&	0.035	&	+0.177	&	0.003	&	0.50	&	+0.76	&	0.37	\\
173	&	J040229.46+571905.4	&	TYC 3726-229-1     	&	04 02 29.47 	&	 +57 19 05.39	&	B8 IV Si   	&	428	&	9.7835	&	0.0006	&	+1.495	&	0.037	&	+0.002	&	0.002	&	0.60	&	+0.06	&	0.07	\\
174	&	J040245.51+485337.9	&	Gaia DR2 247227942587888768	&	04 02 45.53 	&	 +48 53 37.91	&	B9 IV Si   	&	102	&	12.9425	&	0.0004	&	+0.643	&	0.034	&	+0.002	&	0.002	&	1.52	&	+0.46	&	0.12	\\
175	&	J040305.44+444943.7	&	TYC 2889-238-1     	&	04 03 05.44 	&	 +44 49 43.80	&	B9 III$-$IV Si   	&	225	&	11.4741	&	0.0010	&	+0.720	&	0.036	&	+0.035	&	0.003	&	1.01	&	$-$0.24	&	0.12	\\
176	&	J040314.83+534717.4	&	TYC 3718-527-1     	&	04 03 14.83 	&	 +53 47 17.46	&	B9 II$-$III Si   	&	212	&	11.5850	&	0.0006	&	+0.556	&	0.031	&	$-$0.064	&	0.001	&	1.76	&	$-$1.46	&	0.13	\\
177	&	J040452.41+484841.1	&	Gaia DR2 247167984843330304	&	04 04 52.42 	&	 +48 48 41.20	&	A2 IV CrEu   	&	108	&	12.9807	&	0.0006	&	+1.050	&	0.026	&	+0.158	&	0.002	&	1.21	&	+1.87	&	0.07	\\
178	&	J040508.94+484944.7	&	TYC 3335-1943-1    	&	04 05 08.94 	&	 +48 49 44.59	&	B8 III$-$IV EuSi   	&	256	&	10.0895	&	0.0008	&	+1.075	&	0.050	&	$-$0.182	&	0.003	&	1.30	&	$-$1.05	&	0.11	\\
179	&	J040511.96+471903.3	&	Gaia DR2 245917015488631552	&	04 05 11.97 	&	 +47 19 03.46	&	A0 II$-$III Cr   	&	224	&	12.2341	&	0.0002	&	+0.600	&	0.040	&	+0.109	&	0.002	&	1.26	&	$-$0.13	&	0.15	\\
180	&	J040642.34+454640.8 $^{a}$	&	HD 25706   	&	04 06 42.35 	&	 +45 46 40.89	&	kB9hA2mA6 CrSi   	&	774	&	10.1112	&	0.0004	&	+1.769	&	0.049	&	+0.026	&	0.001	&	0.60	&	+0.75	&	0.08	\\
181	&	J040654.80+210519.9	&	HD 284143  	&	04 06 54.80 	&	 +21 05 19.99	&	B9 V CrEu   	&	505	&	10.7260	&	0.0010	&	+1.717	&	0.076	&	$-$0.001	&	0.002	&	0.68	&	+1.22	&	0.11	\\
182	&	J040715.76+310107.8	&	HD 281582  	&	04 07 15.76 	&	 +31 01 07.97	&	kA5hA8mF0 CrEuSi   	&	485	&	9.9505	&	0.0006	&	+2.677	&	0.145	&	+0.184	&	0.002	&	0.74	&	+1.35	&	0.13	\\
183	&	J040723.45+553340.0	&	TYC 3722-782-1     	&	04 07 23.45 	&	 +55 33 40.03	&	A0 IV$-$V Cr   	&	294	&	11.1627	&	0.0010	&	+0.779	&	0.038	&	+0.130	&	0.002	&	1.43	&	$-$0.81	&	0.12	\\
184	&	J040728.92+415903.9	&	Gaia DR2 231940401634554112	&	04 07 28.93 	&	 +41 59 04.10	&	B9 IV$-$V CrSi   	&	119	&	12.7525	&	0.0006	&	+0.674	&	0.042	&	+0.193	&	0.004	&	0.91	&	+0.98	&	0.15	\\
185	&	J040755.56+534052.7	&	Gaia DR2 275908565958715776	&	04 07 55.57 	&	 +53 40 52.74	&	B8 III$-$IV Si   	&	123	&	12.7288	&	0.0007	&	+0.465	&	0.030	&	$-$0.116	&	0.003	&	1.43	&	$-$0.37	&	0.15	\\
186	&	J040809.35+485644.1	&	TYC 3336-1400-1    	&	04 08 09.34 	&	 +48 56 44.04	&	B8 III$-$IV Si   	&	124	&	11.5957	&	0.0009	&	+0.580	&	0.041	&	+0.015	&	0.002	&	1.36	&	$-$0.95	&	0.16	\\
187	&	J040821.16+540119.2	&	TYC 3718-2-1       	&	04 08 21.17 	&	 +54 01 19.32	&	A2 IV$-$V SrCrEu   	&	266	&	10.7801	&	0.0006	&	+2.044	&	0.043	&	+0.007	&	0.002	&	1.08	&	+1.25	&	0.07	\\
188	&	J040832.16+243436.8	&	HD 283428  	&	04 08 32.17 	&	 +24 34 36.77	&	A9 V SrCrEuSi   	&	219	&	11.1143	&	0.0016	&	+0.959	&	0.104	&	+0.349	&	0.003	&	0.77	&	+0.26	&	0.24	\\
189	&	J041056.36+534031.0	&	Gaia DR2 275238001306676480	&	04 10 56.37 	&	 +53 40 31.31	&	kB9.5hA3mA3 SrCrSi   	&	146	&	12.2253	&	0.0003	&	+1.245	&	0.034	&	+0.120	&	0.002	&	1.00	&	+1.70	&	0.08	\\
190	&	J041107.47+433904.3	&	Gaia DR2 232179030014210048	&	04 11 07.47 	&	 +43 39 04.39	&	B8 IV \textit{SiCr}*	&	132	&	12.6631	&	0.0006	&	+0.386	&	0.043	&	+0.009	&	0.003	&	1.01	&	$-$0.41	&	0.24	\\
191	&	J041121.82+545540.8	&	TYC 3722-714-1     	&	04 11 21.74 	&	 +54 55 40.82	&	B8 III$-$IV \textit{(Cr)}*	&	297	&	10.9217	&	0.0006	&	+1.690	&	0.067	&	+0.246	&	0.002	&	1.90	&	+0.16	&	0.10	\\
192	&	J041207.40+235857.2	&	HD 283516  	&	04 12 07.43 	&	 +23 58 57.45	&	B9 V CrEu   	&	474	&	9.8901	&	0.0010	&	+2.825	&	0.048	&	+0.188	&	0.003	&	0.67	&	+1.48	&	0.06	\\
193	&	J041222.48+504917.9	&	Gaia DR2 271636757126607104	&	04 12 22.49 	&	 +50 49 17.99	&	B9.5 IV Eu   	&	102	&	12.9139	&	0.0003	&	+0.799	&	0.032	&	+0.039	&	0.001	&	1.45	&	+0.97	&	0.10	\\
194	&	J041403.70+494110.7 $^{a}$	&	TYC 3336-982-1     	&	04 14 03.71 	&	 +49 41 10.76	&	B9.5 II$-$III Si   	&	185	&	11.8594	&	0.0009	&	+0.864	&	0.043	&	+0.100	&	0.004	&	1.07	&	+0.47	&	0.12	\\
195	&	J041431.97+480234.3	&	TYC 3332-443-1     	&	04 14 31.98 	&	 +48 02 34.36	&	B8 III$-$IV Si   	&	195	&	11.9233	&	0.0034	&	+0.892	&	0.043	&	+0.161	&	0.014	&	1.43	&	+0.25	&	0.12	\\
196	&	J041448.39+484252.7	&	TYC 3332-853-1     	&	04 14 48.39 	&	 +48 42 52.82	&	B9.5 II$-$III  bl4130  	&	272	&	11.0196	&	0.0022	&	+0.889	&	0.049	&	$-$0.152	&	0.006	&	1.57	&	$-$0.81	&	0.13	\\
197	&	J041459.31+532400.8	&	TYC 3719-75-1      	&	04 14 59.32 	&	 +53 24 00.80	&	B9 III$-$IV Si   	&	110	&	11.9190	&	0.0003	&	+0.883	&	0.042	&	+0.015	&	0.002	&	1.13	&	+0.52	&	0.12	\\
198	&	J041506.01+414635.1	&	Gaia DR2 228782565577949952	&	04 15 06.02 	&	 +41 46 35.11	&	A0 III$-$IV bl4077 bl4130  	&	120	&	12.7046	&	0.0004	&	+0.587	&	0.131	&	+0.288	&	0.002	&	1.28	&	+0.27	&	0.49	\\
199	&	J041530.27+511305.8	&	TYC 3340-685-1     	&	04 15 30.28 	&	 +51 13 05.80	&	B9 V Cr   	&	270	&	11.6996	&	0.0008	&	+0.944	&	0.048	&	+0.060	&	0.003	&	0.61	&	+0.96	&	0.12	\\
200	&	J041550.17+494027.7	&	Gaia DR2 270578717700916096	&	04 15 50.18 	&	 +49 40 27.79	&	A4 III$-$IV bl4077   	&	151	&	12.1442	&	0.0003	&	+1.101	&	0.040	&	+0.201	&	0.002	&	0.94	&	+1.42	&	0.09	\\
\hline
\end{tabular}                                                                                                                                                                   
\end{adjustbox}
\end{center}                                                                                                                                             
\end{sidewaystable*}
\setcounter{table}{0}  
\begin{sidewaystable*}
\caption{Essential data for our sample stars, sorted by increasing right ascension. The columns denote: (1) Internal identification number. (2) LAMOST identifier. (3) Alternativ identifier (HD number, TYC identifier or GAIA DR2 number). (4) Right ascension (J2000; GAIA DR2). (5) Declination (J2000; GAIA DR2). (6) Spectral type, as derived in this study. (7) Sloan $g$ band S/N ratio of the analysed spectrum. (8) $G$\,mag (GAIA DR2). (9) $G$\,mag error. (10) Parallax (GAIA DR2). (11) Parallax error. (12) Dereddened colour index $(BP-RP){_0}$ (GAIA DR2). (13) Colour index error. (14) Absorption in the $G$ band, $A_G$. (15) Intrinsic absolute magnitude in the $G$ band, $M_{\mathrm{G,0}}$. (16) Absolute magnitude error.}
\label{table_master5}
\begin{center}
\begin{adjustbox}{max width=\textwidth}
\begin{tabular}{lllcclcccccccccc}
\hline
\hline
(1) & (2) & (3) & (4) & (5) & (6) & (7) & (8) & (9) & (10) & (11) & (12) & (13) & (14) & (15) & (16) \\
No.	&	ID\_LAMOST	&	ID\_alt	&	RA(J2000) 	&	 Dec(J2000)    	&	SpT\_final	&	S/N\,$g$	&	$G$\,mag	&	e\_$G$\,mag	&	pi (mas)	&	e\_pi	&	$(BP-RP){_0}$	&	e\_$(BP-RP){_0}$	&	$A_G$	&	$M_{\mathrm{G,0}}$	&	e\_$M_{\mathrm{G,0}}$	\\
\hline 
201	&	J041607.68+511955.5	&	TYC 3340-1041-1    	&	04 16 07.69 	&	 +51 19 55.51	&	B8 IV Si   	&	250	&	11.7336	&	0.0004	&	+0.794	&	0.040	&	$-$0.122	&	0.002	&	0.81	&	+0.42	&	0.12	\\
202	&	J041619.00+291523.7	&	HD 281818  	&	04 16 19.00 	&	 +29 15 23.79	&	B9 V EuSi   	&	366	&	10.0161	&	0.0005	&	+1.715	&	0.052	&	$-$0.031	&	0.001	&	1.08	&	+0.11	&	0.08	\\
203	&	J041641.15+511253.2	&	TYC 3340-629-1     	&	04 16 41.15 	&	 +51 12 53.21	&	B8 IV CrSi   	&	230	&	12.1156	&	0.0007	&	+0.780	&	0.042	&	$-$0.006	&	0.003	&	0.78	&	+0.80	&	0.13	\\
204	&	J041706.53+462439.2	&	TYC 3328-2095-1    	&	04 17 06.53 	&	 +46 24 39.27	&	B9 IV  bl4130  	&	122	&	12.1136	&	0.0009	&	+1.021	&	0.043	&	$-$0.104	&	0.004	&	0.93	&	+1.23	&	0.10	\\
205	&	J041744.73+403036.1	&	Gaia DR2 227872964522959616	&	04 17 44.73 	&	 +40 30 36.22	&	A0 IV SrCrEu   	&	198	&	11.9723	&	0.0007	&	+1.144	&	0.066	&	+0.052	&	0.002	&	1.15	&	+1.11	&	0.14	\\
206	&	J041748.47+534201.4	&	TYC 3719-604-1     	&	04 17 48.48 	&	 +53 42 01.60	&	B9 IV$-$V Si   	&	108	&	11.7587	&	0.0012	&	+1.201	&	0.035	&	$-$0.015	&	0.005	&	2.10	&	+0.05	&	0.08	\\
207	&	J041813.52+525506.4	&	TYC 3719-599-1     	&	04 18 13.39 	&	 +52 55 06.39	&	B8 III$-$IV  bl4130  	&	267	&	10.8985	&	0.0008	&	+1.365	&	0.109	&	$-$0.036	&	0.003	&	1.11	&	+0.46	&	0.18	\\
208	&	J041819.79+414611.3	&	Gaia DR2 228744980321896448	&	04 18 19.79 	&	 +41 46 11.37	&	kB9.5hA1mA3 Si   	&	112	&	14.2355	&	0.0015	&	+0.273	&	0.039	&	+0.049	&	0.007	&	1.06	&	+0.36	&	0.31	\\
209	&	J041827.05+204500.5	&	HD 284335  	&	04 18 27.05 	&	 +20 45 00.57	&	B9 IV$-$V Sr   	&	290	&	11.4504	&	0.0011	&	+0.958	&	0.096	&	+0.008	&	0.003	&	0.81	&	+0.55	&	0.22	\\
210	&	J041834.97+224205.5	&	HD 284308  	&	04 18 35.08 	&	 +22 42 04.40	&	kA3hA5mA9 SrCrEu   	&	381	&	10.5200	&	0.0005	&	+1.627	&	0.048	&	+0.205	&	0.002	&	0.54	&	+1.04	&	0.08	\\
211	&	J041920.40+530457.5	&	Gaia DR2 275289094232992384	&	04 19 20.41 	&	 +53 04 57.52	&	B9.5 III$-$IV Si   	&	111	&	12.9156	&	0.0006	&	+0.447	&	0.040	&	+0.085	&	0.003	&	1.56	&	$-$0.40	&	0.20	\\
212	&	J041946.03+471655.3	&	Gaia DR2 257920070391100032	&	04 19 46.04 	&	 +47 16 55.41	&	B9.5 III$-$IV Si   	&	108	&	12.5751	&	0.0010	&	+0.532	&	0.044	&	+0.050	&	0.005	&	1.77	&	$-$0.57	&	0.19	\\
213	&	J041952.45+401551.1	&	TYC 2882-1513-1    	&	04 19 52.45 	&	 +40 15 51.20	&	A0 IV$-$V SrCrEu   	&	168	&	11.8203	&	0.0006	&	+0.917	&	0.051	&	+0.098	&	0.002	&	1.19	&	+0.44	&	0.13	\\
214	&	J042104.15+033648.2	&	Gaia DR2 3283347848906203520	&	04 21 04.15 	&	 +03 36 48.20	&	B8 III$-$IV Si   	&	189	&	13.1678	&	0.0005	&	+0.397	&	0.049	&	$-$0.114	&	0.004	&	0.62	&	+0.54	&	0.27	\\
215	&	J042113.88+441145.3	&	TYC 2891-1625-1    	&	04 21 13.88 	&	 +44 11 45.21	&	A1 V SrCrEu   	&	135	&	10.6491	&	0.0004	&	+1.561	&	0.048	&	+0.117	&	0.001	&	0.67	&	+0.94	&	0.08	\\
216	&	J042200.91+402507.7	&	HD 276286  	&	04 22 00.92 	&	 +40 25 07.89	&	B9 V Eu   	&	246	&	11.1562	&	0.0013	&	+1.747	&	0.064	&	$-$0.045	&	0.003	&	1.11	&	+1.26	&	0.09	\\
217	&	J042235.21+515224.6	&	TYC 3341-92-1      	&	04 22 35.22 	&	 +51 52 24.60	&	B8 III  bl4130  	&	303	&	10.5483	&	0.0003	&	+0.300	&	0.449	&	$-$	&	$-$	&	$-$	&	$-$	&	$-$	\\
218	&	J042235.95+411448.9	&	TYC 2883-63-1      	&	04 22 35.95 	&	 +41 14 48.97	&	A1 IV SrCrEuSi   	&	178	&	12.4050	&	0.0003	&	+0.883	&	0.046	&	+0.235	&	0.002	&	0.82	&	+1.31	&	0.12	\\
219	&	J042248.93+474138.0	&	TYC 3333-1433-1    	&	04 22 48.92 	&	 +47 41 38.03	&	A0 II Eu   	&	308	&	11.4005	&	0.0033	&	$-$0.520	&	0.272	&	$-$	&	$-$	&	$-$	&	$-$	&	$-$	\\
220	&	J042304.38+405451.5	&	HD 276275  	&	04 23 04.39 	&	 +40 54 51.60	&	B7 IV Si   	&	494	&	10.6030	&	0.0004	&	+1.035	&	0.052	&	$-$0.045	&	0.001	&	0.78	&	$-$0.10	&	0.12	\\
221	&	J042344.60+350740.4	&	HD 279836  	&	04 23 44.60 	&	 +35 07 40.35	&	A2 IV SrEu   	&	347	&	10.2024	&	0.0005	&	+2.755	&	0.053	&	+0.249	&	0.003	&	0.76	&	+1.64	&	0.07	\\
222	&	J042410.70+540958.6	&	Gaia DR2 275497417325069696	&	04 24 10.70 	&	 +54 09 58.66	&	B9 IV bl4077 bl4130  	&	113	&	12.7648	&	0.0008	&	+0.777	&	0.036	&	+0.013	&	0.004	&	1.55	&	+0.66	&	0.11	\\
223	&	J042441.72+424439.2	&	Gaia DR2 228514770078270976	&	04 24 41.74 	&	 +42 44 39.13	&	B9 IV$-$V CrEu  (He-wk) 	&	117	&	12.5875	&	0.0007	&	+0.738	&	0.058	&	+0.203	&	0.005	&	0.78	&	+1.15	&	0.18	\\
224	&	J042521.99+232323.1	&	HD 284427  	&	04 25 21.99 	&	 +23 23 21.54	&	A0 IV$-$V SrCrEu   	&	516	&	10.6442	&	0.0005	&	+1.508	&	0.051	&	+0.134	&	0.001	&	1.03	&	+0.51	&	0.09	\\
225	&	J042602.49+540804.0	&	TYC 3719-1063-1    	&	04 26 02.49 	&	 +54 08 04.14	&	B9.5 IV$-$V CrEu   	&	236	&	10.7222	&	0.0006	&	+1.105	&	0.041	&	$-$0.155	&	0.002	&	1.20	&	$-$0.26	&	0.09	\\
226	&	J042613.51+435333.7	&	Gaia DR2 253399226476443136	&	04 26 13.52 	&	 +43 53 33.71	&	B9.5 IV$-$V Cr   	&	101	&	14.7966	&	0.0014	&	+0.297	&	0.035	&	+0.255	&	0.006	&	1.17	&	+0.99	&	0.26	\\
227	&	J042616.93+331258.7	&	HD 282086  	&	04 26 16.93 	&	 +33 12 58.76	&	B9 III$-$IV CrEuSi   	&	266	&	11.2353	&	0.0016	&	+0.841	&	0.082	&	+0.021	&	0.006	&	0.66	&	+0.20	&	0.22	\\
228	&	J042736.18+063643.1	&	HD 28238   	&	04 27 35.96 	&	 +06 36 43.13	&	A2 IV$-$V SrCrEu   	&	506	&	9.0822	&	0.0005	&	+2.864	&	0.059	&	+0.188	&	0.002	&	0.23	&	+1.13	&	0.07	\\
229	&	J042920.21+501901.0	&	Gaia DR2 270820296720663168	&	04 29 20.22 	&	 +50 19 01.07	&	B9 IV$-$V CrEu   	&	125	&	12.5741	&	0.0007	&	+0.893	&	0.043	&	+0.055	&	0.003	&	1.25	&	+1.08	&	0.12	\\
230	&	J043045.19+484742.5	&	TYC 3350-370-1     	&	04 30 45.19 	&	 +48 47 42.58	&	B8 IV$-$V Si  (He-wk) 	&	225	&	11.4279	&	0.0011	&	+0.882	&	0.034	&	+0.029	&	0.003	&	1.01	&	+0.15	&	0.10	\\
231	&	J043150.95+220725.7	&	Gaia DR2 145046852381708288	&	04 31 50.95 	&	 +22 07 25.72	&	A4 V SrCrEu   	&	106	&	13.8300	&	0.0003	&	+0.578	&	0.032	&	+0.199	&	0.002	&	1.04	&	+1.60	&	0.13	\\
232	&	J043201.64+471447.8	&	TYC 3346-66-1      	&	04 32 01.64 	&	 +47 14 48.00	&	\textit{A2} V SrCrEu\textit{(Si)}* $^{d}$  	&	419	&	10.2635	&	0.0005	&	+2.500	&	0.061	&	+0.257	&	0.004	&	0.86	&	+1.39	&	0.07	\\
233	&	J043315.40+404158.8	&	Gaia DR2 179844097595682944	&	04 33 15.41 	&	 +40 41 58.88	&	B6 IV   (He-wk) 	&	110	&	14.2200	&	0.0004	&	+0.286	&	0.034	&	+0.042	&	0.003	&	1.02	&	+0.47	&	0.26	\\
234	&	J043459.73+222805.5	&	Gaia DR2 145126399472110848	&	04 34 59.73 	&	 +22 28 05.72	&	kB8hA3mA6 CrSi   	&	106	&	13.0972	&	0.0006	&	+0.687	&	0.024	&	+0.016	&	0.004	&	0.98	&	+1.30	&	0.09	\\
235	&	J043752.46+533259.6	&	Gaia DR2 273642674351866240	&	04 37 52.47 	&	 +53 32 59.65	&	B9.5 V CrSi   	&	323	&	12.4731	&	0.0003	&	+1.247	&	0.041	&	+0.027	&	0.002	&	1.65	&	+1.30	&	0.09	\\
236	&	J043814.39+542341.7	&	Gaia DR2 273828667913353728	&	04 38 14.40 	&	 +54 23 41.77	&	B9 IV$-$V CrSi   	&	219	&	12.5861	&	0.0007	&	+0.862	&	0.036	&	+0.054	&	0.005	&	1.24	&	+1.03	&	0.10	\\
237	&	J044015.93+215631.5	&	HD 284592  	&	04 40 15.93 	&	 +21 56 31.52	&	B9 V CrEu   	&	472	&	11.0040	&	0.0013	&	+1.057	&	0.049	&	+0.053	&	0.004	&	0.85	&	+0.28	&	0.11	\\
238	&	J044108.28+481631.3	&	Gaia DR2 257474944280221824	&	04 41 08.29 	&	 +48 16 31.38	&	B8 IV CrSi  (He-wk) 	&	213	&	11.8758	&	0.0007	&	+0.834	&	0.052	&	+0.117	&	0.002	&	1.51	&	$-$0.03	&	0.15	\\
239	&	J044214.27+353207.5	&	Gaia DR2 174309701521387008	&	04 42 14.28 	&	 +35 32 07.61	&	B8 III  bl4130  	&	103	&	12.4460	&	0.0004	&	+1.007	&	0.039	&	+0.076	&	0.002	&	2.11	&	+0.35	&	0.10	\\
240	&	J044313.88+532544.2	&	TYC 3733-622-1     	&	04 43 13.88 	&	 +53 25 44.22	&	B8 IV$-$V \textit{Cr(Si)}*	&	548	&	11.2241	&	0.0012	&	+1.166	&	0.044	&	+0.107	&	0.002	&	1.14	&	+0.42	&	0.10	\\
241	&	J044407.32-005639.0	&	TYC 4735-654-1     	&	04 44 07.16 	&	 -00 56 39.02	&	F0 V SrEuSi   	&	346	&	10.1724	&	0.0003	&	+1.015	&	0.053	&	+0.098	&	0.001	&	0.11	&	+0.09	&	0.12	\\
242	&	J044433.39+540720.9	&	TYC 3733-47-1      	&	04 44 33.40 	&	 +54 07 20.92	&	A1 II Si   	&	280	&	11.4320	&	0.0009	&	+1.153	&	0.041	&	$-$0.043	&	0.003	&	0.91	&	+0.83	&	0.09	\\
243	&	J044441.74+263924.0	&	Gaia DR2 154343223195896704	&	04 44 41.74 	&	 +26 39 24.09	&	B9 IV$-$V Cr   	&	112	&	11.5960	&	0.0009	&	+1.623	&	0.049	&	+0.218	&	0.002	&	1.24	&	+1.41	&	0.08	\\
244	&	J044446.24+513129.2	&	TYC 3355-141-1     	&	04 44 46.15 	&	 +51 31 29.31	&	A9 V SrEu   	&	376	&	10.9987	&	0.0020	&	+1.882	&	0.181	&	+0.411	&	0.004	&	0.65	&	+1.72	&	0.21	\\
245	&	J044605.51+435344.1	&	TYC 2905-208-1     	&	04 46 05.52 	&	 +43 53 44.09	&	B9.5 IV$-$V CrEu   	&	236	&	11.2435	&	0.0014	&	+1.161	&	0.058	&	+0.079	&	0.003	&	0.93	&	+0.64	&	0.12	\\
246	&	J044713.47+540515.5	&	TYC 3733-133-1     	&	04 47 13.51 	&	 +54 05 15.68	&	A2 IV CrEu   	&	465	&	9.6125	&	0.0005	&	+1.718	&	0.051	&	$-$0.068	&	0.003	&	0.70	&	+0.08	&	0.08	\\
247	&	J044715.48+573746.4	&	TYC 3741-236-1     	&	04 47 15.45 	&	 +57 37 45.45	&	B9.5 IV$-$V Si   	&	370	&	10.9459	&	0.0015	&	+0.953	&	0.060	&	+0.020	&	0.004	&	0.83	&	+0.01	&	0.15	\\
248	&	J044741.43+230820.9	&	TYC 1831-1868-1    	&	04 47 41.44 	&	 +23 08 20.96	&	B8 III$-$IV EuSi   	&	126	&	12.1933	&	0.0013	&	+0.457	&	0.046	&	+0.065	&	0.005	&	0.81	&	$-$0.32	&	0.23	\\
249	&	J044758.11+070009.4	&	TYC 96-157-1       	&	04 47 58.12 	&	 +07 00 09.43	&	B9 V CrSi  (He-wk) 	&	196	&	11.0805	&	0.0006	&	+1.197	&	0.057	&	$-$0.098	&	0.003	&	0.23	&	+1.24	&	0.11	\\
250	&	J045038.83+544917.1	&	Gaia DR2 273444865338468736	&	04 50 38.84 	&	 +54 49 17.23	&	B8 IV$-$V Si  (He-wk) 	&	120	&	12.3142	&	0.0005	&	+0.686	&	0.043	&	+0.250	&	0.003	&	1.46	&	+0.03	&	0.14	\\
\hline
\end{tabular}                                                                                                                                                                   
\end{adjustbox}
\end{center}                                                                                                                                             
\end{sidewaystable*}
\setcounter{table}{0}  
\begin{sidewaystable*}
\caption{Essential data for our sample stars, sorted by increasing right ascension. The columns denote: (1) Internal identification number. (2) LAMOST identifier. (3) Alternativ identifier (HD number, TYC identifier or GAIA DR2 number). (4) Right ascension (J2000; GAIA DR2). (5) Declination (J2000; GAIA DR2). (6) Spectral type, as derived in this study. (7) Sloan $g$ band S/N ratio of the analysed spectrum. (8) $G$\,mag (GAIA DR2). (9) $G$\,mag error. (10) Parallax (GAIA DR2). (11) Parallax error. (12) Dereddened colour index $(BP-RP){_0}$ (GAIA DR2). (13) Colour index error. (14) Absorption in the $G$ band, $A_G$. (15) Intrinsic absolute magnitude in the $G$ band, $M_{\mathrm{G,0}}$. (16) Absolute magnitude error.}
\label{table_master6}
\begin{center}
\begin{adjustbox}{max width=\textwidth}
\begin{tabular}{lllcclcccccccccc}
\hline
\hline
(1) & (2) & (3) & (4) & (5) & (6) & (7) & (8) & (9) & (10) & (11) & (12) & (13) & (14) & (15) & (16) \\
No.	&	ID\_LAMOST	&	ID\_alt	&	RA(J2000) 	&	 Dec(J2000)    	&	SpT\_final	&	S/N\,$g$	&	$G$\,mag	&	e\_$G$\,mag	&	pi (mas)	&	e\_pi	&	$(BP-RP){_0}$	&	e\_$(BP-RP){_0}$	&	$A_G$	&	$M_{\mathrm{G,0}}$	&	e\_$M_{\mathrm{G,0}}$	\\
\hline 
251	&	J045109.07+433556.5	&	TYC 2906-2149-1    	&	04 51 09.11 	&	 +43 35 56.83	&	A1 II$-$III Si   	&	188	&	11.7949	&	0.0015	&	+0.663	&	0.071	&	+0.030	&	0.029	&	0.60	&	+0.30	&	0.24	\\
252	&	J045121.11+093555.8	&	HD 287173  	&	04 51 21.12 	&	 +09 35 55.81	&	A0 V SrCrEuSi   	&	564	&	9.9255	&	0.0003	&	+2.219	&	0.061	&	+0.135	&	0.001	&	0.54	&	+1.12	&	0.08	\\
253	&	J045131.08+395823.5	&	HD 277044  	&	04 51 31.08 	&	 +39 58 23.66	&	B9 V  bl4130  	&	102	&	11.3158	&	0.0011	&	+0.770	&	0.052	&	+0.058	&	0.002	&	0.59	&	+0.16	&	0.15	\\
254	&	J045148.66+435755.1	&	TYC 2906-368-1     	&	04 51 48.67 	&	 +43 57 55.18	&	B9.5 II$-$III Si   	&	170	&	12.3508	&	0.0008	&	+0.421	&	0.035	&	+0.095	&	0.004	&	0.72	&	$-$0.25	&	0.19	\\
255	&	J045150.13+594602.0	&	TYC 3745-269-1     	&	04 51 50.14 	&	 +59 46 02.00	&	B9 V CrEu   	&	250	&	11.4013	&	0.0008	&	+1.577	&	0.035	&	+0.005	&	0.003	&	1.26	&	+1.13	&	0.07	\\
256	&	J045220.67+405034.9	&	HD 277028  	&	04 52 20.69 	&	 +40 50 34.94	&	B9 V SrCr   	&	210	&	10.7310	&	0.0007	&	+0.526	&	0.258	&	$-$	&	$-$	&	$-$	&	$-$	&	$-$	\\
257	&	J045231.56+521715.1	&	TYC 3356-1085-1    	&	04 52 31.56 	&	 +52 17 15.04	&	A5 IV SrCrEu   	&	193	&	10.3405	&	0.0004	&	+2.302	&	0.049	&	+0.314	&	0.002	&	0.62	&	+1.53	&	0.07	\\
258	&	J045316.36+244251.8	&	HD 283952  	&	04 53 16.37 	&	 +24 42 51.77	&	B9.5 V \textit{Cr}*	&	216	&	10.7801	&	0.0005	&	+1.271	&	0.045	&	+0.123	&	0.002	&	0.70	&	+0.60	&	0.09	\\
259	&	J045508.24+204943.7	&	Gaia DR2 3411724520166949120	&	04 55 08.24 	&	 +20 49 43.69	&	kA2hA4mA7 SrCrEu   	&	147	&	14.0406	&	0.0003	&	+0.756	&	0.037	&	+0.353	&	0.002	&	0.62	&	+2.82	&	0.12	\\
260	&	J045538.72+243458.6	&	HD 283960  	&	04 55 38.72 	&	 +24 34 58.70	&	A0 IV CrEu   	&	124	&	11.3335	&	0.0008	&	+1.584	&	0.044	&	+0.294	&	0.002	&	0.76	&	+1.57	&	0.08	\\
261	&	J045637.95+380626.6	&	HD 280235  	&	04 56 37.95 	&	 +38 06 26.74	&	kB9hA1mA4  bl4130  	&	322	&	10.0614	&	0.0004	&	+1.265	&	0.047	&	+0.143	&	0.002	&	1.09	&	$-$0.52	&	0.09	\\
262	&	J045708.30+543307.3	&	Gaia DR2 273280733165163264	&	04 57 08.30 	&	 +54 33 07.13	&	A0 IV Si   	&	151	&	12.2942	&	0.0003	&	+0.835	&	0.040	&	+0.224	&	0.002	&	0.84	&	+1.06	&	0.12	\\
263	&	J045713.44+383741.1	&	HD 280230  	&	04 57 13.44 	&	 +38 37 41.23	&	B9.5 IV bl4077 bl4130  	&	128	&	11.3683	&	0.0009	&	+1.293	&	0.063	&	$-$0.042	&	0.003	&	1.21	&	+0.71	&	0.12	\\
264	&	J045741.74+405440.1	&	HD 277264  	&	04 57 41.74 	&	 +40 54 40.26	&	B8 III$-$IV Si   	&	150	&	10.6503	&	0.0007	&	+0.947	&	0.064	&	$-$0.148	&	0.003	&	0.55	&	$-$0.02	&	0.15	\\
265	&	J045747.66+560121.5	&	TYC 3738-1376-1    	&	04 57 47.65 	&	 +56 01 21.53	&	B9.5 IV CrEu   	&	245	&	10.8983	&	0.0023	&	+1.538	&	0.045	&	+0.018	&	0.005	&	1.74	&	+0.09	&	0.08	\\
266	&	J045808.28+465625.9	&	TYC 3348-2776-1    	&	04 58 08.28 	&	 +46 56 25.90	&	B9 IV$-$V CrEuSi   	&	176	&	11.5823	&	0.0006	&	+0.998	&	0.042	&	+0.109	&	0.003	&	0.80	&	+0.78	&	0.10	\\
267	&	J045914.82+410335.6	&	HD 277256  	&	04 59 14.82 	&	 +41 03 35.61	&	B7 III  bl4130  	&	427	&	9.2353	&	0.0005	&	+0.787	&	0.043	&	$-$0.076	&	0.003	&	0.44	&	$-$1.73	&	0.13	\\
268	&	J045926.29+535030.4	&	TYC 3743-1043-1    	&	04 59 26.17 	&	 +53 50 30.51	&	B9 V CrEu   	&	292	&	10.9138	&	0.0009	&	+1.475	&	0.050	&	+0.081	&	0.003	&	0.64	&	+1.12	&	0.09	\\
269	&	J045933.63+395715.8	&	TYC 2898-2642-1    	&	04 59 33.58 	&	 +39 57 15.94	&	A0 IV$-$V SrCr   	&	168	&	11.9263	&	0.0006	&	+0.734	&	0.044	&	+0.114	&	0.019	&	0.93	&	+0.32	&	0.14	\\
270	&	J050143.90+440946.2	&	Gaia DR2 205566588012912896	&	05 01 43.91 	&	 +44 09 46.35	&	B9 IV Cr   	&	109	&	12.1255	&	0.0005	&	+1.844	&	0.915	&	$-$	&	$-$	&	$-$	&	$-$	&	$-$	\\
271	&	J050144.01+584923.4	&	TYC 3746-1604-1    	&	05 01 44.02 	&	 +58 49 23.39	&	B8 III$-$IV Si   	&	573	&	9.9912	&	0.0005	&	+1.061	&	0.045	&	$-$0.117	&	0.002	&	0.98	&	$-$0.86	&	0.10	\\
272	&	J050146.85+383500.8	&	HD 280281  	&	05 01 46.85 	&	 +38 35 00.81	&	\textit{B8} V Si* $^{c}$	&	104	&	10.9424	&	0.0010	&	+1.253	&	0.041	&	+0.109	&	0.004	&	0.90	&	+0.53	&	0.09	\\
273	&	J050205.88+403800.1	&	Gaia DR2 200809138640147072	&	05 02 05.89 	&	 +40 38 00.15	&	kA1hA3mA3 CrSi   	&	102	&	14.6777	&	0.0006	&	+0.279	&	0.035	&	+0.039	&	0.004	&	0.84	&	+1.07	&	0.28	\\
274	&	J050209.22+395941.4	&	HD 277411  	&	05 02 09.23 	&	 +39 59 41.35	&	B7 IV Si  (He-wk) (He-st)	&	186	&	11.1530	&	0.0027	&	+0.897	&	0.070	&	$-$0.045	&	0.009	&	0.46	&	+0.46	&	0.18	\\
275	&	J050210.72+464600.0	&	TYC 3344-79-1      	&	05 02 10.72 	&	 +46 46 00.07	&	B8 III$-$IV Si   	&	241	&	9.7607	&	0.0006	&	+1.177	&	0.049	&	$-$0.164	&	0.002	&	0.87	&	$-$0.76	&	0.10	\\
276	&	J050230.72+582514.3	&	TYC 3746-2118-1    	&	05 02 30.72 	&	 +58 25 14.33	&	A0 IV CrEu   	&	193	&	11.0443	&	0.0005	&	+1.357	&	0.043	&	+0.214	&	0.002	&	0.80	&	+0.90	&	0.08	\\
277	&	J050435.61+530749.4	&	TYC 3734-376-1     	&	05 04 35.61 	&	 +53 07 49.42	&	B7 V Si   	&	189	&	11.1216	&	0.0014	&	+1.259	&	0.039	&	+0.336	&	0.005	&	1.20	&	+0.42	&	0.08	\\
278	&	J050453.61+403735.7	&	TYC 2899-621-1     	&	05 04 53.62 	&	 +40 37 35.71	&	B5 IV$-$V  bl4130 (He-wk) 	&	135	&	11.2878	&	0.0014	&	+0.832	&	0.055	&	$-$0.089	&	0.004	&	0.55	&	+0.34	&	0.15	\\
279	&	J050519.16+412323.4	&	Gaia DR2 201277766811826176	&	05 05 19.16 	&	 +41 23 23.42	&	B8 IV$-$V Si  (He-wk) 	&	136	&	11.6903	&	0.0011	&	+0.731	&	0.043	&	$-$0.080	&	0.006	&	0.46	&	+0.55	&	0.14	\\
280	&	J050537.89+554553.8	&	TYC 3738-363-1     	&	05 05 37.98 	&	 +55 45 52.99	&	B8 IV$-$V  bl4130  	&	217	&	10.9355	&	0.0009	&	+1.375	&	0.041	&	+0.188	&	0.003	&	1.03	&	+0.60	&	0.08	\\
281	&	J050731.60+570017.2	&	TYC 3743-386-1     	&	05 07 31.62 	&	 +57 00 17.15	&	kB9.5hA7mA6 CrEu   	&	316	&	10.4902	&	0.0012	&	$-$3.275	&	1.147	&	$-$	&	$-$	&	$-$	&	$-$	&	$-$	\\
282	&	J050748.68+385804.4	&	HD 277595  	&	05 07 48.69 	&	 +38 58 04.42	&	B8 V  bl4130  	&	296	&	9.5146	&	0.0006	&	+1.347	&	0.067	&	$-$0.124	&	0.003	&	0.57	&	$-$0.41	&	0.12	\\
283	&	J050813.24+404513.2	&	HD 277537  	&	05 08 13.26 	&	 +40 45 13.17	&	\textit{B9 III Si}*	&	340	&	10.3210	&	0.0005	&	+1.075	&	0.050	&	$-$0.157	&	0.001	&	0.59	&	$-$0.11	&	0.11	\\
284	&	J050839.79+422835.6	&	TYC 2903-225-1     	&	05 08 39.80 	&	 +42 28 35.59	&	B7 V Si   	&	96	&	11.7575	&	0.0006	&	+0.835	&	0.041	&	+0.031	&	0.004	&	0.47	&	+0.90	&	0.12	\\
285	&	J050900.01+493815.1	&	TYC 3353-1331-1    	&	05 09 00.01 	&	 +49 38 15.12	&	B9 V Cr  (He-wk) 	&	249	&	11.6302	&	0.0009	&	+0.840	&	0.047	&	$-$0.006	&	0.004	&	0.99	&	+0.26	&	0.13	\\
286	&	J050925.66+512444.9	&	TYC 3357-906-1     	&	05 09 25.66 	&	 +51 24 44.98	&	A3 III$-$IV CrEu   	&	164	&	11.1055	&	0.0019	&	+1.398	&	0.051	&	+0.108	&	0.004	&	0.59	&	+1.25	&	0.09	\\
287	&	J050952.91+424828.1	&	TYC 2903-279-1     	&	05 09 52.91 	&	 +42 48 28.22	&	A0 V SrCrEu   	&	81	&	11.8222	&	0.0009	&	+0.973	&	0.049	&	$-$0.035	&	0.003	&	0.66	&	+1.10	&	0.12	\\
288	&	J051003.60+382125.9	&	TYC 2896-2008-1    	&	05 10 03.61 	&	 +38 21 26.03	&	A0 V SrCrEu   	&	110	&	12.2482	&	0.0005	&	+0.903	&	0.041	&	+0.077	&	0.004	&	0.73	&	+1.30	&	0.11	\\
289	&	J051013.62+420718.3	&	HD 277634  	&	05 10 13.63 	&	 +42 07 18.31	&	kA0hA1mA3 Eu   	&	573	&	9.6076	&	0.0009	&	+3.674	&	0.703	&	+0.146	&	0.003	&	0.09	&	+2.35	&	0.42	\\
290	&	J051205.75+491307.8	&	TYC 3353-348-1     	&	05 12 05.75 	&	 +49 13 07.81	&	A2 V Sr   	&	189	&	11.3847	&	0.0007	&	+1.528	&	0.044	&	$-$0.039	&	0.003	&	0.66	&	+1.64	&	0.08	\\
291	&	J051322.65+385958.7	&	HD 277821  	&	05 13 22.65 	&	 +38 59 58.78	&	B9 IV$-$V bl4077 bl4130  	&	219	&	11.4385	&	0.0007	&	+0.815	&	0.065	&	$-$0.096	&	0.004	&	0.67	&	+0.32	&	0.18	\\
292	&	J051327.91+573333.0	&	TYC 3743-2168-1    	&	05 13 27.87 	&	 +57 33 33.11	&	B9 V Cr   	&	179	&	11.4714	&	0.0005	&	+0.940	&	0.036	&	+0.093	&	0.002	&	0.88	&	+0.46	&	0.10	\\
293	&	J051506.10+321121.3	&	Gaia DR2 180672678389101952	&	05 15 06.11 	&	 +32 11 21.39	&	A0 IV$-$V SrCr   	&	121	&	11.9559	&	0.0004	&	+0.677	&	0.043	&	+0.291	&	0.002	&	0.81	&	+0.29	&	0.15	\\
294	&	J051510.44+330047.9 $^{a}$	&	HD 241843  	&	05 15 10.44 	&	 +33 00 47.98	&	B9 IV$-$V CrEu   	&	226	&	10.2085	&	0.0010	&	+1.345	&	0.051	&	+0.070	&	0.003	&	0.58	&	+0.27	&	0.10	\\
295	&	J051523.41+183519.5	&	TYC 1287-1240-1    	&	05 15 23.42 	&	 +18 35 19.61	&	B8 IV Sr   	&	114	&	12.8452	&	0.0005	&	+0.398	&	0.046	&	$-$0.138	&	0.003	&	1.05	&	$-$0.20	&	0.25	\\
296	&	J051539.57+360321.9	&	Gaia DR2 186913304529270016	&	05 15 39.57 	&	 +36 03 22.00	&	A0 V SrCr   	&	101	&	14.2619	&	0.0004	&	+0.408	&	0.042	&	+0.176	&	0.005	&	1.39	&	+0.92	&	0.23	\\
297	&	J051540.30+370025.0	&	HD 280682  	&	05 15 40.31 	&	 +37 00 25.17	&	B8 IV Si   	&	198	&	10.9237	&	0.0012	&	+1.427	&	0.103	&	+0.064	&	0.007	&	0.53	&	+1.17	&	0.16	\\
298	&	J051709.99+214859.0	&	Gaia DR2 3414598991456199168	&	05 17 09.99 	&	 +21 48 59.20	&	B9 V Cr   	&	108	&	14.5263	&	0.0010	&	+0.325	&	0.031	&	+0.127	&	0.004	&	1.15	&	+0.94	&	0.21	\\
299	&	J051711.10+365831.6	&	TYC 2402-795-1     	&	05 17 11.11 	&	 +36 58 31.75	&	B9.5 IV$-$V SrCr   	&	187	&	11.9807	&	0.0007	&	+0.832	&	0.043	&	+0.072	&	0.005	&	0.63	&	+0.95	&	0.12	\\
300	&	J051717.29+362615.5	&	HD 280802  	&	05 17 17.21 	&	 +36 26 15.55	&	B8 IV Si  (He-wk) 	&	360	&	10.7389	&	0.0008	&	+0.927	&	0.062	&	$-$0.130	&	0.005	&	0.56	&	+0.01	&	0.15	\\
\hline
\end{tabular}                                                                                                                                                                   
\end{adjustbox}
\end{center}                                                                                                                                             
\end{sidewaystable*}
\setcounter{table}{0}  
\begin{sidewaystable*}
\caption{Essential data for our sample stars, sorted by increasing right ascension. The columns denote: (1) Internal identification number. (2) LAMOST identifier. (3) Alternativ identifier (HD number, TYC identifier or GAIA DR2 number). (4) Right ascension (J2000; GAIA DR2). (5) Declination (J2000; GAIA DR2). (6) Spectral type, as derived in this study. (7) Sloan $g$ band S/N ratio of the analysed spectrum. (8) $G$\,mag (GAIA DR2). (9) $G$\,mag error. (10) Parallax (GAIA DR2). (11) Parallax error. (12) Dereddened colour index $(BP-RP){_0}$ (GAIA DR2). (13) Colour index error. (14) Absorption in the $G$ band, $A_G$. (15) Intrinsic absolute magnitude in the $G$ band, $M_{\mathrm{G,0}}$. (16) Absolute magnitude error.}
\label{table_master7}
\begin{center}
\begin{adjustbox}{max width=\textwidth}
\begin{tabular}{lllcclcccccccccc}
\hline
\hline
(1) & (2) & (3) & (4) & (5) & (6) & (7) & (8) & (9) & (10) & (11) & (12) & (13) & (14) & (15) & (16) \\
No.	&	ID\_LAMOST	&	ID\_alt	&	RA(J2000) 	&	 Dec(J2000)    	&	SpT\_final	&	S/N\,$g$	&	$G$\,mag	&	e\_$G$\,mag	&	pi (mas)	&	e\_pi	&	$(BP-RP){_0}$	&	e\_$(BP-RP){_0}$	&	$A_G$	&	$M_{\mathrm{G,0}}$	&	e\_$M_{\mathrm{G,0}}$	\\
\hline 
301	&	J051725.85+364904.8	&	HD 280795  	&	05 17 25.86 	&	 +36 49 04.87	&	A0 IV$-$V SrCr   	&	204	&	11.4545	&	0.0007	&	+0.696	&	0.041	&	+0.139	&	0.003	&	0.69	&	$-$0.03	&	0.14	\\
302	&	J051800.77+184058.1	&	TYC 1287-1110-1    	&	05 18 00.77 	&	 +18 40 58.19	&	B9 V Cr   	&	333	&	11.3538	&	0.0010	&	+0.970	&	0.059	&	+0.325	&	0.003	&	0.93	&	+0.36	&	0.14	\\
303	&	J051812.81+370758.9	&	TYC 2402-1131-1    	&	05 18 12.82 	&	 +37 07 58.99	&	B9.5 IV$-$V CrEuSi   	&	305	&	11.7380	&	0.0008	&	+0.864	&	0.044	&	$-$0.021	&	0.003	&	0.53	&	+0.89	&	0.12	\\
304	&	J051813.83+374532.4	&	Gaia DR2 187175946071857408	&	05 18 13.84 	&	 +37 45 32.51	&	kA1hA5mA3 SrCrSi   	&	132	&	12.9961	&	0.0005	&	+0.691	&	0.073	&	+0.166	&	0.002	&	0.51	&	+1.68	&	0.23	\\
305	&	J051816.22+380429.4	&	HD 280761  	&	05 18 16.24 	&	 +38 04 29.53	&	B9 V CrSi   	&	160	&	11.8640	&	0.0006	&	+0.943	&	0.042	&	$-$0.044	&	0.004	&	0.44	&	+1.29	&	0.11	\\
306	&	J051844.95+380605.3	&	Gaia DR2 187564730808766848	&	05 18 44.96 	&	 +38 06 05.33	&	B9 \textit{V} SrCr\textit{EuSi}* $^{d}$   	&	116	&	14.0374	&	0.0011	&	+0.418	&	0.028	&	+0.187	&	0.005	&	1.28	&	+0.87	&	0.15	\\
307	&	J051852.91+133356.6	&	TYC 711-1863-1     	&	05 18 52.92 	&	 +13 34 01.78	&	B9.5 IV$-$V Cr   	&	991	&	7.4031	&	0.0009	&	+5.273	&	0.107	&	$-$0.071	&	0.006	&	0.31	&	+0.70	&	0.07	\\
308	&	J051854.92+124446.8	&	Gaia DR2 3387793447726005376	&	05 18 54.93 	&	 +12 44 46.91	&	A1 IV$-$V CrEu   	&	120	&	14.4398	&	0.0006	&	+0.476	&	0.030	&	+0.215	&	0.004	&	1.34	&	+1.49	&	0.15	\\
309	&	J051857.75+412255.9	&	Gaia DR2 195073398794823040	&	05 18 57.75 	&	 +41 22 55.85	&	B9 IV CrSi   	&	105	&	14.1977	&	0.0004	&	+0.290	&	0.039	&	$-$0.063	&	0.004	&	1.21	&	+0.30	&	0.30	\\
310	&	J051922.65+451350.7	&	TYC 3358-659-1     	&	05 19 22.66 	&	 +45 13 50.77	&	B8 III$-$IV \textit{(CrSi)}*	&	252	&	11.1642	&	0.0012	&	+0.565	&	0.059	&	$-$0.139	&	0.005	&	0.71	&	$-$0.78	&	0.23	\\
311	&	J051956.20+392118.6	&	Gaia DR2 187964609445409536	&	05 19 56.21 	&	 +39 21 18.67	&	B9.5 III$-$IV Cr   	&	133	&	14.3460	&	0.0012	&	+0.329	&	0.030	&	+0.101	&	0.007	&	0.96	&	+0.97	&	0.20	\\
312	&	J051957.65+334802.5	&	Gaia DR2 181253976444979456	&	05 19 57.64 	&	 +33 48 02.55	&	B8 V  bl4130  	&	149	&	11.9283	&	0.0004	&	+0.378	&	0.063	&	+0.054	&	0.003	&	0.94	&	$-$1.12	&	0.36	\\
313	&	J052001.14+490915.4	&	HD 34439   	&	05 20 01.18 	&	 +49 09 18.40	&	B8 IV EuSi  (He-wk) 	&	531	&	9.0140	&	0.0014	&	+2.172	&	0.056	&	$-$0.067	&	0.006	&	0.28	&	+0.42	&	0.07	\\
314	&	J052016.52+351301.6	&	HD 280950  	&	05 20 16.53 	&	 +35 13 01.89	&	B5 V  bl4130 (He-wk) (He-st)	&	189	&	10.8196	&	0.0010	&	+0.901	&	0.063	&	$-$0.311	&	0.005	&	0.94	&	$-$0.35	&	0.16	\\
315	&	J052043.33+380212.5	&	Gaia DR2 187568815318517760	&	05 20 43.34 	&	 +38 02 12.55	&	A1 II Si   	&	103	&	13.4429	&	0.0006	&	+0.495	&	0.029	&	+0.022	&	0.003	&	1.40	&	+0.52	&	0.14	\\
316	&	J052043.38+335022.0	&	Gaia DR2 181243187487128960	&	05 20 43.39 	&	 +33 50 21.17	&	B8 IV Si   	&	153	&	12.0607	&	0.0011	&	+0.813	&	0.049	&	$-$0.084	&	0.006	&	0.68	&	+0.94	&	0.14	\\
317	&	J052059.29+351123.5	&	HD 280948  	&	05 20 59.29 	&	 +35 11 23.58	&	B9.5 III  bl4130  	&	200	&	10.2693	&	0.0016	&	+1.571	&	0.056	&	+0.074	&	0.007	&	0.55	&	+0.70	&	0.09	\\
318	&	J052118.97+320805.7	&	HD 242764  	&	05 21 18.98 	&	 +32 08 05.85	&	B4 V\textit{pn}*	&	488	&	9.8145	&	0.0022	&	+0.885	&	0.066	&	$-$0.174	&	0.009	&	0.50	&	$-$0.95	&	0.17	\\
319	&	J052128.61+401445.8	&	HD 278068  	&	05 21 28.62 	&	 +40 14 45.81	&	B8 IV$-$V \textit{SiCr}*	&	286	&	11.2981	&	0.0015	&	+0.849	&	0.111	&	$-$0.177	&	0.007	&	0.70	&	+0.24	&	0.29	\\
320	&	J052143.28+354353.0	&	Gaia DR2 183754437683161216	&	05 21 43.28 	&	 +35 43 53.04	&	A0 III Eu   	&	199	&	12.4640	&	0.0017	&	+0.550	&	0.052	&	+0.011	&	0.007	&	1.38	&	$-$0.21	&	0.21	\\
321	&	J052220.03+132145.8	&	TYC 712-2078-1     	&	05 22 20.03 	&	 +13 21 45.89	&	B9 V SrCr   	&	269	&	10.7849	&	0.0007	&	+1.147	&	0.047	&	+0.135	&	0.003	&	0.80	&	+0.28	&	0.10	\\
322	&	J052237.24+324258.3	&	TYC 2394-214-1     	&	05 22 37.25 	&	 +32 42 58.40	&	B9 IV$-$V CrSi   	&	149	&	12.2183	&	0.0006	&	+0.724	&	0.041	&	$-$0.039	&	0.005	&	0.59	&	+0.93	&	0.13	\\
323	&	J052244.67+345827.3	&	TYC 2398-315-1     	&	05 22 44.68 	&	 +34 58 27.34	&	B6 IV Si   	&	372	&	11.4327	&	0.0017	&	+0.777	&	0.058	&	$-$0.204	&	0.008	&	0.96	&	$-$0.07	&	0.17	\\
324	&	J052254.93+134638.7 $^{a}$	&	HD 243096  	&	05 22 54.93 	&	 +13 46 38.79	&	A1 IV  bl4130  	&	250	&	11.0763	&	0.0008	&	+1.185	&	0.047	&	+0.189	&	0.002	&	0.67	&	+0.78	&	0.10	\\
325	&	J052259.54+343944.9	&	HD 281056  	&	05 22 59.55 	&	 +34 39 44.94	&	B9.5 V Cr   	&	191	&	10.6012	&	0.0010	&	+0.794	&	0.047	&	+0.052	&	0.003	&	0.58	&	$-$0.48	&	0.14	\\
326	&	J052329.37+423212.0	&	Gaia DR2 195256707998564480	&	05 23 29.38 	&	 +42 32 12.11	&	B9 V CrEu   	&	109	&	14.6622	&	0.0008	&	+0.238	&	0.032	&	$-$0.087	&	0.005	&	1.12	&	+0.43	&	0.30	\\
327	&	J052422.19+383106.2	&	Gaia DR2 187644204884340480	&	05 24 22.19 	&	 +38 31 06.29	&	B8 III Si  (He-wk) 	&	103	&	14.2274	&	0.0014	&	+0.373	&	0.031	&	+0.109	&	0.006	&	1.16	&	+0.93	&	0.19	\\
328	&	J052454.62+434145.3	&	Gaia DR2 207372260984327936	&	05 24 54.64 	&	 +43 41 45.36	&	B9.5 III$-$IV  bl4130  	&	107	&	12.8217	&	0.0004	&	+0.672	&	0.043	&	+0.158	&	0.003	&	0.86	&	+1.10	&	0.15	\\
329	&	J052454.90+370932.6	&	TYC 2415-423-1     	&	05 24 54.91 	&	 +37 09 32.67	&	B8 III$-$IV EuSi   	&	116	&	11.3825	&	0.0008	&	+0.677	&	0.043	&	+0.054	&	0.003	&	1.18	&	$-$0.65	&	0.15	\\
330	&	J052459.98-064651.8	&	TYC 4765-1018-1    	&	05 24 59.99 	&	 -06 46 51.83	&	A0 IV$-$V Si   	&	227	&	11.8359	&	0.0004	&	+1.310	&	0.048	&	+0.233	&	0.002	&	0.25	&	+2.17	&	0.09	\\
331	&	J052510.60+440918.0	&	TYC 2921-1065-1    	&	05 25 10.60 	&	 +44 09 18.02	&	B9.5 IV$-$V Eu   	&	148	&	12.3953	&	0.0004	&	+0.874	&	0.060	&	$-$0.078	&	0.003	&	0.96	&	+1.14	&	0.16	\\
332	&	J052552.24+344817.3	&	Gaia DR2 182850540345232256	&	05 25 52.24 	&	 +34 48 17.36	&	A0 V CrEu   	&	136	&	13.1048	&	0.0004	&	+0.728	&	0.029	&	+0.143	&	0.003	&	0.87	&	+1.55	&	0.10	\\
333	&	J052602.25+075254.0	&	TYC 700-1067-1     	&	05 26 02.26 	&	 +07 52 54.03	&	B9 V CrEu   	&	342	&	11.2510	&	0.0013	&	+1.319	&	0.052	&	$-$0.045	&	0.005	&	0.62	&	+1.23	&	0.10	\\
334	&	J052616.48+331544.2	&	HD 243492  	&	05 26 16.49 	&	 +33 15 44.26	&	B8 IV$-$V  bl4130  	&	256	&	10.6016	&	0.0005	&	+0.842	&	0.042	&	+0.044	&	0.003	&	0.58	&	$-$0.35	&	0.12	\\
335	&	J052621.99+491939.9	&	TYC 3367-968-1     	&	05 26 22.00 	&	 +49 19 39.89	&	kA1hA3mA7 SrCrEu   	&	299	&	10.6318	&	0.0012	&	+1.512	&	0.146	&	+0.235	&	0.005	&	0.40	&	+1.13	&	0.22	\\
336	&	J052643.64+341450.5	&	Gaia DR2 182629916466102784	&	05 26 43.64 	&	 +34 14 50.55	&	B7 IV$-$V Si   	&	136	&	13.0085	&	0.0012	&	+0.452	&	0.042	&	$-$0.139	&	0.005	&	0.80	&	+0.49	&	0.21	\\
337	&	J052658.34+155131.6	&	Gaia DR2 3390885033905554176	&	05 26 58.34 	&	 +15 51 31.72	&	A7 V SrCrEu   	&	149	&	14.3652	&	0.0016	&	+0.382	&	0.035	&	+0.291	&	0.006	&	1.07	&	+1.21	&	0.21	\\
338	&	J052718.48-012726.1	&	Gaia DR2 3220192038446213760	&	05 27 18.48 	&	 -01 27 26.16	&	A0 IV$-$V CrEu   	&	108	&	13.3183	&	0.0005	&	+0.571	&	0.034	&	+0.033	&	0.003	&	0.29	&	+1.81	&	0.14	\\
339	&	J052739.41+533935.4	&	TYC 3748-1827-1    	&	05 27 39.32 	&	 +53 39 36.63	&	A0 III$-$IV SrCrEu   	&	385	&	10.7348	&	0.0007	&	+1.074	&	0.045	&	+0.020	&	0.004	&	0.86	&	+0.03	&	0.10	\\
340	&	J052748.30+453546.1	&	TYC 3359-1802-1    	&	05 27 48.30 	&	 +45 35 46.14	&	B8 IV Si   	&	308	&	10.1747	&	0.0011	&	+1.363	&	0.067	&	$-$0.169	&	0.005	&	0.66	&	+0.19	&	0.12	\\
341	&	J052800.13+351644.0	&	Gaia DR2 183262612388074496	&	05 28 00.14 	&	 +35 16 44.03	&	A2 III$-$IV SrCrEu   	&	124	&	12.8995	&	0.0009	&	+0.677	&	0.038	&	+0.171	&	0.004	&	1.01	&	+1.05	&	0.13	\\
342	&	J052805.24+343638.8	&	Gaia DR2 182664997758510848	&	05 28 05.26 	&	 +34 36 38.91	&	B8 III$-$IV Si   	&	102	&	13.3388	&	0.0015	&	+0.493	&	0.034	&	+0.142	&	0.008	&	0.98	&	+0.82	&	0.16	\\
343	&	J052812.16+415006.4	&	HD 278204  	&	05 28 11.99 	&	 +41 50 06.46	&	kA0hA3mA7 Si   	&	247	&	10.3252	&	0.0007	&	+1.173	&	0.066	&	$-$0.141	&	0.002	&	0.39	&	+0.28	&	0.13	\\
344	&	J052816.11-063820.1	&	TYC 4765-708-1     	&	05 28 16.10 	&	 -06 38 20.07	&	kA1\textit{hA9}mA9 Sr\textit{Cr}Eu* $^{d}$   	&	358	&	10.7380	&	0.0005	&	+2.685	&	0.041	&	+0.444	&	0.007	&	0.06	&	+2.82	&	0.06	\\
345	&	J052818.84+414932.6	&	HD 278203  	&	05 28 18.85 	&	 +41 49 32.69	&	A0 IV$-$V CrEu   	&	186	&	10.3076	&	0.0006	&	+0.917	&	0.294	&	$-$	&	$-$	&	$-$	&	$-$	&	$-$	\\
346	&	J052823.82+491141.0	&	Gaia DR2 213505989678368640	&	05 28 23.83 	&	 +49 11 41.06	&	A0 II$-$III  bl4130  	&	109	&	13.2444	&	0.0004	&	+0.273	&	0.022	&	+0.027	&	0.004	&	0.76	&	$-$0.33	&	0.18	\\
347	&	J052900.00+403426.0	&	Gaia DR2 193909630159512576	&	05 29 00.01 	&	 +40 34 26.05	&	B9.5 IV$-$V Cr   	&	171	&	11.3729	&	0.0007	&	+0.444	&	0.658	&	$-$	&	$-$	&	$-$	&	$-$	&	$-$	\\
348	&	J052945.57+440142.2	&	Gaia DR2 195762479048137344	&	05 29 45.57 	&	 +44 01 42.23	&	A2 IV$-$V SrCrEu   	&	194	&	12.3760	&	0.0002	&	+0.784	&	0.051	&	+0.178	&	0.002	&	0.69	&	+1.16	&	0.15	\\
349	&	J052947.73+420240.7	&	HD 278201  	&	05 29 47.74 	&	 +42 02 40.67	&	B9 V Cr   	&	283	&	10.9859	&	0.0010	&	+0.900	&	0.056	&	$-$0.432	&	0.004	&	1.39	&	$-$0.63	&	0.14	\\
350	&	J052950.88+172629.8	&	TYC 1301-1038-1    	&	05 29 50.88 	&	 +17 26 29.80	&	B9 IV bl4077 bl4130  	&	272	&	11.8268	&	0.0005	&	+0.725	&	0.041	&	+0.126	&	0.003	&	0.69	&	+0.44	&	0.13	\\
\hline
\end{tabular}                                                                                                                                                                   
\end{adjustbox}
\end{center}                                                                                                                                             
\end{sidewaystable*}
\setcounter{table}{0}  
\begin{sidewaystable*}
\caption{Essential data for our sample stars, sorted by increasing right ascension. The columns denote: (1) Internal identification number. (2) LAMOST identifier. (3) Alternativ identifier (HD number, TYC identifier or GAIA DR2 number). (4) Right ascension (J2000; GAIA DR2). (5) Declination (J2000; GAIA DR2). (6) Spectral type, as derived in this study. (7) Sloan $g$ band S/N ratio of the analysed spectrum. (8) $G$\,mag (GAIA DR2). (9) $G$\,mag error. (10) Parallax (GAIA DR2). (11) Parallax error. (12) Dereddened colour index $(BP-RP){_0}$ (GAIA DR2). (13) Colour index error. (14) Absorption in the $G$ band, $A_G$. (15) Intrinsic absolute magnitude in the $G$ band, $M_{\mathrm{G,0}}$. (16) Absolute magnitude error.}
\label{table_master8}
\begin{center}
\begin{adjustbox}{max width=\textwidth}
\begin{tabular}{lllcclcccccccccc}
\hline
\hline
(1) & (2) & (3) & (4) & (5) & (6) & (7) & (8) & (9) & (10) & (11) & (12) & (13) & (14) & (15) & (16) \\
No.	&	ID\_LAMOST	&	ID\_alt	&	RA(J2000) 	&	 Dec(J2000)    	&	SpT\_final	&	S/N\,$g$	&	$G$\,mag	&	e\_$G$\,mag	&	pi (mas)	&	e\_pi	&	$(BP-RP){_0}$	&	e\_$(BP-RP){_0}$	&	$A_G$	&	$M_{\mathrm{G,0}}$	&	e\_$M_{\mathrm{G,0}}$	\\
\hline 
351	&	J053025.32+332639.6	&	HD 244168  	&	05 30 25.32 	&	 +33 26 39.77	&	A6 V SrEu   	&	161	&	10.9158	&	0.0003	&	+3.246	&	0.432	&	+0.329	&	0.002	&	0.26	&	+3.21	&	0.29	\\
352	&	J053049.15+393224.7	&	Gaia DR2 190785028567285888	&	05 30 49.16 	&	 +39 32 24.77	&	A2 IV CrEu   	&	134	&	12.7920	&	0.0007	&	+0.729	&	0.056	&	+0.113	&	0.004	&	0.83	&	+1.27	&	0.17	\\
353	&	J053049.19+195827.1	&	Gaia DR2 3401277205495091200	&	05 30 49.20 	&	 +19 58 27.18	&	A1 IV$-$V CrEu   	&	111	&	14.3678	&	0.0006	&	+0.383	&	0.037	&	+0.283	&	0.004	&	0.71	&	+1.58	&	0.22	\\
354	&	J053125.46+414645.6	&	TYC 2918-328-1     	&	05 31 25.47 	&	 +41 46 45.59	&	B9.5 V SrCr   	&	197	&	11.1567	&	0.0008	&	+0.833	&	0.063	&	$-$0.065	&	0.002	&	0.76	&	+0.00	&	0.17	\\
355	&	J053139.66+380231.5	&	TYC 2910-488-1     	&	05 31 39.67 	&	 +38 02 31.45	&	B8 III$-$IV Si   	&	451	&	10.1420	&	0.0011	&	+1.601	&	0.113	&	$-$0.143	&	0.002	&	0.74	&	+0.42	&	0.16	\\
356	&	J053204.19+103549.6	&	TYC 705-856-1      	&	05 32 04.19 	&	 +10 35 49.65	&	B9 V Cr   	&	173	&	11.6659	&	0.0009	&	+0.642	&	0.057	&	+0.191	&	0.004	&	0.57	&	+0.13	&	0.20	\\
357	&	J053207.60+351630.3	&	TYC 2411-667-1     	&	05 32 07.60 	&	 +35 16 30.34	&	B8 IV$-$V Si   	&	200	&	11.2139	&	0.0008	&	+0.882	&	0.048	&	$-$0.071	&	0.004	&	0.47	&	+0.47	&	0.13	\\
358	&	J053229.49+171430.2	&	Gaia DR2 3397273059024206720	&	05 32 29.50 	&	 +17 14 30.22	&	B9.5 IV$-$V SrCrEu   	&	108	&	13.2816	&	0.0006	&	+0.592	&	0.027	&	+0.189	&	0.004	&	0.80	&	+1.34	&	0.11	\\
359	&	J053231.77+441954.0	&	Gaia DR2 195863638416486144	&	05 32 31.77 	&	 +44 19 54.12	&	B8 III$-$IV \textit{SiCrEu}*	&	132	&	13.1000	&	0.0005	&	+0.339	&	0.040	&	+0.012	&	0.004	&	0.62	&	+0.13	&	0.26	\\
360	&	J053239.91+434307.5	&	HD 36259   	&	05 32 39.89 	&	 +43 43 04.20	&	B8 IV Si   	&	218	&	9.0375	&	0.0007	&	+1.856	&	0.072	&	$-$0.221	&	0.003	&	0.50	&	$-$0.12	&	0.10	\\
361	&	J053259.46+161128.3	&	Gaia DR2 3396880529075737984	&	05 32 59.47 	&	 +16 11 28.40	&	B8 III$-$IV Si  (He-wk) 	&	200	&	12.4956	&	0.0019	&	+0.641	&	0.040	&	+0.010	&	0.008	&	0.95	&	+0.58	&	0.14	\\
362	&	J053315.94+351856.0	&	TYC 2412-64-1      	&	05 33 15.95 	&	 +35 18 56.08	&	A0 IV$-$V CrEu   	&	145	&	11.8044	&	0.0006	&	+0.903	&	0.055	&	+0.161	&	0.003	&	0.40	&	+1.18	&	0.14	\\
363	&	J053318.54+201332.6	&	TYC 1305-145-1     	&	05 33 18.55 	&	 +20 13 32.66	&	B9.5 IV$-$V SrCr   	&	171	&	11.5245	&	0.0010	&	+1.000	&	0.038	&	+0.049	&	0.006	&	0.68	&	+0.85	&	0.10	\\
364	&	J053320.79+333114.5	&	Gaia DR2 3449103487404273408	&	05 33 20.80 	&	 +33 31 14.51	&	B8 IV$-$V Si   	&	124	&	12.9540	&	0.0012	&	+0.577	&	0.038	&	$-$0.130	&	0.005	&	1.20	&	+0.56	&	0.15	\\
365	&	J053333.97+303009.2	&	Gaia DR2 3446277502000268672	&	05 33 33.97 	&	 +30 30 09.30	&	B8 III$-$IV  bl4130  	&	133	&	12.3690	&	0.0012	&	+0.668	&	0.045	&	$-$0.074	&	0.006	&	1.34	&	+0.16	&	0.15	\\
366	&	J053343.13+360821.5	&	Gaia DR2 183230593411140864	&	05 33 43.13 	&	 +36 08 21.52	&	B9 III$-$IV  bl4130  	&	135	&	14.2961	&	0.0012	&	+0.363	&	0.037	&	+0.055	&	0.004	&	1.42	&	+0.68	&	0.22	\\
367	&	J053347.07+132207.4	&	TYC 713-421-1      	&	05 33 47.07 	&	 +13 22 07.45	&	B9.5 III$-$IV SrEuSi   	&	233	&	11.0293	&	0.0013	&	+1.803	&	0.296	&	+0.103	&	0.004	&	0.87	&	+1.44	&	0.36	\\
368	&	J053355.06+325544.2	&	Gaia DR2 3448859872559385600	&	05 33 55.07 	&	 +32 55 44.17	&	A0 IV CrEu   	&	116	&	12.9827	&	0.0009	&	+0.628	&	0.028	&	$-$0.018	&	0.005	&	1.20	&	+0.78	&	0.11	\\
369	&	J053401.24+284902.2	&	TYC 1860-166-1     	&	05 34 01.25 	&	 +28 49 02.22	&	B9.5 V SrCr   	&	126	&	11.4276	&	0.0005	&	+0.977	&	0.051	&	+0.229	&	0.003	&	1.05	&	+0.32	&	0.12	\\
370	&	J053439.95-015728.1	&	HD 290693  	&	05 34 39.84 	&	 -01 57 28.19	&	A0 III CrEuSi   	&	385	&	10.5331	&	0.0024	&	+1.019	&	0.072	&	+0.024	&	0.01	&	0.38	&	+0.20	&	0.16	\\
371	&	J053504.75-012406.5	&	TYC 4766-330-1     	&	05 35 04.54 	&	 -01 24 06.58	&	kA0hA2mA4 CrEu   	&	559	&	9.4159	&	0.0008	&	+2.288	&	0.058	&	+0.042	&	0.004	&	0.30	&	+0.91	&	0.07	\\
372	&	J053509.30+350554.4	&	TYC 2412-450-1     	&	05 35 09.31 	&	 +35 05 54.47	&	B9 IV$-$V  bl4130  	&	126	&	12.1210	&	0.0009	&	+0.586	&	0.050	&	$-$0.046	&	0.004	&	0.96	&	0.00	&	0.19	\\
373	&	J053510.70+165914.2	&	Gaia DR2 3397038996190150528	&	05 35 10.71 	&	 +16 59 14.29	&	B9.5 IV$-$V SrCr   	&	205	&	12.5486	&	0.0017	&	+0.628	&	0.055	&	+0.271	&	0.004	&	0.85	&	+0.68	&	0.20	\\
374	&	J053526.26+175705.1	&	Gaia DR2 3397700288010818688	&	05 35 26.26 	&	 +17 57 05.17	&	B9 IV bl4077 bl4130  	&	186	&	12.6681	&	0.0007	&	+0.663	&	0.046	&	+0.036	&	0.005	&	0.90	&	+0.88	&	0.16	\\
375	&	J053531.48+305432.1	&	TYC 2404-1071-1    	&	05 35 31.48 	&	 +30 54 32.17	&	B8 IV$-$V Si   	&	180	&	12.0381	&	0.0007	&	+0.906	&	0.076	&	$-$0.149	&	0.005	&	1.08	&	+0.75	&	0.19	\\
376	&	J053531.81+211856.3	&	TYC 1309-1089-1    	&	05 35 31.81 	&	 +21 18 56.30	&	B9 IV CrSi   	&	167	&	11.8327	&	0.0003	&	+0.669	&	0.047	&	+0.024	&	0.003	&	0.80	&	+0.16	&	0.16	\\
377	&	J053542.47+145443.0	&	Gaia DR2 3395789302442225792	&	05 35 42.47 	&	 +14 54 43.00	&	B9.5 III$-$IV Si   	&	128	&	14.4888	&	0.0005	&	+0.374	&	0.033	&	+0.091	&	0.005	&	1.20	&	+1.16	&	0.20	\\
378	&	J053551.11+450901.3	&	TYC 3359-2146-1    	&	05 35 51.10 	&	 +45 09 01.24	&	B9 V Cr   	&	305	&	11.5348	&	0.0008	&	+0.623	&	0.050	&	+0.188	&	0.004	&	0.52	&	$-$0.01	&	0.18	\\
379	&	J053554.49+210229.6	&	Gaia DR2 3402868160162970496	&	05 35 54.50 	&	 +21 02 29.71	&	A3 II$-$III SrCrEuSi   	&	138	&	12.1587	&	0.0005	&	+0.885	&	0.039	&	+0.500	&	0.003	&	0.57	&	+1.32	&	0.11	\\
380	&	J053601.89+324756.5	&	Gaia DR2 3448827853582010112	&	05 36 01.89 	&	 +32 47 56.56	&	B9 III EuSi   	&	112	&	14.3607	&	0.0006	&	+0.315	&	0.032	&	+0.151	&	0.004	&	1.31	&	+0.54	&	0.23	\\
381	&	J053636.38+305520.4 $^{a}$	&	HD 245261  	&	05 36 36.39 	&	 +30 55 20.53	&	B9.5 V CrSi   	&	172	&	11.0049	&	0.0009	&	+1.777	&	0.068	&	$-$0.047	&	0.004	&	0.77	&	+1.48	&	0.10	\\
382	&	J053749.62+541330.7	&	TYC 3749-702-1     	&	05 37 49.63 	&	 +54 13 30.76	&	B9 V Cr   	&	177	&	11.7930	&	0.0008	&	+0.914	&	0.041	&	+0.159	&	0.004	&	0.63	&	+0.97	&	0.11	\\
383	&	J053755.64+493047.3	&	Gaia DR2 214241180701977216	&	05 37 55.64 	&	 +49 30 47.40	&	B9.5 II$-$III Si   	&	132	&	12.0179	&	0.0007	&	+0.395	&	0.039	&	+0.084	&	0.004	&	0.48	&	$-$0.48	&	0.22	\\
384	&	J053815.61+193035.7	&	TYC 1306-837-1     	&	05 38 15.61 	&	 +19 30 35.84	&	B8 IV Si   	&	162	&	11.3036	&	0.0017	&	+0.718	&	0.063	&	$-$0.002	&	0.008	&	0.77	&	$-$0.19	&	0.20	\\
385	&	J053821.15+143839.3	&	Gaia DR2 3347684564918451328	&	05 38 21.15 	&	 +14 38 39.43	&	B9 IV$-$V SrCr   	&	183	&	11.7856	&	0.0003	&	+0.823	&	0.049	&	+0.026	&	0.003	&	0.87	&	+0.50	&	0.14	\\
386	&	J053836.41+242850.6	&	Gaia DR2 3428723146028987392	&	05 38 36.41 	&	 +24 28 50.69	&	kB7hB9mA1 Si   	&	154	&	12.2723	&	0.0005	&	+0.592	&	0.048	&	+0.024	&	0.003	&	1.22	&	$-$0.09	&	0.18	\\
387	&	J053840.17+413754.0	&	Gaia DR2 192886156629653504	&	05 38 40.18 	&	 +41 37 54.05	&	kA1hA7mA9 SrCrEuSi   	&	114	&	12.2659	&	0.0006	&	+0.912	&	0.041	&	+0.256	&	0.003	&	0.44	&	+1.63	&	0.11	\\
388	&	J053901.26+540638.5	&	TYC 3749-888-1     	&	05 39 00.93 	&	 +54 06 38.55	&	A7 V SrCrEuSi   	&	253	&	9.7068	&	0.0006	&	+2.456	&	0.048	&	+0.222	&	0.003	&	0.63	&	+1.02	&	0.07	\\
389	&	J053909.24+461117.3	&	TYC 3360-1477-1    	&	05 39 09.24 	&	 +46 11 17.34	&	B9 IV$-$V Sr   	&	227	&	11.6665	&	0.0006	&	+0.981	&	0.057	&	+0.074	&	0.005	&	0.65	&	+0.97	&	0.14	\\
390	&	J053912.12+480622.4	&	Gaia DR2 210874235579559808	&	05 39 12.12 	&	 +48 06 22.34	&	B9 IV$-$V Cr   	&	132	&	12.2815	&	0.0004	&	+0.627	&	0.034	&	+0.223	&	0.002	&	0.79	&	+0.48	&	0.13	\\
391	&	J054027.66+333102.4	&	Gaia DR2 3448584552275686144	&	05 40 27.66 	&	 +33 31 02.67	&	kB9.5hA2mA4 Eu   	&	116	&	14.6747	&	0.0035	&	+0.322	&	0.035	&	+0.186	&	0.014	&	1.00	&	+1.21	&	0.24	\\
392	&	J054044.78+243154.9	&	Gaia DR2 3428700842260487040	&	05 40 44.79 	&	 +24 31 55.06	&	A5 IV SrCrEu   	&	194	&	11.9913	&	0.0008	&	+0.967	&	0.065	&	+0.056	&	0.004	&	1.12	&	+0.80	&	0.15	\\
393	&	J054102.12+332331.1	&	TYC 2408-1757-1    	&	05 41 02.13 	&	 +33 23 31.10	&	B7 IV Si  (He-wk) 	&	197	&	11.2705	&	0.0025	&	+0.564	&	0.068	&	$-$0.176	&	0.01	&	0.73	&	$-$0.71	&	0.27	\\
394	&	J054123.41+253043.2	&	HD 246255  	&	05 41 23.29 	&	 +25 30 42.93	&	B9.5 IV$-$V SrCr   	&	243	&	10.4934	&	0.0006	&	+1.521	&	0.065	&	+0.078	&	0.004	&	0.30	&	+1.11	&	0.10	\\
395	&	J054301.11+383843.2	&	Gaia DR2 189854704291739648	&	05 43 01.12 	&	 +38 38 43.22	&	B9 IV$-$V Cr   	&	122	&	11.8730	&	0.0010	&	+1.181	&	0.070	&	+0.047	&	0.006	&	0.86	&	+1.38	&	0.14	\\
396	&	J054402.53+181251.3	&	HD 246892  	&	05 44 02.61 	&	 +18 12 51.39	&	B9.5 II$-$III EuSi   	&	329	&	10.8254	&	0.0013	&	+0.817	&	0.082	&	$-$0.025	&	0.006	&	0.47	&	$-$0.08	&	0.22	\\
397	&	J054418.27+382207.0	&	Gaia DR2 189926408770237184	&	05 44 18.28 	&	 +38 22 07.12	&	A0 III$-$IV SrCr   	&	119	&	12.3646	&	0.0007	&	$-$0.291	&	0.124	&	$-$	&	$-$	&	$-$	&	$-$	&	$-$	\\
398	&	J054543.05+350259.7	&	TYC 2413-643-1     	&	05 45 43.05 	&	 +35 02 59.76	&	A0 IV$-$V SrCr   	&	133	&	11.8787	&	0.0006	&	+1.118	&	0.050	&	$-$0.018	&	0.004	&	0.69	&	+1.43	&	0.11	\\
399	&	J054547.46+400703.6	&	TYC 2915-1913-1    	&	05 45 47.46 	&	 +40 07 03.69	&	kA1hA3mA7 SrCrEuSi   	&	192	&	11.9338	&	0.0003	&	+1.011	&	0.037	&	+0.123	&	0.003	&	0.67	&	+1.29	&	0.09	\\
400	&	J054548.96+222854.3	&	HD 247241  	&	05 45 48.96 	&	 +22 28 55.30	&	B9 V SrCrEu   	&	469	&	10.3016	&	0.0005	&	+1.342	&	0.090	&	$-$0.009	&	0.002	&	0.52	&	+0.42	&	0.15	\\
\hline
\end{tabular}                                                                                                                                                                   
\end{adjustbox}
\end{center}                                                                                                                                             
\end{sidewaystable*}
\setcounter{table}{0}  
\begin{sidewaystable*}
\caption{Essential data for our sample stars, sorted by increasing right ascension. The columns denote: (1) Internal identification number. (2) LAMOST identifier. (3) Alternativ identifier (HD number, TYC identifier or GAIA DR2 number). (4) Right ascension (J2000; GAIA DR2). (5) Declination (J2000; GAIA DR2). (6) Spectral type, as derived in this study. (7) Sloan $g$ band S/N ratio of the analysed spectrum. (8) $G$\,mag (GAIA DR2). (9) $G$\,mag error. (10) Parallax (GAIA DR2). (11) Parallax error. (12) Dereddened colour index $(BP-RP){_0}$ (GAIA DR2). (13) Colour index error. (14) Absorption in the $G$ band, $A_G$. (15) Intrinsic absolute magnitude in the $G$ band, $M_{\mathrm{G,0}}$. (16) Absolute magnitude error.}
\label{table_master9}
\begin{center}
\begin{adjustbox}{max width=\textwidth}
\begin{tabular}{lllcclcccccccccc}
\hline
\hline
(1) & (2) & (3) & (4) & (5) & (6) & (7) & (8) & (9) & (10) & (11) & (12) & (13) & (14) & (15) & (16) \\
No.	&	ID\_LAMOST	&	ID\_alt	&	RA(J2000) 	&	 Dec(J2000)    	&	SpT\_final	&	S/N\,$g$	&	$G$\,mag	&	e\_$G$\,mag	&	pi (mas)	&	e\_pi	&	$(BP-RP){_0}$	&	e\_$(BP-RP){_0}$	&	$A_G$	&	$M_{\mathrm{G,0}}$	&	e\_$M_{\mathrm{G,0}}$	\\
\hline 
401	&	J054613.99+245550.4	&	Gaia DR2 3428625182116541952	&	05 46 14.00 	&	 +24 55 50.41	&	A9 V SrCrEuSi   	&	107	&	12.2017	&	0.0002	&	+0.555	&	0.041	&	+0.154	&	0.003	&	0.90	&	+0.02	&	0.17	\\
402	&	J054622.23+204640.9	&	Gaia DR2 3399995204999449984	&	05 46 22.23 	&	 +20 46 40.82	&	B8 IV Si   	&	94	&	12.0309	&	0.0010	&	+0.851	&	0.038	&	+0.155	&	0.005	&	0.40	&	+1.28	&	0.11	\\
403	&	J054630.44+273518.1	&	TYC 1870-1407-1    	&	05 46 30.45 	&	 +27 35 18.16	&	B9 V CrEu   	&	137	&	11.5944	&	0.0004	&	+0.962	&	0.053	&	$-$0.020	&	0.004	&	0.72	&	+0.79	&	0.13	\\
404	&	J054634.75+370200.6	&	TYC 2417-826-1     	&	05 46 34.76 	&	 +37 02 00.61	&	B8 IV Si   	&	140	&	11.9319	&	0.0007	&	+0.625	&	0.051	&	$-$0.053	&	0.004	&	0.83	&	+0.08	&	0.18	\\
405	&	J054724.36+171922.7	&	Gaia DR2 3398036734272024704	&	05 47 24.37 	&	 +17 19 22.81	&	B9 IV$-$V SrCr   	&	195	&	11.9193	&	0.0026	&	+0.582	&	0.052	&	+0.109	&	0.013	&	0.65	&	+0.09	&	0.20	\\
406	&	J054732.61+121608.0	&	TYC 723-950-1      	&	05 47 32.62 	&	 +12 16 08.07	&	\textit{B5 V HeB9}* &	110	&	12.8837	&	0.0014	&	+0.430	&	0.038	&	$-$0.019	&	0.006	&	0.73	&	+0.32	&	0.20	\\
407	&	J054750.69+115742.3	&	TYC 723-750-1      	&	05 47 50.70 	&	 +11 57 42.35	&	B9 IV$-$V CrSi   	&	223	&	11.6513	&	0.0007	&	+0.538	&	0.044	&	+0.033	&	0.004	&	0.58	&	$-$0.27	&	0.18	\\
408	&	J054754.48+413621.3	&	TYC 2919-515-1     	&	05 47 54.49 	&	 +41 36 21.36	&	B9 IV$-$V Cr   	&	321	&	11.4195	&	0.0008	&	+0.920	&	0.048	&	+0.108	&	0.004	&	0.91	&	+0.33	&	0.12	\\
409	&	J054757.12+235011.8	&	TYC 1862-1494-1    	&	05 47 57.13 	&	 +23 50 11.83	&	B9.5 II$-$III EuSi   	&	281	&	10.3040	&	0.0006	&	+1.009	&	0.055	&	$-$0.031	&	0.003	&	0.42	&	$-$0.10	&	0.13	\\
410	&	J054807.07+210619.0	&	HD 247724  	&	05 48 07.08 	&	 +21 06 19.08	&	B7 IV  bl4130  	&	522	&	9.6286	&	0.0005	&	+0.971	&	0.061	&	$-$0.135	&	0.003	&	0.39	&	$-$0.83	&	0.14	\\
411	&	J054819.92+333516.9 $^{a}$	&	TYC 2409-1922-1    	&	05 48 19.77 	&	 +33 35 16.98	&	kA4hA9mF1 SrCrEuSi   	&	200	&	10.4505	&	0.0006	&	+1.729	&	0.086	&	+0.052	&	0.002	&	0.58	&	+1.06	&	0.12	\\
412	&	J054822.35+103530.7	&	TYC 719-303-1      	&	05 48 22.36 	&	 +10 35 30.78	&	B9 V Cr   	&	112	&	11.3847	&	0.0007	&	+0.829	&	0.041	&	+0.129	&	0.005	&	0.67	&	+0.31	&	0.12	\\
413	&	J054826.28+304929.2	&	TYC 2405-268-1     	&	05 48 26.29 	&	 +30 49 29.21	&	B8 III Eu   	&	123	&	11.6907	&	0.0017	&	+0.487	&	0.055	&	$-$0.007	&	0.007	&	0.88	&	$-$0.76	&	0.25	\\
414	&	J054844.18+522027.9	&	TYC 3372-485-1     	&	05 48 44.19 	&	 +52 20 27.92	&	A1 V CrEu   	&	198	&	12.0289	&	0.0003	&	+0.829	&	0.035	&	+0.202	&	0.002	&	0.54	&	+1.08	&	0.10	\\
415	&	J054850.62+350217.4	&	TYC 2413-1-1       	&	05 48 50.46 	&	 +35 02 17.46	&	kB8hA0mA1 CrEu   	&	156	&	10.3198	&	0.0006	&	+1.416	&	0.066	&	$-$0.066	&	0.002	&	0.60	&	+0.48	&	0.11	\\
416	&	J054928.70+404455.7	&	Gaia DR2 191842273418923776	&	05 49 28.70 	&	 +40 44 55.73	&	B9 V SrCr   	&	204	&	12.4174	&	0.0006	&	+0.505	&	0.040	&	+0.036	&	0.004	&	0.57	&	+0.36	&	0.18	\\
417	&	J054952.18+352212.6	&	TYC 2413-441-1     	&	05 49 52.19 	&	 +35 22 12.66	&	B8 IV Si   	&	259	&	10.9698	&	0.0005	&	+0.971	&	0.044	&	$-$0.103	&	0.002	&	0.72	&	+0.18	&	0.11	\\
418	&	J055002.80+234023.9	&	HD 248072  	&	05 50 02.80 	&	 +23 40 23.97	&	B9 IV  bl4130  	&	349	&	10.5259	&	0.0015	&	+1.124	&	0.134	&	$-$0.103	&	0.006	&	0.48	&	+0.30	&	0.26	\\
419	&	J055002.87+235548.0	&	TYC 1862-2027-1    	&	05 50 02.87 	&	 +23 55 48.06	&	A1 IV$-$V Cr   	&	186	&	11.6992	&	0.0006	&	+1.081	&	0.072	&	$-$0.005	&	0.004	&	0.52	&	+1.35	&	0.15	\\
420	&	J055019.83+214937.6	&	Gaia DR2 3424216763262732544	&	05 50 19.83 	&	 +21 49 37.69	&	B9.5 III$-$IV EuSi   	&	240	&	11.5652	&	0.0009	&	+0.950	&	0.038	&	$-$0.050	&	0.005	&	0.67	&	+0.78	&	0.10	\\
421	&	J055023.89+261330.2	&	TYC 1866-861-1     	&	05 50 23.89 	&	 +26 13 30.29	&	\textit{B4: V HeB9}* 	&	79	&	11.5523	&	0.0015	&	+0.612	&	0.052	&	$-$0.182	&	0.009	&	1.03	&	$-$0.55	&	0.19	\\
422	&	J055026.09+314751.1	&	Gaia DR2 3444961073046795392	&	05 50 26.09 	&	 +31 47 51.20	&	\textit{B9 III Si}*	&	62	&	14.1948	&	0.0009	&	+0.252	&	0.051	&	$-$0.045	&	0.006	&	0.79	&	+0.41	&	0.44	\\
423	&	J055045.30+372809.0	&	TYC 2417-126-1     	&	05 50 45.30 	&	 +37 28 09.07	&	kA2hA4mA7 SrCrEu   	&	133	&	11.8874	&	0.0004	&	+1.534	&	0.059	&	+0.279	&	0.003	&	0.89	&	+1.93	&	0.10	\\
424	&	J055054.75+410356.0	&	TYC 2916-1788-1    	&	05 50 54.75 	&	 +41 03 56.09	&	B8 IV Si  (He-wk) 	&	454	&	11.0859	&	0.0008	&	+0.677	&	0.063	&	$-$0.110	&	0.005	&	0.50	&	$-$0.26	&	0.21	\\
425	&	J055105.53+411410.9	&	Gaia DR2 192253357627486592	&	05 51 05.53 	&	 +41 14 10.94	&	A1 V SrCrEu   	&	169	&	13.0171	&	0.0008	&	+0.548	&	0.039	&	+0.186	&	0.003	&	0.55	&	+1.16	&	0.16	\\
426	&	J055108.25+531610.8 $^{a}$	&	TYC 3750-928-1     	&	05 51 08.26 	&	 +53 16 10.89	&	A2 IV$-$V SrCrEu   	&	261	&	11.5413	&	0.0004	&	+1.375	&	0.037	&	+0.250	&	0.002	&	0.35	&	+1.88	&	0.08	\\
427	&	J055110.67+360517.0	&	TYC 2418-790-1     	&	05 51 10.68 	&	 +36 05 17.01	&	B9.5 V SrCr   	&	116	&	12.0787	&	0.0003	&	+0.854	&	0.039	&	+0.314	&	0.003	&	0.78	&	+0.96	&	0.11	\\
428	&	J055121.05+420610.5	&	HD 38943   	&	05 51 20.74 	&	 +42 06 10.53	&	B8 IV Si   	&	352	&	9.0962	&	0.0011	&	+1.422	&	0.057	&	$-$0.046	&	0.007	&	0.36	&	$-$0.50	&	0.10	\\
429	&	J055143.67+205025.1	&	HD 248458  	&	05 51 43.67 	&	 +20 50 25.09	&	B9 IV$-$V  bl4130  	&	288	&	9.9889	&	0.0007	&	+0.566	&	0.277	&	$-$	&	$-$	&	$-$	&	$-$	&	$-$	\\
430	&	J055154.28+350352.9	&	TYC 2414-679-1     	&	05 51 54.29 	&	 +35 03 52.95	&	A3 III$-$IV SrCrEu   	&	105	&	12.4204	&	0.0005	&	+0.964	&	0.042	&	+0.333	&	0.006	&	0.50	&	+1.85	&	0.11	\\
431	&	J055157.54+410439.7	&	TYC 2916-2241-1    	&	05 51 57.55 	&	 +41 04 39.72	&	A1 IV$-$V Cr   	&	119	&	11.5408	&	0.0006	&	+0.708	&	0.059	&	+0.080	&	0.004	&	0.61	&	+0.18	&	0.19	\\
432	&	J055207.96+242953.9	&	Gaia DR2 3428136865813370880	&	05 52 07.97 	&	 +24 29 53.96	&	B9 IV CrSi   	&	116	&	14.4510	&	0.0008	&	+0.198	&	0.041	&	$-$0.406	&	0.006	&	2.06	&	$-$1.12	&	0.45	\\
433	&	J055218.79+331852.7	&	TYC 2410-2403-1    	&	05 52 18.79 	&	 +33 18 52.82	&	B9.5 V Cr   	&	226	&	11.2003	&	0.0006	&	+0.852	&	0.037	&	+0.087	&	0.003	&	0.56	&	+0.30	&	0.11	\\
434	&	J055219.99+261909.7	&	TYC 1871-116-1     	&	05 52 19.99 	&	 +26 19 09.70	&	B9 IV$-$V  bl4130  	&	173	&	11.1083	&	0.0018	&	+1.085	&	0.074	&	$-$0.134	&	0.005	&	0.67	&	+0.62	&	0.16	\\
435	&	J055237.04+154023.8	&	TYC 1312-2862-1    	&	05 52 37.04 	&	 +15 40 23.90	&	B8 IV$-$V Si   	&	184	&	11.7179	&	0.0010	&	+0.721	&	0.040	&	$-$0.099	&	0.009	&	0.36	&	+0.64	&	0.13	\\
436	&	J055237.95+274922.8	&	TYC 1871-1738-1    	&	05 52 37.95 	&	 +27 49 22.78	&	A6 V SrCrEu   	&	120	&	11.4405	&	0.0005	&	+1.285	&	0.045	&	+0.197	&	0.003	&	0.38	&	+1.60	&	0.09	\\
437	&	J055254.41+310240.3	&	Gaia DR2 3444843223443216000	&	05 52 54.41 	&	 +31 02 40.36	&	A0 III$-$IV CrSi   	&	18	&	14.2614	&	0.0005	&	+0.380	&	0.028	&	+0.063	&	0.003	&	0.64	&	+1.52	&	0.17	\\
438	&	J055300.15+411220.9	&	TYC 2916-1620-1    	&	05 53 00.15 	&	 +41 12 20.96	&	A1 IV$-$V SrCr   	&	160	&	11.2609	&	0.0008	&	+0.830	&	0.048	&	+0.218	&	0.002	&	0.52	&	+0.34	&	0.14	\\
439	&	J055308.38+254739.9	&	TYC 1867-2207-1    	&	05 53 08.39 	&	 +25 47 39.95	&	B8 III$-$IV EuSi   	&	116	&	10.9895	&	0.0006	&	+0.875	&	0.074	&	$-$0.066	&	0.006	&	0.63	&	+0.07	&	0.19	\\
440	&	J055321.75+302915.8	&	TYC 2406-1980-1    	&	05 53 21.76 	&	 +30 29 15.91	&	A1 V SrCrEu   	&	103	&	12.0275	&	0.0005	&	+1.333	&	0.048	&	+0.114	&	0.003	&	0.51	&	+2.14	&	0.09	\\
441	&	J055333.92+180202.9	&	HD 248844  	&	05 53 33.92 	&	 +18 02 02.96	&	F2 V SrEu   	&	182	&	10.3666	&	0.0004	&	+0.561	&	0.155	&	$-$	&	$-$	&	$-$	&	$-$	&	$-$	\\
442	&	J055346.74+144708.2	&	Gaia DR2 3347180885513452160	&	05 53 46.74 	&	 +14 47 08.17	&	B9.5 IV$-$V CrEu   	&	111	&	12.2860	&	0.0013	&	+0.649	&	0.043	&	+0.020	&	0.007	&	0.44	&	+0.91	&	0.15	\\
443	&	J055409.86+165428.3	&	HD 248975  	&	05 54 09.87 	&	 +16 54 28.42	&	B8 IV Si   	&	157	&	10.7329	&	0.0007	&	+0.820	&	0.068	&	$-$0.122	&	0.002	&	0.27	&	+0.03	&	0.19	\\
444	&	J055411.43+304113.3	&	Gaia DR2 3444069098537504512	&	05 54 11.43 	&	 +30 41 13.53	&	B8 IV$-$V Si   	&	114	&	11.5208	&	0.0004	&	+0.766	&	0.041	&	+0.011	&	0.003	&	0.56	&	+0.38	&	0.13	\\
445	&	J055422.76+305401.8	&	HD 248874  	&	05 54 22.76 	&	 +30 54 01.90	&	B8 III  bl4130  	&	253	&	10.0389	&	0.0012	&	+0.812	&	0.058	&	$-$0.092	&	0.003	&	0.71	&	$-$1.13	&	0.16	\\
446	&	J055422.96+415825.7	&	TYC 2920-1365-1    	&	05 54 22.96 	&	 +41 58 25.69	&	B9 V CrEu   	&	409	&	9.9528	&	0.0009	&	+1.443	&	0.057	&	+0.014	&	0.003	&	0.29	&	+0.46	&	0.10	\\
447	&	J055426.04+241129.6	&	Gaia DR2 3427914867543706496	&	05 54 26.05 	&	 +24 11 29.74	&	B9 V Cr   	&	118	&	14.1450	&	0.0004	&	+0.378	&	0.032	&	+0.066	&	0.003	&	1.12	&	+0.91	&	0.19	\\
448	&	J055449.32+263732.6	&	TYC 1871-1008-1    	&	05 54 49.33 	&	 +26 37 32.73	&	A0 IV$-$V SrCrEu   	&	135	&	11.5552	&	0.0003	&	+1.153	&	0.051	&	+0.022	&	0.004	&	0.63	&	+1.24	&	0.11	\\
449	&	J055504.38+262725.6	&	TYC 1871-1483-1    	&	05 55 04.38 	&	 +26 27 25.64	&	B8 IV$-$V SrSi   	&	57	&	11.1396	&	0.0008	&	+0.789	&	0.041	&	$-$0.108	&	0.004	&	0.82	&	$-$0.19	&	0.12	\\
450	&	J055507.06+121347.0	&	TYC 724-1180-1     	&	05 55 07.07 	&	 +12 13 47.04	&	B9.5 II$-$III SrSi   	&	260	&	11.3687	&	0.0010	&	+0.574	&	0.063	&	$-$0.090	&	0.005	&	0.36	&	$-$0.20	&	0.24	\\
\hline
\end{tabular}                                                                                                                                                                   
\end{adjustbox}
\end{center}                                                                                                                                             
\end{sidewaystable*}
\setcounter{table}{0}  
\begin{sidewaystable*}
\caption{Essential data for our sample stars, sorted by increasing right ascension. The columns denote: (1) Internal identification number. (2) LAMOST identifier. (3) Alternativ identifier (HD number, TYC identifier or GAIA DR2 number). (4) Right ascension (J2000; GAIA DR2). (5) Declination (J2000; GAIA DR2). (6) Spectral type, as derived in this study. (7) Sloan $g$ band S/N ratio of the analysed spectrum. (8) $G$\,mag (GAIA DR2). (9) $G$\,mag error. (10) Parallax (GAIA DR2). (11) Parallax error. (12) Dereddened colour index $(BP-RP){_0}$ (GAIA DR2). (13) Colour index error. (14) Absorption in the $G$ band, $A_G$. (15) Intrinsic absolute magnitude in the $G$ band, $M_{\mathrm{G,0}}$. (16) Absolute magnitude error.}
\label{table_master10}
\begin{center}
\begin{adjustbox}{max width=\textwidth}
\begin{tabular}{lllcclcccccccccc}
\hline
(1) & (2) & (3) & (4) & (5) & (6) & (7) & (8) & (9) & (10) & (11) & (12) & (13) & (14) & (15) & (16) \\
No.	&	ID\_LAMOST	&	ID\_alt	&	RA(J2000) 	&	 Dec(J2000)    	&	SpT\_final	&	S/N\,$g$	&	$G$\,mag	&	e\_$G$\,mag	&	pi (mas)	&	e\_pi	&	$(BP-RP){_0}$	&	e\_$(BP-RP){_0}$	&	$A_G$	&	$M_{\mathrm{G,0}}$	&	e\_$M_{\mathrm{G,0}}$	\\
\hline
451	&	J055519.77+154335.4	&	TYC 1312-2891-1    	&	05 55 19.77 	&	 +15 43 35.47	&	B9 V Cr   	&	164	&	10.8051	&	0.0010	&	+1.094	&	0.068	&	$-$0.024	&	0.005	&	0.31	&	+0.69	&	0.14	\\
452	&	J055523.13+150002.5	&	HD 249209  	&	05 55 23.13 	&	 +15 00 02.59	&	B8 IV CrSi   	&	227	&	10.9644	&	0.0013	&	+0.499	&	0.075	&	+0.001	&	0.005	&	0.43	&	$-$0.97	&	0.33	\\
453	&	J055529.24+393248.5	&	TYC 2916-1265-1    	&	05 55 29.25 	&	 +39 32 48.53	&	kB8hA3mA3 CrSi   	&	120	&	11.4179	&	0.0005	&	+0.961	&	0.047	&	$-$0.026	&	0.004	&	0.55	&	+0.79	&	0.12	\\
454	&	J055534.89+281813.2	&	Gaia DR2 3431156738919784320	&	05 55 34.90 	&	 +28 18 13.28	&	B9 III$-$IV Si   	&	159	&	12.3021	&	0.0004	&	+0.404	&	0.058	&	+0.016	&	0.004	&	0.80	&	$-$0.46	&	0.32	\\
455	&	J055539.90+171036.9	&	Gaia DR2 3349797452608779520	&	05 55 39.91 	&	 +17 10 36.91	&	A0 IV$-$V CrEu   	&	112	&	14.0679	&	0.0009	&	+0.469	&	0.028	&	+0.189	&	0.005	&	0.69	&	+1.73	&	0.14	\\
456	&	J055545.19+195746.3	&	TYC 1320-1062-1    	&	05 55 45.19 	&	 +19 57 46.32	&	B9.5 III$-$IV Si   	&	161	&	11.4597	&	0.0004	&	+0.778	&	0.040	&	+0.000	&	0.003	&	0.61	&	+0.30	&	0.12	\\
457	&	J055557.19+275828.0	&	TYC 1871-1862-1    	&	05 55 57.19 	&	 +27 58 28.00	&	A0 IV$-$V Cr   	&	109	&	12.2417	&	0.0009	&	+0.645	&	0.069	&	+0.178	&	0.004	&	0.78	&	+0.51	&	0.24	\\
458	&	J055604.42+160937.1	&	TYC 1312-1184-1    	&	05 56 04.42 	&	 +16 09 37.11	&	B8 IV$-$V Si   	&	260	&	11.2583	&	0.0009	&	+0.977	&	0.051	&	$-$0.151	&	0.006	&	0.49	&	+0.71	&	0.12	\\
459	&	J055618.50+210706.3	&	TYC 1324-209-1     	&	05 56 18.51 	&	 +21 07 06.39	&	B5 III$-$IV  bl4130 (He-wk) 	&	125	&	11.5927	&	0.0008	&	+0.561	&	0.050	&	$-$0.028	&	0.004	&	0.86	&	$-$0.52	&	0.20	\\
460	&	J055625.41+295741.2	&	TYC 1875-195-1     	&	05 56 25.42 	&	 +29 57 41.22	&	B6 IV$-$V  bl4130 (He-wk) (He-st)	&	158	&	12.0124	&	0.0004	&	+0.699	&	0.051	&	$-$0.044	&	0.004	&	0.53	&	+0.70	&	0.16	\\
461	&	J055629.24+120153.3	&	TYC 724-1222-1     	&	05 56 29.25 	&	 +12 01 53.34	&	B9 V SrCr   	&	221	&	11.5961	&	0.0008	&	+0.814	&	0.040	&	+0.016	&	0.005	&	0.37	&	+0.77	&	0.12	\\
462	&	J055657.30+265341.3	&	Gaia DR2 3430777686586008576	&	05 56 57.30 	&	 +26 53 41.38	&	B9 III$-$IV EuSi   	&	101	&	12.4593	&	0.0015	&	+0.368	&	0.038	&	+0.012	&	0.006	&	0.83	&	$-$0.54	&	0.23	\\
463	&	J055703.71+463318.7	&	TYC 3361-532-1     	&	05 57 03.68 	&	 +46 33 17.57	&	B9 V Cr   	&	456	&	10.8254	&	0.0009	&	+1.029	&	0.049	&	+0.003	&	0.004	&	0.30	&	+0.58	&	0.12	\\
464	&	J055719.01+124008.6	&	Gaia DR2 3343641424444963072	&	05 57 19.01 	&	 +12 40 08.67	&	A1 III$-$IV CrEuSi   	&	147	&	12.1762	&	0.0002	&	+0.844	&	0.040	&	+0.367	&	0.003	&	0.25	&	+1.56	&	0.11	\\
465	&	J055730.88+295306.9	&	TYC 1875-271-1     	&	05 57 30.88 	&	 +29 53 06.89	&	B9 IV$-$V SrCr   	&	117	&	11.6520	&	0.0013	&	+0.629	&	0.043	&	+0.039	&	0.006	&	0.56	&	+0.08	&	0.16	\\
466	&	J055739.82+160140.8	&	Gaia DR2 3348827404175942656	&	05 57 39.83 	&	 +16 01 40.94	&	A0 IV$-$V Cr   	&	165	&	12.0056	&	0.0009	&	+0.735	&	0.045	&	+0.084	&	0.007	&	0.53	&	+0.81	&	0.14	\\
467	&	J055940.19+284225.7	&	TYC 1875-2529-1    	&	05 59 40.19 	&	 +28 42 25.72	&	B9 III$-$IV Si   	&	118	&	12.1002	&	0.0021	&	+0.503	&	0.059	&	+0.035	&	0.011	&	0.64	&	$-$0.04	&	0.26	\\
468	&	J060020.24+154701.5	&	TYC 1313-1569-1    	&	06 00 20.25 	&	 +15 47 01.54	&	B9 V bl4077 bl4130  	&	152	&	11.8799	&	0.0012	&	+0.825	&	0.043	&	$-$0.073	&	0.005	&	0.61	&	+0.85	&	0.12	\\
469	&	J060040.60+100410.1	&	TYC 721-2348-1     	&	06 00 40.60 	&	 +10 04 10.11	&	B9.5 V CrEu   	&	207	&	11.9664	&	0.0005	&	+0.766	&	0.039	&	+0.028	&	0.004	&	0.43	&	+0.96	&	0.12	\\
470	&	J060045.94-035344.3	&	HD 40759   	&	06 00 45.73 	&	 -03 53 44.41	&	A0 IV$-$V Si   	&	702	&	8.5214	&	0.0006	&	+2.326	&	0.055	&	$-$0.142	&	0.002	&	0.42	&	$-$0.07	&	0.07	\\
471	&	J060106.34-042123.3	&	Gaia DR2 3024207178875434496	&	06 01 06.35 	&	 -04 21 23.33	&	kA1hA5mA8 SrCrEu   	&	152	&	11.7624	&	0.0005	&	+0.726	&	0.041	&	+0.237	&	0.003	&	0.87	&	+0.19	&	0.13	\\
472	&	J060136.47+293812.0	&	Gaia DR2 3437677014671183744	&	06 01 36.47 	&	 +29 38 12.09	&	B8 IV Si   	&	56	&	13.1481	&	0.0014	&	+0.334	&	0.047	&	+0.033	&	0.006	&	0.61	&	+0.16	&	0.31	\\
473	&	J060155.40+280356.1	&	TYC 1872-1819-1    	&	06 01 55.40 	&	 +28 03 56.11	&	B8 IV Si   	&	247	&	11.6929	&	0.0004	&	+0.623	&	0.073	&	$-$0.002	&	0.004	&	0.33	&	+0.33	&	0.26	\\
474	&	J060202.24+282202.0	&	Gaia DR2 3431346134097961216	&	06 02 02.24 	&	 +28 22 02.03	&	B9 IV$-$V SrEu   	&	56	&	12.7638	&	0.0015	&	+0.272	&	0.045	&	$-$0.063	&	0.009	&	0.62	&	$-$0.68	&	0.36	\\
475	&	J060225.92+244628.5	&	HD 40833   	&	06 02 25.92 	&	 +24 46 28.52	&	B8 IV Si   	&	535	&	9.2072	&	0.0019	&	+1.321	&	0.071	&	$-$0.181	&	0.008	&	0.39	&	$-$0.58	&	0.13	\\
476	&	J060227.33+282943.9	&	HD 250515  	&	06 02 27.33 	&	 +28 29 43.95	&	B8 III$-$IV EuSi  (He-wk) 	&	298	&	10.1046	&	0.0017	&	+0.557	&	0.051	&	+0.006	&	0.007	&	0.34	&	$-$1.51	&	0.20	\\
477	&	J060241.10+231013.0	&	TYC 1864-1494-1    	&	06 02 41.11 	&	 +23 10 13.23	&	B9 III$-$IV Si   	&	308	&	11.3684	&	0.0004	&	+0.992	&	0.041	&	$-$0.334	&	0.004	&	1.49	&	$-$0.13	&	0.10	\\
478	&	J060258.76+160557.2	&	HD 250765  	&	06 02 58.76 	&	 +16 05 57.19	&	B9.5 IV Cr   	&	295	&	10.4195	&	0.0008	&	+0.913	&	0.091	&	$-$0.128	&	0.002	&	0.65	&	$-$0.43	&	0.22	\\
479	&	J060315.94+143510.2	&	TYC 729-949-1      	&	06 03 15.95 	&	 +14 35 10.35	&	B9 IV$-$V CrEu   	&	169	&	11.4316	&	0.0010	&	+0.752	&	0.077	&	$-$0.074	&	0.006	&	0.72	&	+0.09	&	0.23	\\
480	&	J060343.72+013555.5	&	TYC 130-1391-1     	&	06 03 43.73 	&	 +01 35 55.60	&	kB7hB9mA0  bl4130 (He-wk) 	&	184	&	11.8978	&	0.0012	&	+0.825	&	0.036	&	+0.027	&	0.004	&	0.87	&	+0.61	&	0.11	\\
481	&	J060347.88+412532.7	&	TYC 2933-1919-1    	&	06 03 47.89 	&	 +41 25 32.80	&	A0 IV$-$V SrCrEuSi   	&	173	&	12.3304	&	0.0005	&	+0.856	&	0.038	&	+0.185	&	0.003	&	0.59	&	+1.40	&	0.11	\\
482	&	J060354.12+162915.5	&	TYC 1313-470-1     	&	06 03 54.11 	&	 +16 29 15.68	&	B9 V Cr   	&	111	&	11.9379	&	0.0007	&	+1.077	&	0.041	&	$-$0.031	&	0.005	&	0.71	&	+1.39	&	0.10	\\
483	&	J060356.06+213033.2	&	TYC 1325-1708-1    	&	06 03 56.06 	&	 +21 30 33.21	&	A8 V SrEu   	&	386	&	10.9242	&	0.0017	&	+1.771	&	0.065	&	+0.113	&	0.004	&	0.36	&	+1.81	&	0.09	\\
484	&	J060410.32+462824.4	&	TYC 3374-1710-1    	&	06 04 10.33 	&	 +46 28 24.46	&	B9 III Si   	&	110	&	12.9083	&	0.0006	&	+0.397	&	0.065	&	+0.019	&	0.003	&	0.37	&	+0.53	&	0.36	\\
485	&	J060418.34+413658.0	&	TYC 2933-1569-1    	&	06 04 18.22 	&	 +41 36 58.04	&	A1 IV$-$V CrSi   	&	244	&	10.7932	&	0.0006	&	+1.123	&	0.055	&	+0.054	&	0.003	&	0.43	&	+0.62	&	0.12	\\
486	&	J060431.41+350633.1	&	TYC 2427-1223-1    	&	06 04 31.41 	&	 +35 06 33.14	&	B8 III CrSi  (He-wk) 	&	194	&	12.4054	&	0.0013	&	+0.405	&	0.042	&	+0.074	&	0.007	&	0.86	&	$-$0.42	&	0.23	\\
487	&	J060453.02+205450.7	&	TYC 1325-259-1     	&	06 04 53.02 	&	 +20 54 50.74	&	B8 III Si   	&	398	&	10.6130	&	0.0007	&	+0.800	&	0.050	&	$-$0.140	&	0.003	&	0.67	&	$-$0.55	&	0.15	\\
488	&	J060453.41+161322.1	&	TYC 1313-1527-1    	&	06 04 53.41 	&	 +16 13 22.18	&	A0 IV Cr   	&	143	&	11.4692	&	0.0008	&	+1.047	&	0.038	&	+0.169	&	0.003	&	0.58	&	+0.99	&	0.09	\\
489	&	J060516.22+120731.9	&	TYC 725-1158-1     	&	06 05 16.23 	&	 +12 07 31.94	&	B9 V SrCr   	&	211	&	11.5435	&	0.0004	&	+0.713	&	0.042	&	+0.075	&	0.003	&	0.45	&	+0.36	&	0.14	\\
490	&	J060518.19+125728.6	&	Gaia DR2 3342783113882464256	&	06 05 18.20 	&	 +12 57 28.63	&	B9 IV$-$V SrCr   	&	149	&	12.8004	&	0.0016	&	+0.510	&	0.053	&	$-$0.127	&	0.007	&	1.04	&	+0.30	&	0.23	\\
491	&	J060543.16+215523.2	&	Gaia DR2 3423437827993940608	&	06 05 43.17 	&	 +21 55 23.33	&	B8 III$-$IV Cr   	&	102	&	14.3544	&	0.0010	&	+0.298	&	0.039	&	+0.004	&	0.006	&	1.30	&	+0.42	&	0.29	\\
492	&	J060620.10+142624.2	&	Gaia DR2 3345279181372314624	&	06 06 20.11 	&	 +14 26 24.28	&	B9 V SrCrEu   	&	108	&	13.1161	&	0.0010	&	+1.091	&	0.062	&	+0.256	&	0.003	&	0.84	&	+2.47	&	0.13	\\
493	&	J060621.79+212433.8 $^{a}$	&	HD 251556  	&	06 06 21.80 	&	 +21 24 35.74	&	B9.5 III$-$IV Cr   	&	386	&	10.2064	&	0.0009	&	+1.455	&	0.080	&	$-$0.172	&	0.003	&	0.36	&	+0.66	&	0.13	\\
494	&	J060707.31+232405.2	&	Gaia DR2 3425366680625960064	&	06 07 07.32 	&	 +23 24 05.25	&	A0 IV$-$V CrEu   	&	112	&	14.2870	&	0.0005	&	+0.529	&	0.032	&	+0.187	&	0.004	&	0.99	&	+1.91	&	0.14	\\
495	&	J060732.16+221707.0	&	Gaia DR2 3423564130094727552	&	06 07 32.17 	&	 +22 17 06.95	&	B8 II$-$III EuSi   	&	183	&	12.7663	&	0.0023	&	+0.569	&	0.173	&	$-$	&	$-$	&	$-$	&	$-$	&	$-$	\\
496	&	J060809.53+240945.0	&	HD 252026  	&	06 08 09.53 	&	 +24 09 45.05	&	B8 III$-$IV Si   	&	449	&	9.9815	&	0.0007	&	+1.122	&	0.047	&	$-$0.104	&	0.003	&	0.51	&	$-$0.28	&	0.10	\\
497	&	J060815.12+045107.6	&	TYC 139-1180-1     	&	06 08 15.12 	&	 +04 51 07.68	&	B9 IV$-$V Eu   	&	282	&	11.4279	&	0.0017	&	+1.153	&	0.081	&	$-$0.037	&	0.008	&	0.43	&	+1.30	&	0.16	\\
498	&	J060820.34+005411.7	&	Gaia DR2 3122985452386853632	&	06 08 20.35 	&	 +00 54 11.73	&	B5 IV Si  (He-wk) 	&	185	&	12.6323	&	0.0006	&	+0.270	&	0.043	&	+0.011	&	0.018	&	0.46	&	$-$0.68	&	0.35	\\
499	&	J060820.77-025507.5	&	HD 294640  	&	06 08 20.77 	&	 -02 55 07.60	&	A1 IV$-$V Sr   	&	248	&	10.6258	&	0.0004	&	+1.571	&	0.040	&	+0.137	&	0.006	&	0.29	&	+1.32	&	0.07	\\
500	&	J060821.41+502149.8	&	HD 233211  	&	06 08 21.10 	&	 +50 21 49.80	&	kB9hA3mA2 bl4077 bl4130  	&	676	&	9.3215	&	0.0009	&	+2.094	&	0.051	&	+0.017	&	0.007	&	0.33	&	+0.60	&	0.07	\\
\hline
\end{tabular}                                                                                                                                                                   
\end{adjustbox}
\end{center}                                                                                                                                             
\end{sidewaystable*}
\setcounter{table}{0}  
\begin{sidewaystable*}
\caption{Essential data for our sample stars, sorted by increasing right ascension. The columns denote: (1) Internal identification number. (2) LAMOST identifier. (3) Alternativ identifier (HD number, TYC identifier or GAIA DR2 number). (4) Right ascension (J2000; GAIA DR2). (5) Declination (J2000; GAIA DR2). (6) Spectral type, as derived in this study. (7) Sloan $g$ band S/N ratio of the analysed spectrum. (8) $G$\,mag (GAIA DR2). (9) $G$\,mag error. (10) Parallax (GAIA DR2). (11) Parallax error. (12) Dereddened colour index $(BP-RP){_0}$ (GAIA DR2). (13) Colour index error. (14) Absorption in the $G$ band, $A_G$. (15) Intrinsic absolute magnitude in the $G$ band, $M_{\mathrm{G,0}}$. (16) Absolute magnitude error.}
\label{table_master11}
\begin{center}
\begin{adjustbox}{max width=\textwidth}
\begin{tabular}{lllcclcccccccccc}
\hline
\hline
(1) & (2) & (3) & (4) & (5) & (6) & (7) & (8) & (9) & (10) & (11) & (12) & (13) & (14) & (15) & (16) \\
No.	&	ID\_LAMOST	&	ID\_alt	&	RA(J2000) 	&	 Dec(J2000)    	&	SpT\_final	&	S/N\,$g$	&	$G$\,mag	&	e\_$G$\,mag	&	pi (mas)	&	e\_pi	&	$(BP-RP){_0}$	&	e\_$(BP-RP){_0}$	&	$A_G$	&	$M_{\mathrm{G,0}}$	&	e\_$M_{\mathrm{G,0}}$	\\
\hline 
501	&	J060822.66+233735.4	&	TYC 1877-408-1     	&	06 08 22.67 	&	 +23 37 35.48	&	B9.5 IV$-$V CrSi   	&	475	&	10.6377	&	0.0010	&	+1.183	&	0.208	&	$-$0.088	&	0.003	&	0.45	&	+0.56	&	0.38	\\
502	&	J060827.84+204832.0	&	Gaia DR2 3375272518547111680	&	06 08 27.84 	&	 +20 48 32.12	&	B9 IV CrEuSi   	&	119	&	12.3687	&	0.0003	&	+0.890	&	0.071	&	$-$0.082	&	0.003	&	0.93	&	+1.18	&	0.18	\\
503	&	J060851.09+424158.2	&	TYC 2933-1518-1    	&	06 08 50.81 	&	 +42 41 58.24	&	B9 III$-$IV  bl4130  	&	444	&	9.8528	&	0.0009	&	+0.838	&	0.054	&	$-$0.089	&	0.003	&	0.49	&	$-$1.02	&	0.15	\\
504	&	J060853.81+465924.7	&	Gaia DR2 963466378309113856	&	06 08 53.81 	&	 +46 59 24.79	&	A1 IV$-$V Cr   	&	130	&	12.9117	&	0.0017	&	+0.487	&	0.081	&	+0.181	&	0.007	&	0.24	&	+1.11	&	0.37	\\
505	&	J060905.54+114858.8	&	TYC 738-220-1      	&	06 09 05.54 	&	 +11 48 58.81	&	B9 V Cr   	&	285	&	11.9801	&	0.0004	&	+1.015	&	0.060	&	+0.043	&	0.004	&	0.48	&	+1.54	&	0.14	\\
506	&	J060940.01+035028.3	&	Gaia DR2 3317252389461940480	&	06 09 40.02 	&	 +03 50 28.34	&	A0 IV SrCrEu   	&	127	&	13.2614	&	0.0004	&	+0.609	&	0.052	&	+0.170	&	0.002	&	0.64	&	+1.54	&	0.19	\\
507	&	J061001.31+325315.5	&	TYC 2424-327-1     	&	06 10 01.31 	&	 +32 53 15.58	&	B9 V CrEu   	&	129	&	9.6521	&	0.0005	&	+1.997	&	0.043	&	$-$0.001	&	0.003	&	0.26	&	+0.89	&	0.07	\\
508	&	J061029.25-053152.4	&	TYC 4791-807-1     	&	06 10 29.25 	&	 -05 31 52.45	&	B9 IV$-$V bl4077 bl4130  	&	284	&	11.2667	&	0.0011	&	+1.209	&	0.044	&	$-$0.476	&	0.003	&	1.54	&	+0.14	&	0.09	\\
509	&	J061030.89+032858.7	&	TYC 135-1133-1     	&	06 10 30.68 	&	 +03 28 58.71	&	B9 V CrEu   	&	391	&	9.0991	&	0.0015	&	+2.092	&	0.084	&	$-$0.377	&	0.007	&	0.84	&	$-$0.13	&	0.10	\\
510	&	J061054.63+240801.2	&	TYC 1877-112-1     	&	06 10 54.63 	&	 +24 08 01.29	&	A0 IV CrEu   	&	259	&	11.3950	&	0.0007	&	+0.996	&	0.042	&	+0.012	&	0.004	&	0.71	&	+0.67	&	0.10	\\
511	&	J061058.53+165733.8	&	TYC 1318-495-1     	&	06 10 58.53 	&	 +16 57 33.83	&	B9 IV$-$V Eu   	&	203	&	11.6046	&	0.0008	&	+1.264	&	0.051	&	$-$0.066	&	0.004	&	1.62	&	+0.49	&	0.10	\\
512	&	J061102.36+244423.8	&	TYC 1881-1149-1    	&	06 11 02.36 	&	 +24 44 23.80	&	B9 IV$-$V Si   	&	394	&	10.9938	&	0.0019	&	+1.123	&	0.061	&	$-$0.019	&	0.007	&	0.63	&	+0.62	&	0.13	\\
513	&	J061114.05+035146.8	&	HD 42510   	&	06 11 13.84 	&	 +03 51 46.57	&	B9.5 II$-$III EuSi   	&	436	&	9.0517	&	0.0022	&	+2.119	&	0.040	&	$-$0.209	&	0.012	&	0.68	&	+0.01	&	0.06	\\
514	&	J061124.59+124504.9	&	TYC 738-699-1      	&	06 11 24.60 	&	 +12 45 05.01	&	A0 IV$-$V CrEu   	&	635	&	10.4730	&	0.0013	&	+1.465	&	0.049	&	+0.036	&	0.008	&	0.46	&	+0.84	&	0.09	\\
515	&	J061143.19+152832.1	&	Gaia DR2 3345733932510678656	&	06 11 43.20 	&	 +15 28 32.13	&	B9.5 V CrEu   	&	129	&	13.0875	&	0.0012	&	+0.720	&	0.038	&	$-$0.052	&	0.005	&	1.07	&	+1.30	&	0.13	\\
516	&	J061143.58+025133.3	&	Gaia DR2 3316930743654985344	&	06 11 43.59 	&	 +02 51 33.35	&	A9 V SrSi   	&	132	&	13.0664	&	0.0006	&	+0.898	&	0.030	&	+0.324	&	0.005	&	1.15	&	+1.68	&	0.09	\\
517	&	J061143.92+254734.6	&	Gaia DR2 3426811541984219520	&	06 11 43.92 	&	 +25 47 34.58	&	B9 IV$-$V Cr   	&	109	&	13.2418	&	0.0009	&	+0.356	&	0.034	&	+0.034	&	0.005	&	0.79	&	+0.21	&	0.21	\\
518	&	J061222.36+240733.9	&	Gaia DR2 3425476524411862144	&	06 12 22.37 	&	 +24 07 33.91	&	A8 V SrCrEu   	&	137	&	12.1788	&	0.0006	&	+0.762	&	0.077	&	+0.330	&	0.004	&	0.82	&	+0.77	&	0.22	\\
519	&	J061229.08+184051.4	&	TYC 1318-680-1     	&	06 12 29.08 	&	 +18 40 51.35	&	B9.5 V CrSi   	&	328	&	11.7449	&	0.0003	&	+0.627	&	0.037	&	+0.082	&	0.003	&	0.45	&	+0.29	&	0.14	\\
520	&	J061308.61+342342.1	&	TYC 2428-1351-1    	&	06 13 08.60 	&	 +34 23 42.02	&	B8 IV$-$V Si   	&	349	&	11.4852	&	0.0011	&	+1.688	&	0.302	&	+0.177	&	0.007	&	0.32	&	+2.31	&	0.39	\\
521	&	J061312.69+013350.3	&	Gaia DR2 3122955177158513920	&	06 13 12.70 	&	 +01 33 50.39	&	A0 V Cr   	&	149	&	13.1708	&	0.0010	&	+1.066	&	0.409	&	$-$	&	$-$	&	$-$	&	$-$	&	$-$	\\
522	&	J061328.54+405514.0	&	TYC 2930-111-1     	&	06 13 28.54 	&	 +40 55 14.07	&	B9.5 II$-$III \textit{SiCrSr}*	&	306	&	11.2009	&	0.0006	&	+0.537	&	0.047	&	$-$0.040	&	0.004	&	0.30	&	$-$0.45	&	0.20	\\
523	&	J061329.24+041213.9	&	TYC 139-2019-1     	&	06 13 29.24 	&	 +04 12 13.94	&	B9.5 V SrCr   	&	267	&	11.8412	&	0.0004	&	+0.995	&	0.067	&	+0.199	&	0.003	&	0.81	&	+1.02	&	0.15	\\
524	&	J061331.98+135352.9	&	Gaia DR2 3344392730186495744	&	06 13 31.99 	&	 +13 53 52.94	&	A0 II Eu   	&	151	&	13.4168	&	0.0008	&	+0.316	&	0.023	&	$-$0.282	&	0.006	&	1.76	&	$-$0.85	&	0.17	\\
525	&	J061338.27-085719.7	&	TYC 5362-1436-1    	&	06 13 38.27 	&	 -08 57 19.67	&	A7 IV$-$V SrCrEu   	&	286	&	10.4517	&	0.0007	&	+1.598	&	0.071	&	+0.193	&	0.002	&	0.26	&	+1.21	&	0.11	\\
526	&	J061341.68+114751.7	&	Gaia DR2 3331844210134383104	&	06 13 41.68 	&	 +11 47 51.69	&	B9 III  bl4130  	&	106	&	12.6200	&	0.0004	&	+0.610	&	0.067	&	$-$0.049	&	0.005	&	0.87	&	+0.68	&	0.24	\\
527	&	J061342.52+263829.6	&	TYC 1885-92-1      	&	06 13 42.52 	&	 +26 38 29.65	&	B8 III$-$IV Si   	&	102	&	10.8684	&	0.0011	&	+0.985	&	0.056	&	+0.109	&	0.006	&	0.56	&	+0.27	&	0.13	\\
528	&	J061350.63+203015.3	&	TYC 1322-443-1     	&	06 13 50.63 	&	 +20 30 15.49	&	kA0hA2mA4  bl4130  	&	156	&	11.9459	&	0.0004	&	$-$1.529	&	0.211	&	$-$	&	$-$	&	$-$	&	$-$	&	$-$	\\
529	&	J061413.62+112022.5	&	TYC 738-1510-1     	&	06 14 13.63 	&	 +11 20 22.53	&	kA2hA3mA6 SrCrEu   	&	104	&	11.9757	&	0.0007	&	+0.998	&	0.049	&	+0.000	&	0.004	&	0.73	&	+1.25	&	0.12	\\
530	&	J061434.73+050332.4	&	TYC 139-1241-1     	&	06 14 34.73 	&	 +05 03 32.42	&	B9.5 IV$-$V Cr   	&	208	&	12.1693	&	0.0013	&	+0.744	&	0.040	&	+0.191	&	0.006	&	0.76	&	+0.77	&	0.13	\\
531	&	J061436.66+142939.5	&	Gaia DR2 3344825319290434176	&	06 14 36.67 	&	 +14 29 39.62	&	B9 V Cr   	&	227	&	12.2346	&	0.0003	&	+0.580	&	0.068	&	+0.002	&	0.004	&	0.80	&	+0.26	&	0.26	\\
532	&	J061444.86+234440.8	&	TYC 1877-792-1     	&	06 14 44.86 	&	 +23 44 40.89	&	A8 V SrCrEu   	&	622	&	9.7436	&	0.0005	&	+2.486	&	0.050	&	+0.263	&	0.004	&	0.14	&	+1.58	&	0.07	\\
533	&	J061455.11+245214.3	&	TYC 1881-912-1     	&	06 14 55.13 	&	 +24 52 14.32	&	A6 V SrEuSi   	&	274	&	11.6812	&	0.0076	&	$-$7.091	&	1.137	&	$-$	&	$-$	&	$-$	&	$-$	&	$-$	\\
534	&	J061455.58+032317.5	&	TYC 135-1270-1     	&	06 14 55.58 	&	 +03 23 17.54	&	B8 IV Si   	&	212	&	12.0688	&	0.0007	&	+0.711	&	0.041	&	$-$0.083	&	0.003	&	1.50	&	$-$0.17	&	0.13	\\
535	&	J061502.38+184511.7	&	Gaia DR2 3373847718978934656	&	06 15 02.39 	&	 +18 45 11.77	&	B9 IV$-$V Cr   	&	178	&	12.6224	&	0.0003	&	+0.518	&	0.046	&	+0.110	&	0.003	&	1.23	&	$-$0.04	&	0.20	\\
536	&	J061558.32+135157.3	&	Gaia DR2 3344570095153338880	&	06 15 58.33 	&	 +13 51 57.37	&	B5 III$-$IV  bl4130 (He-wk) 	&	148	&	13.2180	&	0.0028	&	+0.487	&	0.030	&	$-$0.032	&	0.013	&	1.03	&	+0.62	&	0.14	\\
537	&	J061609.42+265703.2	&	TYC 1885-208-1     	&	06 16 09.42 	&	 +26 57 03.15	&	B8 IV$-$V  bl4130  	&	192	&	11.5715	&	0.0009	&	+17.732	&	1.692	&	+0.840	&	0.004	&	0.03	&	+7.78	&	0.21	\\
538	&	J061623.52+330303.3	&	TYC 2424-369-1     	&	06 16 23.52 	&	 +33 03 03.38	&	B8 IV$-$V Si  (He-wk) 	&	253	&	10.7561	&	0.0008	&	+1.035	&	0.051	&	$-$0.120	&	0.005	&	0.61	&	+0.22	&	0.12	\\
539	&	J061638.39+125441.1	&	Gaia DR2 3332085625956573696	&	06 16 38.40 	&	 +12 54 41.16	&	B8 III$-$IV Si   	&	114	&	13.5954	&	0.0005	&	+0.267	&	0.028	&	$-$0.020	&	0.003	&	1.39	&	$-$0.66	&	0.23	\\
540	&	J061713.35+232351.5	&	TYC 1878-263-1     	&	06 17 13.36 	&	 +23 23 51.52	&	A1 V Cr   	&	259	&	11.7869	&	0.0005	&	+0.867	&	0.039	&	+0.171	&	0.004	&	0.71	&	+0.77	&	0.11	\\
541	&	J061734.94+252253.8	&	Gaia DR2 3426033706226654848	&	06 17 34.96 	&	 +25 22 53.92	&	B8 IV$-$V CrSi   	&	103	&	14.4724	&	0.0009	&	+0.218	&	0.033	&	+0.041	&	0.005	&	1.00	&	+0.17	&	0.33	\\
542	&	J061743.05+595315.3	&	TYC 3776-815-1     	&	06 17 43.07 	&	 +59 53 15.31	&	A1 IV Eu   	&	148	&	11.3202	&	0.0006	&	+1.339	&	0.054	&	+0.260	&	0.002	&	0.18	&	+1.77	&	0.10	\\
543	&	J061802.68+263156.6	&	Gaia DR2 3432912006155737216	&	06 18 02.69 	&	 +26 31 56.64	&	A0 IV SrCrEu   	&	207	&	12.3734	&	0.0003	&	+0.497	&	0.036	&	+0.295	&	0.003	&	0.59	&	+0.27	&	0.17	\\
544	&	J061817.58+265109.6	&	HD 254625  	&	06 18 17.58 	&	 +26 51 09.74	&	A1 II Si   	&	462	&	10.5653	&	0.0012	&	+0.999	&	0.053	&	$-$0.122	&	0.005	&	0.68	&	$-$0.12	&	0.12	\\
545	&	J061826.24+211412.3	&	TYC 1327-1335-1    	&	06 18 26.24 	&	 +21 14 12.45	&	B8 IV$-$V Si   	&	269	&	11.3062	&	0.0008	&	+0.752	&	0.067	&	$-$0.129	&	0.006	&	0.63	&	+0.06	&	0.20	\\
546	&	J061845.15+190925.3	&	Gaia DR2 3373922661868123136	&	06 18 45.15 	&	 +19 09 25.37	&	B9 IV  bl4130  	&	135	&	14.0601	&	0.0008	&	+0.484	&	0.031	&	+0.203	&	0.005	&	0.79	&	+1.70	&	0.15	\\
547	&	J061909.30+503741.3	&	TYC 3387-281-1     	&	06 19 09.15 	&	 +50 37 41.34	&	A1 IV$-$V SrCr   	&	335	&	10.7271	&	0.0006	&	+1.050	&	0.039	&	+0.079	&	0.003	&	0.25	&	+0.58	&	0.10	\\
548	&	J061911.91+224143.5	&	Gaia DR2 3377043763061518592	&	06 19 11.92 	&	 +22 41 43.59	&	B4 IV Si   	&	101	&	14.1343	&	0.0005	&	+0.558	&	0.035	&	+0.349	&	0.004	&	1.62	&	+1.25	&	0.15	\\
549	&	J061913.90+185953.3	&	Gaia DR2 3373905516358940160	&	06 19 13.91 	&	 +18 59 53.45	&	B9 IV$-$V Cr   	&	133	&	14.3685	&	0.0011	&	+0.296	&	0.033	&	+0.067	&	0.007	&	0.99	&	+0.74	&	0.24	\\
550	&	J061914.21+362331.9	&	TYC 2433-657-1     	&	06 19 14.21 	&	 +36 23 31.78	&	B8 IV Si  (He-wk) 	&	292	&	11.6573	&	0.0006	&	+0.667	&	0.061	&	$-$0.082	&	0.006	&	0.92	&	$-$0.14	&	0.20	\\
\hline
\end{tabular}                                                                                                                                                                   
\end{adjustbox}
\end{center}                                                                                                                                             
\end{sidewaystable*}
\setcounter{table}{0}  
\begin{sidewaystable*}
\caption{Essential data for our sample stars, sorted by increasing right ascension. The columns denote: (1) Internal identification number. (2) LAMOST identifier. (3) Alternativ identifier (HD number, TYC identifier or GAIA DR2 number). (4) Right ascension (J2000; GAIA DR2). (5) Declination (J2000; GAIA DR2). (6) Spectral type, as derived in this study. (7) Sloan $g$ band S/N ratio of the analysed spectrum. (8) $G$\,mag (GAIA DR2). (9) $G$\,mag error. (10) Parallax (GAIA DR2). (11) Parallax error. (12) Dereddened colour index $(BP-RP){_0}$ (GAIA DR2). (13) Colour index error. (14) Absorption in the $G$ band, $A_G$. (15) Intrinsic absolute magnitude in the $G$ band, $M_{\mathrm{G,0}}$. (16) Absolute magnitude error.}
\label{table_master12}
\begin{center}
\begin{adjustbox}{max width=\textwidth}
\begin{tabular}{lllcclcccccccccc}
\hline
\hline
(1) & (2) & (3) & (4) & (5) & (6) & (7) & (8) & (9) & (10) & (11) & (12) & (13) & (14) & (15) & (16) \\
No.	&	ID\_LAMOST	&	ID\_alt	&	RA(J2000) 	&	 Dec(J2000)    	&	SpT\_final	&	S/N\,$g$	&	$G$\,mag	&	e\_$G$\,mag	&	pi (mas)	&	e\_pi	&	$(BP-RP){_0}$	&	e\_$(BP-RP){_0}$	&	$A_G$	&	$M_{\mathrm{G,0}}$	&	e\_$M_{\mathrm{G,0}}$	\\
\hline
551	&	J061925.97+204122.2	&	TYC 1327-257-1     	&	06 19 25.96 	&	 +20 41 22.36	&	B9 IV$-$V Si   	&	253	&	11.6463	&	0.0005	&	+0.760	&	0.046	&	$-$0.093	&	0.004	&	0.74	&	+0.31	&	0.14	\\
552	&	J061951.56+283916.8	&	Gaia DR2 3433772034704740736	&	06 19 51.57 	&	 +28 39 16.95	&	B9 III$-$IV  bl4130  	&	120	&	14.6401	&	0.0010	&	+0.292	&	0.050	&	+0.062	&	0.008	&	0.55	&	+1.42	&	0.38	\\
553	&	J062020.91+252606.8	&	HD 255238  	&	06 20 20.92 	&	 +25 26 06.84	&	kB9.5hA1mA4 Si   	&	173	&	10.6242	&	0.0010	&	+0.905	&	0.060	&	$-$0.163	&	0.006	&	0.46	&	$-$0.06	&	0.15	\\
554	&	J062024.70+212536.0	&	Gaia DR2 3375964901637225728	&	06 20 24.71 	&	 +21 25 36.05	&	B9.5 IV Si   	&	102	&	14.4288	&	0.0006	&	+0.299	&	0.033	&	+0.012	&	0.004	&	1.11	&	+0.70	&	0.24	\\
555	&	J062031.37+300340.5	&	Gaia DR2 3437010199527628544	&	06 20 31.38 	&	 +30 03 40.52	&	B9.5 III$-$IV CrEu   	&	142	&	14.2571	&	0.0012	&	+0.223	&	0.046	&	+0.017	&	0.01	&	0.77	&	+0.22	&	0.45	\\
556	&	J062040.02+131611.9	&	HD 255467  	&	06 20 40.02 	&	 +13 16 11.91	&	kB9hA9mA8 Si   	&	726	&	9.9884	&	0.0006	&	+1.450	&	0.064	&	$-$0.174	&	0.003	&	0.31	&	+0.48	&	0.11	\\
557	&	J062044.63+023859.6	&	Gaia DR2 3124629943825342464	&	06 20 44.64 	&	 +02 38 59.72	&	B8 III Si   	&	105	&	12.4587	&	0.0003	&	+0.425	&	0.036	&	+0.114	&	0.004	&	1.05	&	$-$0.45	&	0.19	\\
558	&	J062119.45+050555.7	&	TYC 140-743-1      	&	06 21 19.46 	&	 +05 05 55.71	&	B6 IV Si   	&	247	&	11.5244	&	0.0007	&	+0.700	&	0.052	&	$-$0.072	&	0.006	&	0.49	&	+0.26	&	0.17	\\
559	&	J062136.92+181258.0	&	Gaia DR2 3370712190994249984	&	06 21 36.93 	&	 +18 12 58.10	&	B9 IV  bl4130  	&	111	&	14.7557	&	0.0011	&	+0.197	&	0.036	&	$-$0.249	&	0.006	&	1.79	&	$-$0.57	&	0.40	\\
560	&	J062155.55+001812.2	&	HD 44456   	&	06 21 55.42 	&	 +00 18 15.17	&	B8 III Si   	&	128	&	8.4953	&	0.0014	&	+2.162	&	0.187	&	$-$0.158	&	0.008	&	0.41	&	$-$0.24	&	0.19	\\
561	&	J062221.82+595613.0	&	TYC 3776-269-1     	&	06 22 21.84 	&	 +59 56 13.02	&	\textit{kA5hA7:mF2} SrCrEu\textit{Si:}* $^{d}$   	&	343	&	10.4744	&	0.0004	&	+1.221	&	0.051	&	+0.313	&	0.001	&	0.19	&	+0.72	&	0.10	\\
562	&	J062226.08+443003.8	&	TYC 2939-1176-1    	&	06 22 26.39 	&	 +44 30 03.31	&	B8 IV Si   	&	384	&	9.2839	&	0.0019	&	+1.835	&	0.073	&	$-$0.110	&	0.009	&	0.26	&	+0.35	&	0.10	\\
563	&	J062248.54+144850.5	&	Gaia DR2 3368707433400422784	&	06 22 48.56 	&	 +14 48 50.48	&	B9 IV$-$V Cr   	&	109	&	14.4736	&	0.0010	&	+0.365	&	0.031	&	+0.004	&	0.004	&	1.38	&	+0.91	&	0.19	\\
564	&	J062257.61+231625.8	&	TYC 1878-882-1     	&	06 22 57.62 	&	 +23 16 25.80	&	A1 IV$-$V SrCrEu   	&	355	&	10.4559	&	0.0010	&	+1.351	&	0.062	&	+0.038	&	0.004	&	0.31	&	+0.80	&	0.11	\\
565	&	J062307.91+264642.0	&	Gaia DR2 3432273606513132544	&	06 23 07.92 	&	 +26 46 42.04	&	B4 V \textit{HeB7}*	&	127	&	14.1786	&	0.0009	&	+0.235	&	0.029	&	$-$0.108	&	0.005	&	0.58	&	+0.46	&	0.27	\\
566	&	J062348.44+043007.6	&	Gaia DR2 3131479759532254720	&	06 23 48.45 	&	 +04 30 07.63	&	A8 V SrCrEu   	&	105	&	11.9647	&	0.0003	&	+0.830	&	0.049	&	+0.170	&	0.002	&	0.70	&	+0.86	&	0.14	\\
567	&	J062348.46+034201.1	&	HD 256582  	&	06 23 48.53 	&	 +03 42 03.78	&	B5 \textit{V HeB9}*	&	174	&	9.8823	&	0.0024	&	+1.416	&	0.046	&	+0.070	&	0.008	&	1.33	&	$-$0.69	&	0.09	\\
568	&	J062407.76+264155.1	&	Gaia DR2 3432227461384464640	&	06 24 07.76 	&	 +26 41 55.20	&	B9 II$-$III \textit{(Mg 4481 wk)}* $^{b}$	&	158	&	14.2335	&	0.0015	&	+0.128	&	0.033	&	$-$	&	$-$	&	$-$	&	$-$	&	$-$	\\
569	&	J062427.61+213544.0	&	Gaia DR2 3376311041641501056	&	06 24 27.62 	&	 +21 35 44.07	&	A0 V Cr   	&	134	&	14.5286	&	0.0005	&	+0.311	&	0.035	&	+0.163	&	0.003	&	0.80	&	+1.19	&	0.25	\\
570	&	J062449.08+190854.0	&	Gaia DR2 3372372797150109184	&	06 24 49.09 	&	 +19 08 54.09	&	kA2hA3mA7 SrCrEu   	&	142	&	12.6717	&	0.0008	&	+0.790	&	0.050	&	+0.340	&	0.003	&	0.51	&	+1.65	&	0.15	\\
571	&	J062449.21+161325.7	&	Gaia DR2 3368971595367574016	&	06 24 49.21 	&	 +16 13 25.76	&	A0 V Cr   	&	138	&	13.4502	&	0.0008	&	+0.283	&	0.032	&	+0.260	&	0.006	&	0.84	&	$-$0.13	&	0.25	\\
572	&	J062451.14+152611.6	&	Gaia DR2 3368854978416307712	&	06 24 51.14 	&	 +15 26 11.77	&	B9 V CrEu   	&	140	&	14.4909	&	0.0010	&	+0.280	&	0.034	&	$-$0.041	&	0.005	&	1.19	&	+0.54	&	0.27	\\
573	&	J062511.12+160439.7	&	Gaia DR2 3368966334029514368	&	06 25 11.13 	&	 +16 04 39.83	&	B9.5 II \textit{(Si) (Mg 4481 wk)}*	&	112	&	14.6802	&	0.0006	&	+0.183	&	0.055	&	$-$	&	$-$	&	$-$	&	$-$	&	$-$	\\
574	&	J062513.47+151050.2	&	Gaia DR2 3368834603091499520	&	06 25 13.49 	&	 +15 10 50.30	&	B9 V Cr   	&	101	&	14.9111	&	0.0005	&	+0.260	&	0.032	&	$-$0.114	&	0.003	&	1.47	&	+0.51	&	0.27	\\
575	&	J062529.19+160620.2	&	Gaia DR2 3368966166529044864	&	06 25 29.20 	&	 +16 06 20.32	&	kA2hA3mA5 SrCrEu   	&	103	&	15.2312	&	0.0013	&	+0.286	&	0.067	&	+0.177	&	0.007	&	0.92	&	+1.59	&	0.51	\\
576	&	J062529.84-032411.9	&	TYC 4789-2924-1    	&	06 25 29.85 	&	 -03 24 11.99	&	 B9 IV SrCrEuSi   	&	123	&	12.1564	&	0.0007	&	+0.657	&	0.037	&	+0.095	&	0.004	&	0.41	&	+0.84	&	0.13	\\
577	&	J062551.25+240212.8	&	Gaia DR2 3383424984952243712	&	06 25 51.26 	&	 +24 02 12.82	&	A5 IV$-$V SrCrEu   	&	174	&	14.2458	&	0.0006	&	$-$0.032	&	0.050	&	$-$	&	$-$	&	$-$	&	$-$	&	$-$	\\
578	&	J062614.15+164334.9	&	Gaia DR2 3369553099579973760	&	06 26 14.16 	&	 +16 43 34.93	&	B8 IV EuSi  (He-wk) 	&	170	&	14.2085	&	0.0004	&	+0.356	&	0.031	&	+0.144	&	0.004	&	0.64	&	+1.33	&	0.19	\\
579	&	J062620.51+221542.7	&	Gaia DR2 3376407420706899712	&	06 26 20.52 	&	 +22 15 42.79	&	B9 IV  bl4130  	&	114	&	14.8829	&	0.0011	&	+0.133	&	0.063	&	$-$	&	$-$	&	$-$	&	$-$	&	$-$	\\
580	&	J062630.83+171904.2	&	Gaia DR2 3369749151948057088	&	06 26 30.84 	&	 +17 19 04.35	&	B8 III$-$IV $^{b}$	&	104	&	15.0172	&	0.0014	&	+0.146	&	0.046	&	$-$	&	$-$	&	$-$	&	$-$	&	$-$	\\
581	&	J062638.98+233156.4	&	Gaia DR2 3377346055741625728	&	06 26 38.99 	&	 +23 31 56.49	&	B9.5 III \textit{SiCrEu}*	&	175	&	14.4967	&	0.0004	&	+0.287	&	0.054	&	+0.074	&	0.003	&	0.86	&	+0.93	&	0.41	\\
582	&	J062650.31+244510.6	&	HD 45148   	&	06 26 50.33 	&	 +24 45 13.62	&	B9.5 IV Cr  (He-wk) 	&	772	&	9.2370	&	0.0010	&	+1.762	&	0.058	&	+0.005	&	0.004	&	0.21	&	+0.26	&	0.09	\\
583	&	J062700.73+153645.9	&	Gaia DR2 3368872776760586880	&	06 27 00.74 	&	 +15 36 45.99	&	A0 II$-$III Si   	&	119	&	14.5403	&	0.0013	&	+0.232	&	0.036	&	$-$0.109	&	0.006	&	1.33	&	+0.04	&	0.34	\\
584	&	J062708.44+174403.3	&	Gaia DR2 3369777468670711936	&	06 27 08.45 	&	 +17 44 03.38	&	\textit{B5 Vp HeB9}*	&	112	&	14.4438	&	0.0013	&	+0.168	&	0.031	&	+0.039	&	0.006	&	0.89	&	$-$0.31	&	0.40	\\
585	&	J062710.47+260440.0	&	Gaia DR2 3431947665737506432	&	06 27 10.47 	&	 +26 04 40.06	&	kA1hA7mA3 CrEu   	&	125	&	12.8741	&	0.0007	&	+0.401	&	0.036	&	+0.163	&	0.004	&	0.56	&	+0.33	&	0.20	\\
586	&	J062715.76+174930.0	&	Gaia DR2 3369802340824496768	&	06 27 15.77 	&	 +17 49 30.08	&	B9.5 IV  bl4130  	&	111	&	14.9602	&	0.0005	&	+0.271	&	0.034	&	+0.050	&	0.004	&	0.80	&	+1.32	&	0.28	\\
587	&	J062723.33+212713.3	&	HD 257383  	&	06 27 23.25 	&	 +21 27 13.40	&	B9 V Sr   	&	627	&	10.8046	&	0.0008	&	+1.063	&	0.038	&	$-$0.092	&	0.004	&	0.41	&	+0.53	&	0.09	\\
588	&	J062753.43+360317.1	&	TYC 2434-1050-1    	&	06 27 53.43 	&	 +36 03 17.17	&	A0 IV$-$V SrCr   	&	301	&	11.5878	&	0.0008	&	+0.778	&	0.054	&	+0.009	&	0.006	&	0.40	&	+0.64	&	0.16	\\
589	&	J062759.86+172007.8	&	Gaia DR2 3369705210141630848	&	06 27 59.87 	&	 +17 20 07.91	&	B9.5 III$-$IV Cr   	&	115	&	14.8257	&	0.0005	&	+0.221	&	0.041	&	+0.119	&	0.005	&	1.14	&	+0.41	&	0.40	\\
590	&	J062811.89+551959.8	&	Gaia DR2 997612807125070080	&	06 28 11.89 	&	 +55 19 59.94	&	B9.5 V Cr   	&	32	&	12.7973	&	0.0004	&	+0.604	&	0.040	&	+0.029	&	0.004	&	0.23	&	+1.47	&	0.15	\\
591	&	J062829.47+233423.0	&	Gaia DR2 3383342418501312384	&	06 28 29.48 	&	 +23 34 23.01	&	B9 V SrCr   	&	101	&	15.1717	&	0.0010	&	+0.180	&	0.040	&	+0.056	&	0.007	&	0.76	&	+0.68	&	0.48	\\
592	&	J062832.29+150924.2	&	Gaia DR2 3356789109612111360	&	06 28 32.30 	&	 +15 09 24.34	&	A3 V SrEu   	&	109	&	14.3911	&	0.0009	&	+0.458	&	0.036	&	+0.221	&	0.005	&	0.75	&	+1.95	&	0.18	\\
593	&	J062833.01+162806.1	&	Gaia DR2 3369363674342613632	&	06 28 33.01 	&	 +16 28 06.14	&	B9 IV$-$V Cr   	&	141	&	14.5703	&	0.0007	&	+0.187	&	0.033	&	+0.029	&	0.004	&	0.73	&	+0.19	&	0.38	\\
594	&	J062838.72+355224.3	&	Gaia DR2 942803599885897856	&	06 28 38.73 	&	 +35 52 24.36	&	B9.5 II$-$III EuSi   	&	107	&	14.3985	&	0.0046	&	+0.165	&	0.048	&	$-$	&	$-$	&	$-$	&	$-$	&	$-$	\\
595	&	J062842.31+162852.9	&	Gaia DR2 3369363914860764032	&	06 28 42.32 	&	 +16 28 53.00	&	B9.5 II$-$III EuSi   	&	168	&	14.0127	&	0.0015	&	+0.260	&	0.028	&	+0.074	&	0.008	&	0.64	&	+0.44	&	0.24	\\
596	&	J062909.51+023823.8	&	Gaia DR2 3124242606494984064	&	06 29 09.52 	&	 +02 38 23.87	&	B8 \textit{V} Si* $^{d}$  	&	103	&	13.3277	&	0.0005	&	+0.492	&	0.031	&	$-$0.008	&	0.003	&	1.05	&	+0.74	&	0.14	\\
597	&	J062914.34+004257.0	&	HD 291674  	&	06 29 14.34 	&	 +00 42 57.04	&	B7 III$-$IV Si   	&	516	&	9.7606	&	0.0021	&	+0.900	&	0.174	&	$-$0.220	&	0.005	&	0.74	&	$-$1.21	&	0.42	\\
598	&	J062921.80+165315.3	&	Gaia DR2 3369474686357680256	&	06 29 21.81 	&	 +16 53 15.45	&	B9.5 II$-$III \textit{Si}*	&	106	&	14.9111	&	0.0010	&	+0.136	&	0.058	&	$-$	&	$-$	&	$-$	&	$-$	&	$-$	\\
599	&	J062951.75+151823.7	&	HD 258289  	&	06 29 51.67 	&	 +15 18 23.65	&	B9 V SrCr   	&	364	&	10.4909	&	0.0008	&	+1.079	&	0.073	&	+0.144	&	0.003	&	0.55	&	+0.11	&	0.15	\\
600	&	J062951.85+285010.3	&	TYC 1891-1115-1    	&	06 29 51.85 	&	 +28 50 10.34	&	A8 V SrCrEu   	&	154	&	11.6738	&	0.0009	&	+0.908	&	0.050	&	+0.177	&	0.004	&	0.36	&	+1.10	&	0.13	\\
\hline
\end{tabular}                                                                                                                                                                   
\end{adjustbox}
\end{center}                                                                                                                                             
\end{sidewaystable*}
\setcounter{table}{0}  
\begin{sidewaystable*}
\caption{Essential data for our sample stars, sorted by increasing right ascension. The columns denote: (1) Internal identification number. (2) LAMOST identifier. (3) Alternativ identifier (HD number, TYC identifier or GAIA DR2 number). (4) Right ascension (J2000; GAIA DR2). (5) Declination (J2000; GAIA DR2). (6) Spectral type, as derived in this study. (7) Sloan $g$ band S/N ratio of the analysed spectrum. (8) $G$\,mag (GAIA DR2). (9) $G$\,mag error. (10) Parallax (GAIA DR2). (11) Parallax error. (12) Dereddened colour index $(BP-RP){_0}$ (GAIA DR2). (13) Colour index error. (14) Absorption in the $G$ band, $A_G$. (15) Intrinsic absolute magnitude in the $G$ band, $M_{\mathrm{G,0}}$. (16) Absolute magnitude error.}
\label{table_master13}
\begin{center}
\begin{adjustbox}{max width=\textwidth}
\begin{tabular}{lllcclcccccccccc}
\hline
\hline
(1) & (2) & (3) & (4) & (5) & (6) & (7) & (8) & (9) & (10) & (11) & (12) & (13) & (14) & (15) & (16) \\
No.	&	ID\_LAMOST	&	ID\_alt	&	RA(J2000) 	&	 Dec(J2000)    	&	SpT\_final	&	S/N\,$g$	&	$G$\,mag	&	e\_$G$\,mag	&	pi (mas)	&	e\_pi	&	$(BP-RP){_0}$	&	e\_$(BP-RP){_0}$	&	$A_G$	&	$M_{\mathrm{G,0}}$	&	e\_$M_{\mathrm{G,0}}$	\\
\hline 
601	&	J063007.41+283333.3	&	TYC 1891-1405-1    	&	06 30 07.42 	&	 +28 33 33.38	&	B9 IV$-$V bl4077 bl4130  	&	172	&	11.5882	&	0.0009	&	+0.748	&	0.042	&	$-$0.060	&	0.007	&	0.34	&	+0.62	&	0.13	\\
602	&	J063021.79+201821.4	&	TYC 1336-731-1     	&	06 30 21.80 	&	 +20 18 21.47	&	A7 V SrCrEuSi   	&	338	&	11.7963	&	0.0003	&	+0.812	&	0.042	&	+0.237	&	0.002	&	0.22	&	+1.12	&	0.12	\\
603	&	J063033.69+014424.1	&	HD 288751  	&	06 30 33.69 	&	 +01 44 24.14	&	B8 IV$-$V EuSi   	&	245	&	10.9133	&	0.0011	&	+0.927	&	0.040	&	$-$0.076	&	0.004	&	0.90	&	$-$0.15	&	0.11	\\
604	&	J063035.50+035245.3	&	TYC 154-966-1      	&	06 30 35.66 	&	 +03 52 46.65	&	A1 IV$-$V SrCr   	&	237	&	9.7386	&	0.0013	&	+1.207	&	0.121	&	$-$0.057	&	0.003	&	0.30	&	$-$0.15	&	0.22	\\
605	&	J063037.14+234154.8	&	Gaia DR2 3382612239700477568	&	06 30 37.14 	&	 +23 41 54.88	&	B9 III$-$IV Cr   	&	134	&	14.3692	&	0.0006	&	+0.103	&	0.041	&	$-$	&	$-$	&	$-$	&	$-$	&	$-$	\\
606	&	J063114.46+184731.8	&	HD 258682  	&	06 31 14.46 	&	 +18 47 31.87	&	kB9.5hA7mA8 SrCrEuSi   	&	256	&	10.7497	&	0.0016	&	+0.698	&	0.045	&	+0.207	&	0.006	&	0.38	&	$-$0.41	&	0.15	\\
607	&	J063115.49+235418.3	&	Gaia DR2 3382717964615683456	&	06 31 15.49 	&	 +23 54 18.31	&	A0 IV$-$V CrEu   	&	145	&	14.2161	&	0.0006	&	+0.292	&	0.051	&	+0.057	&	0.004	&	0.41	&	+1.13	&	0.38	\\
608	&	J063143.15+222551.2	&	TYC 1340-1953-1    	&	06 31 43.25 	&	 +22 25 50.45	&	B9 IV$-$V Si   	&	378	&	10.6750	&	0.0007	&	+0.558	&	0.097	&	+0.013	&	0.003	&	0.11	&	$-$0.70	&	0.38	\\
609	&	J063201.93+185013.4	&	Gaia DR2 3371457591156975232	&	06 32 01.94 	&	 +18 50 13.44	&	B9 III$-$IV Si   	&	105	&	13.4379	&	0.0005	&	+0.289	&	0.026	&	$-$0.030	&	0.003	&	0.50	&	+0.25	&	0.20	\\
610	&	J063204.44+013000.1	&	HD 46203   	&	06 32 04.44 	&	 +01 30 00.11	&	B9 V \textit{SiCrSr(Eu)}*	&	430	&	9.8422	&	0.0007	&	+1.840	&	0.090	&	$-$0.069	&	0.003	&	0.37	&	+0.79	&	0.12	\\
611	&	J063218.45+032146.3	&	HD 259273  	&	06 32 18.45 	&	 +03 21 46.43	&	B9 III$-$IV Si  (He-wk) 	&	196	&	9.7120	&	0.0011	&	+1.479	&	0.088	&	$-$0.140	&	0.004	&	0.29	&	+0.27	&	0.14	\\
612	&	J063232.70+155949.8	&	TYC 1329-606-1     	&	06 32 32.71 	&	 +15 59 49.81	&	B9 IV bl4077 bl4130  	&	341	&	11.5922	&	0.0007	&	+0.969	&	0.051	&	$-$0.011	&	0.004	&	0.64	&	+0.89	&	0.12	\\
613	&	J063238.50+205939.7	&	HD 259117  	&	06 32 38.50 	&	 +20 59 39.79	&	A0 IV$-$V SrCr   	&	251	&	10.0636	&	0.0005	&	+0.829	&	0.036	&	+0.298	&	0.003	&	0.22	&	$-$0.57	&	0.11	\\
614	&	J063256.06+370834.3	&	TYC 2434-33-1      	&	06 32 56.07 	&	 +37 08 34.33	&	kB9.5hA2mA5 SrCr   	&	216	&	11.4857	&	0.0011	&	+0.883	&	0.032	&	+0.081	&	0.005	&	0.29	&	+0.92	&	0.09	\\
615	&	J063257.79+203008.2	&	Gaia DR2 3372764154568111744	&	06 32 57.79 	&	 +20 30 08.21	&	A0 IV$-$V SrCr   	&	154	&	14.0661	&	0.0007	&	+0.191	&	0.050	&	$-$	&	$-$	&	$-$	&	$-$	&	$-$	\\
616	&	J063343.72+582308.2	&	TYC 3777-1866-1    	&	06 33 43.73 	&	 +58 23 08.22	&	B9.5 IV  bl4130  	&	232	&	11.9755	&	0.0007	&	+0.865	&	0.059	&	$-$0.079	&	0.004	&	0.11	&	+1.55	&	0.16	\\
617	&	J063351.27+184418.9	&	TYC 1333-644-1     	&	06 33 51.27 	&	 +18 44 18.89	&	B9 IV$-$V EuSi   	&	322	&	9.7858	&	0.0008	&	+1.461	&	0.100	&	$-$0.222	&	0.003	&	0.34	&	+0.26	&	0.16	\\
618	&	J063355.94+263943.7	&	HD 259452  	&	06 33 55.95 	&	 +26 39 43.74	&	A0 V SrCrEu   	&	172	&	9.3243	&	0.0011	&	+2.073	&	0.080	&	+0.046	&	0.006	&	0.19	&	+0.72	&	0.10	\\
619	&	J063516.46+215714.9	&	TYC 1341-349-1     	&	06 35 16.46 	&	 +21 57 15.02	&	B9 IV$-$V Eu   	&	287	&	9.5388	&	0.0018	&	+1.243	&	0.082	&	$-$0.080	&	0.011	&	0.15	&	$-$0.14	&	0.15	\\
620	&	J063522.08+335132.9 $^{a}$	&	TYC 2430-1205-1    	&	06 35 22.08 	&	 +33 51 32.79	&	A6 V SrEu   	&	162	&	10.9600	&	0.0072	&	$-$	&	$-$	&	$-$	&	$-$	&	$-$	&	$-$	&	$-$	\\
621	&	J063525.23-011456.2	&	HD 291918  	&	06 35 25.24 	&	 -01 14 56.22	&	B9 V CrEu   	&	141	&	11.0443	&	0.0010	&	+1.251	&	0.058	&	$-$0.006	&	0.004	&	0.42	&	+1.11	&	0.11	\\
622	&	J063546.95+061914.9	&	Gaia DR2 3131961453001491584	&	06 35 46.95 	&	 +06 19 14.96	&	B9 IV  bl4130  	&	137	&	11.3376	&	0.0010	&	+0.595	&	0.039	&	+0.053	&	0.005	&	0.53	&	$-$0.32	&	0.15	\\
623	&	J063627.06+014655.4	&	TYC 146-1299-1     	&	06 36 27.06 	&	 +01 46 55.39	&	B9 V Cr   	&	350	&	11.2774	&	0.0008	&	+1.304	&	0.186	&	+0.034	&	0.005	&	0.41	&	+1.44	&	0.31	\\
624	&	J063640.63+315450.6	&	TYC 2439-607-1     	&	06 36 40.63 	&	 +31 54 50.63	&	B9.5 III$-$IV SrCrEuSi   	&	242	&	10.2499	&	0.0007	&	+0.243	&	0.080	&	$-$	&	$-$	&	$-$	&	$-$	&	$-$	\\
625	&	J063711.23+220811.9	&	HD 260562  	&	06 37 11.24 	&	 +22 08 11.94	&	kA0hA1mA3 CrEu   	&	429	&	11.2945	&	0.0006	&	+0.741	&	0.039	&	+0.085	&	0.004	&	0.16	&	+0.49	&	0.13	\\
626	&	J063739.78+241634.6	&	Gaia DR2 3383059191177247488	&	06 37 39.78 	&	 +24 16 34.59	&	A3 IV$-$V SrCrEu   	&	215	&	14.2018	&	0.0012	&	+0.467	&	0.080	&	+0.191	&	0.01	&	0.21	&	+2.34	&	0.37	\\
627	&	J063744.29+195655.1 $^{a}$	&	TYC 1337-1539-1    	&	06 37 44.07 	&	 +19 56 55.16	&	kA1hA7mF4 SrCrSi   	&	811	&	9.1709	&	0.0004	&	+2.662	&	0.044	&	+0.128	&	0.002	&	0.12	&	+1.18	&	0.06	\\
628	&	J063747.15+053115.8	&	HD 260964  	&	06 37 47.16 	&	 +05 31 15.83	&	A0 II$-$III Eu   	&	141	&	11.4906	&	0.0040	&	+0.820	&	0.041	&	+0.001	&	0.018	&	0.38	&	+0.68	&	0.12	\\
629	&	J063748.65+160915.0	&	Gaia DR2 3358634738659193344	&	06 37 48.65 	&	 +16 09 14.98	&	A7 V SrCrEuSi   	&	212	&	12.2557	&	0.0005	&	+0.959	&	0.048	&	+0.357	&	0.004	&	0.48	&	+1.69	&	0.12	\\
630	&	J063752.90+091516.7	&	HD 260958  	&	06 37 52.91 	&	 +09 15 16.75	&	B8 IV Si   	&	185	&	10.0152	&	0.0008	&	+1.380	&	0.064	&	$-$0.397	&	0.004	&	0.62	&	+0.10	&	0.11	\\
631	&	J063800.46+310256.5	&	TYC 2435-109-1     	&	06 38 00.46 	&	 +31 02 56.49	&	B9.5 V SrCr   	&	192	&	10.5260	&	0.0007	&	+0.995	&	0.068	&	+0.025	&	0.003	&	0.20	&	+0.32	&	0.16	\\
632	&	J063814.60+274002.2	&	TYC 1888-847-1     	&	06 38 14.61 	&	 +27 40 02.29	&	B9 IV Si   	&	127	&	12.6032	&	0.0010	&	+0.403	&	0.043	&	$-$0.020	&	0.007	&	0.22	&	+0.41	&	0.24	\\
633	&	J063840.72+203356.8	&	HD 261041  	&	06 38 40.63 	&	 +20 33 56.81	&	A0 IV$-$V Cr   	&	561	&	10.7763	&	0.0015	&	+1.295	&	0.153	&	+0.036	&	0.003	&	0.22	&	+1.12	&	0.26	\\
634	&	J063844.48+172022.5	&	Gaia DR2 3359006648466147968	&	06 38 44.49 	&	 +17 20 22.59	&	B9 III$-$IV Si   	&	162	&	12.9577	&	0.0012	&	+0.324	&	0.052	&	+0.021	&	0.005	&	0.70	&	$-$0.19	&	0.35	\\
635	&	J063851.43+252003.8	&	TYC 1884-766-1     	&	06 38 51.44 	&	 +25 20 03.89	&	B9 IV$-$V Sr   	&	122	&	12.2508	&	0.0008	&	+0.541	&	0.085	&	$-$0.002	&	0.009	&	0.18	&	+0.74	&	0.34	\\
636	&	J063853.85-002541.6	&	TYC 4799-1663-1    	&	06 38 53.85 	&	 -00 25 41.64	&	B8 III$-$IV Si  (He-wk) 	&	214	&	11.6562	&	0.0004	&	+0.773	&	0.039	&	+0.015	&	0.003	&	0.73	&	+0.37	&	0.12	\\
637	&	J063920.04+211735.5	&	Gaia DR2 3378859881392215424	&	06 39 20.04 	&	 +21 17 35.55	&	B9.5 V SrCr   	&	121	&	13.1980	&	0.0009	&	+0.394	&	0.028	&	+0.146	&	0.005	&	0.25	&	+0.92	&	0.16	\\
638	&	J063940.37+032443.2	&	HD 261594  	&	06 39 40.50 	&	 +03 24 42.93	&	A2 IV$-$V SrCrEu   	&	393	&	10.4761	&	0.0004	&	+1.249	&	0.044	&	+0.140	&	0.002	&	0.17	&	+0.79	&	0.09	\\
639	&	J064012.77+082519.3	&	HD 261712  	&	06 40 12.72 	&	 +08 25 19.19	&	B8 IV Si   	&	252	&	10.6409	&	0.0014	&	+1.080	&	0.045	&	+0.037	&	0.006	&	0.66	&	+0.15	&	0.10	\\
640	&	J064013.40+061645.0	&	HD 261715  	&	06 40 13.41 	&	 +06 16 45.02	&	B8 IV$-$V Si  (He-wk) 	&	127	&	11.2513	&	0.0010	&	+0.821	&	0.038	&	$-$0.200	&	0.005	&	0.56	&	+0.26	&	0.11	\\
641	&	J064039.28+264137.2	&	TYC 1888-159-1     	&	06 40 39.28 	&	 +26 41 37.23	&	B9 IV Si   	&	188	&	10.5715	&	0.0009	&	+0.760	&	0.048	&	$-$0.057	&	0.004	&	0.19	&	$-$0.22	&	0.14	\\
642	&	J064112.55+240506.5 $^{a}$	&	TYC 1880-251-1     	&	06 41 12.69 	&	 +24 05 03.41	&	A0 III$-$IV CrEu   	&	921	&	8.6497	&	0.0004	&	+1.632	&	0.405	&	$-$0.030	&	0.005	&	0.10	&	$-$0.38	&	0.54	\\
643	&	J064158.41+223927.2	&	Gaia DR2 3379294635162916864	&	06 41 58.41 	&	 +22 39 27.25	&	A5 IV SrCrEu   	&	111	&	13.5967	&	0.0005	&	+0.553	&	0.122	&	+0.371	&	0.004	&	0.30	&	+2.01	&	0.48	\\
644	&	J064226.95+181244.1	&	Gaia DR2 3359462365976011648	&	06 42 26.95 	&	 +18 12 44.23	&	A2 IV$-$V SrCrEu   	&	113	&	14.4136	&	0.0011	&	+0.328	&	0.034	&	+0.223	&	0.006	&	0.39	&	+1.60	&	0.23	\\
645	&	J064233.27+032500.7	&	Gaia DR2 3127497500212384512	&	06 42 33.27 	&	 +03 25 00.75	&	B8 III$-$IV Si	&	154	&	12.2114	&	0.0003	&	+0.606	&	0.065	&	+0.051	&	0.003	&	1.30	&	$-$0.17	&	0.24	\\
646	&	J064257.32+521951.3	&	TYC 3389-862-1     	&	06 42 57.33 	&	 +52 19 51.34	&	kA4hA6mF2 SrCrEu   	&	213	&	10.6500	&	0.0005	&	+1.194	&	0.038	&	+0.200	&	0.002	&	0.11	&	+0.92	&	0.08	\\
647	&	J064259.18+270608.9	&	HD 48119   	&	06 42 59.19 	&	 +27 06 08.97	&	B8 IV Si   	&	486	&	9.5911	&	0.0014	&	+0.974	&	0.090	&	$-$0.158	&	0.002	&	0.11	&	$-$0.58	&	0.21	\\
648	&	J064310.76+484123.3	&	TYC 3381-73-1      	&	06 43 10.78 	&	 +48 41 23.27	&	B9 V CrSi  (He-wk) 	&	104	&	11.5991	&	0.0004	&	+0.453	&	0.080	&	+0.167	&	0.004	&	0.16	&	$-$0.28	&	0.38	\\
649	&	J064351.19+233534.8	&	TYC 1893-186-1     	&	06 43 51.19 	&	 +23 35 34.79	&	B9.5 III$-$IV Si   	&	165	&	11.2900	&	0.0010	&	+0.384	&	0.328	&	$-$	&	$-$	&	$-$	&	$-$	&	$-$	\\
650	&	J064358.45+033447.0	&	Gaia DR2 3127517738096930432	&	06 43 58.46 	&	 +03 34 47.06	&	B8 IV$-$V Si   	&	126	&	12.7791	&	0.0009	&	+0.513	&	0.036	&	$-$0.003	&	0.005	&	0.67	&	+0.66	&	0.16	\\
\hline
\end{tabular}                                                                                                                                                                   
\end{adjustbox}
\end{center}                                                                                                                                             
\end{sidewaystable*}
\setcounter{table}{0}  
\begin{sidewaystable*}
\caption{Essential data for our sample stars, sorted by increasing right ascension. The columns denote: (1) Internal identification number. (2) LAMOST identifier. (3) Alternativ identifier (HD number, TYC identifier or GAIA DR2 number). (4) Right ascension (J2000; GAIA DR2). (5) Declination (J2000; GAIA DR2). (6) Spectral type, as derived in this study. (7) Sloan $g$ band S/N ratio of the analysed spectrum. (8) $G$\,mag (GAIA DR2). (9) $G$\,mag error. (10) Parallax (GAIA DR2). (11) Parallax error. (12) Dereddened colour index $(BP-RP){_0}$ (GAIA DR2). (13) Colour index error. (14) Absorption in the $G$ band, $A_G$. (15) Intrinsic absolute magnitude in the $G$ band, $M_{\mathrm{G,0}}$. (16) Absolute magnitude error.}
\label{table_master14}
\begin{center}
\begin{adjustbox}{max width=\textwidth}
\begin{tabular}{lllcclcccccccccc}
\hline
\hline
(1) & (2) & (3) & (4) & (5) & (6) & (7) & (8) & (9) & (10) & (11) & (12) & (13) & (14) & (15) & (16) \\
No.	&	ID\_LAMOST	&	ID\_alt	&	RA(J2000) 	&	 Dec(J2000)    	&	SpT\_final	&	S/N\,$g$	&	$G$\,mag	&	e\_$G$\,mag	&	pi (mas)	&	e\_pi	&	$(BP-RP){_0}$	&	e\_$(BP-RP){_0}$	&	$A_G$	&	$M_{\mathrm{G,0}}$	&	e\_$M_{\mathrm{G,0}}$	\\
\hline 
651	&	J064438.75+315333.6	&	Gaia DR2 937202240976723072	&	06 44 38.75 	&	 +31 53 33.76	&	A1 V SrCrEuSi   	&	141	&	12.4186	&	0.0010	&	+0.619	&	0.051	&	+0.143	&	0.005	&	0.21	&	+1.17	&	0.18	\\
652	&	J064441.84+213750.5	&	Gaia DR2 3378235530585937280	&	06 44 41.85 	&	 +21 37 50.54	&	A0 II EuSi   	&	132	&	13.9601	&	0.0020	&	+0.202	&	0.043	&	$-$0.088	&	0.01	&	0.46	&	+0.03	&	0.46	\\
653	&	J064452.82+055428.4	&	TYC 159-3043-1     	&	06 44 52.83 	&	 +05 54 28.73	&	B8 IV$-$V Si   	&	129	&	11.2374	&	0.0011	&	+0.749	&	0.107	&	$-$0.094	&	0.006	&	0.74	&	$-$0.13	&	0.31	\\
654	&	J064505.60+122253.2	&	Gaia DR2 3352393605721639424	&	06 45 05.61 	&	 +12 22 53.30	&	A0 IV$-$V SrCrEu   	&	167	&	14.0446	&	0.0010	&	+0.354	&	0.048	&	+0.152	&	0.012	&	0.51	&	+1.28	&	0.30	\\
655	&	J064510.75+134157.8	&	Gaia DR2 3352818979283285504	&	06 45 10.76 	&	 +13 41 57.85	&	B9 III Si   	&	107	&	13.4362	&	0.0005	&	+0.270	&	0.032	&	+0.017	&	0.003	&	0.42	&	+0.17	&	0.26	\\
656	&	J064511.33+035210.9	&	TYC 156-1661-1     	&	06 45 11.34 	&	 +03 52 10.93	&	A3 IV$-$V SrCr   	&	157	&	10.3271	&	0.0006	&	+1.176	&	0.047	&	+0.128	&	0.002	&	0.14	&	+0.54	&	0.10	\\
657	&	J064514.05+371344.9	&	TYC 2448-182-1     	&	06 45 14.06 	&	 +37 13 44.93	&	B9 V Cr   	&	364	&	10.0004	&	0.0006	&	+1.596	&	0.039	&	$-$0.014	&	0.003	&	0.23	&	+0.79	&	0.07	\\
658	&	J064524.45-020257.8	&	HD 292348  	&	06 45 24.46 	&	 -02 02 57.87	&	B9 II$-$III Si   	&	106	&	14.3338	&	0.0011	&	+0.336	&	0.040	&	$-$0.104	&	0.007	&	1.63	&	+0.33	&	0.26	\\
659	&	J064529.41+072552.2	&	TYC 160-109-1      	&	06 45 29.42 	&	 +07 25 52.30	&	B8 IV Si   	&	211	&	11.3481	&	0.0016	&	+0.883	&	0.081	&	+0.035	&	0.007	&	0.69	&	+0.39	&	0.20	\\
660	&	J064534.06+234628.3	&	HD 263149  	&	06 45 34.07 	&	 +23 46 28.33	&	A1 IV SrCrEu   	&	238	&	11.4281	&	0.0015	&	+0.835	&	0.050	&	+0.066	&	0.009	&	0.17	&	+0.86	&	0.14	\\
661	&	J064540.13+112441.7	&	TYC 754-688-1      	&	06 45 40.14 	&	 +11 24 41.80	&	B9.5 II$-$III Si   	&	202	&	12.4162	&	0.0012	&	+0.437	&	0.040	&	+0.093	&	0.007	&	0.82	&	$-$0.20	&	0.20	\\
662	&	J064545.94+133602.0	&	HD 48806   	&	06 45 46.04 	&	 +13 36 04.80	&	B9 IV$-$V  bl4130  	&	315	&	9.2788	&	0.0010	&	+1.375	&	0.081	&	$-$0.157	&	0.004	&	0.29	&	$-$0.32	&	0.14	\\
663	&	J064549.02+484330.2	&	TYC 3394-359-1     	&	06 45 49.02 	&	 +48 43 30.22	&	B8 IV$-$V  bl4130  	&	157	&	12.0159	&	0.0005	&	+0.460	&	0.046	&	$-$0.038	&	0.004	&	0.24	&	+0.09	&	0.22	\\
664	&	J064559.17+213252.6	&	HD 263301  	&	06 45 59.18 	&	 +21 32 52.66	&	B8 III Si   	&	391	&	10.4364	&	0.0009	&	+0.531	&	0.054	&	$-$0.192	&	0.004	&	0.23	&	$-$1.16	&	0.23	\\
665	&	J064601.44+221654.4	&	Gaia DR2 3379399634225338112	&	06 46 01.45 	&	 +22 16 54.42	&	A0 IV CrEu   	&	107	&	14.2917	&	0.0004	&	+0.326	&	0.033	&	+0.289	&	0.002	&	0.58	&	+1.28	&	0.22	\\
666	&	J064614.76+072231.1	&	Gaia DR2 3133202419371320832	&	06 46 14.77 	&	 +07 22 31.28	&	B8 IV Si   	&	151	&	11.9800	&	0.0008	&	+0.619	&	0.053	&	$-$0.152	&	0.005	&	0.83	&	+0.11	&	0.19	\\
667	&	J064616.91+131936.6	&	Gaia DR2 3352605777105469696	&	06 46 16.92 	&	 +13 19 36.70	&	A0 III \textit{CrEu}*	&	100	&	14.5164	&	0.0005	&	+0.189	&	0.067	&	$-$	&	$-$	&	$-$	&	$-$	&	$-$	\\
668	&	J064617.26+082858.8	&	HD 263556  	&	06 46 17.26 	&	 +08 28 58.91	&	B9 II$-$III bl4077 bl4130  	&	153	&	11.5982	&	0.0022	&	+0.894	&	0.049	&	+0.018	&	0.009	&	0.82	&	+0.54	&	0.13	\\
669	&	J064628.76+150619.4	&	HD 263549  	&	06 46 28.76 	&	 +15 06 19.50	&	B9 III$-$IV Si   	&	267	&	9.7127	&	0.0005	&	+1.003	&	0.088	&	$-$0.195	&	0.003	&	0.31	&	$-$0.59	&	0.20	\\
670	&	J064633.16+020802.9	&	Gaia DR2 3126413484825266560	&	06 46 33.16 	&	 +02 08 02.95	&	B9 IV$-$V  bl4130  	&	110	&	12.9308	&	0.0043	&	+0.393	&	0.109	&	$-$	&	$-$	&	$-$	&	$-$	&	$-$	\\
671	&	J064637.59+051616.2	&	TYC 156-871-1      	&	06 46 37.59 	&	 +05 16 16.21	&	B8 IV  bl4130  	&	110	&	12.9161	&	0.0004	&	+0.796	&	0.040	&	+0.075	&	0.003	&	0.51	&	+1.91	&	0.12	\\
672	&	J064641.00+005717.1	&	Gaia DR2 3125709934822430720	&	06 46 41.00 	&	 +00 57 17.14	&	B8 IV Si   	&	101	&	12.8960	&	0.0007	&	+0.461	&	0.036	&	$-$0.034	&	0.005	&	1.19	&	+0.02	&	0.18	\\
673	&	J064701.81+102355.5	&	Gaia DR2 3350836834693773312	&	06 47 01.82 	&	 +10 23 55.53	&	B8 III$-$IV  bl4130  	&	106	&	12.7947	&	0.0004	&	+0.505	&	0.040	&	+0.026	&	0.004	&	0.42	&	+0.89	&	0.18	\\
674	&	J064719.79+013513.6	&	TYC 148-641-1      	&	06 47 19.79 	&	 +01 35 13.71	&	A0 II$-$III EuSi   	&	143	&	12.6268	&	0.0010	&	+0.533	&	0.039	&	+0.040	&	0.006	&	0.73	&	+0.53	&	0.17	\\
675	&	J064741.02+072458.8	&	TYC 160-321-1      	&	06 47 41.04 	&	 +07 24 58.84	&	B9 III$-$IV Si   	&	150	&	11.2725	&	0.0020	&	+0.575	&	0.036	&	+0.001	&	0.009	&	0.83	&	$-$0.77	&	0.15	\\
676	&	J064745.41+583506.1	&	HD 48560   	&	06 47 45.40 	&	 +58 35 06.25	&	kA1hA2mA5 CrEu   	&	614	&	9.6569	&	0.0008	&	+3.559	&	0.294	&	+0.100	&	0.002	&	0.10	&	+2.31	&	0.19	\\
677	&	J064748.51+160757.8	&	HD 263921  	&	06 47 48.51 	&	 +16 07 57.83	&	B9 IV$-$V Sr   	&	201	&	10.3096	&	0.0036	&	$-$4.029	&	0.497	&	$-$	&	$-$	&	$-$	&	$-$	&	$-$	\\
678	&	J064757.48+105648.2	&	Gaia DR2 3350927819280584448	&	06 47 57.50 	&	 +10 56 48.31	&	kB9.5hA1mA2 Cr   	&	159	&	15.7093	&	0.0009	&	+2.669	&	0.050	&	+1.721	&	0.007	&	0.09	&	+7.75	&	0.06	\\
679	&	J064758.50+283022.6	&	TYC 1905-1123-1    	&	06 47 58.50 	&	 +28 30 22.64	&	B9 V Cr   	&	615	&	10.4109	&	0.0009	&	+1.137	&	0.055	&	$-$0.037	&	0.003	&	0.11	&	+0.58	&	0.12	\\
680	&	J064758.79+105621.5	&	Gaia DR2 3350927784920847872	&	06 47 58.80 	&	 +10 56 21.61	&	kA0hA1mA2 bl4077 bl4130  	&	132	&	12.8648	&	0.0007	&	+0.516	&	0.032	&	+0.235	&	0.003	&	0.53	&	+0.90	&	0.15	\\
681	&	J064826.33+203755.5	&	TYC 1343-2249-1    	&	06 48 26.33 	&	 +20 37 55.60	&	B9 IV$-$V  bl4130  	&	376	&	11.4274	&	0.0005	&	+0.764	&	0.063	&	$-$0.067	&	0.005	&	0.13	&	+0.71	&	0.19	\\
682	&	J064838.53+025528.9	&	Gaia DR2 3126572299835661568	&	06 48 38.53 	&	 +02 55 28.98	&	B8 IV  bl4130  	&	118	&	12.8083	&	0.0012	&	+0.460	&	0.053	&	+0.073	&	0.007	&	0.69	&	+0.43	&	0.26	\\
683	&	J064843.13+115633.2	&	Gaia DR2 3351576084465603072	&	06 48 43.13 	&	 +11 56 33.21	&	A0 IV$-$V SrCrEuSi   	&	203	&	12.2557	&	0.0005	&	+0.857	&	0.048	&	+0.217	&	0.005	&	0.23	&	+1.69	&	0.13	\\
684	&	J064844.97+131201.4	&	HD 264269  	&	06 48 44.98 	&	 +13 12 01.46	&	B8 IV$-$V Si   	&	272	&	11.5207	&	0.0007	&	+0.687	&	0.043	&	$-$0.016	&	0.005	&	0.31	&	+0.40	&	0.14	\\
685	&	J064851.39+211245.6	&	TYC 1343-2670-1    	&	06 48 51.39 	&	 +21 12 45.65	&	B7 III$-$IV  bl4130 (He-wk) 	&	212	&	11.8405	&	0.0011	&	+0.548	&	0.050	&	$-$0.138	&	0.008	&	0.10	&	+0.44	&	0.20	\\
686	&	J064853.79+203905.3	&	Gaia DR2 3377882205098370304	&	06 48 53.79 	&	 +20 39 05.28	&	kB9.5hA2mA5 SrCrEu   	&	143	&	13.1255	&	0.0011	&	+0.374	&	0.049	&	+0.143	&	0.004	&	0.09	&	+0.90	&	0.29	\\
687	&	J064859.90+381609.0	&	HD 49198   	&	06 49 00.10 	&	 +38 16 11.27	&	A0 III$-$IV CrSi   	&	344	&	9.2945	&	0.0010	&	+2.106	&	0.077	&	$-$0.004	&	0.005	&	0.19	&	+0.72	&	0.09	\\
688	&	J064901.28+034633.2	&	TYC 156-582-1      	&	06 49 01.29 	&	 +03 46 33.27	&	B9 V Cr   	&	114	&	12.4737	&	0.0003	&	+0.600	&	0.036	&	$-$0.020	&	0.003	&	0.71	&	+0.66	&	0.14	\\
689	&	J064907.51+114600.1	&	Gaia DR2 3351558114322467968	&	06 49 07.52 	&	 +11 46 00.17	&	kA1hA7mA8 SrCr   	&	134	&	13.5457	&	0.0022	&	$-$1.385	&	0.314	&	$-$	&	$-$	&	$-$	&	$-$	&	$-$	\\
690	&	J064947.96+202510.8	&	TYC 1339-558-1     	&	06 49 47.97 	&	 +20 25 10.82	&	kA3hA5mA7 bl4077 bl4130  	&	330	&	9.9263	&	0.0005	&	+1.510	&	0.049	&	+0.344	&	0.002	&	0.04	&	+0.78	&	0.09	\\
691	&	J064959.12+125329.4	&	TYC 755-771-1      	&	06 49 59.13 	&	 +12 53 29.38	&	A0 IV$-$V CrSi  (He-wk) 	&	292	&	9.9404	&	0.0006	&	+1.715	&	0.069	&	$-$0.109	&	0.002	&	0.18	&	+0.94	&	0.10	\\
692	&	J065000.25+293255.8	&	HD 49522   	&	06 50 00.02 	&	 +29 32 57.67	&	A0 V CrEuSi   	&	142	&	8.8861	&	0.0008	&	+2.934	&	0.084	&	$-$0.026	&	0.003	&	0.07	&	+1.16	&	0.08	\\
693	&	J065021.64+433020.7	&	TYC 2954-959-1     	&	06 50 21.64 	&	 +43 30 20.76	&	B9.5 IV Cr   	&	304	&	10.7962	&	0.0006	&	+0.704	&	0.064	&	+0.119	&	0.003	&	0.23	&	$-$0.19	&	0.20	\\
694	&	J065047.40+183257.1	&	TYC 1335-1780-1    	&	06 50 47.40 	&	 +18 32 57.17	&	B9 III$-$IV Si   	&	414	&	11.1338	&	0.0008	&	+0.825	&	0.059	&	$-$0.105	&	0.008	&	0.08	&	+0.63	&	0.16	\\
695	&	J065119.09+102848.9	&	Gaia DR2 3158687866251327872	&	06 51 19.09 	&	 +10 28 49.03	&	B9.5 V CrEu   	&	102	&	14.1592	&	0.0006	&	+0.335	&	0.036	&	+0.153	&	0.005	&	0.34	&	+1.44	&	0.24	\\
696	&	J065134.66+343547.3	&	Gaia DR2 940138761657217536	&	06 51 34.66 	&	 +34 35 47.41	&	B9.5 II$-$III  bl4130  	&	235	&	14.0506	&	0.0006	&	+0.192	&	0.050	&	$-$	&	$-$	&	$-$	&	$-$	&	$-$	\\
697	&	J065135.29+205612.8	&	HD 265031  	&	06 51 35.12 	&	 +20 56 13.26	&	kB9.5hA2mA5 CrEu   	&	543	&	9.8695	&	0.0007	&	+1.745	&	0.065	&	$-$0.019	&	0.002	&	0.14	&	+0.94	&	0.10	\\
698	&	J065141.48+102538.0	&	TYC 751-2305-1     	&	06 51 41.49 	&	 +10 25 38.07	&	B9 V Cr   	&	276	&	11.7306	&	0.0006	&	+0.597	&	0.049	&	+0.079	&	0.004	&	0.20	&	+0.41	&	0.18	\\
699	&	J065154.51+104428.0	&	Gaia DR2 3159071973767160704	&	06 51 54.52 	&	 +10 44 28.07	&	B9 IV$-$V  bl4130  	&	100	&	13.4866	&	0.0009	&	+0.426	&	0.047	&	+0.061	&	0.008	&	0.13	&	+1.51	&	0.24	\\
700	&	J065200.81+121952.8	&	TYC 755-1591-1     	&	06 52 00.82 	&	 +12 19 52.80	&	A0 IV CrEu   	&	312	&	10.6482	&	0.0005	&	+1.385	&	0.068	&	+0.101	&	0.003	&	0.12	&	+1.24	&	0.12	\\
\hline
\end{tabular}                                                                                                                                                                   
\end{adjustbox}
\end{center}                                                                                                                                             
\end{sidewaystable*}
\setcounter{table}{0}  
\begin{sidewaystable*}
\caption{Essential data for our sample stars, sorted by increasing right ascension. The columns denote: (1) Internal identification number. (2) LAMOST identifier. (3) Alternativ identifier (HD number, TYC identifier or GAIA DR2 number). (4) Right ascension (J2000; GAIA DR2). (5) Declination (J2000; GAIA DR2). (6) Spectral type, as derived in this study. (7) Sloan $g$ band S/N ratio of the analysed spectrum. (8) $G$\,mag (GAIA DR2). (9) $G$\,mag error. (10) Parallax (GAIA DR2). (11) Parallax error. (12) Dereddened colour index $(BP-RP){_0}$ (GAIA DR2). (13) Colour index error. (14) Absorption in the $G$ band, $A_G$. (15) Intrinsic absolute magnitude in the $G$ band, $M_{\mathrm{G,0}}$. (16) Absolute magnitude error.}
\label{table_master15}
\begin{center}
\begin{adjustbox}{max width=\textwidth}
\begin{tabular}{lllcclcccccccccc}
\hline
\hline
(1) & (2) & (3) & (4) & (5) & (6) & (7) & (8) & (9) & (10) & (11) & (12) & (13) & (14) & (15) & (16) \\
No.	&	ID\_LAMOST	&	ID\_alt	&	RA(J2000) 	&	 Dec(J2000)    	&	SpT\_final	&	S/N\,$g$	&	$G$\,mag	&	e\_$G$\,mag	&	pi (mas)	&	e\_pi	&	$(BP-RP){_0}$	&	e\_$(BP-RP){_0}$	&	$A_G$	&	$M_{\mathrm{G,0}}$	&	e\_$M_{\mathrm{G,0}}$	\\
\hline 
701	&	J065205.09+065214.6	&	Gaia DR2 3132980592900051200	&	06 52 05.10 	&	 +06 52 14.67	&	B9 V Cr   	&	123	&	12.5078	&	0.0004	&	+0.418	&	0.049	&	+0.077	&	0.004	&	0.30	&	+0.32	&	0.26	\\
702	&	J065209.98+102634.8	&	Gaia DR2 3158640655970656256	&	06 52 09.98 	&	 +10 26 34.93	&	kA1hA8mA8 SrEuSi   	&	141	&	14.1261	&	0.0014	&	+0.300	&	0.037	&	+0.075	&	0.008	&	0.23	&	+1.28	&	0.27	\\
703	&	J065216.71+205412.6	&	Gaia DR2 3366259508199396864	&	06 52 16.71 	&	 +20 54 12.63	&	kA0hA5mA6 CrSi   	&	125	&	13.2474	&	0.0005	&	+0.375	&	0.033	&	+0.052	&	0.004	&	0.15	&	+0.97	&	0.20	\\
704	&	J065226.21+024001.8	&	HD 289244  	&	06 52 26.22 	&	 +02 40 01.90	&	kB9.5hA3mA3 CrEu   	&	322	&	11.2022	&	0.0010	&	+0.926	&	0.035	&	$-$0.058	&	0.004	&	0.44	&	+0.60	&	0.10	\\
705	&	J065235.59+423715.7	&	HD 49884   	&	06 52 35.20 	&	 +42 37 10.07	&	B9.5 III Eu   	&	198	&	8.2392	&	0.0010	&	+2.741	&	0.066	&	$-$0.010	&	0.003	&	0.13	&	+0.29	&	0.07	\\
706	&	J065302.61+403553.6	&	TYC 2946-305-1     	&	06 53 02.61 	&	 +40 35 53.63	&	A5 IV$-$V SrEu   	&	172	&	11.9586	&	0.0003	&	+1.045	&	0.041	&	+0.253	&	0.003	&	0.17	&	+1.88	&	0.10	\\
707	&	J065310.66+071446.8	&	TYC 161-1956-1     	&	06 53 10.67 	&	 +07 14 46.86	&	kB9hA1mA2 CrEuSi   	&	161	&	12.2648	&	0.0023	&	+0.315	&	0.041	&	+0.110	&	0.018	&	0.38	&	$-$0.63	&	0.29	\\
708	&	J065313.89+103303.3	&	Gaia DR2 3159021155714729472	&	06 53 13.90 	&	 +10 33 03.39	&	kA0hA2mA5 Eu   	&	119	&	13.5618	&	0.0024	&	+0.328	&	0.044	&	+0.023	&	0.011	&	0.21	&	+0.93	&	0.29	\\
709	&	J065318.97+131319.7	&	Gaia DR2 3353293349829167744	&	06 53 18.98 	&	 +13 13 19.81	&	kA0hA3mA6 Cr   	&	123	&	14.4544	&	0.0006	&	+0.155	&	0.048	&	$-$	&	$-$	&	$-$	&	$-$	&	$-$	\\
710	&	J065340.20+562055.0	&	TYC 3775-178-1     	&	06 53 40.19 	&	 +56 20 55.05	&	B9 IV$-$V Cr   	&	159	&	12.8888	&	0.0004	&	+0.358	&	0.046	&	$-$0.012	&	0.002	&	0.12	&	+0.54	&	0.29	\\
711	&	J065356.37+103302.0	&	Gaia DR2 3159010156299346048	&	06 53 56.38 	&	 +10 33 02.06	&	kB9.5hA1mA3 CrEu   	&	147	&	13.1461	&	0.0018	&	+0.385	&	0.033	&	+0.176	&	0.009	&	0.24	&	+0.83	&	0.19	\\
712	&	J065358.16+141845.0	&	TYC 760-381-1      	&	06 53 58.17 	&	 +14 18 45.01	&	B9.5 IV$-$V SrCrEu   	&	211	&	10.8393	&	0.0007	&	+1.168	&	0.046	&	+0.020	&	0.004	&	0.08	&	+1.09	&	0.10	\\
713	&	J065400.61+063645.2	&	Gaia DR2 3129955905132736768	&	06 54 00.62 	&	 +06 36 45.16	&	A0 IV$-$V SrCrEu   	&	153	&	12.2854	&	0.0006	&	+0.856	&	0.063	&	+0.147	&	0.004	&	0.26	&	+1.69	&	0.17	\\
714	&	J065401.04+213810.3	&	TYC 1343-673-1     	&	06 54 01.04 	&	 +21 38 10.40	&	kA1hA3mA7 CrEu   	&	229	&	12.1451	&	0.0004	&	+0.565	&	0.044	&	+0.199	&	0.003	&	0.09	&	+0.81	&	0.18	\\
715	&	J065403.63+221545.2	&	HD 50403   	&	06 54 03.63 	&	 +22 15 45.16	&	A6 IV SrCrEu   	&	386	&	9.2694	&	0.0007	&	+3.141	&	0.045	&	+0.181	&	0.005	&	0.05	&	+1.71	&	0.06	\\
716	&	J065404.66+113512.6	&	HD 265946  	&	06 54 04.60 	&	 +11 35 09.65	&	A0 V Cr   	&	638	&	9.4015	&	0.0003	&	+1.797	&	0.061	&	$-$0.021	&	0.002	&	0.12	&	+0.56	&	0.09	\\
717	&	J065414.93+083325.5	&	TYC 748-1845-1     	&	06 54 14.98 	&	 +08 33 23.77	&	B9.5 II$-$III Si   	&	271	&	10.4132	&	0.0012	&	+0.920	&	0.039	&	$-$0.045	&	0.005	&	0.13	&	+0.10	&	0.10	\\
718	&	J065419.05+053003.4	&	Gaia DR2 3129541049951592576	&	06 54 19.06 	&	 +05 30 03.45	&	B9 IV$-$V SrCr   	&	108	&	12.6897	&	0.0003	&	+1.515	&	0.320	&	+0.297	&	0.003	&	0.14	&	+3.45	&	0.46	\\
719	&	J065444.54+202931.2	&	Gaia DR2 3366195672104620544	&	06 54 44.55 	&	 +20 29 31.27	&	B9 V CrEu   	&	117	&	13.5516	&	0.0007	&	+0.256	&	0.042	&	+0.077	&	0.005	&	0.19	&	+0.41	&	0.36	\\
720	&	J065444.94+135455.5	&	HD 266119  	&	06 54 44.95 	&	 +13 54 55.56	&	A1 IV$-$V Cr   	&	177	&	10.6473	&	0.0006	&	+0.943	&	0.033	&	+0.133	&	0.001	&	0.16	&	+0.36	&	0.09	\\
721	&	J065458.31+040826.9 $^{a}$	&	HD 266311  	&	06 54 58.17 	&	 +04 08 27.55	&	kA1hA3mA6 SrCrEu   	&	488	&	9.7499	&	0.0006	&	+1.985	&	0.049	&	$-$0.022	&	0.003	&	0.19	&	+1.04	&	0.07	\\
722	&	J065505.83+384426.4	&	TYC 2942-279-1     	&	06 55 05.83 	&	 +38 44 26.49	&	B8 IV  bl4130  	&	177	&	12.1167	&	0.0007	&	+0.374	&	0.065	&	+0.005	&	0.005	&	0.14	&	$-$0.16	&	0.38	\\
723	&	J065509.34+044322.6	&	TYC 157-2479-1     	&	06 55 09.35 	&	 +04 43 22.67	&	B8 III Si   	&	165	&	12.3571	&	0.0005	&	+0.404	&	0.046	&	+0.086	&	0.003	&	0.91	&	$-$0.52	&	0.25	\\
724	&	J065511.76+115158.3	&	HD 266267  	&	06 55 11.66 	&	 +11 51 56.24	&	A7 V Sr\textit{Cr}Eu* $^{d}$  	&	566	&	10.0083	&	0.0007	&	+2.020	&	0.049	&	+0.044	&	0.004	&	0.15	&	+1.38	&	0.07	\\
725	&	J065518.21+022530.3	&	Gaia DR2 3126252028414767360	&	06 55 18.21 	&	 +02 25 30.24	&	B9 II$-$III Si   	&	195	&	12.3149	&	0.0014	&	+0.363	&	0.069	&	$-$0.041	&	0.026	&	1.11	&	$-$1.00	&	0.42	\\
726	&	J065520.87+120622.7	&	Gaia DR2 3351433903867696640	&	06 55 20.88 	&	 +12 06 22.71	&	A5 IV$-$V SrCrEu   	&	115	&	12.2659	&	0.0005	&	+0.828	&	0.039	&	+0.175	&	0.008	&	0.43	&	+1.43	&	0.11	\\
727	&	J065544.52+563703.5	&	HD 50243   	&	06 55 44.53 	&	 +56 37 03.52	&	B9.5 IV$-$V Cr   	&	592	&	9.4571	&	0.0004	&	+1.766	&	0.054	&	$-$0.008	&	0.004	&	0.09	&	+0.61	&	0.08	\\
728	&	J065623.57+281013.1	&	TYC 1906-481-1     	&	06 56 23.57 	&	 +28 10 13.15	&	A9 V SrEu   	&	196	&	11.3235	&	0.0005	&	+1.313	&	0.049	&	+0.378	&	0.003	&	0.10	&	+1.81	&	0.09	\\
729	&	J065627.80+031055.9	&	TYC 153-2297-1     	&	06 56 27.82 	&	 +03 10 55.95	&	\textit{B9 III (Cr)}*	&	413	&	11.1237	&	0.0060	&	+1.221	&	0.334	&	$-$	&	$-$	&	$-$	&	$-$	&	$-$	\\
730	&	J065629.15+074030.5	&	TYC 748-2642-1     	&	06 56 29.15 	&	 +07 40 30.49	&	A1 IV$-$V CrEu   	&	241	&	10.4546	&	0.0006	&	+1.068	&	0.062	&	+0.058	&	0.001	&	0.11	&	+0.49	&	0.14	\\
731	&	J065629.87+200101.8	&	TYC 1352-411-1     	&	06 56 29.86 	&	 +20 01 01.81	&	B8 III$-$IV Si   	&	420	&	11.5211	&	0.0009	&	+0.586	&	0.061	&	$-$0.101	&	0.006	&	0.06	&	+0.30	&	0.23	\\
732	&	J065647.94+242958.8	&	HD 266617  	&	06 56 47.78 	&	 +24 29 58.86	& \textit{A0} V \textit{Si}Sr\textit{Cr}* $^{d}$ &	383	&	10.1910	&	0.0011	&	+1.584	&	0.118	&	+0.021	&	0.003	&	0.06	&	+1.13	&	0.17	\\
733	&	J065649.21+541717.5	&	TYC 3767-219-1     	&	06 56 49.21 	&	 +54 17 17.52	&	A0 V CrEu   	&	297	&	11.9272	&	0.0003	&	+0.950	&	0.038	&	+0.142	&	0.002	&	0.07	&	+1.75	&	0.10	\\
734	&	J065707.80+030838.9	&	TYC 153-2323-1     	&	06 57 07.81 	&	 +03 08 38.93	&	B9.5 IV Eu   	&	274	&	11.4200	&	0.0013	&	+0.683	&	0.040	&	+0.081	&	0.006	&	0.63	&	$-$0.04	&	0.14	\\
735	&	J065714.71+073908.9	&	Gaia DR2 3157064639192846720	&	06 57 14.71 	&	 +07 39 08.95	&	B9.5 IV EuSi   	&	122	&	12.1459	&	0.0005	&	+0.461	&	0.040	&	+0.022	&	0.003	&	0.20	&	+0.27	&	0.19	\\
736	&	J065800.69+482950.3	&	HD 50972   	&	06 58 00.79 	&	 +48 29 46.52	&	B9 V SrCr   	&	468	&	8.0500	&	0.0006	&	+4.334	&	0.083	&	+0.001	&	0.002	&	0.06	&	+1.17	&	0.06	\\
737	&	J065815.39+021919.6	&	Gaia DR2 3115727022077258112	&	06 58 15.40 	&	 +02 19 19.63	&	A5 IV$-$V SrCrEu   	&	263	&	11.2521	&	0.0009	&	+1.325	&	0.048	&	+0.179	&	0.003	&	0.22	&	+1.64	&	0.09	\\
738	&	J065826.55+123932.0	&	TYC 756-1181-1     	&	06 58 26.56 	&	 +12 39 32.08	&	B9 IV bl4077 bl4130  	&	363	&	10.8823	&	0.0011	&	+0.624	&	0.042	&	$-$0.022	&	0.005	&	0.11	&	$-$0.25	&	0.15	\\
739	&	J065834.63+282752.9	&	Gaia DR2 887381960372791296	&	06 58 34.63 	&	 +28 27 52.88	&	A0 IV Cr   	&	110	&	14.1264	&	0.0007	&	+0.233	&	0.031	&	+0.243	&	0.004	&	0.11	&	+0.86	&	0.29	\\
740	&	J065846.62+305342.0	&	TYC 2437-446-1     	&	06 58 46.65 	&	 +30 53 41.09	&	B8 IV \textit{SiCr} (He-wk)*	&	368	&	10.7901	&	0.0007	&	+0.773	&	0.059	&	$-$0.128	&	0.006	&	0.17	&	+0.07	&	0.17	\\
741	&	J065847.10+005843.6	&	TYC 149-2573-1     	&	06 58 47.11 	&	 +00 58 43.69	&	A0 III$-$IV SrCrEuSi   	&	194	&	11.7281	&	0.0007	&	+0.869	&	0.039	&	+0.013	&	0.004	&	0.42	&	+1.00	&	0.11	\\
742	&	J065941.01+143418.8	&	TYC 760-1081-1     	&	06 59 41.00 	&	 +14 34 18.83	&	B9 IV EuSi   	&	174	&	11.7104	&	0.0013	&	+0.577	&	0.067	&	$-$0.046	&	0.007	&	0.12	&	+0.40	&	0.26	\\
743	&	J070004.95-010025.7	&	HD 292968  	&	07 00 04.96 	&	 -01 00 25.77	&	B9 V Cr   	&	205	&	10.9741	&	0.0007	&	+0.819	&	0.063	&	$-$0.042	&	0.004	&	0.28	&	+0.26	&	0.17	\\
744	&	J070127.57+142617.6	&	TYC 761-1749-1     	&	07 01 27.58 	&	 +14 26 17.70	&	B8 III$-$IV EuSi   	&	192	&	11.4818	&	0.0006	&	+0.675	&	0.073	&	$-$0.047	&	0.007	&	0.06	&	+0.57	&	0.24	\\
745	&	J070132.95+033625.8	&	TYC 166-2723-1     	&	07 01 32.95 	&	 +03 36 25.82	&	B9 IV$-$V Si   	&	243	&	11.4821	&	0.0010	&	+0.933	&	0.042	&	+0.028	&	0.004	&	0.43	&	+0.90	&	0.11	\\
746	&	J070133.22+061017.5	&	TYC 174-2230-1     	&	07 01 33.23 	&	 +06 10 17.43	&	B9 IV  bl4130  	&	167	&	11.2555	&	0.0011	&	+0.455	&	0.047	&	+0.133	&	0.004	&	0.36	&	$-$0.81	&	0.23	\\
747	&	J070144.69-013353.6	&	TYC 4814-2326-1    	&	07 01 44.70 	&	 -01 33 53.59	&	B9.5 IV$-$V SrCrEu   	&	162	&	12.2465	&	0.0010	&	+0.402	&	0.043	&	+0.129	&	0.006	&	0.36	&	$-$0.09	&	0.24	\\
748	&	J070155.26+010932.0	&	Gaia DR2 3113112903122735744	&	07 01 55.27 	&	 +01 09 32.08	&	B9 V Cr   	&	124	&	12.8682	&	0.0003	&	+0.305	&	0.041	&	+0.151	&	0.002	&	0.58	&	$-$0.29	&	0.29	\\
749	&	J070200.78+161738.2	&	HD 52475   	&	07 02 00.78 	&	 +16 17 38.36	&	B8 IV$-$V Si   	&	288	&	10.7487	&	0.0013	&	+0.757	&	0.063	&	$-$0.171	&	0.007	&	0.07	&	+0.07	&	0.19	\\
750	&	J070211.70+202807.9	&	TYC 1352-1191-1    	&	07 02 11.73 	&	 +20 28 08.91	&	B9.5 V SrCrEu   	&	380	&	10.8125	&	0.0005	&	+0.976	&	0.060	&	+0.037	&	0.002	&	0.05	&	+0.71	&	0.14	\\
\hline
\end{tabular}                                                                                                                                                                   
\end{adjustbox}
\end{center}                                                                                                                                             
\end{sidewaystable*}
\setcounter{table}{0}  
\begin{sidewaystable*}
\caption{Essential data for our sample stars, sorted by increasing right ascension. The columns denote: (1) Internal identification number. (2) LAMOST identifier. (3) Alternativ identifier (HD number, TYC identifier or GAIA DR2 number). (4) Right ascension (J2000; GAIA DR2). (5) Declination (J2000; GAIA DR2). (6) Spectral type, as derived in this study. (7) Sloan $g$ band S/N ratio of the analysed spectrum. (8) $G$\,mag (GAIA DR2). (9) $G$\,mag error. (10) Parallax (GAIA DR2). (11) Parallax error. (12) Dereddened colour index $(BP-RP){_0}$ (GAIA DR2). (13) Colour index error. (14) Absorption in the $G$ band, $A_G$. (15) Intrinsic absolute magnitude in the $G$ band, $M_{\mathrm{G,0}}$. (16) Absolute magnitude error.}
\label{table_master16}
\begin{center}
\begin{adjustbox}{max width=\textwidth}
\begin{tabular}{lllcclcccccccccc}
\hline
\hline
(1) & (2) & (3) & (4) & (5) & (6) & (7) & (8) & (9) & (10) & (11) & (12) & (13) & (14) & (15) & (16) \\
No.	&	ID\_LAMOST	&	ID\_alt	&	RA(J2000) 	&	 Dec(J2000)    	&	SpT\_final	&	S/N\,$g$	&	$G$\,mag	&	e\_$G$\,mag	&	pi (mas)	&	e\_pi	&	$(BP-RP){_0}$	&	e\_$(BP-RP){_0}$	&	$A_G$	&	$M_{\mathrm{G,0}}$	&	e\_$M_{\mathrm{G,0}}$	\\
\hline 
751	&	J070234.35+152247.8	&	Gaia DR2 3354031500087825792	&	07 02 34.36 	&	 +15 22 47.89	&	A8 IV$-$V Eu   	&	104	&	12.3900	&	0.0004	&	+0.619	&	0.054	&	+0.420	&	0.002	&	0.09	&	+1.26	&	0.20	\\
752	&	J070237.06+362235.1	&	Gaia DR2 940040321006131200	&	07 02 37.07 	&	 +36 22 35.22	&	A7 V SrEu   	&	109	&	13.1453	&	0.0005	&	+0.560	&	0.030	&	+0.267	&	0.003	&	0.15	&	+1.74	&	0.13	\\
753	&	J070252.77+023700.0	&	HD 52868   	&	07 02 52.76 	&	 +02 36 57.25	&	kB9hA9mA7 SrSi   	&	524	&	9.4417	&	0.0005	&	+1.504	&	0.085	&	$-$0.071	&	0.004	&	0.12	&	+0.21	&	0.13	\\
754	&	J070305.59-020902.4	&	TYC 4818-265-1     	&	07 03 05.59 	&	 -02 09 02.48	&	B8 II$-$III CrSi   	&	131	&	12.3575	&	0.0010	&	+0.385	&	0.041	&	+0.037	&	0.005	&	0.48	&	$-$0.19	&	0.24	\\
755	&	J070337.20+064533.5	&	TYC 174-29-1       	&	07 03 37.24 	&	 +06 45 33.64	&	B8 III$-$IV Si   	&	135	&	10.9012	&	0.0007	&	+1.115	&	0.059	&	$-$0.075	&	0.005	&	0.16	&	+0.98	&	0.12	\\
756	&	J070337.25+194115.0	&	TYC 1352-451-1     	&	07 03 37.25 	&	 +19 41 15.02	&	kA0hA3mA6 Cr   	&	164	&	12.8146	&	0.0010	&	+0.466	&	0.042	&	+0.195	&	0.005	&	0.08	&	+1.07	&	0.20	\\
757	&	J070343.61+140646.3	&	HD 52959   	&	07 03 43.61 	&	 +14 06 46.33	&	B8 III$-$IV  bl4130 (He-wk) 	&	316	&	10.1297	&	0.0008	&	+0.636	&	0.060	&	+0.174	&	0.002	&	0.08	&	$-$0.93	&	0.21	\\
758	&	J070344.63+183406.1	&	Gaia DR2 3361785015568548608	&	07 03 44.64 	&	 +18 34 06.17	&	A9 V SrCrEu   	&	134	&	13.2150	&	0.0005	&	+0.510	&	0.041	&	+0.253	&	0.004	&	0.05	&	+1.70	&	0.18	\\
759	&	J070439.49+181036.7	&	TYC 1349-141-1     	&	07 04 39.30 	&	 +18 10 36.75	&	B9 IV$-$V CrEuSi   	&	186	&	9.9132	&	0.0008	&	+1.628	&	0.083	&	$-$0.054	&	0.004	&	0.04	&	+0.93	&	0.12	\\
760	&	J070519.30+350140.9	&	Gaia DR2 891667783681590656	&	07 05 19.30 	&	 +35 01 41.03	&	B9 IV$-$V Cr   	&	126	&	12.9313	&	0.0008	&	+0.356	&	0.052	&	$-$0.078	&	0.004	&	0.13	&	+0.56	&	0.32	\\
761	&	J070604.68+145303.0	&	Gaia DR2 3359757408749990016	&	07 06 04.69 	&	 +14 53 03.08	&	kA5hA5mA9 SrCrEu   	&	121	&	14.1791	&	0.0007	&	+0.270	&	0.045	&	+0.240	&	0.003	&	0.08	&	+1.25	&	0.36	\\
762	&	J070613.94+185355.6	&	Gaia DR2 3361893764140100096	&	07 06 13.94 	&	 +18 53 55.60	&	B9.5 II$-$III Si   	&	63	&	14.9322	&	0.0030	&	+0.146	&	0.047	&	$-$	&	$-$	&	$-$	&	$-$	&	$-$	\\
763	&	J070617.23+101601.6	&	HD 53662   	&	07 06 17.01 	&	 +10 16 01.62	&	B9 IV EuSi   	&	433	&	8.6716	&	0.0008	&	+2.050	&	0.060	&	$-$0.076	&	0.006	&	0.03	&	+0.20	&	0.08	\\
764	&	J070709.06+225559.9	&	TYC 1896-1380-1    	&	07 07 09.07 	&	 +22 55 59.91	&	B9 IV$-$V CrEu   	&	457	&	10.5471	&	0.0013	&	+0.847	&	0.044	&	+0.026	&	0.006	&	0.05	&	+0.14	&	0.12	\\
765	&	J070738.03+231201.8	&	TYC 1896-1388-1    	&	07 07 38.03 	&	 +23 12 01.76	&	A0 V SrCrEu   	&	775	&	9.2462	&	0.0007	&	+1.894	&	0.055	&	+0.101	&	0.002	&	0.05	&	+0.58	&	0.08	\\
766	&	J070755.64-001724.5	&	HD 293193  	&	07 07 55.64 	&	 -00 17 24.57	&	B9 IV$-$V Cr   	&	237	&	11.4148	&	0.0006	&	+0.619	&	0.045	&	+0.121	&	0.004	&	0.40	&	$-$0.02	&	0.16	\\
767	&	J070801.31+301921.1	&	TYC 2438-214-1     	&	07 08 01.31 	&	 +30 19 21.16	&	A0 IV CrEu   	&	125	&	11.6315	&	0.0012	&	+1.136	&	0.064	&	+0.229	&	0.008	&	0.17	&	+1.74	&	0.13	\\
768	&	J070832.40+034245.5	&	TYC 167-2919-1     	&	07 08 32.24 	&	 +03 42 45.33	&	B8 III$-$IV Si   	&	128	&	10.0628	&	0.0009	&	+0.977	&	0.068	&	$-$0.224	&	0.003	&	0.27	&	$-$0.25	&	0.16	\\
769	&	J070837.00+164450.5	&	Gaia DR2 3360604552396348800	&	07 08 37.01 	&	 +16 44 50.53	&	B9.5 V Eu   	&	145	&	14.1492	&	0.0008	&	+0.334	&	0.049	&	+0.061	&	0.005	&	0.36	&	+1.41	&	0.32	\\
770	&	J070907.08+441114.7	&	Gaia DR2 953130071357647360	&	07 09 07.09 	&	 +44 11 14.80	&	B9.5 V CrEu   	&	124	&	13.9893	&	0.0007	&	+0.510	&	0.029	&	+0.196	&	0.005	&	0.20	&	+2.32	&	0.13	\\
771	&	J071043.39+095402.9	&	TYC 766-268-1      	&	07 10 43.40 	&	 +09 54 02.97	&	A0 V SrCrEu   	&	166	&	11.2727	&	0.0011	&	+1.297	&	0.043	&	+0.111	&	0.005	&	0.07	&	+1.77	&	0.09	\\
772	&	J071113.33+055418.0	&	TYC 175-4347-1     	&	07 11 13.34 	&	 +05 54 18.04	&	B9 IV Eu   	&	332	&	11.5305	&	0.0008	&	+0.426	&	0.048	&	+0.005	&	0.005	&	0.12	&	$-$0.44	&	0.25	\\
773	&	J071258.59+065952.3	&	Gaia DR2 3153216623376421120	&	07 12 58.60 	&	 +06 59 52.29	&	A2 IV SrCrEu   	&	215	&	12.8113	&	0.0018	&	+0.524	&	0.040	&	+0.357	&	0.016	&	0.06	&	+1.35	&	0.17	\\
774	&	J071337.30+040720.7	&	HD 55585   	&	07 13 37.48 	&	 +04 07 21.46	&	B8 IV CrEuSi   	&	108	&	9.8835	&	0.0024	&	+1.158	&	0.069	&	$-$0.034	&	0.012	&	0.18	&	+0.03	&	0.14	\\
775	&	J071413.88+142449.5	&	Gaia DR2 3167387374048122496	&	07 14 13.89 	&	 +14 24 49.58	&	A3 III$-$IV SrCrEu   	&	111	&	14.0288	&	0.0004	&	+0.336	&	0.032	&	+0.310	&	0.003	&	0.10	&	+1.56	&	0.21	\\
776	&	J071434.25+064339.7	&	Gaia DR2 3153147908195304320	&	07 14 34.25 	&	 +06 43 39.78	&	A0 IV$-$V CrEu   	&	149	&	13.6121	&	0.0005	&	+0.291	&	0.043	&	+0.094	&	0.005	&	0.05	&	+0.88	&	0.32	\\
777	&	J071458.67+125333.8	&	TYC 770-628-1      	&	07 14 58.67 	&	 +12 53 34.09	&	B9 IV$-$V Cr   	&	100	&	11.9075	&	0.0012	&	+0.501	&	0.055	&	+0.107	&	0.005	&	0.22	&	+0.19	&	0.24	\\
778	&	J071535.85+054343.4	&	TYC 176-3306-1     	&	07 15 35.89 	&	 +05 43 42.38	&	B8 IV EuSi   	&	413	&	10.6533	&	0.0005	&	+0.778	&	0.039	&	$-$0.058	&	0.002	&	0.11	&	+0.00	&	0.12	\\
779	&	J071550.38+063655.9	&	TYC 176-2454-1     	&	07 15 50.27 	&	 +06 36 55.95	&	A1 IV SrCrEuSi   	&	390	&	10.5506	&	0.0006	&	+2.092	&	0.053	&	+0.269	&	0.002	&	0.03	&	+2.12	&	0.07	\\
780	&	J071653.67+050509.4	&	TYC 172-1501-1     	&	07 16 53.67 	&	 +05 05 09.40	&	B9 IV bl4077 bl4130  	&	211	&	11.3753	&	0.0003	&	+0.461	&	0.414	&	$-$	&	$-$	&	$-$	&	$-$	&	$-$	\\
781	&	J071706.27+205242.5	&	TYC 1358-1336-1    	&	07 17 06.27 	&	 +20 52 42.50	&	B8 IV Si   	&	638	&	10.4400	&	0.0007	&	+1.058	&	0.077	&	$-$0.068	&	0.003	&	0.16	&	+0.41	&	0.17	\\
782	&	J071750.10+173858.9	&	Gaia DR2 3362024090627348864	&	07 17 50.11 	&	 +17 38 58.94	&	B9.5 V SrCrEu   	&	105	&	14.6387	&	0.0007	&	+0.352	&	0.032	&	+0.176	&	0.005	&	0.22	&	+2.15	&	0.20	\\
783	&	J071752.83+134707.8 $^{a}$	&	TYC 775-1162-1     	&	07 17 52.84 	&	 +13 47 07.82	&	A5 IV$-$V SrCrEu   	&	317	&	9.3263	&	0.0005	&	+1.361	&	0.540	&	$-$	&	$-$	&	$-$	&	$-$	&	$-$	\\
784	&	J071901.76+150939.2	&	Gaia DR2 3167825632511004416	&	07 19 01.76 	&	 +15 09 39.24	&	B8 IV Si   	&	149	&	14.2154	&	0.0011	&	+0.117	&	0.046	&	$-$	&	$-$	&	$-$	&	$-$	&	$-$	\\
785	&	J071954.41+051936.0	&	TYC 172-1153-1     	&	07 19 54.42 	&	 +05 19 36.01	&	B9 V Cr   	&	252	&	12.2588	&	0.0004	&	+0.515	&	0.045	&	+0.018	&	0.005	&	0.20	&	+0.62	&	0.20	\\
786	&	J071958.04+084727.9	&	Gaia DR2 3155230555017333760	&	07 19 58.04 	&	 +08 47 27.99	&	B9 IV$-$V SrCrEu   	&	157	&	13.1321	&	0.0006	&	+0.302	&	0.047	&	+0.029	&	0.005	&	0.06	&	+0.48	&	0.34	\\
787	&	J072000.73-030729.1	&	Gaia DR2 3060095753805333504	&	07 20 00.74 	&	 -03 07 29.12	&	B9 IV$-$V  bl4130  	&	126	&	12.6527	&	0.0004	&	+0.381	&	0.045	&	$-$0.012	&	0.003	&	0.14	&	+0.42	&	0.26	\\
788	&	J072031.89+022923.7	&	TYC 168-38-1       	&	07 20 31.89 	&	 +02 29 23.78	&	B9.5 III$-$IV CrSi   	&	148	&	12.1178	&	0.0009	&	+0.506	&	0.045	&	$-$0.014	&	0.006	&	0.12	&	+0.52	&	0.20	\\
789	&	J072040.31-000827.5	&	Gaia DR2 3110653501771439872	&	07 20 40.31 	&	 -00 08 27.52	&	kA1hA7mA9 CrSi   	&	62	&	13.1563	&	0.0017	&	+0.472	&	0.046	&	+0.087	&	0.007	&	0.15	&	+1.37	&	0.22	\\
790	&	J072052.36-015852.8	&	TYC 4820-842-1     	&	07 20 52.37 	&	 -01 58 52.85	&	B9.5 IV$-$V SrCrEu   	&	152	&	12.6771	&	0.0003	&	+0.366	&	0.031	&	+0.070	&	0.002	&	0.13	&	+0.37	&	0.19	\\
791	&	J072104.20-031507.4	&	Gaia DR2 3059901071523913856	&	07 21 04.21 	&	 -03 15 07.45	&	B9.5 II$-$III bl4077 bl4130  	&	142	&	12.5735	&	0.0054	&	+0.512	&	0.052	&	+0.034	&	0.026	&	0.16	&	+0.96	&	0.23	\\
792	&	J072118.92+223422.7	&	TYC 1909-1687-1    	&	07 21 18.93 	&	 +22 34 22.79	&	B9 V bl4077   	&	544	&	10.3131	&	0.0013	&	+1.522	&	0.061	&	$-$0.098	&	0.005	&	0.15	&	+1.07	&	0.10	\\
793	&	J072134.90-033226.0	&	TYC 4820-3742-1    	&	07 21 34.90 	&	 -03 32 26.00	&	A1 II$-$III Si   	&	107	&	11.8123	&	0.0008	&	+0.492	&	0.038	&	$-$0.081	&	0.005	&	0.12	&	+0.15	&	0.18	\\
794	&	J072243.21+220430.7	&	HD 57590   	&	07 22 43.25 	&	 +22 04 30.52	&	B9 IV$-$V \textit{SrEu(Cr)}*	&	906	&	9.5348	&	0.0008	&	+1.185	&	0.084	&	$-$0.044	&	0.039	&	0.07	&	$-$0.16	&	0.16	\\
795	&	J072318.82+082617.7	&	Gaia DR2 3154962720855355520	&	07 23 18.82 	&	 +08 26 17.78	&	A3 III$-$IV CrEu   	&	114	&	13.3816	&	0.0006	&	+0.305	&	0.031	&	+0.093	&	0.005	&	0.06	&	+0.75	&	0.22	\\
796	&	J072341.49-024106.9	&	TYC 4821-604-1     	&	07 23 41.49 	&	 -02 41 07.02	&	A2 V CrEu   	&	265	&	11.6810	&	0.0010	&	+0.844	&	0.044	&	+0.142	&	0.004	&	0.09	&	+1.23	&	0.12	\\
797	&	J072412.26-050025.2	&	TYC 4825-2009-1    	&	07 24 12.26 	&	 -05 00 25.35	&	B8 II$-$III  bl4130  	&	206	&	11.6433	&	0.0008	&	+0.281	&	0.042	&	$-$0.079	&	0.018	&	0.32	&	$-$1.44	&	0.33	\\
798	&	J072415.37-003247.8	&	TYC 4817-1116-1    	&	07 24 15.38 	&	 -00 32 47.84	&	B8 III  bl4130  	&	138	&	11.4538	&	0.0008	&	+0.368	&	0.055	&	$-$0.059	&	0.004	&	0.08	&	$-$0.80	&	0.33	\\
799	&	J072423.08-003904.7	&	TYC 4817-1061-1    	&	07 24 23.09 	&	 -00 39 04.76	&	B9 IV$-$V bl4077 bl4130  	&	222	&	11.6593	&	0.0007	&	+0.547	&	0.043	&	$-$0.024	&	0.004	&	0.07	&	+0.28	&	0.18	\\
800	&	J072433.03-001416.4	&	TYC 4817-1368-1    	&	07 24 33.05 	&	 -00 14 16.47	&	B9.5 V CrEu   	&	229	&	11.3966	&	0.0007	&	+0.561	&	0.048	&	+0.031	&	0.003	&	0.07	&	+0.07	&	0.19	\\
\hline
\end{tabular}                                                                                                                                                                   
\end{adjustbox}
\end{center}                                                                                                                                             
\end{sidewaystable*}
\setcounter{table}{0}  
\begin{sidewaystable*}
\caption{Essential data for our sample stars, sorted by increasing right ascension. The columns denote: (1) Internal identification number. (2) LAMOST identifier. (3) Alternativ identifier (HD number, TYC identifier or GAIA DR2 number). (4) Right ascension (J2000; GAIA DR2). (5) Declination (J2000; GAIA DR2). (6) Spectral type, as derived in this study. (7) Sloan $g$ band S/N ratio of the analysed spectrum. (8) $G$\,mag (GAIA DR2). (9) $G$\,mag error. (10) Parallax (GAIA DR2). (11) Parallax error. (12) Dereddened colour index $(BP-RP){_0}$ (GAIA DR2). (13) Colour index error. (14) Absorption in the $G$ band, $A_G$. (15) Intrinsic absolute magnitude in the $G$ band, $M_{\mathrm{G,0}}$. (16) Absolute magnitude error.}
\label{table_master17}
\begin{center}
\begin{adjustbox}{max width=\textwidth}
\begin{tabular}{lllcclcccccccccc}
\hline
\hline
(1) & (2) & (3) & (4) & (5) & (6) & (7) & (8) & (9) & (10) & (11) & (12) & (13) & (14) & (15) & (16) \\
No.	&	ID\_LAMOST	&	ID\_alt	&	RA(J2000) 	&	 Dec(J2000)    	&	SpT\_final	&	S/N\,$g$	&	$G$\,mag	&	e\_$G$\,mag	&	pi (mas)	&	e\_pi	&	$(BP-RP){_0}$	&	e\_$(BP-RP){_0}$	&	$A_G$	&	$M_{\mathrm{G,0}}$	&	e\_$M_{\mathrm{G,0}}$	\\
\hline 
801	&	J072535.29+072629.2	&	TYC 177-635-1      	&	07 25 35.30 	&	 +07 26 29.23	&	A0 III$-$IV Cr   	&	370	&	11.5436	&	0.0003	&	+0.095	&	0.198	&	$-$	&	$-$	&	$-$	&	$-$	&	$-$	\\
802	&	J072551.97-041719.5	&	TYC 4825-1586-1    	&	07 25 51.98 	&	 -04 17 19.38	&	B9.5 IV CrSi   	&	179	&	12.1882	&	0.0005	&	+0.386	&	0.044	&	$-$0.021	&	0.004	&	0.25	&	$-$0.13	&	0.25	\\
803	&	J072614.34-004443.5	&	TYC 4817-111-1     	&	07 26 14.34 	&	 -00 44 43.58	&	B5 IV  bl4130  	&	209	&	11.4700	&	0.0011	&	+0.567	&	0.051	&	$-$0.131	&	0.005	&	0.08	&	+0.16	&	0.20	\\
804	&	J072659.14+121919.4	&	TYC 772-1162-1     	&	07 26 59.14 	&	 +12 19 19.44	&	A4 IV$-$V SrCrEu   	&	132	&	11.3252	&	0.0006	&	+0.888	&	0.048	&	+0.171	&	0.003	&	0.08	&	+0.99	&	0.13	\\
805	&	J072725.86-012818.4	&	TYC 4817-1586-1    	&	07 27 25.87 	&	 -01 28 18.42	&	B5 IV Si   	&	154	&	12.1513	&	0.0011	&	+0.146	&	0.041	&	$-$	&	$-$	&	$-$	&	$-$	&	$-$	\\
806	&	J072915.84+003208.5	&	TYC 165-835-1      	&	07 29 15.85 	&	 +00 32 08.58	&	B9 V Cr   	&	312	&	10.4212	&	0.0012	&	+1.207	&	0.054	&	+0.024	&	0.005	&	0.10	&	+0.73	&	0.11	\\
807	&	J072940.44-045552.0	&	TYC 4825-1371-1    	&	07 29 40.29 	&	 -04 55 52.06	&	B9 V Cr   	&	342	&	10.1552	&	0.0005	&	+1.063	&	0.054	&	$-$0.050	&	0.001	&	0.11	&	+0.18	&	0.12	\\
808	&	J072943.39-042433.4	&	TYC 4825-1200-1    	&	07 29 43.46 	&	 -04 24 33.38	&	A0 V SrCr   	&	283	&	10.8594	&	0.0006	&	+0.688	&	0.049	&	+0.191	&	0.002	&	0.05	&	0.00	&	0.16	\\
809	&	J073054.73+101640.1	&	TYC 768-1345-1     	&	07 30 54.73 	&	 +10 16 40.12	&	B9 IV$-$V bl4077 bl4130  	&	219	&	10.9476	&	0.0015	&	+0.859	&	0.060	&	$-$0.064	&	0.004	&	0.05	&	+0.57	&	0.16	\\
810	&	J073102.16+070734.5	&	TYC 190-233-1      	&	07 31 02.17 	&	 +07 07 34.55	&	kA3hA5mA7 SrCrEu   	&	104	&	10.8112	&	0.0004	&	+1.393	&	0.261	&	+0.198	&	0.002	&	0.03	&	+1.50	&	0.41	\\
811	&	J073220.17-024048.1	&	TYC 4834-617-1     	&	07 32 20.18 	&	 -02 40 48.14	&	kA2hA4mA7 CrEuSi   	&	262	&	11.0241	&	0.0010	&	+1.052	&	0.065	&	+0.030	&	0.005	&	0.06	&	+1.08	&	0.14	\\
812	&	J073254.94+111314.5	&	HD 59999   	&	07 32 54.95 	&	 +11 13 14.55	&	B9 V SrCr   	&	289	&	9.5405	&	0.0006	&	+1.814	&	0.057	&	+0.018	&	0.003	&	0.04	&	+0.80	&	0.08	\\
813	&	J073351.72+242221.8	&	Gaia DR2 868363467228923904	&	07 33 51.73 	&	 +24 22 21.93	&	B8 IV$-$V  bl4130 (He-wk) 	&	106	&	11.5540	&	0.0006	&	+0.838	&	0.048	&	$-$0.094	&	0.007	&	0.11	&	+1.07	&	0.13	\\
814	&	J073548.83+123225.1	&	TYC 773-931-1      	&	07 35 48.83 	&	 +12 32 25.19	&	kA1hA9mA8 SrEu   	&	197	&	11.2239	&	0.0016	&	+1.309	&	0.116	&	+0.357	&	0.003	&	0.04	&	+1.77	&	0.20	\\
815	&	J073949.90-030558.0	&	TYC 4835-1400-1    	&	07 39 49.96 	&	 -03 05 59.49	&	B7 \textit{V Si (He-wk)}*	&	214	&	10.4654	&	0.0006	&	+0.710	&	0.053	&	$-$0.102	&	0.002	&	0.14	&	$-$0.41	&	0.17	\\
816	&	J073950.01+201812.7	&	Gaia DR2 672828861965623040	&	07 39 50.01 	&	 +20 18 12.75	&	B9 IV$-$V SrCrEu   	&	80	&	15.4476	&	0.0010	&	+0.196	&	0.048	&	$-$0.067	&	0.006	&	0.06	&	+1.84	&	0.53	\\
817	&	J073953.64-002625.6	&	TYC 4831-2071-1    	&	07 39 53.65 	&	 -00 26 25.63	&	A8 V SrCrEu   	&	124	&	11.9413	&	0.0003	&	+0.624	&	0.041	&	+0.297	&	0.002	&	0.10	&	+0.82	&	0.15	\\
818	&	J073953.72+080221.4	&	TYC 778-1580-1     	&	07 39 53.59 	&	 +08 02 21.47	&	B8 III$-$IV  bl4130  	&	433	&	10.3616	&	0.0008	&	+0.867	&	0.052	&	$-$0.116	&	0.003	&	0.07	&	$-$0.01	&	0.14	\\
819	&	J074104.88-003828.3	&	TYC 4831-885-1     	&	07 41 04.77 	&	 -00 38 28.30	&	B9 V CrEu  (He-wk) 	&	283	&	10.5274	&	0.0006	&	+1.013	&	0.047	&	+0.079	&	0.002	&	0.05	&	+0.51	&	0.11	\\
820	&	J074354.81+022855.1	&	TYC 183-266-1      	&	07 43 54.82 	&	 +02 28 55.05	&	A0 V Cr   	&	109	&	11.3431	&	0.0008	&	+0.927	&	0.064	&	+0.137	&	0.002	&	0.05	&	+1.13	&	0.16	\\
821	&	J074414.88+051124.7	&	TYC 187-2124-1     	&	07 44 14.89 	&	 +05 11 24.73	&	B8 III$-$IV Si   	&	294	&	11.4865	&	0.0008	&	+0.614	&	0.062	&	$-$0.148	&	0.003	&	0.06	&	+0.37	&	0.23	\\
822	&	J074417.67-043551.0	&	TYC 4839-894-1     	&	07 44 17.67 	&	 -04 35 51.05	&	B9 V CrEu   	&	477	&	10.1625	&	0.0008	&	+1.007	&	0.053	&	$-$0.094	&	0.003	&	0.17	&	+0.01	&	0.12	\\
823	&	J074419.88+523638.9	&	HD 233446  	&	07 44 19.88 	&	 +52 36 38.86	&	A0 V SrCrEu   	&	497	&	10.1707	&	0.0008	&	+1.824	&	0.075	&	+0.083	&	0.003	&	0.11	&	+1.36	&	0.10	\\
824	&	J074738.21+052329.6	&	TYC 188-171-1      	&	07 47 38.22 	&	 +05 23 29.70	&	B9 V Cr   	&	319	&	11.5441	&	0.0008	&	+0.437	&	0.050	&	$-$0.033	&	0.003	&	0.06	&	$-$0.31	&	0.25	\\
825	&	J074744.39+052243.8	&	TYC 188-247-1      	&	07 47 44.39 	&	 +05 22 43.84	&	B9.5 V CrEuSi   	&	150	&	10.4210	&	0.0010	&	+1.661	&	0.048	&	+0.144	&	0.006	&	0.04	&	+1.48	&	0.08	\\
826	&	J074830.59+002713.2	&	TYC 180-954-1      	&	07 48 30.48 	&	 +00 27 13.23	&	B9.5 V SrCrEu   	&	224	&	10.4835	&	0.0008	&	+1.344	&	0.068	&	+0.038	&	0.003	&	0.06	&	+1.07	&	0.12	\\
827	&	J074851.40+001619.1	&	TYC 180-1290-1     	&	07 48 51.41 	&	 +00 16 19.07	&	kB8hA3mA3 CrEu   	&	229	&	9.9138	&	0.0005	&	+1.394	&	0.053	&	+0.027	&	0.001	&	0.13	&	+0.51	&	0.10	\\
828	&	J074919.49+051551.8	&	TYC 188-700-1      	&	07 49 19.50 	&	 +05 15 51.83	&	kA1hA9mA9 SrCrEu   	&	368	&	11.1328	&	0.0007	&	+1.260	&	0.052	&	+0.275	&	0.002	&	0.04	&	+1.60	&	0.10	\\
829	&	J074959.61+013517.8	&	TYC 180-2568-1     	&	07 49 59.62 	&	 +01 35 17.83	&	kA1hA3mA7 SrCrEuSi   	&	261	&	9.9910	&	0.0007	&	+0.884	&	0.062	&	+0.049	&	0.001	&	0.07	&	$-$0.35	&	0.16	\\
830	&	J075041.80-060338.3	&	HD 63843   	&	07 50 41.81 	&	 -06 03 38.29	&	A2 IV SrCrEu   	&	130	&	10.2188	&	0.0004	&	+0.814	&	0.045	&	+0.076	&	0.002	&	0.34	&	$-$0.57	&	0.13	\\
831	&	J075220.93+113710.6	&	Gaia DR2 3150803440165601664	&	07 52 20.93 	&	 +11 37 10.63	&	kA5hA6mA9 SrCrEu   	&	104	&	13.2627	&	0.0003	&	+0.313	&	0.032	&	+0.207	&	0.002	&	0.06	&	+0.68	&	0.23	\\
832	&	J075429.74-074804.3	&	Gaia DR2 3042652727681575936	&	07 54 29.74 	&	 -07 48 04.50	&	B9 III$-$IV Si   	&	109	&	12.7518	&	0.0006	&	$-$2.860	&	0.543	&	$-$	&	$-$	&	$-$	&	$-$	&	$-$	\\
833	&	J075516.94+101951.2	&	TYC 784-149-1      	&	07 55 16.94 	&	 +10 19 51.28	&	A7 V SrEuSi   	&	229	&	12.0348	&	0.0003	&	+0.436	&	0.038	&	+0.315	&	0.002	&	0.05	&	+0.18	&	0.20	\\
834	&	J075656.82-054907.8	&	TYC 4845-1387-1    	&	07 56 56.83 	&	 -05 49 07.89	&	B8 IV$-$V  bl4130 (He-wk) 	&	371	&	11.2779	&	0.0017	&	+0.990	&	0.067	&	$-$0.116	&	0.004	&	0.11	&	+1.15	&	0.16	\\
835	&	J075951.17+021610.9	&	HD 65644   	&	07 59 51.17 	&	 +02 16 10.84	&	B8 IV CrEu   	&	334	&	9.9878	&	0.0009	&	+1.328	&	0.064	&	$-$0.088	&	0.003	&	0.05	&	+0.56	&	0.12	\\
836	&	J080112.84+305547.8	&	TYC 2468-1474-1    	&	08 01 12.85 	&	 +30 55 47.81	&	A3 IV SrCrEu   	&	254	&	11.4569	&	0.0029	&	+1.547	&	0.901	&	$-$	&	$-$	&	$-$	&	$-$	&	$-$	\\
837	&	J080210.08+334820.2	&	TYC 2476-1457-1    	&	08 02 10.09 	&	 +33 48 20.25	&	kA1hA6mA6 Eu   	&	141	&	13.0555	&	0.0011	&	+0.682	&	0.049	&	+0.049	&	0.006	&	0.17	&	+2.05	&	0.16	\\
838	&	J080339.87-082141.0	&	TYC 5413-1871-1    	&	08 03 39.88 	&	 -08 21 41.01	&	kB9hA3mA8 SrCrEu   	&	251	&	9.4489	&	0.0006	&	+1.205	&	0.055	&	+0.039	&	0.002	&	0.20	&	$-$0.34	&	0.11	\\
839	&	J080435.19-031900.7	&	TYC 4850-398-1     	&	08 04 35.19 	&	 -03 19 00.74	&	A0 IV$-$V SrCr   	&	207	&	10.0101	&	0.0008	&	+1.101	&	0.087	&	+0.104	&	0.002	&	0.05	&	+0.17	&	0.18	\\
840	&	J080734.82-044027.2	&	Gaia DR2 3067862055404261120	&	08 07 34.83 	&	 -04 40 27.22	&	kB9.5hA2mA4 SrCrEu   	&	116	&	13.0785	&	0.0009	&	+0.437	&	0.040	&	+0.092	&	0.004	&	0.08	&	+1.20	&	0.21	\\
841	&	J081024.89+382402.2	&	TYC 2973-199-1     	&	08 10 24.68 	&	 +38 24 02.31	&	B8 IV Si   	&	426	&	9.7665	&	0.0010	&	+1.034	&	0.077	&	$-$0.103	&	0.004	&	0.08	&	$-$0.24	&	0.17	\\
842	&	J081025.65-062756.5	&	TYC 4859-583-1     	&	08 10 25.65 	&	 -06 27 56.38	&	B9 IV$-$V Cr   	&	175	&	11.3055	&	0.0007	&	+0.526	&	0.052	&	$-$0.002	&	0.003	&	0.10	&	$-$0.19	&	0.22	\\
843	&	J081148.45-075357.2	&	TYC 5426-304-1     	&	08 11 48.45 	&	 -07 53 57.21	&	B8 IV Si  (He-wk) 	&	168	&	11.1818	&	0.0013	&	+0.751	&	0.067	&	$-$0.006	&	0.003	&	0.42	&	+0.14	&	0.20	\\
844	&	J081342.91-053547.1	&	TYC 4855-1556-1    	&	08 13 42.91 	&	 -05 35 47.23	&	kB9.5hA3mA3 SrCrEu   	&	217	&	10.6418	&	0.0009	&	+1.488	&	0.106	&	+0.062	&	0.003	&	0.08	&	+1.43	&	0.16	\\
845	&	J081445.01+172256.9	&	TYC 1381-753-1     	&	08 14 44.86 	&	 +17 22 57.00	&	kB9hA5mA5 CrEu   	&	402	&	10.2896	&	0.0011	&	+0.662	&	0.070	&	$-$0.008	&	0.004	&	0.14	&	$-$0.75	&	0.24	\\
846	&	J081930.22+001232.3	&	TYC 196-522-1      	&	08 19 30.23 	&	 +00 12 32.29	&	B6 III$-$IV Si   	&	132	&	11.4510	&	0.0019	&	+0.606	&	0.093	&	$-$0.219	&	0.008	&	0.06	&	+0.30	&	0.34	\\
847	&	J082002.55+000809.3	&	Gaia DR2 3077512228440068352	&	08 20 02.50 	&	 +00 08 09.40	&	kA2hA4mA7 CrEu   	&	225	&	11.5727	&	0.0006	&	+0.596	&	0.062	&	+0.157	&	0.003	&	0.08	&	+0.37	&	0.23	\\
848	&	J082137.47+064401.0	&	Gaia DR2 3096854940075299712	&	08 21 37.48 	&	 +06 44 01.06	&	A1 IV SrCrEu   	&	118	&	13.3369	&	0.0003	&	+0.613	&	0.027	&	+0.262	&	0.002	&	0.06	&	+2.21	&	0.11	\\
849	&	J082235.26+161149.3	&	TYC 1378-930-1     	&	08 22 35.27 	&	 +16 11 49.30	&	B8 IV Si  (He-wk) 	&	227	&	11.8453	&	0.0009	&	+0.443	&	0.088	&	$-$0.204	&	0.005	&	0.08	&	0.00	&	0.43	\\
850	&	J082326.74+072116.4	&	TYC 209-1570-1     	&	08 23 26.70 	&	 +07 21 16.02	&	kA3hA7mF1 SrCrEu   	&	411	&	11.1021	&	0.0008	&	+1.228	&	0.067	&	+0.203	&	0.002	&	0.08	&	+1.47	&	0.13	\\
\hline
\end{tabular}                                                                                                                                                                   
\end{adjustbox}
\end{center}                                                                                                                                             
\end{sidewaystable*}
\setcounter{table}{0}  
\begin{sidewaystable*}
\caption{Essential data for our sample stars, sorted by increasing right ascension. The columns denote: (1) Internal identification number. (2) LAMOST identifier. (3) Alternativ identifier (HD number, TYC identifier or GAIA DR2 number). (4) Right ascension (J2000; GAIA DR2). (5) Declination (J2000; GAIA DR2). (6) Spectral type, as derived in this study. (7) Sloan $g$ band S/N ratio of the analysed spectrum. (8) $G$\,mag (GAIA DR2). (9) $G$\,mag error. (10) Parallax (GAIA DR2). (11) Parallax error. (12) Dereddened colour index $(BP-RP){_0}$ (GAIA DR2). (13) Colour index error. (14) Absorption in the $G$ band, $A_G$. (15) Intrinsic absolute magnitude in the $G$ band, $M_{\mathrm{G,0}}$. (16) Absolute magnitude error.}
\label{table_master18}
\begin{center}
\begin{adjustbox}{max width=\textwidth}
\begin{tabular}{lllcclcccccccccc}
\hline
\hline
(1) & (2) & (3) & (4) & (5) & (6) & (7) & (8) & (9) & (10) & (11) & (12) & (13) & (14) & (15) & (16) \\
No.	&	ID\_LAMOST	&	ID\_alt	&	RA(J2000) 	&	 Dec(J2000)    	&	SpT\_final	&	S/N\,$g$	&	$G$\,mag	&	e\_$G$\,mag	&	pi (mas)	&	e\_pi	&	$(BP-RP){_0}$	&	e\_$(BP-RP){_0}$	&	$A_G$	&	$M_{\mathrm{G,0}}$	&	e\_$M_{\mathrm{G,0}}$	\\
\hline 
851	&	J082331.87+491533.9	&	TYC 3418-1363-1    	&	08 23 31.88 	&	 +49 15 33.89	&	B9 V SrCr   	&	245	&	10.1705	&	0.0005	&	+1.211	&	0.045	&	+0.041	&	0.001	&	0.11	&	+0.47	&	0.09	\\
852	&	J082706.99+453552.9	&	HD 71047   	&	08 27 06.71 	&	 +45 35 52.93	&	A5 III$-$IV Sr   	&	362	&	9.6733	&	0.0007	&	+2.785	&	0.095	&	+0.143	&	0.003	&	0.06	&	+1.84	&	0.09	\\
853	&	J083539.12+002150.4	&	TYC 210-1694-1     	&	08 35 39.15 	&	 +00 21 50.63	&	kA2hA3mA6 SrCrEu   	&	220	&	11.6695	&	0.0006	&	+0.605	&	0.044	&	+0.150	&	0.002	&	0.14	&	+0.43	&	0.17	\\
854	&	J084232.90+211043.9	&	TYC 1398-1797-1    	&	08 42 32.90 	&	 +21 10 43.90	&	A3 IV SrCrEu   	&	246	&	12.1062	&	0.0004	&	+0.611	&	0.045	&	+0.235	&	0.002	&	0.11	&	+0.93	&	0.17	\\
855	&	J084546.89+054048.1	&	HD 74722   	&	08 45 46.71 	&	 +05 40 48.09	&	A8 V SrEu   	&	192	&	9.4740	&	0.0004	&	+1.823	&	0.058	&	+0.150	&	0.002	&	0.06	&	+0.72	&	0.09	\\
856	&	J084613.37+191508.9	&	HD 74719   	&	08 46 13.20 	&	 +19 15 08.96	&	B9.5 V SrCr   	&	431	&	9.7889	&	0.0007	&	+1.380	&	0.065	&	+0.045	&	0.003	&	0.06	&	+0.43	&	0.11	\\
857	&	J085148.85+434402.2	&	Gaia DR2 913924411585612928	&	08 51 48.86 	&	 +43 44 02.29	&	B8 IV$-$V $^{b}$	&	155	&	12.3875	&	0.0003	&	+0.564	&	0.050	&	$-$0.171	&	0.003	&	0.08	&	+1.06	&	0.20	\\
858	&	J091022.04+281832.0	&	Gaia DR2 698186451960307712	&	09 10 21.97 	&	 +28 18 32.27	&	B9 V CrEu   	&	158	&	13.0315	&	0.0008	&	+0.273	&	0.088	&	$-$	&	$-$	&	$-$	&	$-$	&	$-$	\\
859	&	J091053.70+285032.7	&	Gaia DR2 698486305101018240	&	09 10 53.71 	&	 +28 50 32.81	&	A9 V SrCrEu   	&	139	&	12.5636	&	0.0005	&	+0.375	&	0.069	&	+0.262	&	0.003	&	0.06	&	+0.37	&	0.40	\\
860	&	J091451.94+082606.5	&	TYC 819-736-1      	&	09 14 51.95 	&	 +08 26 06.73	&	kA2hA7mA6 SrEu   	&	272	&	9.8052	&	0.0007	&	+2.704	&	0.062	&	+0.282	&	0.008	&	0.09	&	+1.87	&	0.07	\\
861	&	J092233.23+072519.4	&	TYC 234-81-1       	&	09 22 33.01 	&	 +07 25 19.39	&	kA3hA6mA9 SrCrEu   	&	422	&	9.2254	&	0.0006	&	+2.718	&	0.044	&	+0.412	&	0.009	&	0.06	&	+1.33	&	0.06	\\
862	&	J092615.37+280323.0	&	HD 81416   	&	09 26 15.20 	&	 +28 03 23.00	&	B8 IV SrCrEu   	&	498	&	10.1268	&	0.0008	&	+1.106	&	0.075	&	$-$0.073	&	0.005	&	0.08	&	+0.26	&	0.16	\\
863	&	J093551.74+093917.8	&	TYC 821-367-1      	&	09 35 51.64 	&	 +09 39 17.54	&	B9.5 IV$-$V CrEuSi   	&	121	&	11.5541	&	0.0010	&	+1.095	&	0.059	&	+0.040	&	0.003	&	0.08	&	+1.67	&	0.13	\\
864	&	J093915.30-041535.0	&	Gaia DR2 3824575537075330816	&	09 39 15.29 	&	 -04 15 35.11	&	A4 IV$-$V SrCr   	&	138	&	13.3655	&	0.0005	&	+0.429	&	0.062	&	+0.193	&	0.004	&	0.09	&	+1.44	&	0.32	\\
865	&	J094438.61+312641.5	&	TYC 2501-808-1     	&	09 44 38.63 	&	 +31 26 41.56	&	B8 IV Si   	&	438	&	11.2249	&	0.0014	&	+0.386	&	0.108	&	$-$	&	$-$	&	$-$	&	$-$	&	$-$	\\
866	&	J094532.52-062723.0	&	TYC 4901-537-1     	&	09 45 32.52 	&	 -06 27 23.04	&	kB8hA4mA4 Cr   	&	137	&	12.7837	&	0.0011	&	+0.667	&	0.073	&	+0.036	&	0.006	&	0.10	&	+1.81	&	0.24	\\
867	&	J095644.95-021719.5	&	HD 86170   	&	09 56 45.18 	&	 -02 17 20.43	&	kA2hA3mA6 SrCrEu   	&	762	&	8.4143	&	0.0008	&	+3.811	&	0.106	&	+0.128	&	0.006	&	0.09	&	+1.23	&	0.08	\\
868	&	J095855.77-044413.8	&	Gaia DR2 3822254609763252864	&	09 58 55.77 	&	 -04 44 13.74	&	B9.5 III$-$IV Eu   	&	105	&	13.2069	&	0.0011	&	+0.167	&	0.052	&	$-$	&	$-$	&	$-$	&	$-$	&	$-$	\\
869	&	J102027.58+280919.4	&	Gaia DR2 740907873177225088	&	10 20 27.63 	&	 +28 09 19.32	&	B9 V CrEu   	&	127	&	13.7089	&	0.0007	&	+0.173	&	0.049	&	$-$	&	$-$	&	$-$	&	$-$	&	$-$	\\
870	&	J104323.95+045214.3	&	Gaia DR2 3858264955602552320	&	10 43 23.95 	&	 +04 52 14.38	&	A1 IV$-$V CrEu   	&	169	&	12.7714	&	0.0005	&	+0.585	&	0.057	&	+0.133	&	0.003	&	0.05	&	+1.55	&	0.22	\\
871	&	J105734.90+260846.0	&	TYC 1978-545-1     	&	10 57 34.89 	&	 +26 08 45.97	&	A0 V CrEu   	&	282	&	11.9675	&	0.0007	&	+0.838	&	0.064	&	$-$0.031	&	0.005	&	0.10	&	+1.49	&	0.17	\\
872	&	J113813.46+441229.5	&	TYC 3015-595-1     	&	11 38 13.46 	&	 +44 12 29.53	&	A7 V Sr   	&	460	&	10.9897	&	0.0006	&	+1.419	&	0.065	&	+0.204	&	0.002	&	0.05	&	+1.70	&	0.11	\\
873	&	J114130.23+403822.7	&	TYC 3014-2468-1    	&	11 41 30.23 	&	 +40 38 22.80	&	B9 IV$-$V CrEuSi   	&	201	&	11.6523	&	0.0011	&	+0.620	&	0.105	&	$-$0.150	&	0.004	&	0.07	&	+0.55	&	0.37	\\
874	&	J120632.73+521507.9	&	TYC 3457-908-1     	&	12 06 32.66 	&	 +52 15 07.91	&	B9 V Cr  (He-wk) 	&	472	&	10.8068	&	0.0009	&	+0.921	&	0.086	&	$-$0.064	&	0.003	&	0.09	&	+0.54	&	0.21	\\
875	&	J122139.23+383309.5	&	Gaia DR2 1532535594973538688	&	12 21 39.23 	&	 +38 33 09.56	&	kA2hA4mA7 bl4077 bl4130  	&	83	&	13.1753	&	0.0004	&	+0.527	&	0.038	&	+0.299	&	0.002	&	0.04	&	+1.75	&	0.16	\\
876	&	J122746.05+113635.3	&	Gaia DR2 3907547639444408064	&	12 27 46.01 	&	 +11 36 35.95	&	B8 IV Si  (He-wk) 	&	124	&	12.6834	&	0.0008	&	+0.359	&	0.061	&	$-$0.226	&	0.005	&	0.08	&	+0.38	&	0.37	\\
877	&	J122855.36+255446.3	&	HD 108662  	&	12 28 54.70 	&	 +25 54 46.27	&	B9 V CrEu   	&	741	&	5.2066	&	0.0033	&	+13.538	&	0.225	&	$-$0.013	&	0.01	&	0.02	&	+0.84	&	0.06	\\
878	&	J133835.32+100716.1	&	TYC 896-860-1      	&	13 38 35.51 	&	 +10 07 16.57	&	kA3hA7mA9 SrCrEu   	&	473	&	9.8044	&	0.0013	&	+2.107	&	0.102	&	+0.320	&	0.004	&	0.09	&	+1.34	&	0.12	\\
879	&	J140422.54+044357.9	&	TYC 319-461-1      	&	14 04 22.55 	&	 +04 43 57.79	&	kA4hA7mF0 SrCrEu   	&	240	&	12.0595	&	0.0004	&	+0.619	&	0.056	&	+0.172	&	0.003	&	0.06	&	+0.95	&	0.20	\\
880	&	J150331.87+093125.4	&	Gaia DR2 1167894108493926016	&	15 03 31.88 	&	 +09 31 25.44	&	A8 V SrCrEu   	&	195	&	12.6710	&	0.0004	&	+0.435	&	0.046	&	+0.288	&	0.002	&	0.08	&	+0.78	&	0.23	\\
881	&	J155549.85+401144.4	&	Gaia DR2 1382933122321062912	&	15 55 49.85 	&	 +40 11 44.44	&	B8 IV Si   	&	113	&	14.0505	&	0.0008	&	+0.126	&	0.028	&	$-$0.213	&	0.005	&	0.03	&	$-$0.49	&	0.49	\\
882	&	J163458.89+004659.8	&	TYC 382-541-1      	&	16 34 58.89 	&	 +00 46 59.82	&	A0 V CrEu   	&	228	&	11.5347	&	0.0008	&	+1.247	&	0.070	&	+0.083	&	0.002	&	0.19	&	+1.83	&	0.13	\\
883	&	J165909.90+262236.7	&	TYC 2067-291-1     	&	16 59 09.91 	&	 +26 22 36.74	&	A6 V CrEu   	&	234	&	11.2186	&	0.0007	&	+1.143	&	0.035	&	+0.144	&	0.003	&	0.09	&	+1.42	&	0.08	\\
884	&	J170210.51+194917.2	&	TYC 1530-836-1     	&	17 02 10.51 	&	 +19 49 17.27	&	B9.5 IV$-$V SrCr   	&	237	&	12.0212	&	0.0005	&	+0.419	&	0.033	&	+0.055	&	0.003	&	0.17	&	$-$0.04	&	0.18	\\
885	&	J172937.90+414015.2	&	TYC 3094-784-1     	&	17 29 37.91 	&	 +41 40 15.21	&	B9 IV$-$V Cr   	&	71	&	12.4394	&	0.0004	&	+0.841	&	0.023	&	+0.099	&	0.002	&	0.12	&	+1.94	&	0.08	\\
886	&	J173152.09+393831.0	&	TYC 3091-119-1     	&	17 31 52.09 	&	 +39 38 31.10	&	B9 V Cr   	&	128	&	11.6335	&	0.0005	&	+0.656	&	0.028	&	$-$0.020	&	0.003	&	0.11	&	+0.60	&	0.10	\\
887	&	J173536.70+033738.9	&	TYC 418-2336-1     	&	17 35 36.71 	&	 +03 37 39.22	&	A2 IV$-$V Sr   	&	219	&	10.8839	&	0.0005	&	+1.681	&	0.049	&	+0.066	&	0.002	&	0.35	&	+1.67	&	0.08	\\
888	&	J173844.78+243856.2	&	TYC 2080-1036-1    	&	17 38 44.79 	&	 +24 38 56.21	&	B8 III Si   	&	118	&	11.8401	&	0.0009	&	+0.431	&	0.034	&	$-$0.122	&	0.005	&	0.12	&	$-$0.10	&	0.18	\\
889	&	J174031.07+102400.8	&	TYC 997-994-1      	&	17 40 31.07 	&	 +10 24 00.86	&	A6 V SrCrEu   	&	224	&	11.6382	&	0.0007	&	+1.254	&	0.097	&	+0.215	&	0.003	&	0.36	&	+1.77	&	0.18	\\
890	&	J175134.06+263903.1	&	TYC 2085-2086-1    	&	17 51 34.06 	&	 +26 39 03.11	&	A0 V Cr   	&	125	&	12.0794	&	0.0004	&	+0.541	&	0.025	&	+0.076	&	0.002	&	0.09	&	+0.66	&	0.11	\\
891	&	J175822.90+065649.5	&	Gaia DR2 4475629258450293120	&	17 58 22.91 	&	 +06 56 49.53	&	kA4hA7mF0 SrCrEu   	&	127	&	12.9984	&	0.0005	&	+0.564	&	0.031	&	+0.173	&	0.002	&	0.47	&	+1.29	&	0.13	\\
892	&	J175919.34+053509.5	&	Gaia DR2 4474449821775956608	&	17 59 19.33 	&	 +05 35 09.52	&	A1 IV$-$V CrEu   	&	256	&	12.1289	&	0.0009	&	+0.819	&	0.044	&	0.000	&	0.015	&	0.47	&	+1.23	&	0.13	\\
893	&	J180512.65+125948.8	&	TYC 1016-588-1     	&	18 05 12.61 	&	 +12 59 51.41	&	A0 IV SrCr   	&	392	&	10.0765	&	0.0008	&	+1.261	&	0.046	&	+0.095	&	0.003	&	0.36	&	+0.23	&	0.09	\\
894	&	J180615.12+023732.9 $^{a}$	&	TYC 434-1758-1     	&	18 06 15.07 	&	 +02 37 30.78	&	kB9hA3mA7 CrEuSi   	&	578	&	10.1759	&	0.0012	&	+1.803	&	0.070	&	$-$0.003	&	0.004	&	0.18	&	+1.27	&	0.10	\\
895	&	J181156.38+523411.4	&	Gaia DR2 2125305094015307520	&	18 11 56.40 	&	 +52 34 11.48	&	B7 IV$-$V Si   	&	119	&	13.4058	&	0.0012	&	+0.170	&	0.035	&	$-$0.237	&	0.009	&	0.08	&	$-$0.52	&	0.45	\\
896	&	J181409.49+053445.1	&	TYC 439-340-1      	&	18 14 09.49 	&	 +05 34 45.13	&	A0 IV$-$V SrCr   	&	155	&	11.6393	&	0.0008	&	+0.577	&	0.042	&	+0.123	&	0.005	&	0.43	&	+0.02	&	0.17	\\
897	&	J184021.60+273858.7	&	TYC 2115-1062-1    	&	18 40 21.61 	&	 +27 38 58.63	&	B9.5 V Cr   	&	115	&	12.4871	&	0.0003	&	+0.341	&	0.031	&	+0.041	&	0.002	&	0.30	&	$-$0.15	&	0.20	\\
898	&	J184217.44+283421.1	&	TYC 2120-220-1     	&	18 42 17.44 	&	 +28 34 20.95	&	A1 IV CrEu   	&	122	&	12.2758	&	0.0005	&	+0.839	&	0.028	&	+0.179	&	0.002	&	0.36	&	+1.53	&	0.09	\\
899	&	J185127.99+262012.1	&	TYC 2117-2336-1    	&	18 51 28.00 	&	 +26 20 12.13	&	A0 IV$-$V Cr   	&	175	&	12.5515	&	0.0004	&	+0.543	&	0.026	&	+0.100	&	0.002	&	0.46	&	+0.76	&	0.12	\\
900	&	J185444.63+482024.8	&	Gaia DR2 2131718510982169728	&	18 54 44.62 	&	 +48 20 24.80	&	A4 IV$-$V CrEu   	&	107	&	12.7484	&	0.0004	&	+0.593	&	0.034	&	+0.081	&	0.003	&	0.15	&	+1.47	&	0.13	\\
\hline
\end{tabular}                                                                                                                                                                   
\end{adjustbox}
\end{center}                                                                                                                                             
\end{sidewaystable*}
\setcounter{table}{0}  
\begin{sidewaystable*}
\caption{Essential data for our sample stars, sorted by increasing right ascension. The columns denote: (1) Internal identification number. (2) LAMOST identifier. (3) Alternativ identifier (HD number, TYC identifier or GAIA DR2 number). (4) Right ascension (J2000; GAIA DR2). (5) Declination (J2000; GAIA DR2). (6) Spectral type, as derived in this study. (7) Sloan $g$ band S/N ratio of the analysed spectrum. (8) $G$\,mag (GAIA DR2). (9) $G$\,mag error. (10) Parallax (GAIA DR2). (11) Parallax error. (12) Dereddened colour index $(BP-RP){_0}$ (GAIA DR2). (13) Colour index error. (14) Absorption in the $G$ band, $A_G$. (15) Intrinsic absolute magnitude in the $G$ band, $M_{\mathrm{G,0}}$. (16) Absolute magnitude error.}
\label{table_master19}
\begin{center}
\begin{adjustbox}{max width=\textwidth}
\begin{tabular}{lllcclcccccccccc}
\hline
\hline
(1) & (2) & (3) & (4) & (5) & (6) & (7) & (8) & (9) & (10) & (11) & (12) & (13) & (14) & (15) & (16) \\
No.	&	ID\_LAMOST	&	ID\_alt	&	RA(J2000) 	&	 Dec(J2000)    	&	SpT\_final	&	S/N\,$g$	&	$G$\,mag	&	e\_$G$\,mag	&	pi (mas)	&	e\_pi	&	$(BP-RP){_0}$	&	e\_$(BP-RP){_0}$	&	$A_G$	&	$M_{\mathrm{G,0}}$	&	e\_$M_{\mathrm{G,0}}$	\\
\hline 
901	&	J185821.27+434926.5	&	TYC 3131-1074-1    	&	18 58 21.29 	&	 +43 49 26.42	&	A7 V SrCrEu   	&	244	&	11.4667	&	0.0006	&	+1.016	&	0.029	&	+0.215	&	0.003	&	0.09	&	+1.41	&	0.08	\\
902	&	J190132.50+415158.9 $^{a}$	&	HD 177128  	&	19 01 32.49 	&	 +41 51 59.33	&	kA1hA4mA6 SrCrEu   	&	629	&	9.1646	&	0.0004	&	+3.155	&	0.043	&	+0.114	&	0.003	&	0.06	&	+1.60	&	0.06	\\
903	&	J191305.80+500013.3	&	TYC 3550-977-1     	&	19 13 05.80 	&	 +50 00 13.39	&	A4 IV SrCrEu   	&	204	&	11.8878	&	0.0005	&	+0.965	&	0.024	&	+0.090	&	0.003	&	0.12	&	+1.69	&	0.07	\\
904	&	J191430.53+402714.1	&	HD 180374  	&	19 14 30.37 	&	 +40 27 16.66	&	B9 IV$-$V CrEu   	&	660	&	8.7984	&	0.0007	&	+2.529	&	0.061	&	+0.021	&	0.004	&	0.08	&	+0.74	&	0.07	\\
905	&	J191837.96+394726.4 $^{a}$	&	TYC 3125-282-1     	&	19 18 37.97 	&	 +39 47 26.33	&	B9.5 IV$-$V Cr  (He-wk) 	&	254	&	11.0248	&	0.0007	&	+0.839	&	0.208	&	+0.145	&	0.004	&	0.14	&	+0.50	&	0.54	\\
906	&	J191920.17+444706.6	&	TYC 3133-70-1      	&	19 19 20.18 	&	 +44 47 06.71	&	B8 IV$-$V Si   	&	185	&	11.7006	&	0.0006	&	+0.288	&	0.039	&	$-$0.059	&	0.004	&	0.18	&	$-$1.18	&	0.30	\\
907	&	J192135.14+440302.4	&	TYC 3146-839-1     	&	19 21 35.15 	&	 +44 03 02.27	&	A0 II EuSi   	&	399	&	10.3730	&	0.0017	&	+1.113	&	0.031	&	$-$0.072	&	0.008	&	0.15	&	+0.46	&	0.08	\\
908	&	J192151.34+450655.3	&	TYC 3543-971-1     	&	19 21 51.35 	&	 +45 06 55.31	&	B8 IV Si   	&	213	&	12.3466	&	0.0006	&	+0.271	&	0.029	&	$-$0.076	&	0.004	&	0.22	&	$-$0.71	&	0.23	\\
909	&	J192524.14+431911.5	&	TYC 3146-198-1     	&	19 25 24.12 	&	 +43 19 11.12	&	kA3hA7mA9 SrCrEu   	&	161	&	11.8109	&	0.0004	&	+1.023	&	0.029	&	+0.196	&	0.003	&	0.19	&	+1.67	&	0.08	\\
910	&	J192630.17+380251.8	&	Gaia DR2 2052567864358948736	&	19 26 30.17 	&	 +38 02 51.73	&	B9 V Si   	&	107	&	13.6254	&	0.0004	&	+0.261	&	0.030	&	$-$0.039	&	0.003	&	0.18	&	+0.53	&	0.25	\\
911	&	J192943.77+422930.6	&	Gaia DR2 2125752389095257088	&	19 29 43.77 	&	 +42 29 30.68	&	B9 V Cr   	&	170	&	12.6734	&	0.0004	&	+0.371	&	0.026	&	+0.001	&	0.003	&	0.20	&	+0.32	&	0.16	\\
912	&	J193021.81+465043.2	&	Gaia DR2 2128347270896190208	&	19 30 21.78 	&	 +46 50 43.22	&	B9 V Cr   	&	198	&	13.0596	&	0.0004	&	+0.329	&	0.031	&	$-$0.024	&	0.003	&	0.12	&	+0.52	&	0.21	\\
913	&	J193128.34+470548.2	&	TYC 3560-2289-1    	&	19 31 28.35 	&	 +47 05 48.26	&	F0 V SrEu   	&	395	&	10.7146	&	0.0003	&	+2.156	&	0.024	&	+0.361	&	0.001	&	0.10	&	+2.28	&	0.06	\\
914	&	J193307.57+270643.1	&	HD 338514  	&	19 33 07.57 	&	 +27 06 42.92	&	kB9.5hA2mA4 CrEu   	&	102	&	11.1459	&	0.0010	&	+1.146	&	0.031	&	$-$0.031	&	0.004	&	0.77	&	+0.67	&	0.08	\\
915	&	J193927.58+471826.3	&	TYC 3560-3168-1    	&	19 39 27.59 	&	 +47 18 26.45	&	B9 V CrEuSi   	&	661	&	10.3174	&	0.0010	&	+1.265	&	0.032	&	$-$0.031	&	0.005	&	0.15	&	+0.68	&	0.07	\\
916	&	J194154.63+380153.1	&	HD 225429  	&	19 41 54.65 	&	 +38 01 53.21	&	kB8hA3mA6 CrEu   	&	297	&	9.9751	&	0.0005	&	+1.624	&	0.032	&	$-$0.072	&	0.002	&	0.17	&	+0.86	&	0.07	\\
917	&	J194334.90+300004.0	&	HD 332508  	&	19 43 34.86 	&	 +30 00 01.96	&	kA0hA1mA3 EuSi   	&	275	&	10.3491	&	0.0010	&	+0.772	&	0.035	&	$-$0.076	&	0.004	&	0.35	&	$-$0.56	&	0.11	\\
918	&	J194611.22+391333.3	&	HD 225728  	&	19 46 11.21 	&	 +39 13 34.93	&	B8 V bl4077 bl4130  	&	376	&	10.3882	&	0.0006	&	+1.011	&	0.155	&	$-$0.091	&	0.002	&	0.28	&	+0.14	&	0.34	\\
919	&	J194629.20+473750.0 $^{a}$	&	TYC 3561-1857-1    	&	19 46 29.21 	&	 +47 37 50.05	&	kA5hA7mA9 SrCrEu   	&	352	&	11.3822	&	0.0009	&	+1.144	&	0.044	&	+0.368	&	0.003	&	0.13	&	+1.55	&	0.10	\\
920	&	J194650.18+280638.9	&	HD 332756  	&	19 46 50.18 	&	 +28 06 38.98	&	B8 III$-$IV  bl4130  	&	106	&	11.1576	&	0.0006	&	+0.623	&	0.040	&	$-$0.081	&	0.002	&	0.51	&	$-$0.38	&	0.15	\\
921	&	J194925.91+470218.0	&	Gaia DR2 2086363381465025024	&	19 49 25.95 	&	 +47 02 16.98	&	B9.5 II$-$III EuSi   	&	123	&	12.8462	&	0.0003	&	+0.115	&	0.028	&	$-$0.012	&	0.002	&	0.24	&	$-$2.09	&	0.54	\\
922	&	J195042.19+483639.4	&	TYC 3561-781-1     	&	19 50 42.19 	&	 +48 36 39.52	&	kA5hA6mA9 SrCrEu   	&	307	&	11.4833	&	0.0011	&	+1.042	&	0.029	&	+0.403	&	0.008	&	0.10	&	+1.47	&	0.08	\\
923	&	J195251.15+403621.4	&	HD 226339  	&	19 52 51.08 	&	 +40 36 21.48	&	B9 IV Si   	&	412	&	10.8132	&	0.0007	&	+0.970	&	0.038	&	$-$0.195	&	0.003	&	0.42	&	+0.33	&	0.10	\\
924	&	J195314.26+445712.4	&	Gaia DR2 2079301664955979392	&	19 53 14.26 	&	 +44 57 12.41	&	kB9.5hA5mA2 SrEu   	&	117	&	13.7325	&	0.0004	&	+0.293	&	0.019	&	+0.069	&	0.003	&	0.71	&	+0.36	&	0.15	\\
925	&	J195341.40+441939.9	&	Gaia DR2 2079071931443788672	&	19 53 41.39 	&	 +44 19 39.86	&	A0 V CrEu   	&	210	&	11.9820	&	0.0004	&	+0.805	&	0.041	&	+0.041	&	0.002	&	0.53	&	+0.98	&	0.12	\\
926	&	J195344.32+414104.1	&	HD 226421  	&	19 53 44.32 	&	 +41 41 04.26	&	B8 V Si   	&	582	&	10.1422	&	0.0006	&	+0.939	&	0.032	&	$-$0.205	&	0.002	&	0.48	&	$-$0.48	&	0.09	\\
927	&	J195506.34+442900.7	&	TYC 3149-428-1     	&	19 55 06.35 	&	 +44 29 00.69	&	A5 IV$-$V SrEu   	&	297	&	11.9201	&	0.0002	&	+1.189	&	0.022	&	+0.267	&	0.002	&	0.41	&	+1.89	&	0.06	\\
928	&	J195551.27+480759.8	&	TYC 3562-1452-1    	&	19 55 51.23 	&	 +48 07 59.42	&	B9.5 V Cr   	&	363	&	10.9518	&	0.0010	&	+0.946	&	0.031	&	$-$0.040	&	0.004	&	0.31	&	+0.52	&	0.09	\\
929	&	J195631.74+253407.8	&	Gaia DR2 2026771741029840128	&	19 56 31.75 	&	 +25 34 07.84	&	B8 \textit{V} Si\textit{Cr}*   	&	134	&	12.2836	&	0.0010	&	+0.765	&	0.032	&	$-$0.119	&	0.004	&	1.76	&	$-$0.06	&	0.10	\\
930	&	J195644.95+432951.5	&	TYC 3149-1303-1    	&	19 56 44.96 	&	 +43 29 51.42	&	kB9hA8mA6 Si   	&	261	&	11.8054	&	0.0005	&	+0.709	&	0.026	&	$-$0.137	&	0.003	&	0.41	&	+0.65	&	0.09	\\
931	&	J195801.13+251550.6	&	Gaia DR2 1834595113014285824	&	19 58 01.13 	&	 +25 15 50.65	&	B9 V Cr   	&	108	&	13.2872	&	0.0004	&	+0.700	&	0.019	&	$-$0.022	&	0.002	&	1.66	&	+0.86	&	0.08	\\
932	&	J200227.66+463007.0	&	TYC 3558-1813-1    	&	20 02 27.66 	&	 +46 30 06.91	&	A2 IV SrEu   	&	469	&	11.3210	&	0.0006	&	+1.643	&	0.023	&	+0.290	&	0.002	&	0.19	&	+2.21	&	0.06	\\
933	&	J200529.70+440656.7	&	TYC 3162-162-1     	&	20 05 29.68 	&	 +44 06 56.39	&	A1 III$-$IV bl4077 bl4130  	&	178	&	11.8068	&	0.0004	&	+0.603	&	0.033	&	$-$0.078	&	0.002	&	0.70	&	+0.01	&	0.13	\\
934	&	J200544.86+434645.9	&	TYC 3162-1336-1    	&	20 05 44.85 	&	 +43 46 45.18	&	B9 V  bl4130  	&	188	&	11.8986	&	0.0002	&	+0.635	&	0.026	&	+0.086	&	0.002	&	0.39	&	+0.53	&	0.10	\\
935	&	J200814.38+464916.2	&	TYC 3559-1471-1    	&	20 08 14.38 	&	 +46 49 15.75	&	A0 V SrCrEu   	&	444	&	11.0934	&	0.0009	&	+0.940	&	0.025	&	+0.104	&	0.003	&	0.82	&	+0.14	&	0.08	\\
936	&	J200911.86+445359.0	&	TYC 3162-277-1     	&	20 09 11.87 	&	 +44 53 59.14	&	B9 III Si   	&	214	&	11.5085	&	0.0006	&	+0.690	&	0.036	&	$-$0.220	&	0.002	&	0.98	&	$-$0.27	&	0.12	\\
937	&	J201138.29+273639.6	&	HD 333914  	&	20 11 38.26 	&	 +27 36 38.24	&	B7 IV$-$V Si   	&	308	&	10.7394	&	0.0010	&	+1.032	&	0.034	&	$-$0.055	&	0.004	&	0.94	&	$-$0.13	&	0.09	\\
938	&	J201237.90+513759.1	&	TYC 3571-735-1     	&	20 12 37.90 	&	 +51 37 59.20	&	B8 IV CrSi   	&	220	&	11.3409	&	0.0013	&	+0.541	&	0.104	&	$-$0.005	&	0.003	&	0.22	&	$-$0.22	&	0.42	\\
939	&	J202056.13+380407.6	&	HD 229032  	&	20 20 56.23 	&	 +38 04 07.29	&	B7 V Si   	&	279	&	10.7593	&	0.0004	&	+1.156	&	0.030	&	$-$0.405	&	0.002	&	1.32	&	$-$0.24	&	0.08	\\
940	&	J202416.52+434141.9	&	TYC 3164-1108-1    	&	20 24 16.52 	&	 +43 41 41.92	&	A0 II$-$III bl4077   	&	231	&	11.8897	&	0.0008	&	+0.804	&	0.025	&	$-$1.219	&	0.003	&	3.13	&	$-$1.72	&	0.08	\\
941	&	J202636.19+424251.0	&	TYC 3160-232-1     	&	20 26 36.19 	&	 +42 42 51.07	&	B9.5 II Si   	&	136	&	12.6712	&	0.0004	&	+0.498	&	0.026	&	$-$0.095	&	0.003	&	1.86	&	$-$0.70	&	0.13	\\
942	&	J202924.63+411507.6	&	Gaia DR2 2068000609561250816	&	20 29 24.63 	&	 +41 15 07.68	&	B8 V bl4077 bl4130  	&	201	&	12.0036	&	0.0004	&	+0.939	&	0.023	&	$-$0.502	&	0.002	&	2.03	&	$-$0.16	&	0.07	\\
943	&	J202943.73+384756.6	&	Gaia DR2 2064117095834734464	&	20 29 43.73 	&	 +38 47 56.62	&	B9 II$-$III Si   	&	171	&	12.3830	&	0.0014	&	+1.898	&	0.218	&	+1.180	&	0.003	&	0.17	&	+3.60	&	0.25	\\
944	&	J203012.93+413750.4	&	TYC 3161-1091-1    	&	20 30 12.93 	&	 +41 37 50.38	&	B9 III$-$IV CrSi   	&	286	&	11.4399	&	0.0007	&	+0.901	&	0.031	&	+0.004	&	0.002	&	0.90	&	+0.32	&	0.09	\\
945	&	J203252.92+440742.5	&	TYC 3165-381-1     	&	20 32 52.63 	&	 +44 07 42.74	&	kB9hA0mA2 Si   	&	161	&	9.5781	&	0.0004	&	+1.393	&	0.040	&	$-$0.139	&	0.002	&	0.42	&	$-$0.13	&	0.08	\\
946	&	J203337.82+480113.3	&	TYC 3577-1410-1    	&	20 33 37.83 	&	 +48 01 13.40	&	B9 III Si   	&	343	&	11.2524	&	0.0009	&	+0.459	&	0.058	&	$-$0.358	&	0.003	&	1.23	&	$-$1.67	&	0.28	\\
947	&	J203813.43+442217.0	&	TYC 3165-370-1     	&	20 38 13.68 	&	 +44 22 18.33	&	B9 IV$-$V Cr   	&	405	&	9.7620	&	0.0008	&	+1.839	&	0.043	&	$-$0.103	&	0.003	&	0.30	&	+0.78	&	0.07	\\
948	&	J204216.10+360057.0	&	TYC 2699-2845-1    	&	20 42 16.10 	&	 +36 00 57.00	&	B8 IV  bl4130  	&	107	&	11.3592	&	0.0007	&	+0.955	&	0.039	&	$-$0.085	&	0.002	&	0.89	&	+0.37	&	0.10	\\
949	&	J204308.71+504709.0	&	Gaia DR2 2180453470538367488	&	20 43 08.72 	&	 +50 47 09.04	&	B9.5 III$-$IV CrSi   	&	141	&	12.8618	&	0.0007	&	+0.615	&	0.023	&	+0.052	&	0.003	&	1.09	&	+0.71	&	0.09	\\
950	&	J204922.38+372816.3	&	TYC 2699-981-1     	&	20 49 22.38 	&	 +37 28 16.30	&	B9.5 IV$-$V SrCr   	&	107	&	12.4261	&	0.0002	&	+0.634	&	0.027	&	+0.052	&	0.001	&	0.90	&	+0.54	&	0.11	\\
\hline
\end{tabular}                                                                                                                                                                   
\end{adjustbox}
\end{center}                                                                                                                                             
\end{sidewaystable*}
\setcounter{table}{0}  
\begin{sidewaystable*}
\caption{Essential data for our sample stars, sorted by increasing right ascension. The columns denote: (1) Internal identification number. (2) LAMOST identifier. (3) Alternativ identifier (HD number, TYC identifier or GAIA DR2 number). (4) Right ascension (J2000; GAIA DR2). (5) Declination (J2000; GAIA DR2). (6) Spectral type, as derived in this study. (7) Sloan $g$ band S/N ratio of the analysed spectrum. (8) $G$\,mag (GAIA DR2). (9) $G$\,mag error. (10) Parallax (GAIA DR2). (11) Parallax error. (12) Dereddened colour index $(BP-RP){_0}$ (GAIA DR2). (13) Colour index error. (14) Absorption in the $G$ band, $A_G$. (15) Intrinsic absolute magnitude in the $G$ band, $M_{\mathrm{G,0}}$. (16) Absolute magnitude error.}
\label{table_master20}
\begin{center}
\begin{adjustbox}{max width=\textwidth}
\begin{tabular}{lllcclcccccccccc}
\hline
\hline
(1) & (2) & (3) & (4) & (5) & (6) & (7) & (8) & (9) & (10) & (11) & (12) & (13) & (14) & (15) & (16) \\
No.	&	ID\_LAMOST	&	ID\_alt	&	RA(J2000) 	&	 Dec(J2000)    	&	SpT\_final	&	S/N\,$g$	&	$G$\,mag	&	e\_$G$\,mag	&	pi (mas)	&	e\_pi	&	$(BP-RP){_0}$	&	e\_$(BP-RP){_0}$	&	$A_G$	&	$M_{\mathrm{G,0}}$	&	e\_$M_{\mathrm{G,0}}$	\\
\hline 
951	&	J205711.99+362112.8	&	TYC 2700-3067-1    	&	20 57 12.00 	&	 +36 21 12.81	&	B8 IV Si   	&	165	&	11.0595	&	0.0006	&	+0.589	&	0.045	&	$-$0.005	&	0.003	&	0.53	&	$-$0.62	&	0.17	\\
952	&	J205820.49+340411.3	&	TYC 2696-1529-1    	&	20 58 20.50 	&	 +34 04 11.39	&	B8 IV$-$V Si   	&	101	&	11.4910	&	0.0006	&	+0.680	&	0.035	&	$-$0.063	&	0.002	&	0.35	&	+0.31	&	0.12	\\
953	&	J210327.21+013808.7	&	Gaia DR2 1729657826308369792	&	21 03 27.22 	&	 +01 38 08.55	&	A2 IV$-$V SrEu   	&	252	&	12.5182	&	0.0003	&	+0.950	&	0.052	&	+0.177	&	0.003	&	0.29	&	+2.12	&	0.13	\\
954	&	J212802.14+395946.6	&	TYC 3186-2260-1    	&	21 28 02.15 	&	 +39 59 46.71	&	B8 III$-$IV Si   	&	172	&	11.8125	&	0.0004	&	+0.433	&	0.031	&	+0.003	&	0.002	&	0.64	&	$-$0.65	&	0.16	\\
955	&	J213646.10+494739.4	&	Gaia DR2 1979156634085728384	&	21 36 46.11 	&	 +49 47 39.45	&	A2 IV$-$V SrCr   	&	137	&	14.4306	&	0.0003	&	+0.244	&	0.027	&	$-$0.109	&	0.002	&	1.52	&	$-$0.15	&	0.25	\\
956	&	J213710.42+402056.7	&	TYC 3187-943-1     	&	21 37 10.42 	&	 +40 20 56.75	&	kA1hA6mA7 SrCrEu   	&	298	&	11.0760	&	0.0013	&	$-$2.540	&	0.506	&	$-$	&	$-$	&	$-$	&	$-$	&	$-$	\\
957	&	J213851.08+414514.0	&	TYC 3191-1090-1    	&	21 38 50.98 	&	 +41 45 16.21	&	B8 IV Si   	&	846	&	9.9430	&	0.0005	&	+1.545	&	0.054	&	$-$0.206	&	0.001	&	0.55	&	+0.34	&	0.09	\\
958	&	J213940.97-044004.9	&	TYC 5217-1367-1    	&	21 39 40.98 	&	 -04 40 04.84	&	B8 IV Si   	&	143	&	10.8596	&	0.0008	&	+0.608	&	0.097	&	$-$0.172	&	0.004	&	0.13	&	$-$0.35	&	0.35	\\
959	&	J214029.69+432658.6 $^{a}$	&	TYC 3196-875-1     	&	21 40 29.61 	&	 +43 26 58.63	&	kA1hA7mA8 SrCrEuSi   	&	270	&	10.7689	&	0.0004	&	+1.566	&	0.038	&	$-$0.753	&	0.002	&	1.75	&	$-$0.01	&	0.07	\\
960	&	J214045.49+120849.2	&	TYC 1128-1649-1    	&	21 40 45.37 	&	 +12 08 49.21	&	B9 V SrCr   	&	429	&	10.4149	&	0.0005	&	+1.530	&	0.050	&	$-$0.039	&	0.003	&	0.21	&	+1.13	&	0.09	\\
961	&	J214140.45+082543.9	&	TYC 1120-1719-1    	&	21 41 40.38 	&	 +08 25 43.90	&	B9 IV$-$V CrEu   	&	314	&	10.7778	&	0.0011	&	+0.596	&	0.066	&	+0.015	&	0.006	&	0.09	&	$-$0.44	&	0.24	\\
962	&	J214337.80+494041.9	&	Gaia DR2 1978967483714383616	&	21 43 37.81 	&	 +49 40 42.02	&	kA0hA3mA7 CrSi   	&	122	&	14.5697	&	0.0006	&	+0.398	&	0.021	&	$-$0.199	&	0.003	&	1.82	&	+0.75	&	0.13	\\
963	&	J214358.57+430044.4	&	TYC 3192-676-1     	&	21 43 58.54 	&	 +43 00 43.65	&	kB9.5hA2mA5 CrEu   	&	229	&	10.8782	&	0.0007	&	+1.599	&	0.030	&	+0.028	&	0.003	&	0.44	&	+1.46	&	0.06	\\
964	&	J214815.98+083232.4	&	Gaia DR2 2701846248105636480	&	21 48 15.98 	&	 +08 32 32.42	&	B8 IV$-$V \textit{(Cr)}*	&	117	&	14.4455	&	0.0008	&	+0.277	&	0.073	&	$-$	&	$-$	&	$-$	&	$-$	&	$-$	\\
965	&	J220052.34+411659.6	&	TYC 3206-1198-1    	&	22 00 52.35 	&	 +41 16 59.66	&	B9 III$-$IV Si   	&	139	&	9.6347	&	0.0005	&	+1.656	&	0.040	&	$-$0.071	&	0.002	&	0.27	&	+0.46	&	0.07	\\
966	&	J220500.91+043900.2	&	TYC 561-1768-1     	&	22 05 00.91 	&	 +04 39 00.30	&	kA2hA8mA9 SrCrEu   	&	197	&	11.7202	&	0.0003	&	+0.745	&	0.065	&	+0.289	&	0.003	&	0.15	&	+0.93	&	0.20	\\
967	&	J222549.96+343851.0	&	HD 212714  	&	22 25 49.67 	&	 +34 38 51.06	&	B8 IV EuSi   	&	181	&	8.7177	&	0.0005	&	+1.459	&	0.064	&	$-$0.046	&	0.003	&	0.18	&	$-$0.64	&	0.11	\\
968	&	J222640.42+545105.7	&	Gaia DR2 2004936299100544128	&	22 26 40.43 	&	 +54 51 05.87	&	kB9hA0mA2  bl4130  	&	110	&	12.8230	&	0.0004	&	+0.485	&	0.027	&	$-$0.092	&	0.002	&	0.80	&	+0.45	&	0.13	\\
969	&	J223811.04+533238.4	&	TYC 3983-3022-1    	&	22 38 11.05 	&	 +53 32 38.43	&	A0 II$-$III \textit{CrEu}*	&	224	&	12.1422	&	0.0007	&	+0.652	&	0.040	&	$-$0.015	&	0.008	&	0.50	&	+0.71	&	0.14	\\
970	&	J224146.84+521705.7	&	TYC 3633-2112-1    	&	22 41 46.59 	&	 +52 17 05.77	&	B8 III$-$IV Si   	&	330	&	10.1168	&	0.0008	&	+0.729	&	0.033	&	$-$0.078	&	0.002	&	0.48	&	$-$1.05	&	0.11	\\
971	&	J224704.08+343418.9	&	TYC 2744-2279-1    	&	22 47 04.08 	&	 +34 34 18.92	&	A8 V SrEuSi   	&	167	&	11.7583	&	0.0006	&	+1.252	&	0.042	&	+0.206	&	0.002	&	0.18	&	+2.07	&	0.09	\\
972	&	J224927.99+584130.6	&	TYC 3996-813-1     	&	22 49 27.99 	&	 +58 41 30.63	&	B8 IV Si  (He-wk) 	&	118	&	11.7313	&	0.0008	&	+0.966	&	0.136	&	+0.101	&	0.003	&	0.97	&	+0.68	&	0.31	\\
973	&	J224939.77+341256.5	&	TYC 2757-2035-1    	&	22 49 39.78 	&	 +34 12 56.60	&	A9 V SrCrEuSi   	&	205	&	11.6114	&	0.0007	&	+1.036	&	0.033	&	+0.344	&	0.002	&	0.15	&	+1.54	&	0.09	\\
974	&	J225245.93+531333.8	&	TYC 3984-1798-1    	&	22 52 45.93 	&	 +53 13 33.81	&	B9 II$-$III EuSi   	&	235	&	11.9652	&	0.0006	&	+0.408	&	0.035	&	$-$0.038	&	0.003	&	0.49	&	$-$0.47	&	0.19	\\
975	&	J225518.40+560355.8	&	TYC 3989-187-1     	&	22 55 18.40 	&	 +56 03 55.81	&	B9.5 IV Cr   	&	298	&	11.0244	&	0.0008	&	+1.326	&	0.034	&	+0.064	&	0.003	&	0.57	&	+1.06	&	0.07	\\
976	&	J225656.56+551017.2	&	TYC 3989-1590-1    	&	22 56 56.56 	&	 +55 10 17.29	&	B8 III Si   	&	174	&	11.7430	&	0.0005	&	+0.340	&	0.030	&	$-$0.171	&	0.004	&	0.91	&	$-$1.52	&	0.20	\\
977	&	J225737.31+492342.7	&	Gaia DR2 1985841767843891200	&	22 57 37.33 	&	 +49 23 42.83	&	B8 IV$-$V Cr   	&	130	&	13.3784	&	0.0006	&	+0.329	&	0.025	&	$-$0.071	&	0.003	&	0.41	&	+0.55	&	0.17	\\
978	&	J225918.13+554247.1	&	Gaia DR2 2008961233205915776	&	22 59 18.13 	&	 +55 42 47.13	&	A1 III Cr   	&	127	&	12.5886	&	0.0005	&	+0.598	&	0.030	&	+0.104	&	0.003	&	0.61	&	+0.86	&	0.12	\\
979	&	J230159.71+575548.7	&	TYC 3993-58-1      	&	23 01 59.71 	&	 +57 55 48.72	&	B9 V Eu   	&	235	&	11.5508	&	0.0008	&	+0.953	&	0.031	&	$-$0.011	&	0.004	&	0.58	&	+0.87	&	0.09	\\
980	&	J230905.79+523711.2	&	TYC 3998-2321-1    	&	23 09 05.44 	&	 +52 37 11.29	&	B8 IV EuSi   	&	786	&	9.3047	&	0.0005	&	+1.298	&	0.034	&	$-$0.033	&	0.003	&	0.33	&	$-$0.45	&	0.08	\\
981	&	J231037.91+584007.7	&	TYC 4010-80-1      	&	23 10 37.91 	&	 +58 40 07.75	&	B9.5 IV Cr   	&	129	&	11.8740	&	0.0003	&	+0.638	&	0.046	&	$-$0.004	&	0.002	&	1.13	&	$-$0.23	&	0.16	\\
982	&	J231205.86+550220.0	&	TYC 4002-294-1     	&	23 12 05.86 	&	 +55 02 20.02	&	B8 III$-$IV Si   	&	101	&	11.7994	&	0.0007	&	+0.539	&	0.028	&	$-$0.152	&	0.005	&	0.57	&	$-$0.11	&	0.12	\\
983	&	J231211.12+564543.8	&	TYC 4006-501-1     	&	23 12 11.12 	&	 +56 45 43.83	&	B9 IV$-$V CrEu   	&	134	&	11.8741	&	0.0007	&	+0.769	&	0.032	&	+0.060	&	0.003	&	0.65	&	+0.66	&	0.10	\\
984	&	J231412.30+232336.9	&	TYC 2236-1035-1    	&	23 14 12.31 	&	 +23 23 36.98	&	B9.5 V SrCrEu   	&	161	&	11.2083	&	0.0013	&	+0.620	&	0.059	&	+0.112	&	0.002	&	0.23	&	$-$0.06	&	0.21	\\
985	&	J231505.60+020916.4	&	HD 219348  	&	23 15 05.61 	&	 +02 09 16.61	&	A7 V SrCrEu   	&	577	&	9.7839	&	0.0004	&	+2.116	&	0.074	&	+0.325	&	0.004	&	0.16	&	+1.26	&	0.09	\\
986	&	J231553.45+561845.4	&	TYC 4006-1320-1    	&	23 15 53.32 	&	 +56 18 45.56	&	kB9.5hA3mA3 EuSi   	&	168	&	11.0930	&	0.0049	&	+1.242	&	0.136	&	+0.084	&	0.007	&	0.60	&	+0.96	&	0.24	\\
987	&	J232106.34+552607.8	&	TYC 4003-1466-1    	&	23 21 06.35 	&	 +55 26 07.85	&	A0 IV$-$V Cr   	&	248	&	11.3421	&	0.0007	&	+1.227	&	0.035	&	+0.057	&	0.002	&	0.27	&	+1.51	&	0.08	\\
988	&	J232111.01+552314.2	&	TYC 4003-1670-1    	&	23 21 11.12 	&	 +55 23 14.26	&	B9 IV$-$V  bl4130  	&	252	&	10.9044	&	0.0006	&	+0.917	&	0.043	&	$-$0.043	&	0.002	&	0.47	&	+0.24	&	0.11	\\
989	&	J232126.71+551802.8	&	TYC 4003-1920-1    	&	23 21 26.71 	&	 +55 18 02.88	&	A1 II$-$III Eu   	&	221	&	11.6543	&	0.0006	&	+0.737	&	0.032	&	+0.015	&	0.003	&	0.38	&	+0.61	&	0.11	\\
990	&	J232231.75+543933.2	&	Gaia DR2 1996099008739321088	&	23 22 31.76 	&	 +54 39 33.22	&	B9.5 V Cr   	&	128	&	12.2867	&	0.0003	&	+0.711	&	0.027	&	+0.076	&	0.002	&	0.33	&	+1.21	&	0.10	\\
991	&	J232507.50+432535.0	&	TYC 3242-686-1     	&	23 25 07.22 	&	 +43 25 35.58	&	B8 IV Si  (He-wk) 	&	143	&	9.4734	&	0.0005	&	+1.713	&	0.053	&	$-$0.103	&	0.002	&	0.40	&	+0.24	&	0.08	\\
992	&	J232524.99+561822.0	&	TYC 4007-1393-1    	&	23 25 25.11 	&	 +56 18 22.51	&	B9 III$-$IV Si   	&	207	&	10.8588	&	0.0008	&	+1.093	&	0.046	&	$-$0.021	&	0.002	&	0.75	&	+0.30	&	0.10	\\
993	&	J232801.60+555306.9	&	TYC 4003-1156-1    	&	23 28 01.69 	&	 +55 53 06.55	&	kB9hA3mA3 CrEuSi   	&	244	&	10.9460	&	0.0006	&	+1.525	&	0.034	&	+0.049	&	0.002	&	0.43	&	+1.43	&	0.07	\\
994	&	J232808.48+564209.1	&	TYC 4007-911-1     	&	23 28 08.48 	&	 +56 42 09.09	&	kB9hA6mA5 CrEu   	&	138	&	12.0462	&	0.0003	&	+1.010	&	0.032	&	+0.097	&	0.002	&	0.76	&	+1.31	&	0.09	\\
995	&	J233101.92+564608.8	&	TYC 4007-1067-1    	&	23 31 02.06 	&	 +56 46 09.63	&	B9 IV$-$V SrCrEuSi  (He-wk) 	&	182	&	10.7793	&	0.0007	&	+0.783	&	0.032	&	+0.241	&	0.003	&	0.57	&	$-$0.32	&	0.10	\\
996	&	J233123.69+564325.5	&	TYC 4007-1207-1    	&	23 31 23.92 	&	 +56 43 24.50	&	B8 IV  bl4130  	&	169	&	10.2869	&	0.0003	&	+1.056	&	0.039	&	$-$0.039	&	0.002	&	0.49	&	$-$0.09	&	0.10	\\
997	&	J233539.01+555058.3	&	TYC 4004-1005-1    	&	23 35 39.01 	&	 +55 50 58.30	&	B8 III$-$IV CrEuSi  (He-wk) 	&	102	&	11.3755	&	0.0014	&	+0.580	&	0.035	&	+0.046	&	0.005	&	0.48	&	$-$0.28	&	0.14	\\
998	&	J234055.72+565101.4	&	TYC 4008-1298-1    	&	23 40 55.73 	&	 +56 51 01.41	&	kB9hA1mA3 Cr   	&	149	&	11.0877	&	0.0010	&	+0.581	&	0.234	&	$-$	&	$-$	&	$-$	&	$-$	&	$-$	\\
999	&	J234915.69+560102.0	&	TYC 4005-947-1     	&	23 49 15.70 	&	 +56 01 02.02	&	B9 IV$-$V Cr   	&	267	&	10.9879	&	0.0010	&	+1.154	&	0.041	&	$-$0.049	&	0.004	&	0.42	&	+0.88	&	0.09	\\
1000	&	J235351.09+525134.5	&	TYC 4001-1858-1    	&	23 53 51.03 	&	 +52 51 35.48	&	B9 IV$-$V SrCrEu   	&	302	&	10.7189	&	0.0006	&	+0.582	&	0.049	&	+0.124	&	0.003	&	0.44	&	$-$0.89	&	0.19	\\
\hline
\end{tabular}                                                                                                                                                                   
\end{adjustbox}
\end{center}                                                                                                                                             
\end{sidewaystable*}
\setcounter{table}{0}  
\begin{sidewaystable*}
\caption{Essential data for our sample stars, sorted by increasing right ascension. The columns denote: (1) Internal identification number. (2) LAMOST identifier. (3) Alternativ identifier (HD number, TYC identifier or GAIA DR2 number). (4) Right ascension (J2000; GAIA DR2). (5) Declination (J2000; GAIA DR2). (6) Spectral type, as derived in this study. (7) Sloan $g$ band S/N ratio of the analysed spectrum. (8) $G$\,mag (GAIA DR2). (9) $G$\,mag error. (10) Parallax (GAIA DR2). (11) Parallax error. (12) Dereddened colour index $(BP-RP){_0}$ (GAIA DR2). (13) Colour index error. (14) Absorption in the $G$ band, $A_G$. (15) Intrinsic absolute magnitude in the $G$ band, $M_{\mathrm{G,0}}$. (16) Absolute magnitude error.}
\label{table_master21}
\begin{center}
\begin{adjustbox}{max width=\textwidth}
\begin{tabular}{lllcclcccccccccc}
\hline
\hline
(1) & (2) & (3) & (4) & (5) & (6) & (7) & (8) & (9) & (10) & (11) & (12) & (13) & (14) & (15) & (16) \\
No.	&	ID\_LAMOST	&	ID\_alt	&	RA(J2000) 	&	 Dec(J2000)    	&	SpT\_final	&	S/N\,$g$	&	$G$\,mag	&	e\_$G$\,mag	&	pi (mas)	&	e\_pi	&	$(BP-RP){_0}$	&	e\_$(BP-RP){_0}$	&	$A_G$	&	$M_{\mathrm{G,0}}$	&	e\_$M_{\mathrm{G,0}}$	\\
\hline 
1001	&	J235740.51+470001.7	&	TYC 3643-1589-1    	&	23 57 40.27 	&	 +47 00 02.36	&	B9 IV CrSi   	&	316	&	9.8410	&	0.0005	&	+1.215	&	0.048	&	$-$0.076	&	0.002	&	0.21	&	+0.06	&	0.10	\\
1002	&	J235825.56+564224.7	&	TYC 4009-1911-1    	&	23 58 25.56 	&	 +56 42 24.77	&	B9 III$-$IV Si   	&	309	&	11.4611	&	0.0009	&	$-$0.142	&	0.088	&	$-$	&	$-$	&	$-$	&	$-$	&	$-$	\\
\hline
\hline
\multicolumn{16}{l}{Notes:} \\
\multicolumn{16}{l}{$^{a}$ Contained in the sample of strongly magnetic Ap stars of \citet{Scholz2019}.} \\
\multicolumn{16}{l}{$^{b}$ Enhanced metal-lines but no traditional Si, Cr, Sr, or Eu peculiarities present.} \\
\multicolumn{16}{l}{$^{c}$ Spectrum indicative of an SB2 system (cf. Section \ref{SB2_system}).} \\
\multicolumn{16}{l}{$^{d}$ Cf. Table \ref{table_MKCLASS_manual}.} \\
\hline
\hline
\end{tabular}                                                                                                                                                                   
\end{adjustbox}
\end{center}                                                                                                                                             
\end{sidewaystable*}

\section{Masses and fractional ages on the main sequence}

\begin{table*}
\caption{Masses ($M$) and fractional ages on the main sequence ($\tau$) for the 903 sample stars fulfilling our accuracy criteria. Values have been calculated assuming solar metallicity [Z]\,=\,0.020. The columns denote: (1) Internal identification number. (2) $M$\,($M_\odot$). (3) $\sigma(M)-$. (4) $\sigma(M)+$. (5) $\tau$\,(\%). (6) $\sigma(\tau)-$. (7) $\sigma(\tau)+$.}
\label{table_age_mass}
\begin{center}
\begin{adjustbox}{max width=\textwidth}
\begin{tabular}{llllllllll}
\hline
\hline
(1) & (2) & (3) & (4) & (5) & (6) & (7) \\
No.	&	$M$\,($M_\odot$) & $\sigma(M)-$ & $\sigma(M)+$ & $\tau$\,(\%) & $\sigma(\tau)-$ & $\sigma(\tau)+$ \\
\hline
1	&	1.947	&	0.038	&	0.021	&	62	&	0	&	4	\\
2	&	1.697	&	0.023	&	0.023	&	39	&	4	&	7	\\
3	&	3.325	&	0.322	&	0.377	&	99	&	25	&	33	\\
4	&	2.414	&	0.067	&	0.052	&	73	&	4	&	9	\\
5	&	2.730	&	0.092	&	0.008	&	72	&	4	&	9	\\
6	&	2.584	&	0.066	&	0.109	&	74	&	8	&	4	\\
\hline
\end{tabular}
\end{adjustbox}
\end{center}
\end{table*}
\setcounter{table}{0}  
\begin{table*}
\caption{Masses ($M$) and fractional ages on the main sequence ($\tau$) for the 903 sample stars fulfilling our accuracy criteria. Values have been calculated assuming solar metallicity [Z]\,=\,0.020. The columns denote: (1) Internal identification number. (2) $M$\,($M_\odot$). (3) $\sigma(M)-$. (4) $\sigma(M)+$. (5) $\tau$\,(\%). (6) $\sigma(\tau)-$. (7) $\sigma(\tau)+$.}
\label{table_age_mass}
\begin{center}
\begin{adjustbox}{max width=\textwidth}
\begin{tabular}{llllllllll}
\hline
\hline
(1) & (2) & (3) & (4) & (5) & (6) & (7) \\
No.	&	$M$\,($M_\odot$) & $\sigma(M)-$ & $\sigma(M)+$ & $\tau$\,(\%) & $\sigma(\tau)-$ & $\sigma(\tau)+$ \\
\hline
64	&	2.518	&	0.102	&	0.120	&	73	&	8	&	4	\\
65	&	2.356	&	0.137	&	0.062	&	34	&	17	&	23	\\
66	&	2.524	&	0.057	&	0.061	&	65	&	4	&	8	\\
67	&	2.416	&	0.069	&	0.122	&	69	&	7	&	4	\\
68	&	1.699	&	0.025	&	0.026	&	79	&	4	&	0	\\
69	&	2.174	&	0.051	&	0.045	&	57	&	0	&	7	\\
\hline
\end{tabular}
\end{adjustbox}
\end{center}
\end{table*}
\setcounter{table}{0}  
\begin{table*}
\caption{Masses ($M$) and fractional ages on the main sequence ($\tau$) for the 903 sample stars fulfilling our accuracy criteria. Values have been calculated assuming solar metallicity [Z]\,=\,0.020. The columns denote: (1) Internal identification number. (2) $M$\,($M_\odot$). (3) $\sigma(M)-$. (4) $\sigma(M)+$. (5) $\tau$\,(\%). (6) $\sigma(\tau)-$. (7) $\sigma(\tau)+$.}
\label{table_age_mass}
\begin{center}
\begin{adjustbox}{max width=\textwidth}
\begin{tabular}{llllllllll}
\hline
\hline
(1) & (2) & (3) & (4) & (5) & (6) & (7) \\
No.	&	$M$\,($M_\odot$) & $\sigma(M)-$ & $\sigma(M)+$ & $\tau$\,(\%) & $\sigma(\tau)-$ & $\sigma(\tau)+$ \\
\hline
133	&	2.137	&	0.050	&	0.041	&	69	&	4	&	4	\\
135	&	2.936	&	0.173	&	0.178	&	99	&	16	&	19	\\
136	&	2.507	&	0.014	&	0.065	&	51	&	8	&	0	\\
137	&	3.812	&	0.000	&	0.185	&	34	&	12	&	6	\\
138	&	3.361	&	0.035	&	0.086	&	86	&	9	&	5	\\
139	&	2.251	&	0.033	&	0.038	&	71	&	0	&	4	\\
\hline
\end{tabular}
\end{adjustbox}
\end{center}
\end{table*}
\setcounter{table}{0}  
\begin{table*}
\caption{Masses ($M$) and fractional ages on the main sequence ($\tau$) for the 903 sample stars fulfilling our accuracy criteria. Values have been calculated assuming solar metallicity [Z]\,=\,0.020. The columns denote: (1) Internal identification number. (2) $M$\,($M_\odot$). (3) $\sigma(M)-$. (4) $\sigma(M)+$. (5) $\tau$\,(\%). (6) $\sigma(\tau)-$. (7) $\sigma(\tau)+$.}
\label{table_age_mass}
\begin{center}
\begin{adjustbox}{max width=\textwidth}
\begin{tabular}{llllllllll}
\hline
\hline
(1) & (2) & (3) & (4) & (5) & (6) & (7) \\
No.	&	$M$\,($M_\odot$) & $\sigma(M)-$ & $\sigma(M)+$ & $\tau$\,(\%) & $\sigma(\tau)-$ & $\sigma(\tau)+$ \\
\hline
197	&	2.467	&	0.102	&	0.071	&	69	&	4	&	8	\\
198	&	2.246	&	0.172	&	0.392	&	100	&	37	&	26	\\
199	&	2.219	&	0.048	&	0.046	&	61	&	7	&	4	\\
200	&	1.966	&	0.019	&	0.027	&	61	&	3	&	4	\\
201	&	2.902	&	0.064	&	0.066	&	38	&	11	&	10	\\
202	&	2.679	&	0.041	&	0.103	&	77	&	12	&	0	\\
\hline
\end{tabular}
\end{adjustbox}
\end{center}
\end{table*}
\setcounter{table}{0}  
\begin{table*}
\caption{Masses ($M$) and fractional ages on the main sequence ($\tau$) for the 903 sample stars fulfilling our accuracy criteria. Values have been calculated assuming solar metallicity [Z]\,=\,0.020. The columns denote: (1) Internal identification number. (2) $M$\,($M_\odot$). (3) $\sigma(M)-$. (4) $\sigma(M)+$. (5) $\tau$\,(\%). (6) $\sigma(\tau)-$. (7) $\sigma(\tau)+$.}
\label{table_age_mass}
\begin{center}
\begin{adjustbox}{max width=\textwidth}
\begin{tabular}{llllllllll}
\hline
\hline
(1) & (2) & (3) & (4) & (5) & (6) & (7) \\
No.	&	$M$\,($M_\odot$) & $\sigma(M)-$ & $\sigma(M)+$ & $\tau$\,(\%) & $\sigma(\tau)-$ & $\sigma(\tau)+$ \\
\hline
259	&	1.550	&	0.050	&	0.031	&	3	&	0	&	7	\\
260	&	1.874	&	0.025	&	0.025	&	67	&	4	&	0	\\
261	&	2.814	&	0.058	&	0.059	&	99	&	6	&	6	\\
262	&	2.074	&	0.036	&	0.055	&	75	&	4	&	4	\\
263	&	2.462	&	0.021	&	0.110	&	54	&	18	&	3	\\
264	&	3.210	&	0.100	&	0.138	&	54	&	18	&	7	\\
\hline
\end{tabular}
\end{adjustbox}
\end{center}
\end{table*}
\setcounter{table}{0}  
\begin{table*}
\caption{Masses ($M$) and fractional ages on the main sequence ($\tau$) for the 903 sample stars fulfilling our accuracy criteria. Values have been calculated assuming solar metallicity [Z]\,=\,0.020. The columns denote: (1) Internal identification number. (2) $M$\,($M_\odot$). (3) $\sigma(M)-$. (4) $\sigma(M)+$. (5) $\tau$\,(\%). (6) $\sigma(\tau)-$. (7) $\sigma(\tau)+$.}
\label{table_age_mass}
\begin{center}
\begin{adjustbox}{max width=\textwidth}
\begin{tabular}{llllllllll}
\hline
\hline
(1) & (2) & (3) & (4) & (5) & (6) & (7) \\
No.	&	$M$\,($M_\odot$) & $\sigma(M)-$ & $\sigma(M)+$ & $\tau$\,(\%) & $\sigma(\tau)-$ & $\sigma(\tau)+$ \\
\hline
322	&	2.368	&	0.017	&	0.120	&	43	&	14	&	3	\\
323	&	3.629	&	0.130	&	0.189	&	30	&	19	&	15	\\
324	&	2.202	&	0.035	&	0.030	&	79	&	4	&	0	\\
325	&	2.708	&	0.109	&	0.247	&	100	&	25	&	12	\\
326	&	2.720	&	0.125	&	0.182	&	54	&	25	&	7	\\
327	&	2.178	&	0.052	&	0.087	&	69	&	7	&	4	\\
\hline
\end{tabular}
\end{adjustbox}
\end{center}
\end{table*}
\setcounter{table}{0}  
\begin{table*}
\caption{Masses ($M$) and fractional ages on the main sequence ($\tau$) for the 903 sample stars fulfilling our accuracy criteria. Values have been calculated assuming solar metallicity [Z]\,=\,0.020. The columns denote: (1) Internal identification number. (2) $M$\,($M_\odot$). (3) $\sigma(M)-$. (4) $\sigma(M)+$. (5) $\tau$\,(\%). (6) $\sigma(\tau)-$. (7) $\sigma(\tau)+$.}
\label{table_age_mass}
\begin{center}
\begin{adjustbox}{max width=\textwidth}
\begin{tabular}{llllllllll}
\hline
\hline
(1) & (2) & (3) & (4) & (5) & (6) & (7) \\
No.	&	$M$\,($M_\odot$) & $\sigma(M)-$ & $\sigma(M)+$ & $\tau$\,(\%) & $\sigma(\tau)-$ & $\sigma(\tau)+$ \\
\hline
385	&	2.414	&	0.049	&	0.104	&	73	&	8	&	0	\\
386	&	2.720	&	0.120	&	0.130	&	85	&	9	&	10	\\
387	&	1.869	&	0.020	&	0.032	&	59	&	6	&	3	\\
388	&	2.083	&	0.009	&	0.046	&	76	&	4	&	0	\\
389	&	2.219	&	0.061	&	0.046	&	58	&	0	&	11	\\
390	&	2.326	&	0.072	&	0.057	&	87	&	5	&	5	\\
\hline
\end{tabular}
\end{adjustbox}
\end{center}
\end{table*}
\setcounter{table}{0}  
\begin{table*}
\caption{Masses ($M$) and fractional ages on the main sequence ($\tau$) for the 903 sample stars fulfilling our accuracy criteria. Values have been calculated assuming solar metallicity [Z]\,=\,0.020. The columns denote: (1) Internal identification number. (2) $M$\,($M_\odot$). (3) $\sigma(M)-$. (4) $\sigma(M)+$. (5) $\tau$\,(\%). (6) $\sigma(\tau)-$. (7) $\sigma(\tau)+$.}
\label{table_age_mass}
\begin{center}
\begin{adjustbox}{max width=\textwidth}
\begin{tabular}{llllllllll}
\hline
\hline
(1) & (2) & (3) & (4) & (5) & (6) & (7) \\
No.	&	$M$\,($M_\odot$) & $\sigma(M)-$ & $\sigma(M)+$ & $\tau$\,(\%) & $\sigma(\tau)-$ & $\sigma(\tau)+$ \\
\hline
447	&	2.219	&	0.048	&	0.084	&	65	&	10	&	0	\\
448	&	2.169	&	0.000	&	0.050	&	40	&	13	&	2	\\
449	&	3.081	&	0.063	&	0.185	&	72	&	15	&	4	\\
450	&	3.005	&	0.105	&	0.261	&	75	&	15	&	9	\\
451	&	2.462	&	0.058	&	0.076	&	54	&	6	&	7	\\
452	&	3.231	&	0.408	&	0.305	&	97	&	20	&	48	\\
\hline
\end{tabular}
\end{adjustbox}
\end{center}
\end{table*}
\setcounter{table}{0}  
\begin{table*}
\caption{Masses ($M$) and fractional ages on the main sequence ($\tau$) for the 903 sample stars fulfilling our accuracy criteria. Values have been calculated assuming solar metallicity [Z]\,=\,0.020. The columns denote: (1) Internal identification number. (2) $M$\,($M_\odot$). (3) $\sigma(M)-$. (4) $\sigma(M)+$. (5) $\tau$\,(\%). (6) $\sigma(\tau)-$. (7) $\sigma(\tau)+$.}
\label{table_age_mass}
\begin{center}
\begin{adjustbox}{max width=\textwidth}
\begin{tabular}{llllllllll}
\hline
\hline
(1) & (2) & (3) & (4) & (5) & (6) & (7) \\
No.	&	$M$\,($M_\odot$) & $\sigma(M)-$ & $\sigma(M)+$ & $\tau$\,(\%) & $\sigma(\tau)-$ & $\sigma(\tau)+$ \\
\hline
511	&	2.630	&	0.053	&	0.090	&	55	&	14	&	7	\\
512	&	2.471	&	0.060	&	0.067	&	62	&	10	&	8	\\
513	&	3.689	&	0.130	&	0.127	&	16	&	12	&	11	\\
514	&	2.302	&	0.050	&	0.054	&	60	&	3	&	7	\\
515	&	2.334	&	0.069	&	0.022	&	9	&	6	&	11	\\
516	&	1.824	&	0.025	&	0.030	&	65	&	7	&	0	\\
\hline
\end{tabular}
\end{adjustbox}
\end{center}
\end{table*}
\setcounter{table}{0}  
\begin{table*}
\caption{Masses ($M$) and fractional ages on the main sequence ($\tau$) for the 903 sample stars fulfilling our accuracy criteria. Values have been calculated assuming solar metallicity [Z]\,=\,0.020. The columns denote: (1) Internal identification number. (2) $M$\,($M_\odot$). (3) $\sigma(M)-$. (4) $\sigma(M)+$. (5) $\tau$\,(\%). (6) $\sigma(\tau)-$. (7) $\sigma(\tau)+$.}
\label{table_age_mass}
\begin{center}
\begin{adjustbox}{max width=\textwidth}
\begin{tabular}{llllllllll}
\hline
\hline
(1) & (2) & (3) & (4) & (5) & (6) & (7) \\
No.	&	$M$\,($M_\odot$) & $\sigma(M)-$ & $\sigma(M)+$ & $\tau$\,(\%) & $\sigma(\tau)-$ & $\sigma(\tau)+$ \\
\hline
575	&	1.924	&	0.125	&	0.165	&	48	&	31	&	9	\\
576	&	2.219	&	0.045	&	0.079	&	68	&	7	&	4	\\
578	&	2.038	&	0.070	&	0.046	&	53	&	6	&	10	\\
581	&	2.219	&	0.136	&	0.145	&	61	&	10	&	7	\\
582	&	2.584	&	0.066	&	0.054	&	74	&	4	&	4	\\
583	&	3.024	&	0.189	&	0.242	&	61	&	12	&	16	\\
\hline
\end{tabular}
\end{adjustbox}
\end{center}
\end{table*}
\setcounter{table}{0}  
\begin{table*}
\caption{Masses ($M$) and fractional ages on the main sequence ($\tau$) for the 903 sample stars fulfilling our accuracy criteria. Values have been calculated assuming solar metallicity [Z]\,=\,0.020. The columns denote: (1) Internal identification number. (2) $M$\,($M_\odot$). (3) $\sigma(M)-$. (4) $\sigma(M)+$. (5) $\tau$\,(\%). (6) $\sigma(\tau)-$. (7) $\sigma(\tau)+$.}
\label{table_age_mass}
\begin{center}
\begin{adjustbox}{max width=\textwidth}
\begin{tabular}{llllllllll}
\hline
\hline
(1) & (2) & (3) & (4) & (5) & (6) & (7) \\
No.	&	$M$\,($M_\odot$) & $\sigma(M)-$ & $\sigma(M)+$ & $\tau$\,(\%) & $\sigma(\tau)-$ & $\sigma(\tau)+$ \\
\hline
646	&	2.129	&	0.013	&	0.045	&	76	&	4	&	0	\\
647	&	3.570	&	0.152	&	0.240	&	71	&	11	&	9	\\
648	&	2.598	&	0.305	&	0.327	&	100	&	29	&	41	\\
650	&	2.419	&	0.068	&	0.053	&	61	&	10	&	8	\\
651	&	2.084	&	0.077	&	0.055	&	60	&	3	&	7	\\
652	&	2.902	&	0.222	&	0.364	&	68	&	23	&	13	\\
\hline
\end{tabular}
\end{adjustbox}
\end{center}
\end{table*}
\setcounter{table}{0}  
\begin{table*}
\caption{Masses ($M$) and fractional ages on the main sequence ($\tau$) for the 903 sample stars fulfilling our accuracy criteria. Values have been calculated assuming solar metallicity [Z]\,=\,0.020. The columns denote: (1) Internal identification number. (2) $M$\,($M_\odot$). (3) $\sigma(M)-$. (4) $\sigma(M)+$. (5) $\tau$\,(\%). (6) $\sigma(\tau)-$. (7) $\sigma(\tau)+$.}
\label{table_age_mass}
\begin{center}
\begin{adjustbox}{max width=\textwidth}
\begin{tabular}{llllllllll}
\hline
\hline
(1) & (2) & (3) & (4) & (5) & (6) & (7) \\
No.	&	$M$\,($M_\odot$) & $\sigma(M)-$ & $\sigma(M)+$ & $\tau$\,(\%) & $\sigma(\tau)-$ & $\sigma(\tau)+$ \\
\hline
714	&	2.174	&	0.058	&	0.091	&	81	&	9	&	0	\\
715	&	1.899	&	0.025	&	0.020	&	41	&	2	&	8	\\
716	&	2.538	&	0.075	&	0.000	&	59	&	3	&	11	\\
717	&	2.720	&	0.041	&	0.126	&	76	&	12	&	4	\\
719	&	2.455	&	0.190	&	0.145	&	76	&	4	&	14	\\
720	&	2.427	&	0.050	&	0.032	&	83	&	0	&	5	\\
\hline
\end{tabular}
\end{adjustbox}
\end{center}
\end{table*}
\setcounter{table}{0}  
\begin{table*}
\caption{Masses ($M$) and fractional ages on the main sequence ($\tau$) for the 903 sample stars fulfilling our accuracy criteria. Values have been calculated assuming solar metallicity [Z]\,=\,0.020. The columns denote: (1) Internal identification number. (2) $M$\,($M_\odot$). (3) $\sigma(M)-$. (4) $\sigma(M)+$. (5) $\tau$\,(\%). (6) $\sigma(\tau)-$. (7) $\sigma(\tau)+$.}
\label{table_age_mass}
\begin{center}
\begin{adjustbox}{max width=\textwidth}
\begin{tabular}{llllllllll}
\hline
\hline
(1) & (2) & (3) & (4) & (5) & (6) & (7) \\
No.	&	$M$\,($M_\odot$) & $\sigma(M)-$ & $\sigma(M)+$ & $\tau$\,(\%) & $\sigma(\tau)-$ & $\sigma(\tau)+$ \\
\hline
778	&	2.846	&	0.064	&	0.110	&	72	&	8	&	9	\\
779	&	1.742	&	0.018	&	0.013	&	34	&	5	&	6	\\
781	&	2.720	&	0.143	&	0.072	&	54	&	9	&	14	\\
782	&	1.799	&	0.075	&	0.030	&	7	&	4	&	21	\\
785	&	2.420	&	0.078	&	0.052	&	65	&	4	&	8	\\
786	&	2.467	&	0.168	&	0.148	&	69	&	0	&	13	\\
\hline
\end{tabular}
\end{adjustbox}
\end{center}
\end{table*}
\setcounter{table}{0}  
\begin{table*}
\caption{Masses ($M$) and fractional ages on the main sequence ($\tau$) for the 903 sample stars fulfilling our accuracy criteria. Values have been calculated assuming solar metallicity [Z]\,=\,0.020. The columns denote: (1) Internal identification number. (2) $M$\,($M_\odot$). (3) $\sigma(M)-$. (4) $\sigma(M)+$. (5) $\tau$\,(\%). (6) $\sigma(\tau)-$. (7) $\sigma(\tau)+$.}
\label{table_age_mass}
\begin{center}
\begin{adjustbox}{max width=\textwidth}
\begin{tabular}{llllllllll}
\hline
\hline
(1) & (2) & (3) & (4) & (5) & (6) & (7) \\
No.	&	$M$\,($M_\odot$) & $\sigma(M)-$ & $\sigma(M)+$ & $\tau$\,(\%) & $\sigma(\tau)-$ & $\sigma(\tau)+$ \\
\hline
846	&	3.447	&	0.479	&	0.369	&	4	&	1	&	13	\\
847	&	2.408	&	0.139	&	0.078	&	86	&	5	&	10	\\
848	&	1.724	&	0.025	&	0.025	&	25	&	8	&	10	\\
849	&	3.629	&	0.363	&	0.274	&	22	&	18	&	25	\\
850	&	1.947	&	0.040	&	0.046	&	59	&	6	&	3	\\
851	&	2.404	&	0.000	&	0.062	&	76	&	8	&	0	\\
\hline
\end{tabular}
\end{adjustbox}
\end{center}
\end{table*}
\setcounter{table}{0}  
\begin{table*}
\caption{Masses ($M$) and fractional ages on the main sequence ($\tau$) for the 903 sample stars fulfilling our accuracy criteria. Values have been calculated assuming solar metallicity [Z]\,=\,0.020. The columns denote: (1) Internal identification number. (2) $M$\,($M_\odot$). (3) $\sigma(M)-$. (4) $\sigma(M)+$. (5) $\tau$\,(\%). (6) $\sigma(\tau)-$. (7) $\sigma(\tau)+$.}
\label{table_age_mass}
\begin{center}
\begin{adjustbox}{max width=\textwidth}
\begin{tabular}{llllllllll}
\hline
\hline
(1) & (2) & (3) & (4) & (5) & (6) & (7) \\
No.	&	$M$\,($M_\odot$) & $\sigma(M)-$ & $\sigma(M)+$ & $\tau$\,(\%) & $\sigma(\tau)-$ & $\sigma(\tau)+$ \\
\hline
910	&	2.523	&	0.112	&	0.122	&	61	&	23	&	4	\\
911	&	2.538	&	0.071	&	0.100	&	74	&	8	&	0	\\
912	&	2.523	&	0.111	&	0.068	&	61	&	7	&	8	\\
913	&	1.649	&	0.024	&	0.009	&	43	&	9	&	5	\\
914	&	2.463	&	0.051	&	0.075	&	58	&	12	&	3	\\
915	&	2.507	&	0.095	&	0.031	&	51	&	3	&	13	\\
\hline
\end{tabular}
\end{adjustbox}
\end{center}
\end{table*}
\setcounter{table}{0}  
\begin{table*}
\caption{Masses ($M$) and fractional ages on the main sequence ($\tau$) for the 903 sample stars fulfilling our accuracy criteria. Values have been calculated assuming solar metallicity [Z]\,=\,0.020. The columns denote: (1) Internal identification number. (2) $M$\,($M_\odot$). (3) $\sigma(M)-$. (4) $\sigma(M)+$. (5) $\tau$\,(\%). (6) $\sigma(\tau)-$. (7) $\sigma(\tau)+$.}
\label{table_age_mass}
\begin{center}
\begin{adjustbox}{max width=\textwidth}
\begin{tabular}{llllllllll}
\hline
\hline
(1) & (2) & (3) & (4) & (5) & (6) & (7) \\
No.	&	$M$\,($M_\odot$) & $\sigma(M)-$ & $\sigma(M)+$ & $\tau$\,(\%) & $\sigma(\tau)-$ & $\sigma(\tau)+$ \\
\hline
976	&	4.344	&	0.547	&	0.306	&	89	&	14	&	52	\\
977	&	2.626	&	0.088	&	0.108	&	52	&	15	&	6	\\
978	&	2.219	&	0.045	&	0.046	&	68	&	4	&	4	\\
979	&	2.356	&	0.031	&	0.048	&	51	&	8	&	6	\\
980	&	3.022	&	0.090	&	0.067	&	86	&	5	&	10	\\
981	&	2.870	&	0.161	&	0.032	&	83	&	0	&	16	\\
982	&	3.266	&	0.056	&	0.190	&	56	&	14	&	11	\\
983	&	2.352	&	0.054	&	0.012	&	68	&	0	&	8	\\
984	&	2.391	&	0.049	&	0.287	&	100	&	33	&	6	\\
985	&	1.981	&	0.034	&	0.012	&	78	&	0	&	5	\\
986	&	2.219	&	0.107	&	0.084	&	61	&	3	&	7	\\
987	&	2.055	&	0.018	&	0.028	&	29	&	5	&	5	\\
988	&	2.740	&	0.101	&	0.000	&	65	&	0	&	12	\\
989	&	2.420	&	0.064	&	0.052	&	65	&	4	&	8	\\
990	&	2.112	&	0.029	&	0.046	&	53	&	8	&	3	\\
991	&	2.902	&	0.052	&	0.049	&	54	&	6	&	7	\\
992	&	2.639	&	0.101	&	0.004	&	70	&	4	&	9	\\
993	&	2.083	&	0.000	&	0.046	&	34	&	8	&	4	\\
994	&	2.083	&	0.045	&	0.029	&	48	&	3	&	9	\\
995	&	2.693	&	0.055	&	0.056	&	98	&	6	&	6	\\
996	&	2.825	&	0.058	&	0.104	&	80	&	9	&	5	\\
997	&	2.547	&	0.000	&	0.301	&	100	&	33	&	0	\\
999	&	2.418	&	0.050	&	0.120	&	43	&	22	&	3	\\
1001	&	2.854	&	0.057	&	0.103	&	69	&	11	&	4	\\
\hline
\hline
\end{tabular}
\end{adjustbox}
\end{center}
\end{table*}

\section{LAMOST Standard Star Library}

\begin{table*}
\caption{Standard stars of the \textit{liblamost} library. The columns denote: (1) Original LAMOST spectrum FITS file name. (2) Identification number from the $Kepler$ input catalogue \citep{KIC}. (3) 2MASS identifier \citep{2MASS}. (4) Spectral type, as derived by \citet{gray16}. (5) Sloan g band S/N ratio of the spectrum according to \citet{gray16}. (6) Quality flag according to \citet{gray16}. (7) Suitability estimate as an MKK standard star (1 = suitable to a lesser extent; 2 = suitable; 3 = fully suitable).}
\label{table_liblamost}
\begin{center}
\begin{adjustbox}{max width=0.9\textwidth}
\begin{tabular}{llllccr}
\hline
\hline
(1) & (2) & (3) & (4) & (5) & (6) & (7) \\
\textbf{ID\_Spec}	&	\textbf{ID\_KIC}	&	\textbf{ID\_2MASS}	&	\textbf{SpT}	&	\textbf{S/N}	&	\textbf{quality flag}	&	\textbf{suitability}	\\
\hline
spec-56914-KP193637N444141V01\_sp15-029	&	KIC09472174	&	19383260+4603591	&	B3 IV	&	224	&	vgood	& 1 \\
spec-56561-KP195920N454621V01\_sp01-071	&	KIC08324482	&	19570365+4413556	&	B3 V	&	314	&	vgood	& 2-3$^{a}$ \\
\hline
spec-56561-KP195920N454621V02\_sp12-045	&	KIC10501393	&	20064002+4736539	&	B5 III	&	311	&	vgood	& 1-2 \\
spec-56591-KP195920N454621V3\_sp13-174	&	KIC09860322	&	20063327+4637279	&	B5 V	&	265	&	vgood	& 1$^{b}$ \\
\hline
spec-56562-KP192102N424113V02\_sp06-083	&	KIC06780397	&	19305469+4215203	&	B7 III$-$IV	&	413	&	vgood	& 2 \\
spec-56096-kepler08B56096\_2\_sp13-250	&	KIC05380341	&	19454878+4033190	&	B7 V	&	182	&	vgood	& 2.5 \\
\hline
spec-56561-KP195920N454621V02\_sp06-117	&	KIC08530971	&	20111938+4435273	&	B8 III$-$IV	&	242	&	vgood	& 2$^{a}$ \\
spec-56562-KP192102N424113V02\_sp08-216	&	KIC06121547	&	19243929+4124248	&	B8 V	&	382	&	vgood	& 2$^{c}$ \\
\hline
spec-56561-KP195920N454621V02\_sp02-182	&	KIC08577307	&	19502070+4436477	&	B9 III	&	167	&	vgood	& 2 \\
spec-56561-KP195920N454621V01\_sp02-063	&	KIC08189641	&	19553255+4400228	&	B9 V	&	547	&	vgood	& 2-3$^{d}$ \\
\hline
spec-56561-KP195920N454621V01\_sp05-032	&	KIC08583770	&	19565852+4440592	&	A0 III	&	609	&	vgood	& 2-3 \\
spec-56811-KP190339N395439V02\_sp16-205	&	KIC06265185	&	18550325+4140225	&	A0 V	&	278	&	excel	& 3 \\
\hline
spec-56919-KP190651N485531V01\_sp12-144	&	KIC12055345	&	19105861+5030322	&	A1 III	&	589	&	vgood	& 2 \\
spec-56561-KP195920N454621V02\_sp11-197	&	KIC10624050	&	19590430+4749002	&	A1 V	&	395	&	excel	& 3 \\
\hline
spec-56096-kepler08B56096\_2\_sp03-235	&	KIC04564619	&	19301385+3940339	&	A2 III$-$IV	&	114	&	excel	& 2 \\
spec-56914-KP193637N444141V01\_sp15-170	&	KIC09529773	&	19325433+4609042	&	A2 V	&	247	&	excel	& 2-3 \\
\hline
spec-56914-KP193637N444141V02\_sp03-104	&	KIC08887625	&	19302112+4509101	&	A3 III	&	328	&	vgood	& 2 \\
spec-56568-KP195920N454621M01\_sp12-241	&	KIC10628271	&	20035323+4750419	&	A3 V	&	407	&	excel	& 3 \\
\hline
spec-56562-KP192102N424113V02\_sp05-044	&	KIC06199731	&	19203196+4135194	&	A5 III	&	410	&	vgood	& 1$^{e}$ \\
spec-56561-KP195920N454621V02\_sp08-150	&	KIC08916492	&	20050274+4506175	&	A5 V	&	440	&	excel	& 3 \\
\hline
spec-56550-KP194045N483045V02\_sp04-175	&	KIC11090405	&	19391238+4836241	&	A7 III	&	605	&	vgood	& 2 \\
spec-56550-KP194045N483045V02\_sp04-202	&	KIC11252382	&	19422768+4854228	&	A7 V	&	275	&	excel	& 3 \\
\hline
spec-56094-kepler05B56094\_2\_sp10-065	&	KIC03535046	&	19143404+3836017	&	F0 III	&	265	&	vgood	& 1$^{e}$ \\
spec-56562-KP192102N424113V02\_sp15-119	&	KIC07748238	&	19203821+4329037	&	F0 V	&	478	&	excel	& 3 \\
\hline
spec-56798-KP192314N471144V01\_sp04-125	&	KIC10073601	&	19244870+4702411	&	F2 III	&	296	&	vgood	& 2 \\
spec-56919-KP190651N485531V01\_sp11-032	&	KIC12058428	&	19180566+5033358	&	F2 V	&	216	&	excel	& 3 \\
\hline
spec-56918-KP192323N501616V\_sp07-062	&	KIC11139951	&	19322218+4845539	&	F3 III Fe-0.6	&	100	&	vgood	& 1 \\
spec-56780-KP185111N464417V01\_sp15-194	&	KIC10320849	&	18495615+4726035	&	F3 V	&	242	&	excel	& 3 \\
\hline
spec-56798-KP192314N471144V01\_sp03-117	&	KIC10398258	&	19161952+4735107	&	F5 III$-$IV	&	254	&	vgood	& 2 \\
spec-56561-KP195920N454621V01\_sp08-245	&	KIC08985402	&	20043155+4517541	&	F5 V	&	377	&	excel	& 3 \\
\hline
spec-56807-KP185111N464417V03\_sp09-173	&	KIC10256595	&	18531919+4723443	&	F6 III	&	243	&	vgood	& 2 \\
spec-56918-KP192323N501616V\_sp11-087	&	KIC12785394	&	19221083+5203294	&	F6 V	&	287	&	excel	& 3 \\
\hline
spec-56918-KP192323N501616V\_sp01-135	&	KIC11028682	&	19253344+4830538	&	F8 III	&	306	&	vgood	& 2 \\
spec-56550-KP194045N483045V02\_sp10-121	&	KIC10732086	&	19284619+4802316	&	F8 V	&	286	&	excel	& 3 \\
\hline
spec-56811-KP190339N395439V01\_sp16-061	&	KIC06346065	&	18585279+4144275	&	F9 III	&	190	&	vgood	& 1 \\
spec-56919-KP190651N485531V01\_sp01-020	&	KIC09821151	&	19072635+4637318	&	F9 V	&	177	&	excel	& 3 \\
\hline
spec-56550-KP194045N483045V02\_sp08-014	&	KIC10482869	&	19454882+4739317	&	G0 III$-$IV	&	294	&	vgood	& 2 \\
spec-56562-KP192102N424113V02\_sp05-154	&	KIC06600155	&	19195410+4204013	&	G0 V	&	330	&	excel	& 3 \\
\hline
\hline
\multicolumn{7}{l}{Notes:} \\
\multicolumn{7}{l}{$^{a}$ Interstellar contribution to the \ion{Ca}{ii} K line. DIB at 4130\,\AA.} \\
\multicolumn{7}{l}{$^{b}$ Slightly broadened lines -- rapid rotator? Glitch at 4315\,\AA.} \\
\multicolumn{7}{l}{$^{c}$ He lines slightly strong. Glitch at 4307\,\AA.} \\
\multicolumn{7}{l}{$^{d}$ Slightly metal-weak?} \\
\multicolumn{7}{l}{$^{e}$ Characteristica of luminosity class III only weakly expressed.} \\
\end{tabular}
\end{adjustbox}
\end{center}
\end{table*}

\end{document}